\documentclass[twocolumn,twocolappendix]{aastex701}

\usepackage{xcolor}
\usepackage{comment}

\newcommand{\beagle}{\textsc{Beagle}}
\newcommand{\lya}{\hbox{Ly$\alpha$}}

\newcommand{\kms}{\ifmmode\,{\rm km}\,{\rm s}^{-1}\else km$\,$s$^{-1}$\fi}
\newcommand{\ModelMZ}{M25$_{\rm{\dot{M}\propto Z}}$}
\newcommand{\ModelMLMC}{M25$_{\rm{\dot{M},LMC}}$}
\shorttitle{Deep Spectroscopy of GN-z11}
\shortauthors{Chen et al.}

\begin{document}

\title{SPURS: Massive Stars, Dense Gas, and Ly$\alpha$ Escape in GN-z11 at $z = 10.6$}

\author[0000-0002-2178-5471]{Zuyi Chen}
\affiliation{Cosmic Dawn Center (DAWN)}
\affiliation{Niels Bohr Institute, University of Copenhagen, Jagtvej 128, 2200 Copenhagen N, Denmark}
\email[show]{zuyi.chen@nbi.ku.dk}

\author[0000-0001-6106-5172]{Daniel P. Stark}
\affiliation{Department of Astronomy, University of California, Berkeley, Berkeley, CA 94720, USA}
\email{dpstark@berkeley.edu}

\author[0000-0002-3407-1785]{Charlotte A. Mason}
\affiliation{Cosmic Dawn Center (DAWN)}
\affiliation{Niels Bohr Institute, University of Copenhagen, Jagtvej 128, 2200 Copenhagen N, Denmark}
\email{charlotte.mason@nbi.ku.dk}

\author[0000-0002-9132-6561]{Peter Senchyna}
\affiliation{The Observatories of the Carnegie Institution for Science, 813 Santa Barbara Street, Pasadena, CA 91101, USA}
\email{psenchyna@carnegiescience.edu}

\author[0000-0001-5940-338X]{Mengtao Tang}
\affiliation{Tsung-Dao Lee Institute, Shanghai Jiao Tong University, 1 Lisuo Road, Shanghai 201210, People’s Republic of China}
\affiliation{School of Physics and Astronomy, Shanghai Jiao Tong University, 800 Dongchuan Road, Shanghai 200240, People’s Republic of China}
\email{mengtao.tang@sjtu.edu.cn}

\author[0000-0002-2645-679X]{Keerthi Vasan G. C.}
\affiliation{The Observatories of the Carnegie Institution for Science, 813 Santa Barbara Street, Pasadena, CA 91101, USA}
\email{kvch153@gmail.com}

\author[0000-0003-1432-7744]{Lily Whitler}
\affiliation{Kavli Institute for Cosmology, University of Cambridge, Madingley Road, Cambridge, CB3 0HA, UK}
\affiliation{Cavendish Laboratory, University of Cambridge, JJ Thomson Avenue, Cambridge, CB3 0US, UK}
\email{lw851@cam.ac.uk}

\author[0000-0003-0390-0656]{Adele Plat}
\affiliation{Institute of Physics, GalSpec Laboratory, Ecole Polytechnique Federale de Lausanne, Observatoire de Sauverny, Chemin Pegasi 51, 1290
Versoix, Switzerland}
\email{adele.plat@epfl.ch}

\author[0000-0001-5487-0392]{Viola Gelli}
\affiliation{Cosmic Dawn Center (DAWN)}
\affiliation{Niels Bohr Institute, University of Copenhagen, Jagtvej 128, 2200 Copenhagen N, Denmark}
\email{viola.gelli@nbi.ku.dk}

\begin{abstract}
We present ultra-deep {\it JWST} spectroscopy of GN-z11 ($z=10.6$) obtained through the SPURS Cycle 4 Large Program, providing the deepest rest-UV view yet obtained of a galaxy at $z>10$. GN-z11 was previously found to be nitrogen-enhanced  with detectable Ly$\alpha$. The SPURS spectrum reveals P-Cygni stellar wind features and broad He II emission that are jointly reproduced by stellar population models incorporating very massive stars (VMS; $>100\,M_\odot$) at low metallicity and young ages ($\lesssim3$ Myr). We also detect a broad ($\rm FWHM=1670$ km s$^{-1}$) component to N IV] $\lambda1486$, now seen in several nitrogen emitters, potentially arising from dense WN-like winds or LBV-like outbursts associated with a population of 
VMS in a dense environment,
though an AGN-driven wind cannot be excluded. In either scenario, this broad component may trace the gas producing GN-z11's nitrogen enhancement. Rest-UV absorption lines reveal a fast ($\sim500$~km~s$^{-1}$), highly ionized outflow and a negligible neutral gas covering fraction.  We resolve the weak Ly$\alpha$ emission ($\rm EW=5.6$ \AA, $f_{\rm esc,Ly\alpha}=2.7$\%), finding a broad red wing (44\% of flux at $>500$ km s$^{-1}$) that should experience reduced IGM damping wing suppression and help explain Ly$\alpha$ visibility at $z>10$. Fine-structure O I* $\lambda1304$ emission indicates dense neutral gas near a subset of the ionizing sources, which may also scatter Ly$\alpha$ to the large observed velocities.  The weak low-ionization absorption favors a picture in which this dense neutral gas is confined to a compact nuclear region. Together, these results are consistent with a rapid burst of  star formation building up the dense nuclear regions and surrounding  clusters in GN-z11.
\end{abstract}



\section{Introduction} \label{sec:introduction}

\begin{figure*}[t]
    \centering
    $\vcenter{\hbox{\includegraphics[width=0.23\linewidth]{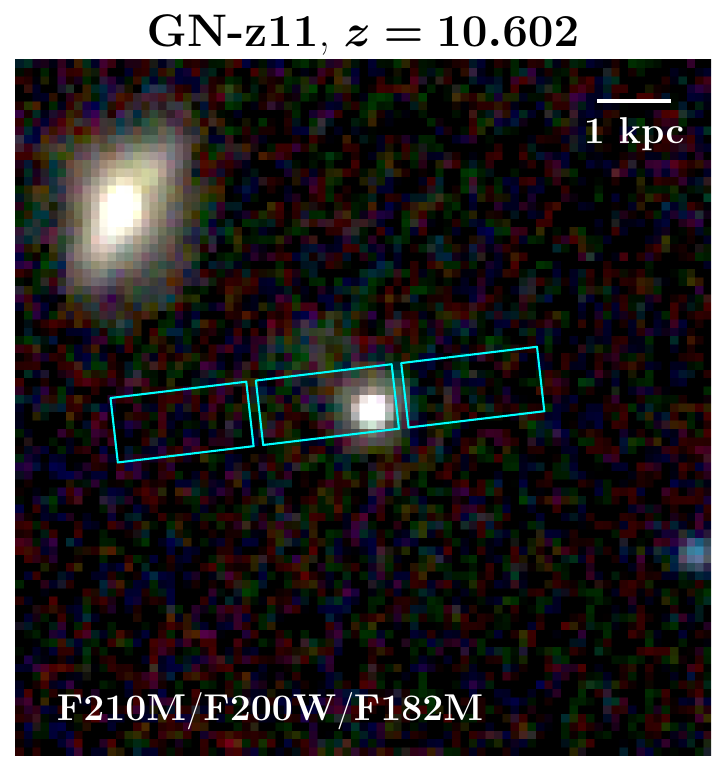}}}$
    $\vcenter{\hbox{\includegraphics[width=0.4\linewidth]{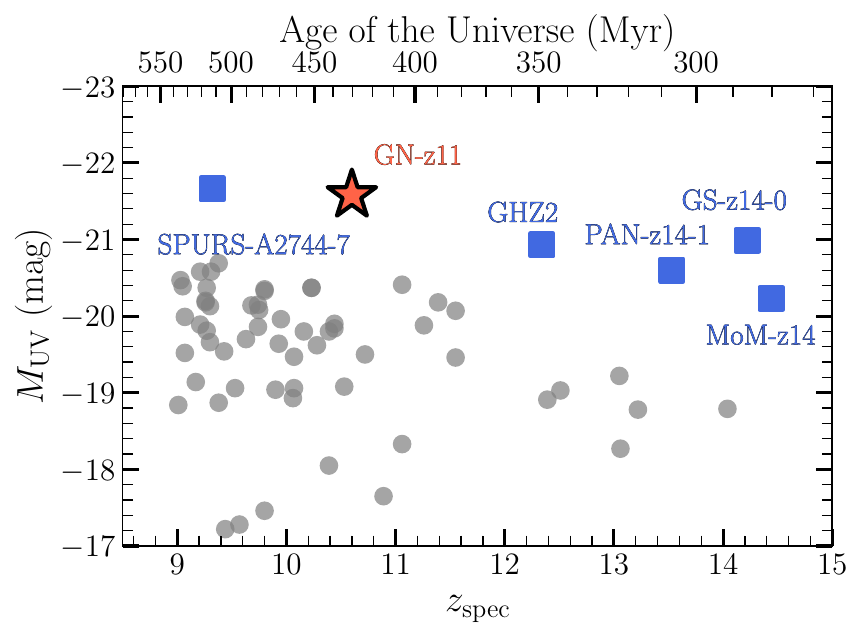}}}$
    \caption{
    \textbf{Left}: NIRCam RGB image ($2\arcsec{}.5\times2\arcsec{}.5$) of GN-z11, with our SPURS NIRSpec shutter in cyan.
    A scale bar of 1~kpc is also shown.
    \textbf{Right}: UV absolute magnitude of GN-z11 compared to other spectroscopically confirmed galaxies at $z>9$.
    We plot the NIRSpec sample compiled in \cite{Tang2025} as gray dots and highlight other notably luminous galaxies at these redshifts, including SPURS-A2744-7 (or Gz9p3; \citealt{Boyett2024_z9p3,Chen2026}), GHZ2 (\citealt{Castellano2024}), PAN-z14-1 (\citealt{Donnan2026}), GS-z14-0 (\citealt{Carniani2024_z14}),  and MoM-z14 (\citealt{Naidu2026_MoMz14}).}
    \label{fig:image}
\end{figure*}

The launch of the {\it James Webb Space Telescope} ({\it JWST}; \citealt{Gardner2023,Rigby2023})has recently pushed the galaxy redshift frontier back beyond $z\simeq 14$ (see \citealt{Ellis2025,Matthee2025,Stark2026} for a review). Prior to {\it JWST}, many expected the galaxy population at $z\gtrsim 10$ to be predominantly extremely faint, reflecting the redshift evolution seen in UV luminosity functions with the {\it Hubble Space Telescope} ({\it HST}) \citep[e.g.,][]{Bouwens2015,Bouwens2021,Finkelstein2015,Oesch2018,Bowler2020}.
One of the biggest surprises from early {\it JWST} operations has been the discovery that very luminous galaxies (M$_{\rm{UV}}<-20$) remain abundant out to $z\simeq 14$ \citep[e.g.,][]{Castellano2022,Naidu2022,Adams2023,Finkelstein2023,Harikane2023a,Donnan2024,Robertson2024,Whitler2025}, with the current record holder now spectroscopically confirmed at $z=14.4$ \citep{Naidu2026_MoMz14}. 

The physics that makes galaxies so luminous at early epochs remains debated. Some argue star formation efficiencies are elevated in the low mass halos that come into view at $z\gtrsim 10$ \citep[e.g.,][]{Dekel2023,Qin2023,Feldmann2025,Somerville2025}, while others suggest stochasticity in star formation histories adds considerable scatter in the UV luminosity at fixed halo mass, with burstiness enabling low mass halos to be luminous for short periods of time \citep[e.g.,][]{Mason2023,Mirocha2023,Shen2023,Kravtsov2024,Gelli2024}. Additionally, if dust is preferentially ejected from early galaxies (i.e., due to their large specific star formation rates), the emergent UV luminosities will be further boosted \citep[e.g.,][]{Ferrara2025}. Finally, the UV output may be further enhanced by a population of early AGN or the presence of top-heavy initial mass functions \citep[e.g.,][]{Chon2022,Harikane2023b,Steinhardt2023,Hegde2024,Ventura2024,Hutter2025}.

Attention is now turning to detailed investigations of the galaxies discovered at these redshifts.
One of the first deep $z\gtrsim 10$ galaxy spectra was of GN-z11 \citep{Oesch2016}, a compact (half-light radius $=64\pm20$~pc; \citealt{Tacchella2023}) UV luminous (M$_{\rm UV}=-21.6$) system at $z=10.6$, observed for 6.9 hours in the prism and 3.45 hours in each of G140M/F070LP, G235M/F170LP, and G395M/F290LP grating/filter combinations \citep{Bunker2023}. The GN-z11 spectrum revealed numerous emission lines throughout the rest-UV and rest-optical.  Ly$\alpha$ emission was detected, potentially suggesting GN-z11 sits in an early ionized region. The detection of both \ion{N}{3}] and \ion{N}{4}] pointed to a nitrogen-enhanced abundance pattern that is similar to globular cluster stars \citep[e.g.,][]{Cameron2023,Senchyna2024}. High ionization lines are detected, revealing the presence of hard ionizing sources, and the various density-sensitive ratios reveal that GN-z11 contains high density gas ($>10^5$~cm$^{-3}$; \citealt{Senchyna2024,Maiolino2024}) not often seen in lower redshift galaxies. MIRI MRS spectroscopy has identified narrow emission from [\ion{O}{3}] and H$\alpha$ in the rest-frame optical \citep{Alvarez-Marquez2025}, and NIRSpec/IFU observations reveal tentative indications of rotation \citep{Xu2024}. The emission line ratios from NIRSpec and MIRI point to low metallicity gas ($0.13-0.17\ Z_\odot$; \citealt{Cameron2023,Alvarez-Marquez2025}), suggesting GN-z11 may be a compact starburst in a metal poor galaxy. 

However, the nature of GN-z11 remains debated. \citet{Maiolino2024} have argued GN-z11 is likely an AGN based on very high ionization nebular emission, fast outflows, and very high gas densities.
Additional evidence in favor of an AGN interpretation for GN-z11 was presented in \cite{Ji2025}. They find a continuum excess at rest-frame $3000-3550$~\AA\ in the NIRSpec prism spectrum, potentially corresponding to the small blue bump often seen in lower redshift quasar spectra. On the other hand, an AGN is not required to explain the narrow emission line ratios, and MIRI/MRS spectroscopy reveals that H$\alpha$ is narrow (albeit not ruling out a fainter broad contribution) \citep[e.g.,][]{Alvarez-Marquez2025}. MIRI imaging has recently extended the GN-z11 SED to $10\ \mu$m \citep{CrespoGomez2026}. The shape of the rest-optical SED does not support the presence of a Type I AGN, but it does reveal excess emission at rest-frame wavelengths of $6600-8600$~\AA, potentially corresponding to hot dust emission that could either be from a Type II AGN or a compact starburst  (\citealt{CrespoGomez2026}). 
Without deeper spectroscopy, the role of an AGN in establishing the spectral characteristics of GN-z11 remains unsettled.

Since the discovery of GN-z11, other nitrogen-enhanced spectra with hard radiation fields and high density ionized gas have been found at $z\gtrsim 10$ \citep{Castellano2024,Naidu2026_MoMz14}, suggesting that GN-z11 may be representative of a phase in early galaxy evolution. On the other hand, only one further galaxy at $z\gtrsim 10$ has been found with Ly$\alpha$ emission (GS-z13-1-LA at $z=13.1$, \citealt{Witstok2025,Witstok2026}). Whether GN-z11 and GS-z13-1-LA have unique physical conditions that facilitate Ly$\alpha$ escape in a highly neutral IGM is not known. Owing to its brightness, GN-z11 is one of the best galaxies  for answering  questions related to nitrogen enhancements and Ly$\alpha$ transfer through the early IGM. 
However, current data are not up to either task. The existing rest-UV spectrum of GN-z11 lacks the sensitivity to robustly characterize the  stellar wind features (e.g., broad He II, \ion{N}{5}, C IV) that would directly identify the massive star population driving its extreme nitrogen enhancement. Similarly, characterizing the gas-phase conditions that permit Ly$\alpha$ escape requires substantially higher signal-to-noise than presently available. A deep rest-UV spectrum reaching sufficient signal-to-noise and resolution is therefore essential to move toward a physical characterization of the detailed nature of GN-z11. 

In this paper, we present new observations of GN-z11 obtained as part of  the SPURS Large Program in Cycle 4 of {\it JWST} operations \citep[GO 9214, PIs C. Mason and D. Stark,][]{Chen2026}. The SPURS data deliver the first ultra-deep view of the rest-frame UV spectrum of GN-z11, providing new insights into the massive stars, ISM conditions, and Ly$\alpha$ profile, while also delivering new constraints on the role of an AGN in  driving its spectroscopic properties.  We note that an independent analysis of the SPURS spectrum of GN-z11 is presented in \cite{Nakane2026}.
The paper is organized as follows.
In \S~\ref{sec:obs}, we describe the new SPURS observations and our measurements of the emission lines in the NIRSpec spectrum.
We then characterize the rest-frame UV and optical emission lines of GN-z11 in \S~\ref{sec:results} and place constraints on the interstellar absorption lines in \S~\ref{sec:ISMabs}.
The physical properties inferred from the emission lines are presented in \S~\ref{sec:phys}.
We model the stellar continuum of the far-UV spectrum in \S~\ref{sec:VMS}.
We discuss the implications for the physical nature of GN-z11 in \S~\ref{sec:discussion} and summarize our conclusions in \S~\ref{sec:summary}. 
Throughout this paper, we adopt a flat $\Lambda$CDM cosmology with $H_0 = 70$ km~s$^{-1}$~Mpc$^{-1}$, $\Omega_{\rm m} = 0.3$, and $\Omega_\Lambda = 0.7$.
All magnitudes are measured in the AB system \citep{Oke1983}.
The line equivalent widths are calculated in the rest frame, with positive values corresponding to emission lines and negative values to absorption lines.

\section{Observations and Analysis} \label{sec:obs}

\begin{figure*}
    \centering
    \includegraphics[width=1\textwidth]{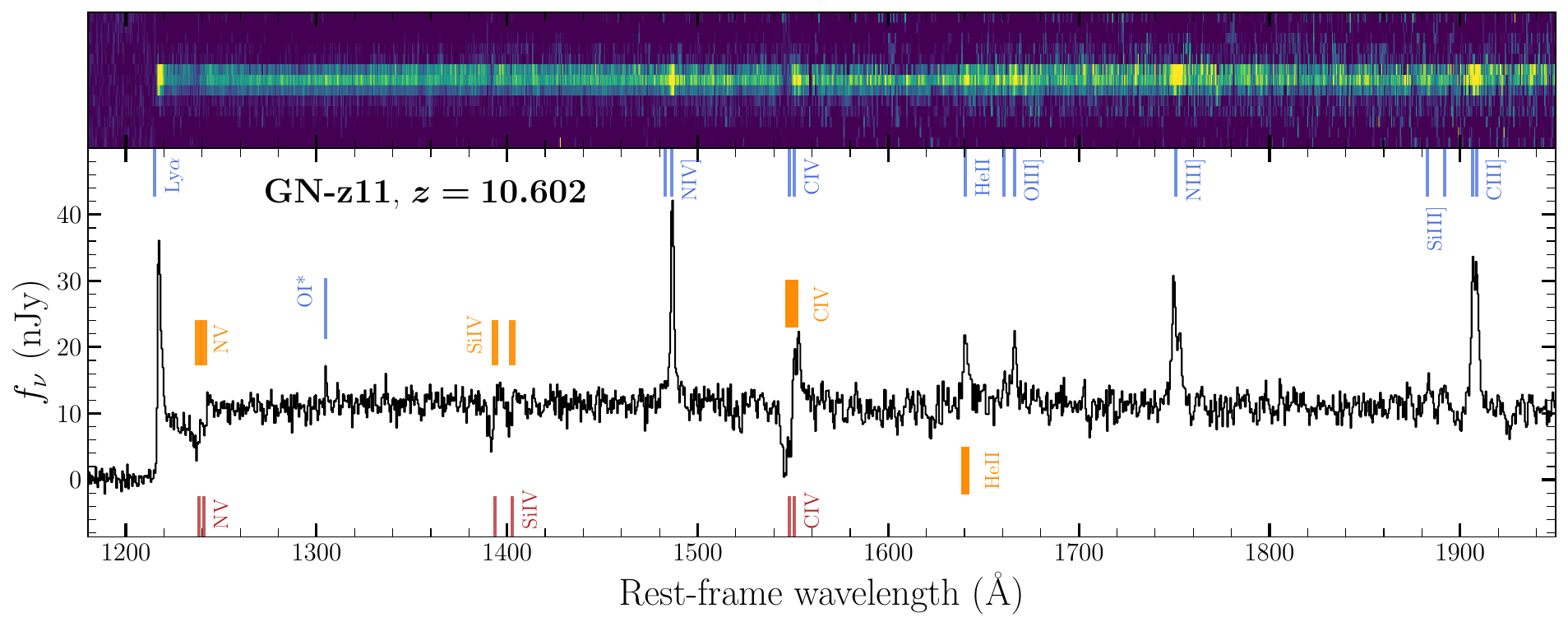}
    \caption{Rest-frame far-UV ($1180-1950$~\AA) spectrum of GN-z11. 
    The main panel shows the final 1D spectrum used in this work, created by stacking the SPURS and JADES observations. The top panel shows the 2D spectrum from the SPURS observations, which dominate the total exposure time.
    We mark the detected emission lines with blue vertical lines. The interstellar absorption lines and stellar wind features are highlighted by red and orange vertical lines, respectively.}
    \label{fig:spectrum}
\end{figure*}

\subsection{SPURS Spectroscopy}\label{subsec:spurs}

Here we present new spectra of GN-z11 obtained as part of the SPURS program.
The full description of the SPURS observations will be presented in a future paper, and we briefly summarize the relevant observations and spectra.
GN-z11 was observed in the SPURS GOODS-N pointing taken over May 12--14, 2026.
Data were acquired in a single pointing using the Multi-Object Spectroscopy (MOS) mode (aperture position angle PA = 276.5$^\circ$), adopting the standard three-shutter nodding pattern.
We use three grating/filter pairs at medium resolution ($R\sim1000$): G140M/F100LP, G235M/F170LP, and G395M/F290LP, with integration times of 29.18, 7.90, and 2.92 hours in each grating, respectively. 
In Figure~\ref{fig:image}, we show the NIRCam image of GN-z11, overlaid with the NIRSpec shutter position of our SPURS observation.

We reduce the spectra using the \texttt{msaexp}\footnote{\url{https://github.com/gbrammer/msaexp/}} package (\texttt{v0.9.14}; \citealt{Brammer2023}), based on the official \textit{JWST} pipeline version 1.16.1 and reference file mapping \texttt{jwst\_1464.pmap}, following the same procedures described in \citet{Chen2026} (also see \citealt{deGraaff2025,Heintz2025,Valentino2025}).
Briefly, we obtain the stage-1 data products (\texttt{*\_rate.fits}) of the standard \textit{JWST} pipeline from the MAST archive, and subsequently use \texttt{msaexp} to apply $1/f$ correction and read noise array scaling based on empty regions of each image.
The resulting files are then processed through the stage-2 pipeline to assign the world coordinate system, flag open microshutters, cut out individual 2D spectra from each exposure of every source, apply flat-field correction, correct for vignetting of the MSA bars, and apply the absolute photometric calibration.
For each source, we subtract the background using the difference spectra taken at the three nod offset positions.
We then rectify and combine individual background-subtracted 2D cutouts on a common wavelength grid to produce the final 2D spectrum of each source.
The final 1D spectra used in our analysis are extracted using a boxcar aperture with the extraction window chosen to match the spatial profile of each source.
Throughout this reduction, we utilize the extended wavelength extraction available as part of \texttt{msaexp} \citep{Valentino2025} to achieve coverage beyond the nominal limits of each grating by extracting the full wavelength range covered by the detector, which is based on the extended flats and photometric calibration derived by \texttt{msaexp}.
While this extended wavelength coverage is susceptible to second-order spectral contamination, in our analysis below, we will exclude the affected spectral regions from our continuum measurements but will consider them for emission line measurements.
We assume a point source pathloss correction, consistent with the compact morphology seen from the NIRCam images of GN-z11.

To maximize the signal-to-noise ratio, we create a composite spectrum for GN-z11 by stacking the SPURS and JADES $R\sim1000$ grating spectra \citep{Bunker2023,Maiolino2024}. 
We utilize the JADES G140M/F070LP, G235M/F170LP, and G395M/F290LP spectra taken by the program GTO 1181 (PI: D. Eisenstein) and released by the DAWN {\it JWST} Archive\footnote{\url{https://dawn-cph.github.io/dja/}} (DJA; \citealt{Brammer2023,deGraaff2024,Heintz2025}), which were also reduced using the \texttt{msaexp} package with the same procedures described above by the DJA team. 
Before stacking, for each spectrum, we first make sure the error spectrum matches the observed continuum fluctuation, derived as the standard deviations in bins of 20 pixels, similar to the approach adopted in previous works (e.g., \citealt{Senchyna2022,deGraaff2025}).
We model the resulting ratio as a function of wavelength with an iteratively fit, outlier-rejected third-order polynomial, which we then apply as a multiplicative correction to the error spectrum.
We then coadd individual SPURS and JADES spectra with inverse variance weighting (i.e., weighted by $1/\sigma^2$ where $\sigma$ is the error spectrum of the individual spectra). 
This adds an additional 5.96 hours to each grating, leading to total integration times of 35.14, 13.86, and 8.88 hours in G140M, G235M, and G395M, respectively.
Finally, we recalibrate the absolute flux of the coadded spectrum to match that of NIRCam by applying a second-order polynomial correction factor as a function of wavelength. 
We derive this correction by fitting the ratio of the NIRCam photometry, where we adopt the CIRC1 photometry available from the JADES DR5 photometric catalog \citep{Robertson2026}, to that synthesized from the coadded spectrum in broad NIRCam filters (from F150W to F444W).
In what follows, we will primarily use this final spectrum for our analysis, but we will also separately examine the low-SNR emission lines in the SPURS spectrum alone for comparison with JADES.
As shown in \cite{Bunker2023}, the JADES spectrum extracted with the 3-pixel (0\arcsec{}.3) window also matches the NIRCam CIRC1 photometry, so our comparison will primarily focus on their values based on the 3-pixel extraction.

Figure~\ref{fig:spectrum} presents the far-UV spectrum of GN-z11, using the redshift determined in \S~\ref{sec:obs_lines}.
The continuum is well detected in the combined G140M grating spectrum, with an SNR of 8.5 per pixel (corresponding to 6~\AA{}) at the observed-frame wavelength of 1.5~$\mu$m, significantly higher than what was achieved with the earlier JADES spectrum alone (SNR = 2.6 per pixel).
The typical $3\sigma$ emission line flux limit is $8.1\times10^{-20}$~erg~s$^{-1}$~cm$^{-2}$, assuming an integration window of 600~km~s$^{-1}$ ($\sim2$ resolution elements).
This translates to a $3\sigma$ rest-frame line EW limit of 0.4~\AA{} in GN-z11 at rest-frame 1400~\AA{}.
We estimate typical $3\sigma$ line flux limits of $7.8\times10^{-20}$~erg~s$^{-1}$~cm$^{-2}$ in G235M and $7.6\times10^{-20}$~erg~s$^{-1}$~cm$^{-2}$ in G395M, corresponding to EW limits of 1.2~\AA{} in G235M and 3.4~\AA{} in G395M for GN-z11.

\subsection{Emission  Line Measurements } \label{sec:obs_lines}

We first derive the systemic redshift by simultaneously fitting the centroids of the available strong narrow rest-frame optical emission lines. 
Since the strong optical lines (H$\beta$, [O~{\small III}]) are shifted out of the NIRSpec spectrum, we utilize 
the [Ne~{\small III}]~$\lambda3869$ and H$\delta$ emission lines. The derived systemic redshift ($z_{\rm sys}=10.6016$)  is consistent with the redshift reported in \citet{Bunker2023}. 

Emission lines are characterized as follows. 
For well-detected lines (S/N $>5$), we measure the line flux, centroid, width, and EW by fitting the line profile and nearby continuum with a Gaussian function. For each line, we visually examine whether there are broad wings that are not captured by our derived profile. We report the small number of lines where this is the case in Section \S~\ref{sec:results}.
For emission lines that are close in wavelength or that show complex profiles, we fit with multiple Gaussians simultaneously. 
We will describe measurements for these lines in more detail in \S~\ref{sec:results}.
If an emission line is detected with lower S/N ($<5$), we compute the line flux using direct integration. 
Finally, the uncertainties of line fluxes, widths, and EWs are evaluated as follows. 
We resample the flux densities of each spectrum $1000$ times by taking the observed value as the mean and the error as the standard deviation. 
Then we compute the line fluxes, widths, and EWs from the resampled spectra of each source using the same approach described above. 
We take the standard deviation of these measurements as the uncertainty. 
When fitting broad emission lines, we convolve the Gaussian models with the instrument resolution ($R=1000$, corresponding to 300~km~s$^{-1}$). 
We obtain consistent results if we adopt the wavelength-dependent line spread function from the JDox, rescaled by a factor of 1.3 to match in-flight estimates of the resolution \citep{Isobe2023,deGraaff2024,Shajib2025}.

\section{Spectroscopic Measurements} \label{sec:results}


\begin{deluxetable}{lcccc}
\tablecaption{Rest-frame far-UV emission line observed-frame wavelengths (\AA), fluxes ($10^{-19}$~erg~s$^{-1}$~cm$^{-2}$), EWs (\AA), and observed FWHMs (km~s$^{-1}$) of GN-z11. Upper limits are provided at the $3\sigma$ level.}
\tablehead{
Line & $\lambda_{\rm obs}$ & Flux & EW & FWHM
}
\startdata
Ly$\alpha_{\rm t}$ & $14120.5$ & $12.8_{-0.3}^{+0.4}$ & $5.6_{-0.2}^{+0.2}$ & -- \\
Ly$\alpha_{\rm n}$ & $14120.4$ & $5.1_{-2.5}^{+2.3}$ & $2.3_{-0.7}^{+0.8}$ & $315^{+9}_{-20}$ \\
Ly$\alpha_{\rm b}$ & $14126.9$ & $7.9_{-2.7}^{+2.2}$ & $3.2_{-0.9}^{+0.7}$ & $592^{+78}_{-79}$ \\
N~{\scriptsize V}~$\lambda1240$\tablenotemark{a} & $14440.3$ & $1.9^{+0.3}_{-0.3}$ & $1.0^{+0.2}_{-0.2}$ & $754^{+253}_{-253}$ \\
O~{\scriptsize I}*~$\lambda1304$ & $15140.8$ & $1.0^{+0.3}_{-0.2}$ & $0.5^{+0.2}_{-0.1}$ & $<300$ \\
N~{\scriptsize IV}]$_{\rm t}$ & -- & $12.2^{+0.6}_{-0.6}$ & $7.9^{+0.4}_{-0.4}$ & -- \\
{[}N~{\scriptsize IV}]~$\lambda1483_{\rm n}$ & -- & $<0.8$ & $<0.5$ & -- \\
N~{\scriptsize IV}]~$\lambda1486_{\rm n}$ & $17248.8$ & $6.3^{+0.6}_{-0.6}$ & $4.0^{+0.4}_{-0.4}$ & $319^{+24}_{-24}$ \\
{[}N~{\scriptsize IV}]~$\lambda1483_{\rm b}$ & -- & $<2.0$ & $<1.3$ & -- \\
N~{\scriptsize IV}]~$\lambda1486_{\rm b}$ & $17248.8$ & $6.0^{+1.8}_{-1.8}$ & $3.8^{+1.1}_{-1.1}$ & $1670^{+292}_{-292}$\tablenotemark{b} \\
C~{\scriptsize IV}$_{\rm t}$ & -- & $4.5^{+0.5}_{-0.5}$ & $3.3^{+0.3}_{-0.3}$ & -- \\
C~{\scriptsize IV}~$\lambda1548$ & $17990.1$ & $1.8^{+0.3}_{-0.3}$ & $1.3^{+0.2}_{-0.2}$ & $347^{+31}_{-31}$ \\
C~{\scriptsize IV}~$\lambda1550$ & $18020.3$ & $2.8^{+0.4}_{-0.4}$ & $2.0^{+0.3}_{-0.3}$ & $347^{+31}_{-31}$ \\
He~{\scriptsize II}~$\lambda1640$ & $19033.6$ & $4.1_{-0.8}^{+0.7}$ & $3.3_{-0.7}^{+0.6}$ & $671_{-178}^{+175}$ \\
O~{\scriptsize III}]~$\lambda1661$ & $19269.3$ & $0.8^{+0.2}_{-0.2}$ & $0.6^{+0.2}_{-0.2}$ & $386^{+45}_{-45}$ \\
O~{\scriptsize III}]~$\lambda1666$ & $19331.9$ & $2.4^{+0.4}_{-0.4}$ & $1.9^{+0.3}_{-0.3}$ & $386^{+45}_{-45}$ \\
N~{\scriptsize III}]$_{\rm t}$ & -- & $7.9^{+0.5}_{-0.5}$ & $7.3^{+0.4}_{-0.4}$ & -- \\
N~{\scriptsize III}]~$\lambda1746$ & $20269.2$ & $0.3^{+0.0}_{-0.0}$ & $0.3^{+0.0}_{-0.0}$ & $347^{+22}_{-22}$ \\
N~{\scriptsize III}]~$\lambda1748$ & $20290.0$ & $0.8^{+0.2}_{-0.2}$ & $0.8^{+0.2}_{-0.2}$ & $347^{+22}_{-22}$ \\
N~{\scriptsize III}]~$\lambda1750$ & $20302.8$ & $3.7^{+0.3}_{-0.3}$ & $3.4^{+0.3}_{-0.3}$ & $347^{+22}_{-22}$ \\
N~{\scriptsize III}]~$\lambda1752$ & $20331.8$ & $2.2^{+0.3}_{-0.3}$ & $2.0^{+0.3}_{-0.3}$ & $347^{+22}_{-22}$ \\
N~{\scriptsize III}]~$\lambda1754$ & $20352.7$ & $0.9^{+0.2}_{-0.2}$ & $0.8^{+0.2}_{-0.2}$ & $347^{+22}_{-22}$ \\
Si~{\scriptsize III}]~$\lambda1883$ & $21843.1$ & $1.0^{+0.3}_{-0.3}$ & $1.1^{+0.4}_{-0.4}$ & $413^{+116}_{-116}$ \\
Si~{\scriptsize III}]~$\lambda1892$ & $21951.0$ & $0.7^{+0.3}_{-0.3}$ & $0.8^{+0.3}_{-0.3}$ & $413^{+116}_{-116}$ \\
C~{\scriptsize III}]$_{\rm t}$ & -- & $7.9^{+0.5}_{-0.5}$ & $8.8^{+0.4}_{-0.4}$ & -- \\
{[}C~{\scriptsize III}]~$\lambda1907$ & $22121.5$ & $3.8^{+0.4}_{-0.4}$ & $4.3^{+0.5}_{-0.5}$ & $442^{+32}_{-32}$ \\
C~{\scriptsize III}]~$\lambda1909$ & $22144.7$ & $4.0^{+0.4}_{-0.4}$ & $4.5^{+0.5}_{-0.5}$ & $442^{+32}_{-32}$ \\
\enddata
\tablecomments{Lines with subscript 't' show the total fluxes and EWs of multiple components. Lines with subscript 'n' and 'b' show the narrow and broad components, respectively. }
\tablenotetext{a}{Unresolved N~{\scriptsize V}~$\lambda\lambda1238,1242$ emission.}
\tablenotetext{b}{Intrinsic FWHM for broad \ion{N}{4}] corrected for instrument resolution assuming $R=1000$.}
\label{tab:fuv_lines}
\end{deluxetable}

\subsection{Ly$\alpha$ Emission} \label{sec:spec_lya}

\begin{figure*}[t]
\centering
\includegraphics[width=\linewidth]{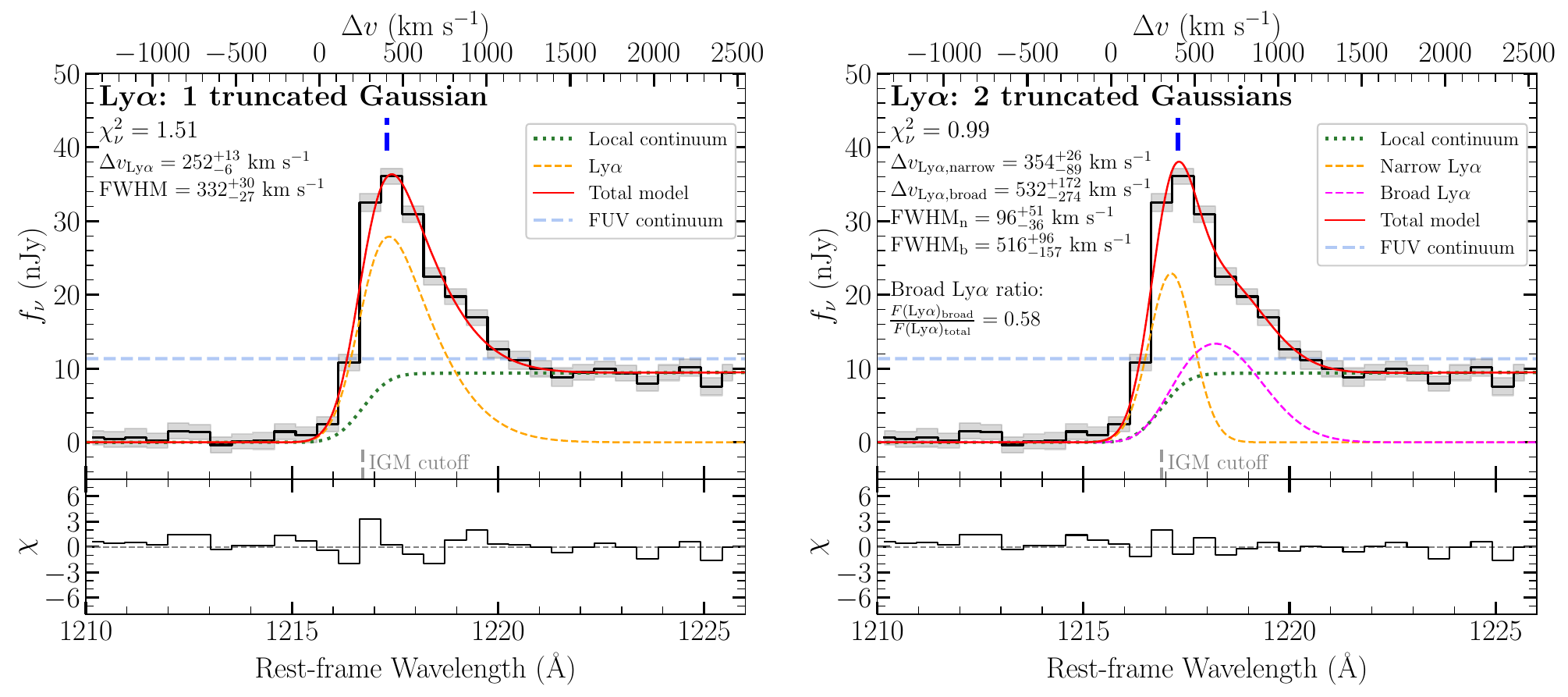}
\caption{Ly$\alpha$ of GN-z11 with best-fit model (red) to the line profile (black).
We present both the fit assuming a single truncated Gaussian (left panel) and that assuming a double truncated Gaussian (right panel).
In both cases, we additionally include a constant component as the continuum.
Both the continuum and the line profile (single or double Gaussians) are truncated at the same IGM cutoff velocity, which is a free parameter in our fit.
We plot the model convolved with instrument resolution for each truncated Gaussian (orange and magenta dashed lines) as well as for the continuum (green dotted line). We also plot the continuum extrapolated from our fit over $1300-1600$\,\AA\ (light blue dashed lines), which we use for our EW estimate.
In each fit, we also plot the residual of the observed spectrum subtracted by the best-fit model. The residual is shown by $\chi\equiv(f-m)/\sigma$, where $f$ is the observed flux density, $m$ is the best-fit model, and $\sigma$ is the uncertainty.
We also show the peak velocity offset and FWHM derived from our intrinsic model (i.e., before convolving with the instrument resolution) in each panel.
}
\label{fig:Lya}
\end{figure*}

GN-z11 is one of only two galaxies above $z\simeq 10$ detected with Ly$\alpha$ emission (see also \citealt{Witstok2025}). The line was first identified in the JADES discovery spectrum \citep{Bunker2023} and subsequently observed with NIRSpec IFU \citep{Scholtz2024}. With the new SPURS observations, we recover \lya{} with improved significance (S/N=24), enabling better characterization of the line properties. 
The observed Ly$\alpha$ profile (Figure~\ref{fig:Lya}) is asymmetric, reaches peak intensity at $418_{-14}^{+12}$\,km~s$^{-1}$
redward of the systemic velocity (uncertainties estimated by resampling the flux densities based on the error spectrum 1000 times), and shows both a narrow core and a broad red wing.
Asymmetric and redshifted Ly$\alpha$ lines are commonly seen in $z\gtrsim5$ galaxies due to resonant scattering by neutral hydrogen in the ISM and IGM \citep[e.g.,][]{Pentericci2018,Hayes2021,Tang2024,PrietoLyon2025}. However, given the $R\approx1000$ resolution of the G140M spectrum, the observed peak velocity may overestimate the intrinsic velocity offset of \lya{} from systemic \citep[as has been noted in sources with both G140H and G140M \lya{} spectroscopy,][]{Whitler2026}.

To characterize the line properties, we thus consider fits to the spectrum with a truncated Gaussian, which we convolve to the instrument resolution, following the approach in previous works \citep[e.g.,][]{Endsley2022_bubble,Tang2026_spursQSO1}. 
In our fits, we truncate the emission below a velocity of $v_{\rm cutoff}$ (a free parameter) and 
we assume a fixed continuum level redwards of the cutoff velocity, using the median flux just redward of the emission line ([1222, 1226]~\AA{}, $1.89\times10^{-19}$~erg~s$^{-1}$~cm$^{-2}$~\AA$^{-1}$).
We fit over the rest-frame wavelength window of [1212, 1226]~\AA{}, corresponding to [$-1000$, 2500]~km~s$^{-1}$ relative to the systemic wavelength of Ly$\alpha$.
The free parameters are thus the amplitude, centroid, and width of the Gaussian component, as well as the cutoff velocity.

We show the resulting Ly$\alpha$ model, convolved to the instrument resolution, in Figure~\ref{fig:Lya}.
The truncated Gaussian profile (left panel) reproduces the peak and the extended red wing of the observed Ly$\alpha$ emission ($\chi^2_{\nu, \rm single}=1.5$). 
The model profile reaches its intrinsic peak intensity at a redshifted velocity of $v_{\rm peak} = 252_{-6}^{+13}$~km~s$^{-1}$, significantly lower than its value once convolved with the instrumental resolution ($417_{-6}^{+1}$~km~s$^{-1}$). 
The intrinsic FWHM of the model profile is $312_{-25}^{+29}$~km~s$^{-1}$, or $511_{-17}^{+17}$ ~km~s$^{-1}$ including the instrumental resolution.  
We find the asymmetric profile is fit with a truncation velocity of $v_{\rm cutoff} = 250_{-9}^{+8}$~km~s$^{-1}$, comparable to the velocity offset. One interpretation of such a high truncation velocity is resonant scattering by neutral hydrogen from IGM gas infalling towards the galaxy, which should produce a sharp cut-off on the blue side of the line \citep[predicted at velocities around $100-300$~km~s$^{-1}$][]{Santos2004,Mason2018,Park2021}, though it may also reflect resonant scattering in the ISM \citep[e.g.,][]{Neufeld1990,Verhamme2006,Verhamme2008}. A higher resolution spectrum should be able to better resolve a sharp cut-off due to IGM resonant scattering.
Integrating the line profile, we measure a total Ly$\alpha$ flux of $1.28_{-0.03}^{+0.04}\times10^{-18}$~erg~s$^{-1}$~cm$^{-2}$. Importantly, we find that $44\%$ of the line flux arises from very large velocities ($>500$~km~s$^{-1}$ relative to line center), where this number is computed from the intrinsic profile, before convolution with the instrumental profile.

Given the very high SNR of the SPURS spectrum, we are now able to better resolve the asymmetric profile of Ly$\alpha$. This results in some differences in our inferred line properties relative to those reported from the earlier JADES spectrum by \citet{Bunker2023} and \citet{Scholtz2024}, which we will now describe.
The \textit{observed} line profile peaks at a velocity consistent with the shallower JADES spectrum \citep[$\approx 420$~km~s$^{-1}$][]{Bunker2023}, and has a consistent observed linewidth (FWHM$=511_{-17}^{+17}$~km~s$^{-1}$, relative to $566\pm 61$~km~s$^{-1}$ in the JADES spectrum).
However, estimates of the \textit{intrinsic} velocity offset and FWHM are sensitive to the shape of the line profile.
Based on our truncated Gaussian fits we recover a lower de-convolved peak velocity offset ($v_{\rm peak} = 252_{-6}^{+13}$~km~s$^{-1}$) and FWHM ($312_{-25}^{+29}$~km~s$^{-1}$) than reported by \citet{Bunker2023} ($v_{\rm peak}$ = 520--555~km~s$^{-1}$, FWHM $= 530\pm65$~km~s$^{-1}$ accounting for the resolution).
We attribute these differences primarily to the assumed line profiles.
\citet{Bunker2023} found a single full (i.e. not truncated) Gaussian profile, without continuum, was a good fit to the Ly$\alpha$ profile in the JADES spectrum.
Fitting the SPURS spectrum with a full Gaussian produces much closer agreement with the results of \citet{Bunker2023}: a full Gaussian fit shifts the inferred peak to higher velocities and requires a broader line profile, relative to our truncated Gaussian model (Figure~\ref{fig:Lya}), to fit the asymmetric profile (and continuum at $\approx 1220-1225$\AA{}). 
However, we find that the full Gaussian fits to the SPURS spectrum are substantially worse ($\chi_\nu^2 > 3$) than the truncated Gaussian ($\chi_\nu^2 = 1.5$), as expected given the asymmetry of the line.
A smaller contribution to the difference in intrinsic FWHM comes from the assumed instrumental resolution: \citet{Bunker2023} adopted $R\approx1500$, whereas we adopt $R=1000$ (see Section~\ref{sec:obs_lines}), which slightly increases our intrinsic FWHM measurement relative to their reported value.
Our results indicate that accounting for asymmetric lineshapes is important for interpreting \lya{} in G140M spectra. Ultimately, a G140H spectrum would provide the most precise estimate of the Ly$\alpha$ velocity offset in GN-z11.

\begin{figure*}[t]
\centering
\includegraphics[width=\linewidth]{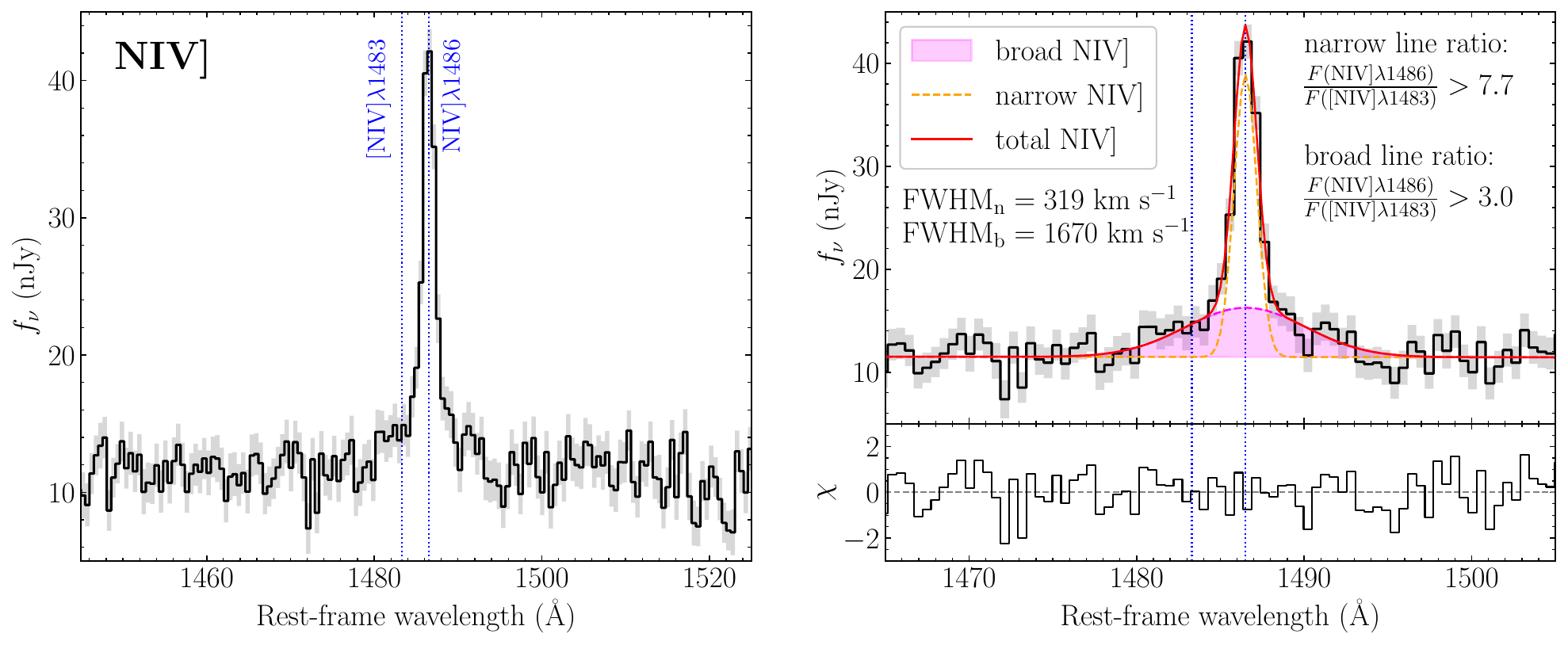}
\caption{Left panel: \ion{N}{4}] emission line of GN-z11. Right panel: Fits to the \ion{N}{4}] profile. Rest-frame wavelength of each individual component of \ion{N}{4}] is marked by blue dotted line. The \ion{N}{4}] profile is best-fitted by a narrow (FWHM $=319$~km~s$^{-1}$; orange dashed line) and a broad (FWHM $=1670$~km~s$^{-1}$; magenta shaded region) \ion{N}{4}]$\lambda1486$ component, while the [\ion{N}{4}]$\lambda1483$ component remains undetected. The red curves show the sum of fits to individual components. In the right panel, we also plot the residual of the observed spectrum subtracted by the best-fit model ($\chi$).}
\label{fig:n4}
\end{figure*}

We measure the Ly$\alpha$ EW, which is a well-established tracer for neutral hydrogen in the reionizing IGM \citep[e.g.,][]{Stark2010,Stark2011,Treu2012,Pentericci2014,Mason2018}.
Our goal is to characterize how the ratio of \lya{} flux relative to the FUV continuum evolves with redshift, as the Ly$\alpha$ damping wing due to neutral hydrogen in the IGM can significantly depress both Ly$\alpha$ emission and the continuum redwards of Ly$\alpha$, out to $\sim1300$~\AA\ \citep[e.g.,][]{Mason2026}.
Thus, to measure the EW, we compute the continuum by fitting a power law ($f_\lambda \propto \lambda^\beta$) to the observed UV continuum over the rest-frame range 1300--1800~\AA, masking regions affected by any emission and absorption features (described in the Sections below). This range also avoids the strong \ion{N}{5} P-Cygni absorption and emission at wavelengths 1226--1240~\AA\ seen in Figure~\ref{fig:spectrum} and described below in Section~\ref{subsec:obs_wind}. 
We extrapolate from the fit to derive the continuum at the Ly$\alpha$ wavelength ($2.30\pm0.02\times10^{-19}$~erg~s$^{-1}$~cm$^{-2}$~\AA$^{-1}$), yielding an integrated EW of $5.6_{-0.2}^{+0.2}$~\AA{}.
We note that this value is  lower than that reported by \cite{Bunker2023}, which is $12$~\AA{}, assuming their line flux determined from the 3-pixel, or 0\arcsec{}.3, extraction aperture. 
As our line flux measurement is only slightly smaller than their value from 3-pixel extraction, we attribute the higher EW reported by \citet{Bunker2023} primarily to differences in the adopted continuum estimate.
\citet{Bunker2023} estimate the continuum directly at the Ly$\alpha$ wavelength (priv. comm.) from the JADES prism spectrum, where the lower spectral resolution blends the continuum with both the Ly$\alpha$ break and the \ion{N}{5} P-Cygni absorption \citep[e.g.,][]{Keating2024,Mason2026}, resulting in a lower observed continuum value relative to our estimate from the high SNR grating spectrum.
Such a low Ly$\alpha$ EW is $\approx3-4 \times$ smaller than the median EW in Lyman-break galaxies at $z\sim5-6$ \citep[][]{Tang2024,PrietoLyon2025}, but comparable values have been reported in similarly UV luminous galaxies at slightly lower redshifts ($z\simeq8\mbox{--}9$; e.g., \citealt{Whitler2026}), and may suggest significant IGM attenuation \citep{Tang2024a,Kageura2025}.
We return to the effect of the IGM in \S~\ref{sec:VMS_IGM}, where we independently model its effect as part of the full UV continuum spectrum fitting. 

Finally, motivated by the high fraction of flux at $>500$~km~s$^{-1}$, we also consider whether the Ly$\alpha$ emission can be better fit including a broad component, to help establish the potential presence of outflows and/or an AGN broad line region which may broaden Ly$\alpha$ without significant resonant scattering. 
We thus also fit the Ly$\alpha$ profile with a double-truncated Gaussian model, which we present in the right panel of Figure~\ref{fig:Lya}.
We find the double-truncated Gaussian model provides a slightly better fit to the data compared to the single-truncated Gaussian model, as it better reproduces the core of the line $\approx200-400$~km~s and the red tail at $\approx800-1000$~km~s, as seen in the residuals in Figure~\ref{fig:Lya} ($\chi^2_{\nu, \rm double}=1.0$ compared to $\chi^2_{\nu, \rm single}=1.5$ at rest-frame wavelengths 1215--1220~\AA{}).
In the double-truncated Gaussian, the narrow core is peaked at $v_{\rm peak, narrow} = 360_{-45}^{+26}$~km~s$^{-1}$ with intrinsic FWHM $=95_{-28}^{+60}$~km~s$^{-1}$, while the broad wing is more redshifted ($v_{\rm peak, broad} = 538_{-255}^{+165}$~km~s$^{-1}$), with intrinsic FWHM = $526_{-155}^{+98}$~km~s$^{-1}$.
We note the total Ly$\alpha$ flux ($1.30_{-0.06}^{+0.05}\times10^{-18}$~erg~s$^{-1}$~cm$^{-2}$) and large fraction ($40\%$) of flux at $>500$~km~s$^{-1}$ are unchanged relative to the single truncated Gaussian model, and these results are unchanged if we fix the broad component centroid to that of the narrow peak.
We will discuss the implications of a potential broad component in Ly$\alpha$ in more detail in Section~\ref{sec:discussion}.

\subsection{\ion{N}{4}] Emission} \label{sec:spec_NIV}

The next strongest emission line in the rest-UV is the \ion{N}{4}]$\lambda\lambda$1483,1486 doublet (Fig.~\ref{fig:n4}).
Emission from the red component of \ion{N}{4}] ($\lambda$1486)  was detected by \citet{Bunker2023}, and it was pointed out that the line ratios of the doublet (i.e., the absence of $\lambda$1483 emission) imply that the nitrogen-emitting gas is very high density ($\gtrsim 10^5$ cm$^{-3}$)  \citep{Senchyna2024,Maiolino2024}. The SPURS measurements reveal a strong $\lambda 1486$ line but    do not unambiguously detect the  $\lambda 1483$ component. 
One of the key findings of the deeper spectrum  is the potential detection of a broad \ion{N}{4}] component, which visually appears centered on the red side of the doublet (see  Fig.~\ref{fig:n4}). 

\begin{figure*}[t]
\centering
\includegraphics[width=\linewidth]{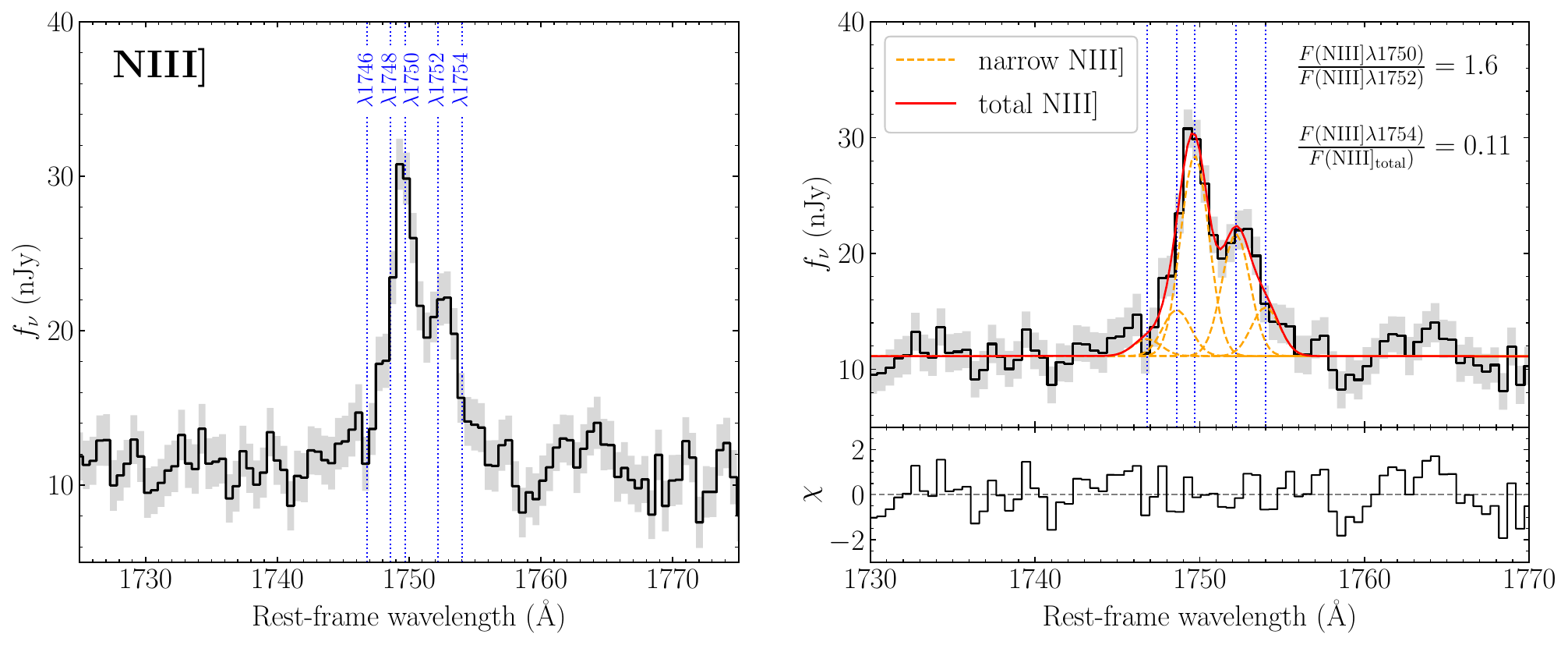}
\caption{Left panel: \ion{N}{3}] emission line of GN-z11. Right panel: Fits to the \ion{N}{3}] profile of GN-z11. The figure is shown in the same way as Figure~\ref{fig:n4}. The \ion{N}{3}] profile is best-fitted by five narrow (FWHM $=347$~km~s$^{-1}$; orange dashed line) Gaussians without a broad component.}
\label{fig:n3}
\end{figure*}

We fit the \ion{N}{4}] complex with a number of model profiles. We first consider only two narrow lines, then we add in one and two broad components. The results are described in Appendix~\ref{app:BIC}. 
This investigation shows that fits including two narrow components and a broad component are preferred to those including only two narrow components. Our main results are not strongly dependent on the chosen profile, but we will adopt the model that allows two narrow lines (centered on $\lambda$1483 and $\lambda$1486) and a broad line (centered on $\lambda$1486). Additional information about the fitting process is provided in Appendix~\ref{app:BIC}.

The \ion{N}{4}] fit is shown in the right panel of Figure~\ref{fig:n4}, and the line measurements are listed in Table~\ref{tab:fuv_lines}. The narrow lines are found with a FWHM of $319^{+24}_{-24}$~km~s$^{-1}$, similar to that found in other nebular emission lines in GN-z11. 
The broad line is fit with a FWHM of $1670^{+292}_{-292}$~km~s$^{-1}$, implying a fast-moving nitrogen-emitting component in GN-z11.
The narrow \ion{N}{4}]$\lambda1486$ line is very strong  (S/N $=10$, EW $=4.0^{+0.4}_{-0.4}$~\AA). We place an upper limit on the narrow [\ion{N}{4}]$\lambda1483$ flux, indicating a large \ion{N}{4}]$\lambda1486$/[\ion{N}{4}]$\lambda1483$ flux ratio ($>7.7$ at $3\sigma$). 
The broad N~{\small IV}] component (S/N $=3$, EW $=3.8^{+1.1}_{-1.1}$~\AA) comprises $49\%$ of the total N~{\small IV}] flux, corresponding to a luminosity of $8.8^{+2.6}_{-2.6}\times10^{41}$~erg~s$^{-1}$. The absence of a broad component centered at [\ion{N}{4}]$\lambda$1483 suggests an \ion{N}{4}]$\lambda1486$/[\ion{N}{4}]$\lambda1483$ ratio of $>3.0$ in the broad line emitting region.

One potential explanation of the broad \ion{N}{4}]$\lambda$1486 component is fast-moving gas associated with stellar winds. The most prominent \ion{N}{4}] wind line in Wolf-Rayet stars is typically \ion{N}{4}]$\lambda$1719 (e.g. \citealt{Hainich2014}), so if the broad \ion{N}{4}]$\lambda$1486 line is powered by stellar winds, we should expect to see even stronger broad \ion{N}{4}]$\lambda$1719 emission.
However, the SPURS observations do not detect any emission at the wavelength of \ion{N}{4}]$\lambda$1719 (Fig.~\ref{fig:spectrum}). 
Assuming a line width the same as the broad N IV]~$\lambda$1486 (FWHM $=1670$~km~s$^{-1}$), we place a $3\sigma$ limiting flux of $<1.6\times10^{-19}$~erg~s$^{-1}$~cm$^{-2}$, corresponding to a broad \ion{N}{4}]$\lambda$1486/\ion{N}{4}]$\lambda$1719 flux ratio of $>3.7$. 
We thus conclude that the stellar winds typically seen in WR stars of the local universe are unlikely to be responsible for the broad \ion{N}{4}]$\lambda$1486 emission, but we will discuss in \S~\ref{sec:discussion_NIV} that this could be nevertheless produced in a dense WR-like wind.

\begin{figure*}[t]
\centering
\includegraphics[width=\linewidth]{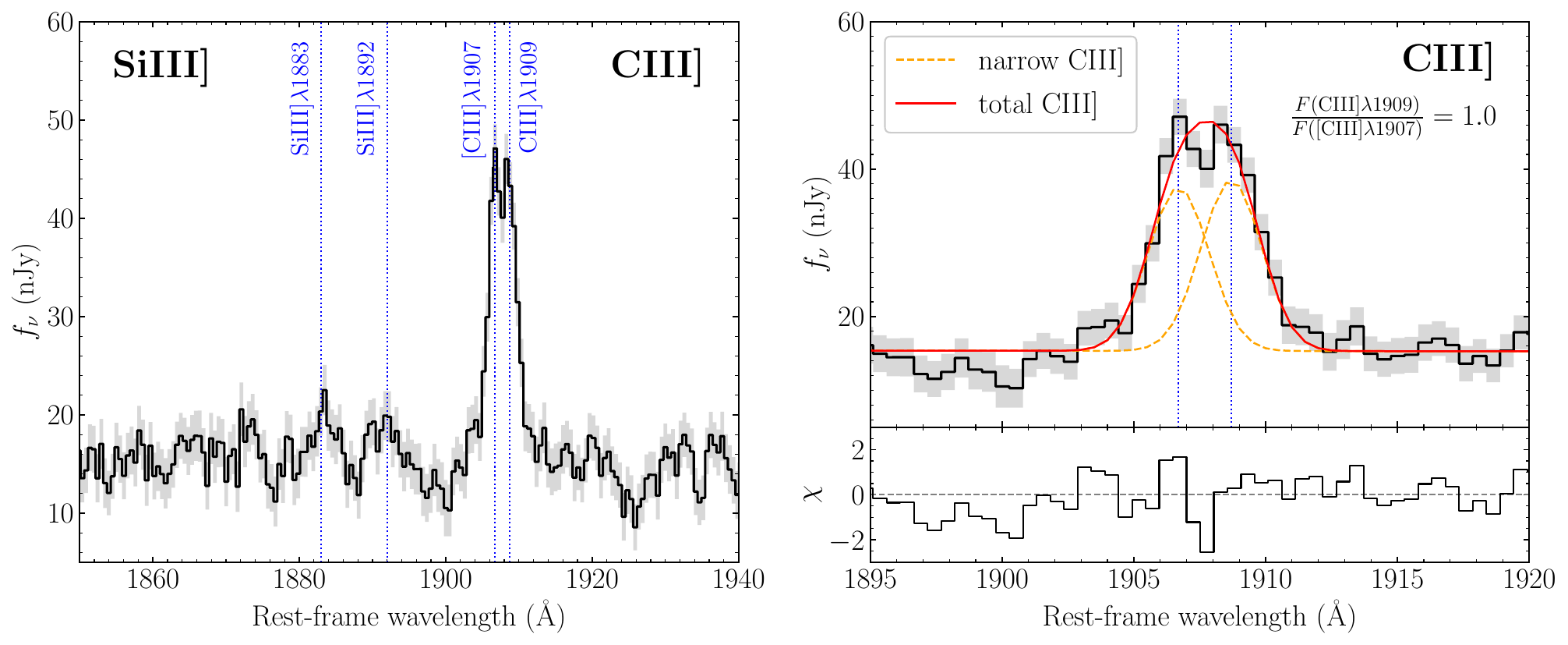}
\caption{Left panel: \ion{Si}{3}] and \ion{C}{3}] emission lines of GN-z11. Right panel: Fits to the \ion{C}{3}] profile of GN-z11. The figure is shown in the same way as Figure~\ref{fig:n3}. The \ion{C}{3}] emission line is best-fitted by two narrow (FWHM $=442$~km~s$^{-1}$) Gaussians centered at the rest-frame wavelengths of [\ion{C}{3}]~$\lambda1907$ and \ion{C}{3}]~$\lambda1909$.}
\label{fig:si3_c3}
\end{figure*}

\subsection{\ion{N}{3}] Emission} \label{sec:spec_NIII}

GN-z11 also powers strong emission from the \ion{N}{3}]$\lambda1750$ quintuplet. This feature was detected in the JADES spectrum \citep{Bunker2023}. The relative quintuplet ratios provide a constraint on the electron density traced by nitrogen-emitting gas in GN-z11 \citep[e.g.,][]{Maiolino2024}.   In particular, the \ion{N}{3}]$\lambda$1750/$\lambda$1752 ratio is very density sensitive, increasing from $1.0-1.7$ at low gas densities ($10^2-10^4$ cm$^{-3}$) to $2.5$ at extremely high gas densities ($10^9$ cm$^{-3}$).
The SPURS spectrum improves the significance of the \ion{N}{3}] line (S/N $=17$; left panel of Figure~\ref{fig:n3}) relative to the earlier data.
The total \ion{N}{3}] flux is $7.9^{+0.5}_{-0.5}\times10^{-19}$~erg~s$^{-1}$~cm$^{-2}$, implying a luminosity of $1.17^{+0.07}_{-0.07}\times10^{42}$~erg~s$^{-1}$.
Unlike \ion{N}{4}], we do not see concrete evidence for a broad profile in \ion{N}{3}] (see Appendix~\ref{app:BIC}). However, it is plausible a broad component may be detected with a deeper spectrum: assuming the same broad/narrow flux ratio as in \ion{N}{4}], a broad component in \ion{N}{3}] would be detected at SNR$\leq3$ in our spectrum.

To assess the relative fluxes of the individual \ion{N}{3}] components, we fit the observed line profile with five Gaussian functions, each centered at the rest-wavelengths of the different quintuplet lines, assuming all five components have the same line width. We fix the \ion{N}{3}]$\lambda1746$/\ion{N}{3}]$\lambda1752$ and \ion{N}{3}]$\lambda1748$/\ion{N}{3}]$\lambda1754$ flux ratios to the respective theoretical values ($f_{\rm NIII]\lambda1746}/f_{\rm NIII]\lambda1752}=0.14$, $f_{\rm NIII]\lambda1748}/f_{\rm NIII]\lambda1754}=0.95$). 
The \ion{N}{3}] fit is shown in the right panel of Figure~\ref{fig:n3}. The derived $\lambda1750$/$\lambda1752$ flux ratio is $1.6^{+0.3}_{-0.3}$. 
As we will discuss in more detail in Section~\ref{sec:phys_density}, this ratio lies in a regime that is consistent with a wide range of gas densities.
The \ion{N}{3}]~$\lambda1754$ to total \ion{N}{3}] flux ratio is also sensitive to electron density. 
Using JADES spectrum, \citet{Maiolino2024} reported a \ion{N}{3}]$\lambda1754$/\ion{N}{3}] flux ratio of $0.25\pm0.04$, consistent with an extremely high density ($\gtrsim10^9$~cm$^{-3}$). 
We find a \ion{N}{3}]~$\lambda1754$/\ion{N}{3}] flux ratio of $0.11^{+0.02}_{-0.02}$, which is lower than that reported in \citet{Maiolino2024}. 
This ratio is also consistent with a broad range of densities (see Section~\ref{sec:phys_density}).

\subsection{\ion{C}{3}], \ion{O}{3}], and \ion{Si}{3}] Emission} \label{sec:spec_CIII}

The [\ion{C}{3}]$\lambda$1907, \ion{C}{3}]$\lambda$1909 doublet is very prominent in the spectrum of GN-z11, providing an important anchor for relative chemical abundance ratios (C/O, N/C) and an additional check on the gas-phase density. 
Both [\ion{C}{3}]$\lambda$1907 and \ion{C}{3}]$\lambda$1909 are detected in the spectrum (Figure~\ref{fig:si3_c3}), with the sum of both components implying a total C~{\small III}] EW of $8.8^{+0.4}_{-0.4}$~\AA. 
The \ion{C}{3}]$\lambda1909$/[\ion{C}{3}]$\lambda1907$ flux ratio varies from $0.6$ in the low density limit ($<10^2$~cm$^{-3}$) to $2$ at high densities ($\simeq10^5$~cm$^{-3}$) \citep[e.g.,][]{Osterbrock2006,Kewley2019}.
The derived $\lambda$1909/$\lambda$1907 flux ratio in our spectrum ($1.0^{+0.2}_{-0.2}$; right panel of Figure~\ref{fig:si3_c3}) suggests that \ion{C}{3}] probes moderately large electron densities in GN-z11. This is consistent with what was concluded based on the lower S/N JADES spectrum \citep{Senchyna2024}.

We also detect Si~{\small III}]~$\lambda1883$ (S/N $=3.1$) and Si~{\small III}]~$\lambda1892$ (S/N $=2.6$) emission (left panel of Figure~\ref{fig:si3_c3}).  
We measure a line flux of $1.0^{+0.3}_{-0.3}\times10^{-19}$~erg~s$^{-1}$~cm$^{-2}$ ($7.4^{+2.8}_{-2.8}\times10^{-20}$~erg~s$^{-1}$~cm$^{-2}$) for Si~{\small III}]~$\lambda1883$ (Si~{\small III}]~$\lambda1892$), indicating an EW of $1.1^{+0.4}_{-0.4}$~\AA\ ($0.8^{+0.3}_{-0.3}$~\AA).
While the doublet is also sensitive to electron density, the S/N of the lines is not large enough for a robust calculation. 

The auroral \ion{O}{3}]$\lambda1666$ emission line (S/N $=6$, EW $=1.9^{+0.3}_{-0.3}$~\AA) is well-detected in the SPURS spectrum (Figure~\ref{fig:heii}). When combined with [\ion{O}{3}]$\lambda$5007 (from MIRI, \citealt{Alvarez-Marquez2025}), the UV auroral line provides a constraint on the electron temperature. 
We also detect the \ion{O}{3}]$\lambda1661$ emission line (S/N $=3.5$, EW $=0.6^{+0.2}_{-0.2}$~\AA). 
This implies an \ion{O}{3}]$\lambda1661$/\ion{O}{3}]$\lambda1666$ flux ratio of $0.34^{+0.11}_{-0.11}$, consistent with that calculated from theoretical transition probabilities ($\simeq0.4$; \citealt{FroeseFischer1985}). 
We will derive the electron temperature and compute a direct method oxygen abundance in section \S~\ref{sec:phys_TeOH}.

\subsection{He II Emission}\label{subsec:heii}

\begin{figure}
\centering
\includegraphics[width=\linewidth]{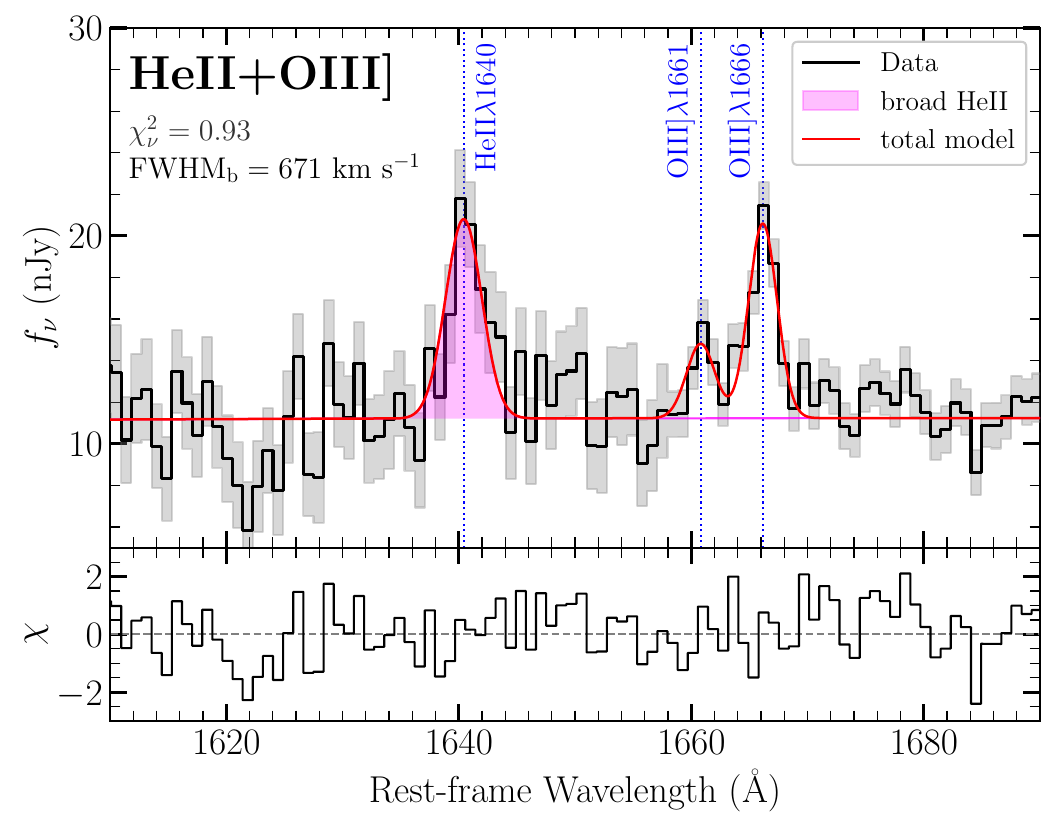}
\caption{\ion{He}{2} and \ion{O}{3}] emission lines of GN-z11 and the best-fit model to the line profiles. The spectrum is shown in the same way as Figure~\ref{fig:n3}.}
\label{fig:heii}
\end{figure}

The He II$\lambda1640$ line was detected (S/N $=4$) in the discovery spectrum of GN-z11. 
He II is a superposition of stellar and nebular emission. The nebular emission is expected to be narrow, with line widths similar to other nebular lines ($\simeq300-400$ km s$^{-1}$, in the case of GN-z11). In contrast, the stellar wind contribution may be significantly broader ($>500$~km~s$^{-1}$), powered by strong winds of WR or VMS populations  \citep[e.g.,][]{Schaerer1996,Chandar2004,Brinchmann2008,Nanayakkara2019,Senchyna2021}.
While He II falls in the detector gap of our deepest G140M grating, the SPURS G235M observation covers this line, so the stacked spectrum still has more than double the previous integration time at this wavelength (13.86~hours in our stack).
To quantify the line width of the observed feature, we fit He II with a single Gaussian, which is shown in Figure~\ref{fig:heii}. The measured FWHM ($671_{-178}^{+175}$ km s$^{-1}$, or $600_{-209}^{+191}$ km s$^{-1}$ after correcting for instrument resolution) is broader than the isolated nebular lines in the rest-UV, consistent with a significant contribution of stellar winds, though we note the FWHM is still consistent with that of other UV nebular lines within $1\sigma$. 
We find a total He II line flux of $4.1_{-0.8}^{+0.7}\times10^{-19}$~erg~s$^{-1}$~cm$^{-2}$, corresponding to an EW of  $3.3_{-0.7}^{+0.6}$~\AA.

\subsection{\ion{C}{4}, \ion{N}{5}, and Si IV Emission and Absorption}\label{subsec:obs_wind}

\begin{figure*}
\centering
\includegraphics[width=\linewidth]{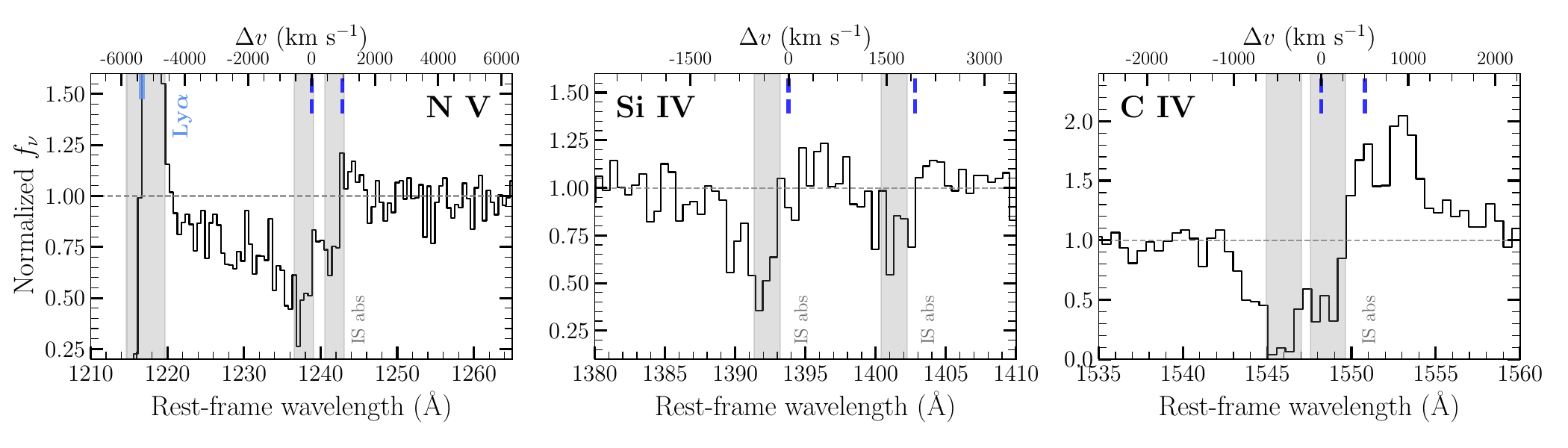}
\caption{Stellar wind P-Cygni features of GN-z11 (left panel: \ion{N}{5}; middle panel: \ion{Si}{4}; right panel: \ion{C}{4}). In each panel, we mark the line center of each component of \ion{N}{5}, \ion{Si}{4}, or \ion{C}{4} as blue dashed vertical line. The regions that are impacted by interstellar absorption are masked by grey shades.}
\label{fig:winds}
\end{figure*}

The resonant P-Cygni absorption in N~V, Si~IV, and C~IV is a key stellar wind diagnostic from young massive stars \citep[e.g.,][]{Leitherer1995,Chisholm2019,Senchyna2021}.
One challenge in characterizing these features is the blending with the narrow interstellar absorption lines.
Here, we describe their profiles qualitatively, while leaving quantitative measurements of interstellar absorption contribution to \S~\ref{sec:ISMabs} and the comparison to stellar population models in \S~\ref{sec:VMS}.

The \ion{C}{4} $\lambda\lambda$1548,1550 complex provides additional diagnostic power on the nature of the ionizing sources. 
The JADES discovery spectrum provided tentative evidence for a stellar P-Cygni  profile for \ion{C}{4}, but the S/N was not adequate to characterize the profile. 
SPURS observations provide a more robust view of \ion{C}{4} (see right panel of Figure~\ref{fig:winds}). We see interstellar absorption lines from both components of the doublet, each blueshifted by $\approx-500$ km s$^{-1}$ (Figure~\ref{fig:ismabs}). There is also a broad blueshifted tail of absorption, extending to $\simeq-1000$ km s$^{-1}$. This is consistent with expectations for strong stellar wind absorption, although we cannot rule out an additional component of fast-moving gas associated with an outflow \citep[e.g.,][]{Maiolino2024}.

The \ion{C}{4} emission  is separated into two peaks, one appearing at rest-frame $1550.7$~\AA\ and the other at $1553.3$~\AA, both redshifted $\simeq+484$ km s$^{-1}$ from the resonance wavelengths of the doublet, which could be consistent with resonant scattering in a fast ($\gtrsim 500$\,km~s$^{-1}$), optically thick outflow (e.g., Lim et al. in prep). The total line flux of the \ion{C}{4} emission is $4.5^{+0.5}_{-0.5}\times 10^{-19}$ erg s$^{-1}$ cm$^{-2}$, and the implied EW is $3.3^{+0.3}_{-0.3}$~\AA. 
The \ion{C}{4} emission is likely a combination of emission from nebular gas and stellar winds. We will estimate the stellar contribution using continuum fitting in section \S~6.

The \ion{N}{5}$\lambda\lambda1238,1242$ doublet is also detected in the GN-z11 spectrum,  revealing a strong P-Cygni profile (left panel of Figure~\ref{fig:winds}).
We detect a broad  (FWHM $754^{+253}_{-253}$ km s$^{-1}$) and redshifted emission component, with a blueshifted absorption wing  extending to $\simeq-3000$ km s$^{-1}$.
The peak flux of the emission line occurs at rest-frame $1244.7$~\AA, and the implied EW is $1.0^{+0.2}_{-0.2}$~\AA. We also detect two narrow absorption components at rest-frame $1237.1$ and $1241.2$~\AA, both blueshifted by $\approx-400$ km s$^{-1}$ from resonance.  
We note that the \ion{N}{5} P-Cygni profile can also be suppressed by the IGM damping wing, which we discuss in more detail in Section~\ref{sec:VMS}.
 
Finally, the Si IV resonant doublet ($\lambda\lambda1393,1402$) also shows a broad P-Cygni profile, consistent with an origin in stellar winds (middle panel of Figure~\ref{fig:winds}), as well as narrower strong interstellar absorption lines.  
We will explore and quantify the ISM absorption in N~V, Si~IV and C~IV profiles in more detail in section \S~\ref{sec:ISMabs}, and the implications for the stellar populations in \S~\ref{sec:VMS}.

\subsection{\ion{O}{1}$^*$ fine structure emission}

\begin{figure}
\centering
\includegraphics[width=\linewidth]{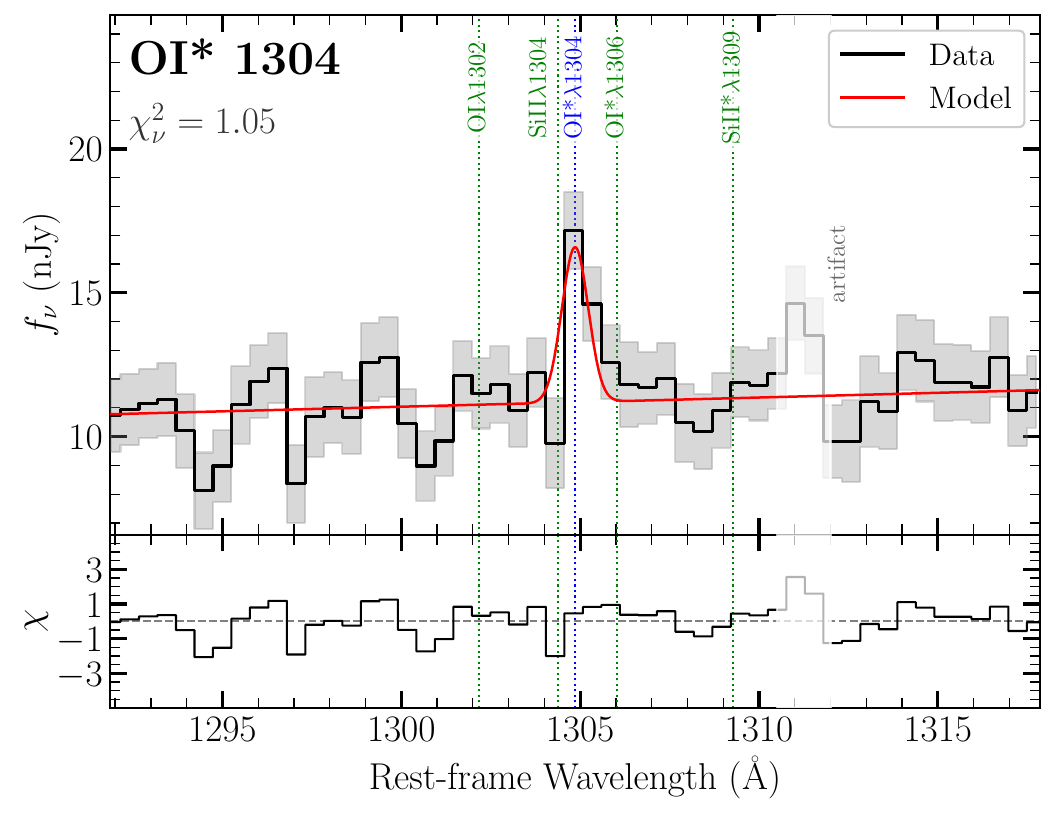}
\caption{Detection of fine structure emission in O~I$^*$~$\lambda1304$.
We present the observed spectrum along with the Gaussian fit, shown in the same way as Figure~\ref{fig:n3}.
We also mark the wavelengths of other emission lines around this region, with the detected emission feature best aligning with that of O~I$^*$~$\lambda1304$. We note a feature at $\approx1311$\,\AA\ corresponds to an artifact in a single exposure.}
\label{fig:oi1304}
\end{figure}

We detect a faint emission feature (S/N = 3.6) at the observed wavelength of 15140.8~\AA{}, translating to a rest-frame wavelength of 1304.8~\AA{}. 
This emission feature is just redshifted of the \ion{O}{1}$\lambda1302$ and \ion{Si}{2}$\lambda1304$ low ionization absorption lines, however we detect no significant absorption (see Section~\ref{sec:ISMabs} below). This wavelength range includes a blend of several potential emission features: resonant \ion{Si}{2} at 1304.37~\AA{} -- which may be seen in emission rather than absorption if resonantly scattered photons escape along an optically thin sightline; and \ion{O}{1}$^*$ fine structure emission at 1304.86~\AA{} and 1306.03~\AA{}.
As we show in Figure~\ref{fig:oi1304}, given the systemic redshift, the emission feature we detect is most consistent with \ion{O}{1}$^*$ fine structure emission at 1304.86~\AA{}.
By fitting a Gaussian profile along with a linear function as the continuum, we derive an observed flux of $1.03^{+0.33}_{-0.24}\times10^{-19}$~erg~s$^{-1}$~cm$^{-2}$, corresponding to a rest-frame EW of $0.52^{+0.17}_{-0.12}$~\AA.

\ion{O}{1}$^*\lambda1304$ emission is rarely detected in galaxies at lower redshifts, but is prominently seen in AGN and has recently been reported in several LRDs \citep{Tripodi2025,Tang2026_spursQSO1}.
In these sources, as well as the Sunburst arc, which shows rest-frame optical \ion{O}{1}$^*$ emission \citep{Pascale2024,Choe2025}, \ion{O}{1}$^*$ flux is thought to be boosted in very neutral dense gas via Ly$\beta$ (Bowen) fluorescence of the $3{\rm d}^3\,{\rm D}^0$ excited state of \ion{O}{1} \citep[][]{Kwan1981,Matsuoka2007}. 
If the line is primarily driven by Ly$\beta$ fluorescence, we may also expect to see \ion{O}{1}$\lambda1302$ in emission: our non-detection suggests the emitting gas is likely optically thick to \ion{O}{1}$\lambda1302$.
We should also expect \ion{O}{1}$\lambda11287$,$\lambda8446$ emission, though these are beyond the wavelength range of NIRSpec.
\ion{O}{1}$^*\lambda1304$ can also be produced by continuum pumping of \ion{O}{1}$\lambda1302$ and subsequent re-emission via fine-structure cascades, analogous to fine-structure lines such as \ion{Si}{2}$^*$, \ion{C}{2}$^*$ and \ion{Fe}{2}$^*$ commonly seen in $z\sim2-4$ star-forming galaxies \citep[e.g.,][]{Shapley2003,Erb2012,Kornei2013,Wang2020}, however we do not detect the low-ionization absorption features which would indicate continuum absorption, nor any other fine-structure emission lines at high significance in GN-z11. This suggests Ly$\beta$ pumping may dominate the \ion{O}{1}$^*\lambda1304$ production in GN-z11.


\begin{deluxetable}{lcccc}
\tablecaption{Rest-frame near-UV emission line observed-frame wavelengths (\AA), fluxes ($10^{-19}$~erg~s$^{-1}$~cm$^{-2}$), EWs (\AA), and FWHMs (km~s$^{-1}$) of GN-z11. Upper limits are provided at the $3\sigma$ level.}
\label{tab:nuv_lines}
\tablehead{
Line & $\lambda_{\rm obs}$ & Flux & EW & FWHM
}
\startdata
{[}Ne~{\scriptsize IV}]~$\lambda2422$ & -- & $<0.8$ & $<1.6$ & -- \\
{[}Ne~{\scriptsize IV}]~$\lambda2424$ & -- & $<0.8$ & $<1.6$ & -- \\
Mg~{\scriptsize II}~$\lambda2796$ & $32446.6$ & $2.1^{+0.3}_{-0.3}$ & $5.8^{+0.8}_{-0.8}$ & $366^{+34}_{-34}$ \\
Mg~{\scriptsize II}~$\lambda2803$ & $32529.0$ & $1.2^{+0.3}_{-0.3}$ & $3.5^{+0.9}_{-0.9}$ & $366^{+34}_{-34}$ \\
{[}Ne~{\scriptsize V}]~$\lambda3427$ & -- & $<0.7$ & $<3.2$ & -- \\
\enddata
\end{deluxetable}

\subsection{[\ion{Ne}{4}] and [\ion{Ne}{5}] Emission}

The near-UV is host to several high ionization lines which offer additional diagnostic power on the presence of an AGN. The JADES spectrum revealed a tentative detection of 
[\ion{Ne}{4}]~$\lambda\lambda2422,2424$ \citep{Maiolino2024,Tang2025}.
We do not find [\ion{Ne}{4}] in the deeper SPURS spectrum (top panel of Figure~\ref{fig:nuv}), placing a $3\sigma$ limiting flux of $<8\times10^{-20}$~erg~s$^{-1}$~cm$^{-2}$ (EW $<1.6$~\AA) for each component. This is smaller than the value of the tentative feature in the JADES spectrum ($3.14\pm0.65\times10^{-19}$~erg~s$^{-1}$~cm$^{-2}$). Either the line is variable, or it is weaker than implied by the tentative feature in the earlier spectrum. 

We also do not detect the [\ion{Ne}{5}] emission line, placing a $3\sigma$ limiting line flux of $<7\times10^{-20}$~erg~s$^{-1}$~cm$^{-2}$. 
This pushes the limit placed in the JADES spectrum ($<1.5\times10^{-19}$~erg~s$^{-1}$~cm$^{-2}$; \citealt{Maiolino2024}) to a deeper sensitivity. 
The corresponding $3\sigma$ upper limit of [\ion{Ne}{5}] EW of GN-z11 is $<3.2$~\AA.
We summarize the measurements for these NUV lines in Table~\ref{tab:nuv_lines}.

\subsection{Mg~{\small II} Emission}

\begin{figure}
\centering
\includegraphics[width=\linewidth]{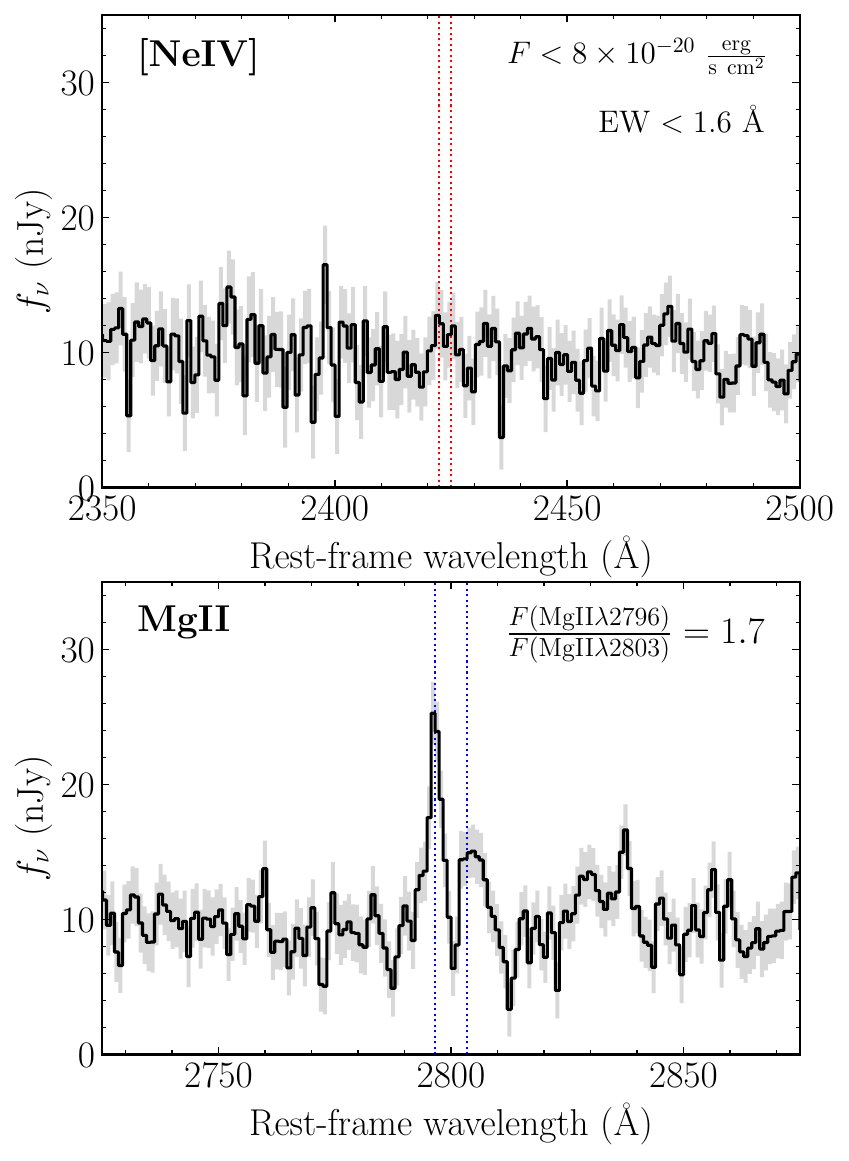}
\caption{\textbf{Top} panel: Non-detection (red dashed lines) of [\ion{Ne}{4}]~$\lambda2422,2424$ emission lines of GN-z11. \textbf{Bottom} panel: \ion{Mg}{2}~$\lambda2796,2803$ emission line detections (blue dashed lines) of GN-z11.}
\label{fig:nuv}
\end{figure}

We also detect the Mg~{\small II}~$\lambda\lambda2796,2803$ doublet in emission in the near UV (bottom panel of Figure~\ref{fig:nuv}), providing an indication of the neutral gas conditions in GN-z11. Mg~{\small II} is a resonant transition tracing cool, $T\sim10^4$\,K gas, and its nebular emission doublet ratio is sensitive to gas kinematics, dust attenuation and the Mg II column density \citep[e.g.,][]{Chang2024}. Empirically, sources with the Mg~{\small II}~$\lambda2796$/Mg~{\small II}~$\lambda2803$ emission flux ratio close to the theoretical value (2:1) have been shown to have high Ly$\alpha$ escape fractions \citep{Henry2018} and Lyman continuum escape fractions \citep{Chisholm2020,Xu2023}.
Dust attenuation, fast outflows ($>700$\,km~s$^{-1}$), and scattering of the stellar continuum can all preferentially suppress the Mg~{\small II}~$\lambda2796$ line relative to Mg~{\small II}~$\lambda2803$, reducing the flux ratio below its intrinsic value of 2 \citep{Chang2024}.

We detect both components (combined S/N $=6$) in the GN-z11 spectrum, with velocity centroids consistent with line center, implying a total Mg~{\small II} EW of $9.3_{-1.2}^{+1.2}$~\AA, in agreement with previous measurements from the JADES spectrum \citep{Bunker2023,Maiolino2024}. 
We measure a Mg~{\small II}~$\lambda2796$/Mg~{\small II}~$\lambda2803$ flux ratio of $1.7^{+0.5}_{-0.5}$, 
which is consistent with the theoretically expected value, indicating optically thin neutral gas, though does not rule out lower values. This is consistent with the ratio reported by \citet{Maiolino2024}, but with improved precision.

\begin{figure*}
\centering
\includegraphics[width=\linewidth]{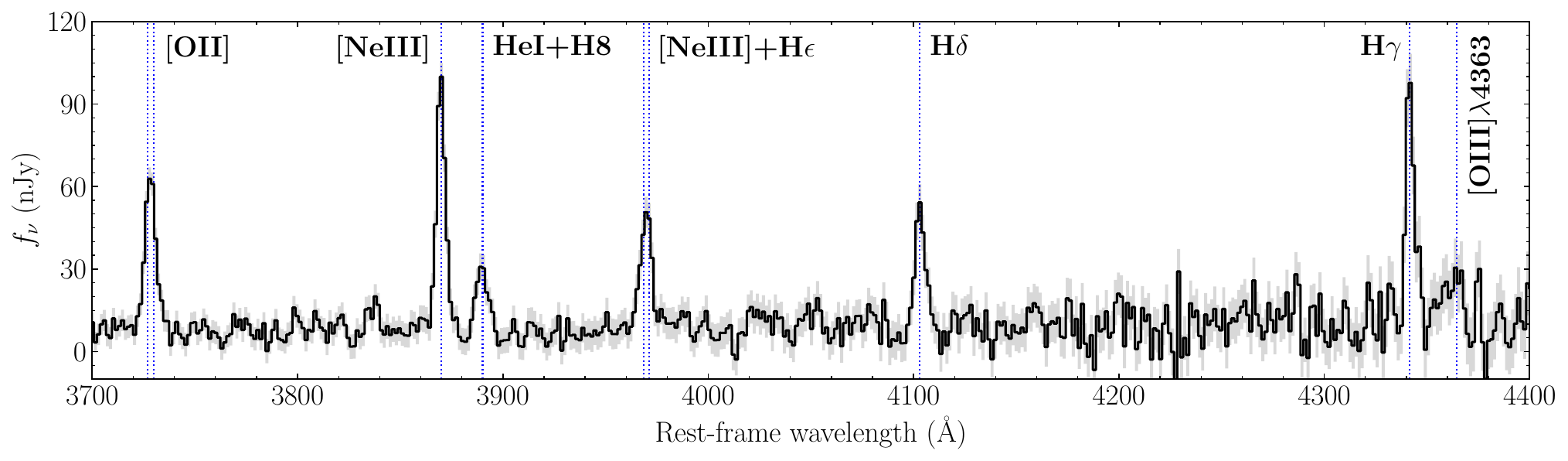}
\caption{Rest-frame optical (G395M) spectrum of GN-z11. The spectrum is created by stacking both SPURS and JADES observations. We detect a suite of rest-frame optical emission lines ([\ion{O}{2}], [\ion{Ne}{3}], \ion{He}{1}, H$\delta$, H$\gamma$, [\ion{O}{3}]~$\lambda4363$).}
\label{fig:opt}
\end{figure*}


\begin{deluxetable}{lcccc}
\tablecaption{Rest-frame optical emission line observed-frame wavelengths (\AA), fluxes ($10^{-19}$~erg~s$^{-1}$~cm$^{-2}$), EWs (\AA), and FWHMs (km~s$^{-1}$) of GN-z11. Upper limits are provided at the $3\sigma$ level.}
\tablehead{
Line & $\lambda_{\rm obs}$ & Flux & EW & FWHM
}
\startdata
{[}O~{\scriptsize II}]~$\lambda3728$\tablenotemark{a} & $43254.2$ & $8.0^{+0.4}_{-0.4}$ & $44^{+2}_{-2}$ & $394^{+24}_{-24}$ \\
{[}Ne~{\scriptsize III}]~$\lambda3869$ & $44895.9$ & $9.7^{+0.4}_{-0.4}$ & $58^{+2}_{-2}$ & $384^{+15}_{-15}$ \\
He~{\scriptsize I}~$\lambda3889$+H8 & $45126.9$ & $2.9^{+0.4}_{-0.4}$ & $18^{+2}_{-2}$ & $468^{+68}_{-68}$ \\
{[}Ne~{\scriptsize III}]~$\lambda3968$+H$\epsilon$ & $46055.3$ & $5.4^{+0.4}_{-0.4}$ & $35^{+3}_{-3}$ & $367^{+42}_{-42}$ \\
H$\delta$ & $47603.3$ & $4.6^{+0.6}_{-0.6}$ & $32^{+4}_{-4}$ & $409^{+43}_{-43}$ \\
H$\gamma$ & $50370.4$ & $7.0^{+0.7}_{-0.7}$ & $55^{+6}_{-6}$ & $317^{+29}_{-29}$ \\
{[}O~{\scriptsize III}]~$\lambda4363$ & $50645.7$ & $3.0^{+0.8}_{-0.8}$ & $24^{+6}_{-6}$ & $459^{+43}_{-43}$ \\
\enddata
\tablenotetext{a}{Unresolved [O~{\scriptsize II}]~$\lambda\lambda3727,3729$ doublet.}
\label{tab:opt_lines}
\end{deluxetable}

\subsection{Rest-Optical Emission Lines} \label{sec:res_optical}

We identify several rest-frame optical emission lines ([\ion{O}{2}], [\ion{Ne}{3}]$\lambda3869$, \ion{He}{1}$\lambda3889$+H8, [\ion{Ne}{3}]$\lambda3968$+H$\epsilon$, H$\delta$, H$\gamma$, [\ion{O}{3}]$\lambda4363$) in the stacked G395M spectrum of SPURS and JADES (Figure~\ref{fig:opt}). 
We present the rest-frame optical emission line fluxes and EWs in Table~\ref{tab:opt_lines}.
The line fluxes are generally consistent with those measured from the JADES spectrum \citep{Bunker2023,Maiolino2024}. 
The Ne3O2 ratio is large ($1.2^{+0.1}_{-0.1}$), indicating extreme ionization conditions in the gas of GN-z11.
The H$\delta$/H$\gamma$ flux ratio ($0.66^{+0.10}_{-0.10}$) is consistent with that expected from case B recombination, suggesting minimal dust attenuation (see \S~\ref{sec:phys_dust}).

We examine whether the Balmer emission lines present broad components.
Visually, no broad emission components are clearly detected in H$\gamma$ or H$\delta$, and their linewidths are consistent with other optical emission lines. However, it is possible a broad component exists below our SNR limit.
Assuming the broad line width of \ion{N}{4}] (FWHM $=1670$~km~s$^{-1}$), we find the $3\sigma$ limiting broad H$\gamma$ line flux is $<3.0\times10^{-19}$~erg~s$^{-1}$~cm$^{-2}$. 
We consider whether a broad H$\gamma$ component would be detected in our data assuming a range of broad-to-total flux ratios given the narrow line flux we measure.
We first assume a broad H$\gamma$ flux ratio similar to that of \ion{N}{4}] ($49\%$), and find an expected broad H$\gamma$ flux of $6.7\times10^{-19}$~erg~s$^{-1}$~cm$^{-2}$. 
This is above the $3\sigma$ limiting flux, indicating a broad component in H$\gamma$ as strong as that seen in \ion{N}{4}] is ruled out.
We also consider a smaller broad H$\gamma$ flux ratio similar to what is found for H$\beta$ in EGSY8p7 ($\simeq0.2$; \citealt{Larson2023,Zamora2025}).
In this case, we expect the broad H$\gamma$ flux to be $1.2\times10^{-19}$~erg~s$^{-1}$~cm$^{-2}$, which is comparable to the $3\sigma$ limiting flux, indicating we cannot rule out the presence of a broad H$\gamma$ component at this level.
We note that the same is true for H$\alpha$, where a similarly broad component is not ruled out by the current MIRI MRS observation of GN-z11 \citep{Alvarez-Marquez2025}\footnote{Using H$\alpha$ flux measured from the MIRI MRS spectrum \citep{Alvarez-Marquez2025},
we would expect the broad H$\alpha$ line flux in GN-z11 to be $1.3\times10^{-18}$~erg~s$^{-1}$~cm$^{-2}$ assuming the broad-to-narrow ratio similar to that in EGSY8p7.
Assuming the broad line width is the same as that of broad N~{\small IV}] (FWHM$_{\rm broad}=1670$~km~s$^{-1}$)
we place a $3\sigma$ flux limit of $<4.8\times10^{-18}$~erg~s$^{-1}$~cm$^{-2}$ for broad H$\alpha$ in the current MIRI MRS spectrum \citep{Alvarez-Marquez2025} -- higher than the expected broad line fluxes.}.

\section{Interstellar Absorption Lines}\label{sec:ISMabs}

The SPURS rest-UV spectrum also enables the detection of multiple metal absorption lines, tracing enriched ISM and CGM gas along the line-of-sight to the UV-emitting region in GN-z11, and providing a critical probe of the kinematics and covering fraction of multi-phase gas \citep[e.g.,][]{Pettini2000,Pettini2002,Shapley2003,steidel2010}.
A key new result from the SPURS spectrum is that the absorbing gas is dominated by highly ionized, outflowing gas.

\begin{figure*}[!t]
    \centering  
    \begin{minipage}{0.495\linewidth}
        \includegraphics[width=\linewidth]{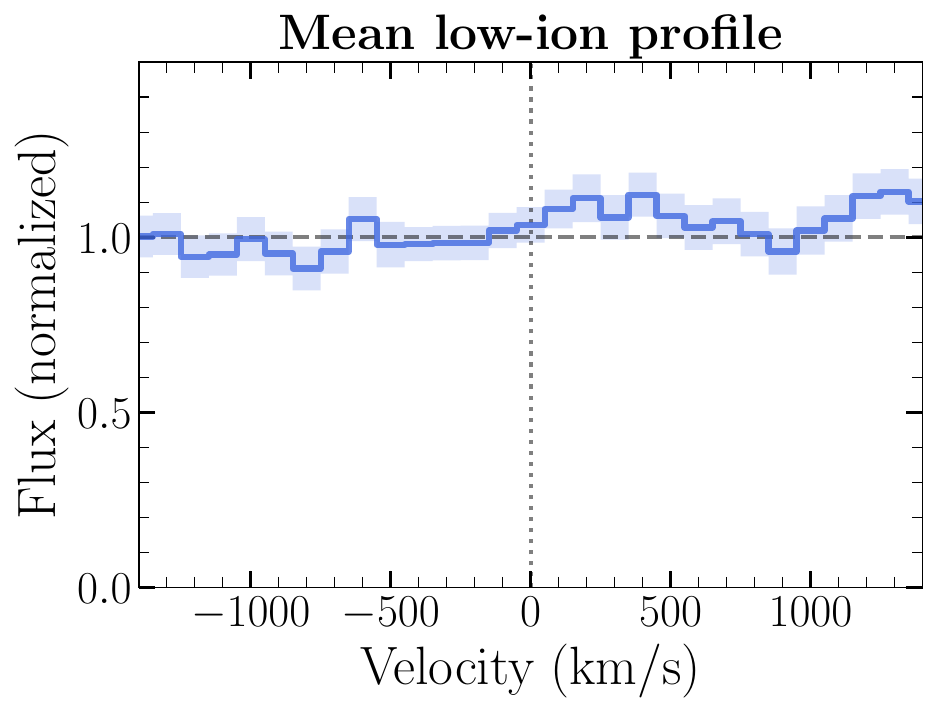} \\
        \includegraphics[width=\linewidth]{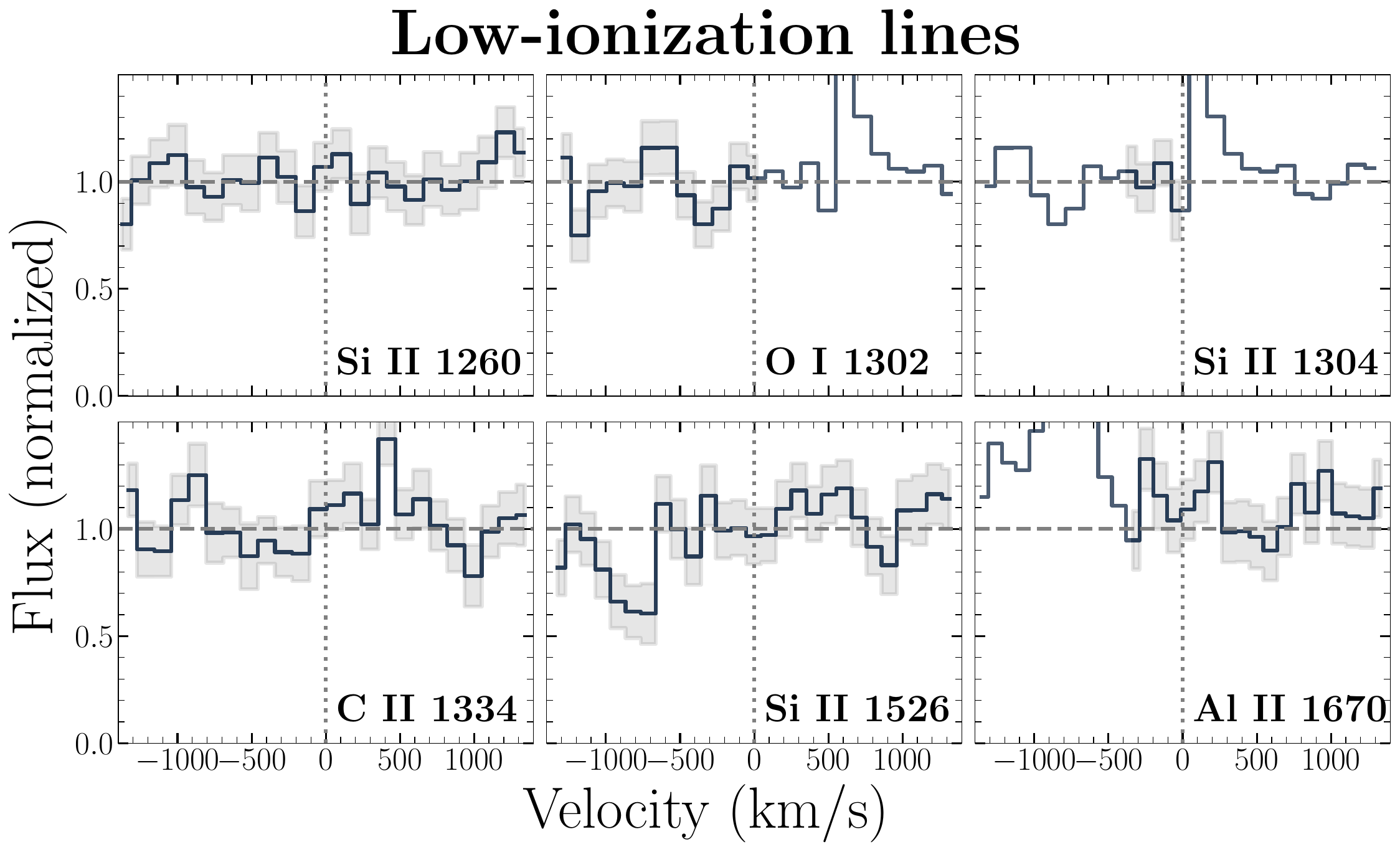}
    \end{minipage}
    \begin{minipage}{0.495\linewidth}
        \includegraphics[width=\linewidth]{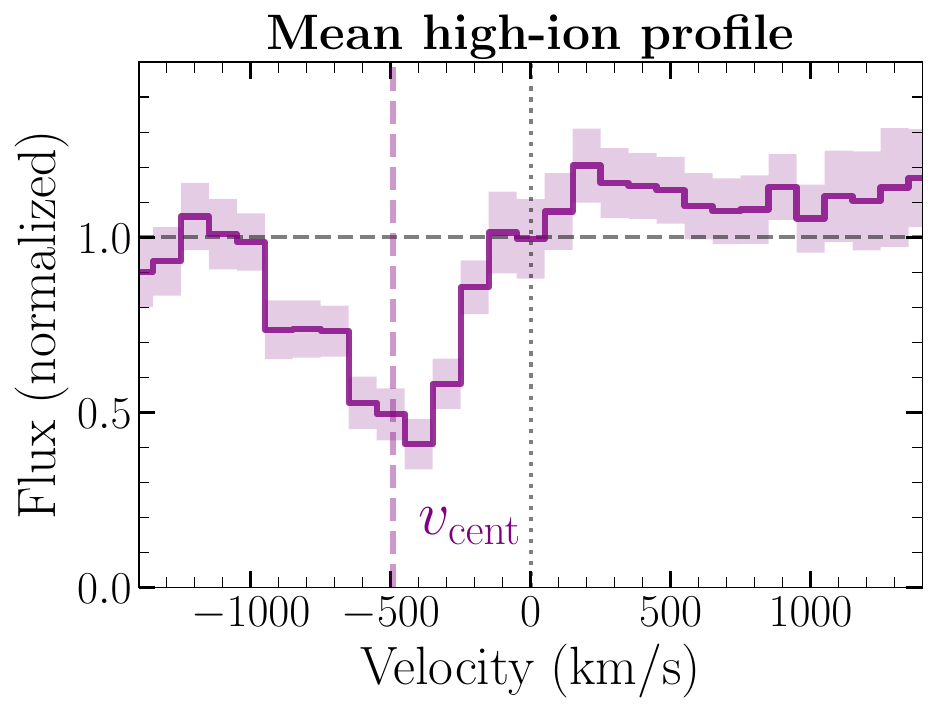} \\
        \includegraphics[width=\linewidth]{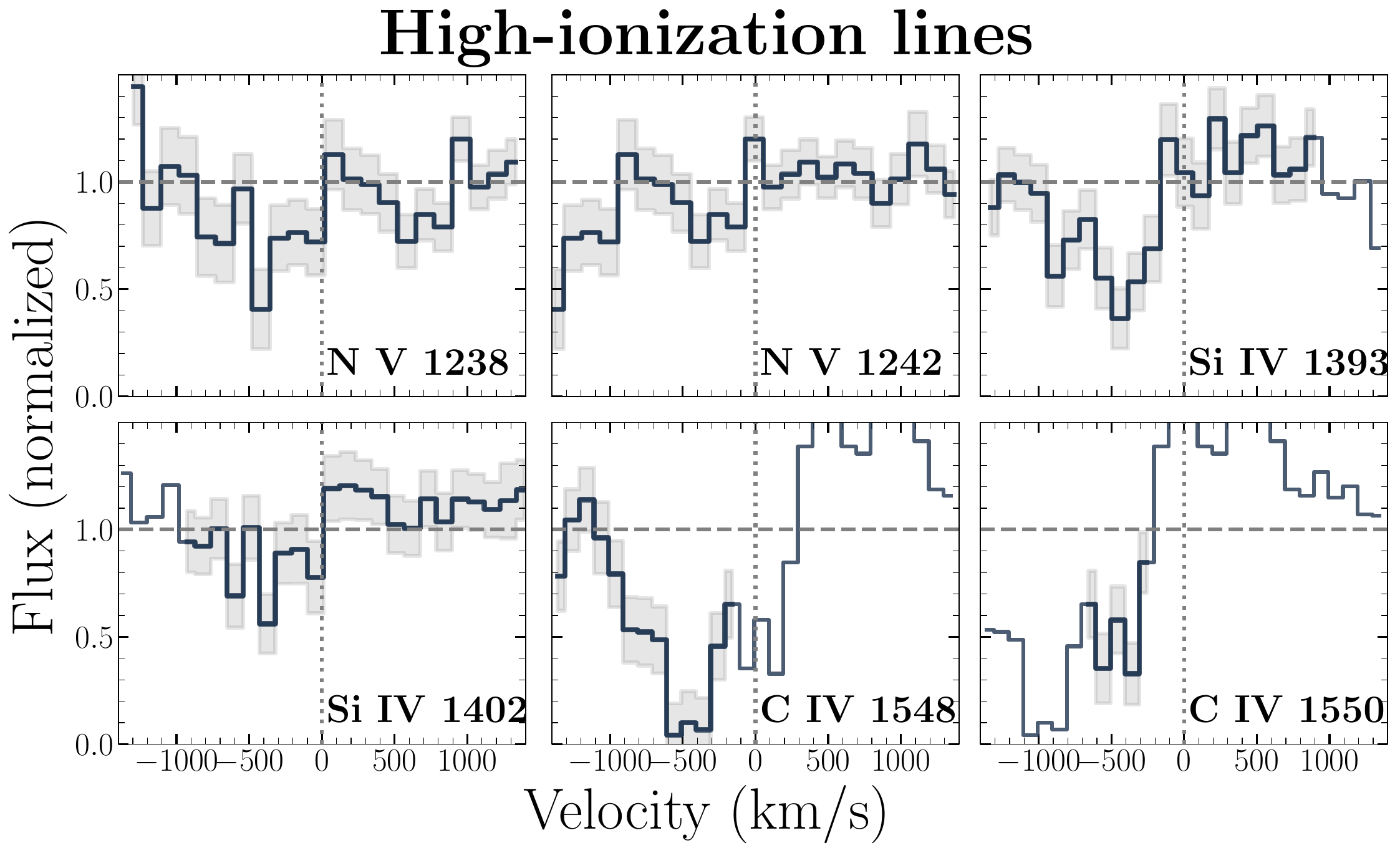}    
    \end{minipage}
    \caption{\emph{Top panels}: Mean low- and high-ionization ISM absorption line profiles for GN-z11. The colored shaded region in each panel denotes the $1\sigma$ uncertainties. The dotted gray vertical line marks the systemic velocity. The velocity centroid ($v_{\rm cent}$) of the high-ionization mean profile is denoted as purple dashed line. \emph{Bottom panels}: Individual low- and high-ionization line profiles used to construct the mean profiles. In each panel, the gray shaded region denotes the $1\sigma$ uncertainties and regions without the uncertainties are masked before combining the lines. For the high-ionization lines, we only use the \ion{Si}{4} and \ion{C}{4} lines but note that the \ion{N}{5} shares a similar velocity structure. We show the absorption line profiles normalized based on the CB19 stellar continuum model presented in \S~\ref{sec:VMS_comparison}, but we note that the main features, i.e., the strength and the velocity centroid, remain unchanged if using the \ModelMZ{} continuum model also considered in \S~\ref{sec:VMS_comparison}.
    }\label{fig:ismabs}
\end{figure*}

In Figure~\ref{fig:ismabs} we show the spectral regions around low-ionization and high-ionization lines with high oscillator strengths, which are typically observed in bright, star-forming galaxies \citep[e.g.,][]{Shapley2003,Steidel2016}.
We detect no significant absorption in the low-ionization lines  (LIS; \ion{Si}{2}$\lambda1260$, \ion{O}{1}$\lambda1302$, \ion{Si}{2}$\lambda1304$, \ion{C}{2}$\lambda1334$, \ion{Si}{2}$\lambda1526$, \ion{Al}{2}$\lambda1670$), implying a low covering fraction of metal-enriched neutral gas.
By contrast, we detect clear absorption in the high-ionization lines (HIS; \ion{Si}{4}$\lambda\lambda1393,1402$, \ion{C}{4}$\lambda\lambda1548,1550$) at blueshifted velocities, indicating a highly ionized outflow.
We also detect narrow absorption features consistent with blueshifted \ion{N}{5}$\lambda\lambda1238,1242$ absorption, at velocities similar to the other high-ionization absorption lines.
Narrow \ion{N}{5} absorption is rarely detected at lower redshifts \citep{Shapley2003,Steidel2016,Berg2022}, but was recently reported in another bright $z>9$ galaxy observed by SPURS \citep{Chen2026}. This could have an interstellar origin, but we will also discuss in \S\ref{sec:discussion} that it may be a feature of unusual stellar winds.

To quantify the differences between the absorption profiles and the outflow kinematics we fit both the individual line profiles, as well as stacks of the mean low- and high-ionization line profiles.
We follow the approach of \citet{vasan2026} and refer readers there for a detailed description of our methodology. 
Briefly, we first normalize the absorption profiles by the continuum. 
To account for stellar wind features in the high ionization lines, we consider both of the continuum models presented in Section~\ref{sec:VMS_comparison} below (CB19 and \ModelMZ{}).
To generate stacks of the low- and high-ionization absorption profiles we interpolate the individual spectral regions shown in Figure~\ref{fig:ismabs} onto a common velocity grid of 100\,km~s$^{-1}$ and combine them using an inverse-variance weighted mean. 
To create the mean profiles, utilize the spectral regions adopted \citet{vasan2026} for $z\simeq5-9$ galaxies. 
We mask wavelength ranges corresponding to rest-UV emission lines (e.g. \ion{C}{4} nebular emission), and in the case of nearby and potentially blended absorption features (e.g. \ion{O}{1}$\lambda1302$, \ion{Si}{2}$\lambda1304$, \ion{C}{4}$\lambda\lambda$1548,50) we mask wavelength ranges expected to be contaminated by the other transition.
The mean profiles are shown in the top panels of Figure~\ref{fig:ismabs}.
Consistent with the individual lines, we detect no significant absorption in the low-ionization stack, while the high-ionization stack shows clear blueshifted absorption.

To characterize the line profiles, we use a single Gaussian fit to the individual and mean normalized line profiles, measuring the velocity centroid ($v_{\rm cent}$), equivalent width (EW), and intrinsic deconvolved line width, accounting for the spectral resolution (assuming $R=1000$, see \S~\ref{sec:obs_lines}).
We find the high-ionization absorption line EWs are strong, though the exact values are slightly sensitive to the choice of continuum model. We find the mean EW of the HIS lines is $-$1.65\,\AA\ assuming the CB19 model, and $-$1.60\,\AA\ assuming the \ModelMZ{} model.
In contrast, for the low-ionization lines, assuming unresolved lines, we measure stringent limits of $\gtrsim-0.34$ to $-0.28$\,\AA\ on the absorption EW, significantly weaker than values typically reported at lower redshifts, and in other bright $z\sim5-9$ galaxies observed by SPURS \citep[$\approx -0.9$\,\AA\ to $-1.5$\,\AA,][]{Shapley2003,Du2018,Pahl2020,vasan2026}.
Comparably low LIS absorption EWs are typically seen in strong \lya-emitters, where they have been interpreted as tracing sightlines with a low covering fraction of neutral gas, facilitating \lya{} transmission through the ISM and CGM \citep{Shapley2003,Trainor2015,Trainor2019,Du2018,Reddy2022,Glazer2025}.
We report all the inferred line properties for the individual high ionization lines and the mean profile, and the EW upper limits on the individual low-ionization lines in Table~\ref{tab:abslines}. 

The high-ionization lines are systematically blueshifted, indicating outflowing ionized gas. 
We measure a velocity centroid of $-492\pm15$\,\kms\ from the mean high-ionization profile, assuming the CB19 continuum model.
Adopting the \ModelMZ{} model results in a marginally lower velocity centroid of $-466\pm15$\,\kms.
We note these velocities are also consistent with the high outflow velocity suggested by the redshifted \ion{C}{4} emission ($\approx500$\,\kms; Section~\ref{sec:results}).
The peak outflow velocities in GN-z11 are significantly faster than the median of the $z\simeq5-9$ sample reported by \citet[][]{vasan2026} ($-211\pm 63$\,\kms).
We note similarly high peak outflow velocities have been seen in a handful of low redshift galaxies with star formation rates and star formation rate surface densities comparable to GN-z11 \citep{Chisholm2015,Heckman2016,Xu2022}, as well as AGN \citep{Hainline2011}. 
In summary, the combination of blueshifted high-ionization absorption and the non-detection of low-ionization species indicates that the absorbing medium in GN-z11 is dominated by outflowing, highly ionized gas.

{\setlength{\tabcolsep}{11pt}
\begin{deluxetable}{lccc}
\tablecaption{High-ionization and low-ionization absorption line properties. Equivalent widths (EW; \AA), centroid velocities ($v_{\rm cent}$; km~s$^{-1}$), and velocity dispersions corrected for instrumental resolution ($\sigma_{\rm deconv}$; km~s$^{-1}$).
For high-ionization lines, we report values based on the absorption profiles considering normalization by two different stellar continuum models (CB19 and \ModelMZ{}; see \S~\ref{sec:VMS_comparison}).}
\label{tab:abslines}
\tablehead{
\colhead{Line} & \colhead{EW} & \colhead{$v_{\rm cent}$} & \colhead{$\sigma_{\rm deconv}$}
}
\startdata
\multicolumn{4}{l}{\textbf{High-ionization lines (CB19) }} \\
\hline 
\ion{Si}{4} $\lambda1393$  				  & $-1.3^{+0.2}_{-0.2}$ & $-500^{+51}_{-51}$ & $214^{+43}_{-43}$ \\
\ion{Si}{4} $\lambda1402$ 				  & $-0.5^{+0.2}_{-0.2}$ & $-387^{+70}_{-70}$ & $157^{+54}_{-54}$ \\
\ion{C}{4} $\lambda1548$\tablenotemark{a} & $-3.0^{+0.2}_{-0.2}$ & $-479^{+22}_{-22}$ & $243^{+20}_{-20}$ \\
\ion{C}{4} $\lambda1550$\tablenotemark{a} & $-1.8^{+0.5}_{-0.5}$ & $-503^{+51}_{-51}$ & $235^{+89}_{-89}$ \\
Mean profile 						      & -- 				   & $-492^{+15}_{-15}$ & $195^{+21}_{-21}$ \\
\hline
\multicolumn{4}{l}{\textbf{High-ionization lines (\ModelMZ{}) }} \\
\hline 
\ion{Si}{4} $\lambda1393$  				  & $-1.3^{+0.2}_{-0.2}$ & $-441^{+36}_{-36}$ & $194^{+61}_{-61}$  \\
\ion{Si}{4} $\lambda1402$ 				  & $-0.6^{+0.1}_{-0.1}$ & $-332^{+72}_{-72}$ & $182^{+36}_{-36}$  \\
\ion{C}{4} $\lambda1548$\tablenotemark{a} & $-2.7^{+0.2}_{-0.2}$ & $-469^{+19}_{-19}$ & $212^{+21}_{-21}$ \\
\ion{C}{4} $\lambda1550$\tablenotemark{a} & $-1.8^{+0.4}_{-0.4}$ & $-501^{+40}_{-40}$ & $217^{+70}_{-70}$ \\
Mean profile 						      & -- 				     & $-466^{+15}_{-15}$ & $188^{+20}_{-20}$  \\
\hline
\multicolumn{4}{l}{\textbf{Low-ionization lines}} \\
\hline 
\ion{Si}{2} $\lambda1260$ & $>-0.31$ & -- & -- \\
\ion{O}{1} $\lambda1302$ & $>-0.29$ & -- & -- \\
\ion{Si}{2} $\lambda1304$ & $>-0.33$ & -- & -- \\
\ion{C}{2} $\lambda1334$ & $>-0.28$ & -- & -- \\
\ion{Si}{2} $\lambda1526$ & $>-0.33$ & -- & -- \\
\ion{Al}{2} $\lambda1670$ & $>-0.34$ & -- & -- \\
\enddata
\tablenotetext{a}{Only the unblended portion of the absorption line was used in the fit. Low-ionization lines are non-detections; limits are $3\sigma$ rest-frame equivalent width lower limits.}
\end{deluxetable}
}

\section{Inferred Physical Properties } \label{sec:phys}


\begin{deluxetable}{cc}
\tablecaption{Emission line flux ratios of GN-z11. Upper limits and lower limits are provided at the $3\sigma$ level.}
\tablehead{
Line ratio & Value
}
\startdata
N~{\scriptsize IV}]~$\lambda1486$/[N~{\scriptsize IV}]~$\lambda1483$ (narrow) & $>7.7$ \\
N~{\scriptsize IV}]~$\lambda1486$/[N~{\scriptsize IV}]~$\lambda1483$ (broad) & $>3.0$ \\
N~{\scriptsize III}]~$\lambda1750$/[N~{\scriptsize III}]~$\lambda1752$ & $1.6^{+0.3}_{-0.3}$ \\
N~{\scriptsize III}]~$\lambda1754$/[N~{\scriptsize III}]$_{\rm total}$ & $0.11^{+0.02}_{-0.02}$ \\
C~{\scriptsize III}]~$\lambda1909$/[C~{\scriptsize III}]~$\lambda1907$ & $1.0^{+0.2}_{-0.2}$ \\
O~{\scriptsize III}]~$\lambda1666$/[O~{\scriptsize III}]~$\lambda5007$ & $0.018^{+0.003}_{-0.003}$ \\
{[}O~{\scriptsize III}]~$\lambda4363$/[O~{\scriptsize III}]~$\lambda5007$ & $0.022^{+0.006}_{-0.006}$ \\
H$\delta$/H$\gamma$ & $0.66^{+0.10}_{-0.10}$ \\
Ne3O2 & $1.2^{+0.1}_{-0.1}$ \\
O32 & $23^{+3}_{-3}$ \\
\enddata
\tablecomments{Line ratios are calculated from narrow line fluxes. The O~{\scriptsize III}]~$\lambda1666$/[O~{\scriptsize III}]~$\lambda5007$, [O~{\scriptsize III}]~$\lambda4363$/[O~{\scriptsize III}]~$\lambda5007$, and O32 ratios of GN-z11 are calculated using the [O~{\scriptsize III}]~$\lambda5007$ flux measured from MIRI spectrum reported in \citet{Alvarez-Marquez2025} and assuming $f_{\rm [OIII]\lambda4959}:f_{\rm [OIII]\lambda5007}=1:3$.}
\label{tab:line_ratios}
\end{deluxetable}


\begin{deluxetable}{cccc}
\tablecaption{Physical properties of GN-z11.}
\tablehead{
Properties & Value
}
\startdata
$\tau_{\rm V}$ & $0.01_{-0.01}^{+0.01}$\\
$\Delta v_{\rm Ly\alpha}/{\rm km~s^{-1}}$ & $252_{-6}^{+13}$\\
$f_{\rm esc,Ly\alpha}$ & $0.027_{-0.003}^{+0.003}$\\
$n_{\rm e}$(N~{\scriptsize IV]})$_{\rm narrow}/{\rm cm}^{-3}$ & $>7.9\times10^5$ \\
$n_{\rm e}$(N~{\scriptsize IV]})$_{\rm broad}/{\rm cm}^{-3}$ & $>2.5\times10^5$ \\
$n_{\rm e}$(N~{\scriptsize III]})$/{\rm cm}^{-3}$ & $1.1\times10^3-1.1\times10^8$ \\
$n_{\rm e}$(C~{\scriptsize III]})$/{\rm cm}^{-3}$ & $2.5^{+0.7}_{-0.9}\times10^4$ \\
$T_{\rm e}$(O$^{2+}$)/K & $1.3^{+0.1}_{-0.1}\times10^4$ \\
$12+\log{\rm (O/H)}$ & $7.94^{+0.14}_{-0.14}$ \\
N/O$_{\rm narrow}$ & $0.54^{+0.10}_{-0.08}$ \\
(N/O)$_{\rm narrow}$/(N/O)$_{\odot}$ & $3.9^{+0.8}_{-0.5}$ \\
\enddata
\tablecomments{The electron temperature and gas-phase oxygen abundance of GN-z11 are calculated using the [O~{\scriptsize III}]~$\lambda5007$ flux measured from MIRI spectrum reported in \citet{Alvarez-Marquez2025}. The dust optical depth in the V-band ($\tau_V$) is inferred from photoionization modeling with \beagle{}.}
\label{tab:properties}
\end{deluxetable}

The deeper spectrum motivates updated measurements of several GN-z11 properties, several of which depend sensitively on the assumed physical conditions of the emitting gas. In this section, we quantify the Balmer decrements, the Ly$\alpha$ escape fraction, the electron density and temperature, and the gas-phase oxygen abundance and N/O ratio.
We summarize the line ratios and inferred physical properties in Table~\ref{tab:line_ratios} and \ref{tab:properties}.

\subsection{Dust attenuation} \label{sec:phys_dust}

To interpret emission lines, we must first constrain the impact of dust reddening on the spectrum.
We therefore consider the Balmer decrement. In our stacked spectrum,  the H$\delta$/H$\gamma$ ratio 
($0.66^{+0.10}_{-0.10}$) provides our best constraint on the Balmer decrement, with both line fluxes measured from the same G395M grating.
While the observed Balmer decrement is slightly elevated above the Case B value at an electron temperature of $10^4$ K (0.55; \citealt{Osterbrock2006}), the measurement is consistent with the intrinsic value within just over 1$\sigma$. 
This suggests negligible dust attenuation, consistent with the blue UV slope ($\beta=-2.4$) reported in \citet{Bunker2023} using the prism spectrum.
This is further supported by the 
H$\alpha$ flux measurement from MIRI.
If we take the H$\delta$ flux and assume case B recombination without dust, we derive a H$\alpha$ flux of $5.1\pm0.6\times10^{-18}$~erg~s$^{-1}$~cm$^{-2}$. This is consistent with that measured from MIRI within $1.6\sigma$ ($6.8\pm0.9\times10^{-18}$~erg~s$^{-1}$~cm$^{-2}$; \citealt{Alvarez-Marquez2025}).
As we will show in \S~\ref{sec:phy_modeling}, the absence of dust attenuation is also supported by our joint photoionization modeling of the rest-optical emission lines with BEAGLE.

\subsection{Ly$\alpha$  Escape Fraction} \label{sec:phys_lya}

The escape fraction of Ly$\alpha$ ($f_{\rm esc, Ly\alpha}$) can be computed using dust-corrected Balmer line fluxes. 
We use the measured H$\gamma$ emission line flux assuming no dust correction, consistent with our measurement of the Balmer decrement.
Under the assumption of Case B recombination (assuming $T_e=1.3\times10^4$~K and $n_e=2.5\times10^4$~cm$^{-3}$ as derived in Section~\ref{sec:phys_density} and \ref{sec:phys_TeOH}), we expect an intrinsic Ly$\alpha$ luminosity that is $69\times$ stronger than H$\gamma$. 
The observed ratio (\lya/H$\gamma=1.87_{-0.20}^{+0.19}$) implies that GN-z11 has a very low Ly$\alpha$ escape fraction ($f_{\rm esc, Ly\alpha}=0.027_{-0.003}^{+0.003}$). We note that 
the Ly$\alpha$ escape fraction would still be small $0.033_{-0.003}^{+0.004}$ if we instead assume the intrinsic Ly$\alpha$/H$\gamma$ ratio (56) found in more typical \ion{H}{2} region conditions ($T_e=10^4$~K and $n_e=10^3$~cm$^{-3}$).
Our value is broadly consistent with that reported by \cite{Bunker2023} ($0.038\pm0.004$); the small differences reflect differences in the assumed gas conditions and a \lya{} flux measurement 1$\sigma$
lower than theirs.

While we expect that some fraction of the intrinsic \lya{} emission may be extended beyond the aperture of the NIRSpec shutter ($0.2^{\prime\prime} \times 0.46^{\prime\prime}$), we note that even if we adopt the integrated Ly$\alpha$ flux as measured from the NIRSpec IFU observations of GN-z11 (Central+NE as reported in Table 1 of \citealt{Scholtz2024}), the resulting Ly$\alpha$ escape fraction is still low ($f_{\rm esc, Ly\alpha}=0.081_{-0.013}^{+0.014}$ assuming the temperature and density derived in Section~\ref{sec:phys_density} and \ref{sec:phys_TeOH}).

\subsection{Electron Density} \label{sec:phys_density}

\begin{figure*}
\centering
\includegraphics[width=\linewidth]{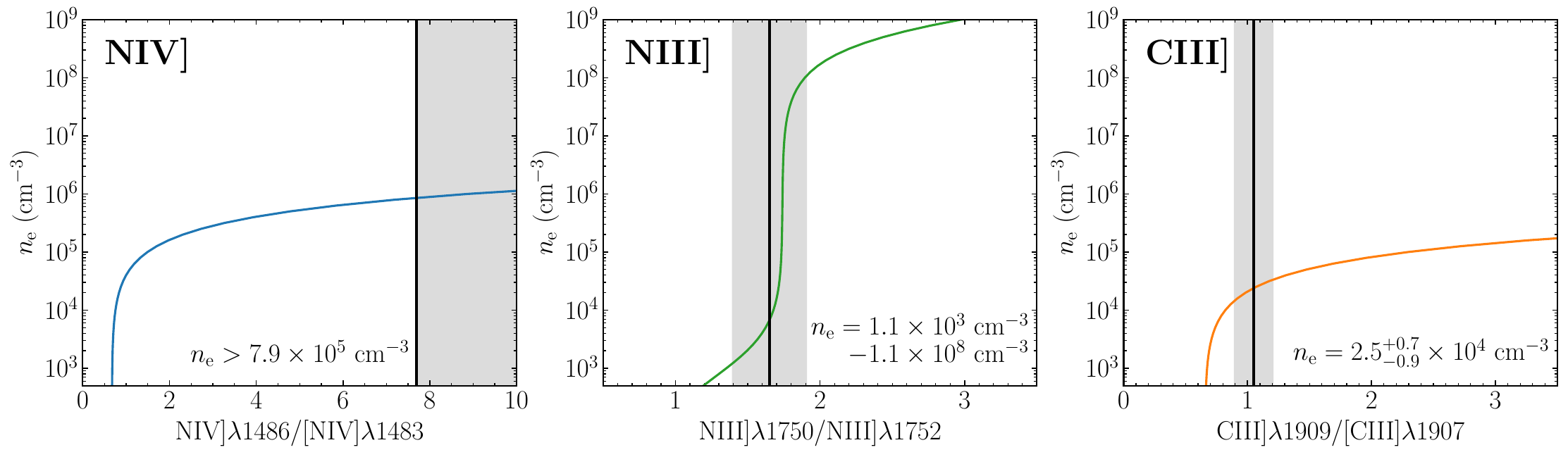}
\caption{Electron densities of GN-z11 derived from density-sensitive line ratios \ion{N}{4}]~$\lambda1486$/[\ion{N}{4}]~$\lambda1483$ (left; narrow line only), \ion{N}{3}]~$\lambda1750$/\ion{N}{3}]~$\lambda1752$ (middle), and \ion{C}{3}]~$\lambda1909$/[\ion{C}{3}]~$\lambda1907$ (right). The colored solid curve in each panel shows the relation between each line ratio and electron density assuming $T=1.3\times10^4$~K (see \S~\ref{sec:phys_TeOH}), which is calculated using \texttt{PyNeb}. The black solid vertical line and grey shaded region present the observed line ratio and $1\sigma$ confidence interval. For \ion{N}{4}] density, we give the $3\sigma$ lower limit of the narrow \ion{N}{4}]~$\lambda1486$/[\ion{N}{4}]~$\lambda1483$ ratio.}
\label{fig:density}
\end{figure*}

We compute the electron densities by jointly fitting the density-sensitive ratios using the \texttt{Python} package \texttt{PyNeb} \citep{Luridiana2015}.
We first investigate the density implied by the \ion{C}{3}] emission, which has an ionization potential of 24.4\,eV and thus constrains the density of gas traced by other ions of similar ionization state, such as [\ion{O}{3}].
In the right panel of Figure~\ref{fig:density}, we plot the ratio--density relation calculated using \texttt{PyNeb}.
We measure a \ion{C}{3}]$\lambda1909$/[\ion{C}{3}]$\lambda1907$ flux ratio of $1.0^{+0.2}_{-0.2}$, which corresponds to an electron density of $2.5^{+0.7}_{-0.9}\times10^4$\,cm$^{-3}$ for the \ion{C}{3}]-emitting gas.
This \ion{C}{3}]-based electron density we infer appears to be lower than those measured in other strong \ion{N}{4}] emitters (e.g., \citealt{Topping2024, Topping2025_demographic}), but it remains $\sim1.8\times$ larger compared to what is typically inferred in \ion{C}{3}] emitting galaxies (e.g., \citealt{Christensen2012,James2014,Maseda2017,Berg2021,Mingozzi2022,Mainali2023,Topping2025b}).
Similar values have also been seen in those with very large H$\beta$ EW, as would be expected if densities are higher in the strongest bursts of star formation \citep{Topping2025_demographic}.

The \ion{N}{4}]$\lambda\lambda$1483,1486 doublet (ionization potential 47.4\,eV) probes gas of considerably higher ionization than \ion{C}{3}], providing a constraint on the density of the nitrogen-emitting gas.
As discussed in \S~3.2, we detect both a broad and a narrow component in \ion{N}{4}]$\lambda1486$, and we constrain their densities separately.
As shown in the left panel of Figure~\ref{fig:density}, the narrow \ion{N}{4}]$\lambda1486$/[\ion{N}{4}]$\lambda1483$ flux ratio ($>7.7$) implies an extremely large electron density of $>7.9\times10^5$\,cm$^{-3}$ ($3\sigma$ lower limit) for the narrow-line-emitting gas.
The broad component likewise indicates a very high density, with its flux ratio ($>3.0$) implying $>2.5\times10^5$\,cm$^{-3}$ for the broad-line-emitting gas.
Both values are lower limits because, even with data $6\times$ deeper than the previous analysis of GN-z11, the blue component of the doublet, [\ion{N}{4}]$\lambda1483$, remains challenging to detect unambiguously.
These densities are more than an order of magnitude above the value inferred from \ion{C}{3}], and also exceed those found in other known \ion{N}{4}]-emitting galaxies at high redshift ($z>5$; \citealt{Ji2024,Marques-Chaves2024, Topping2024, Topping2025_demographic}).
Both the narrow and broad \ion{N}{4}] lines therefore arise in far denser gas than \ion{C}{3}], either because they occupy the densest, innermost zones near the ionizing sources, or because \ion{N}{4}] traces a physically distinct component of the ionized gas.

The final density diagnostic we consider is the \ion{N}{3}] multiplet, which traces gas of intermediate ionization (29.6\,eV) between that probed by \ion{C}{3}] and \ion{N}{4}].
Previous analysis of GN-z11, based on the shallower JADES spectrum, inferred an extremely high density ($>10^9$\,cm$^{-3}$) from the \ion{N}{3}] ratio, as expected for a broad-line region (BLR; \citealt{Maiolino2024}).
We revisit this possibility using the improved constraints on the \ion{N}{3}] ratios afforded by the stacked SPURS+JADES spectrum.
We first consider \ion{N}{3}]~$\lambda1750$/\ion{N}{3}]~$\lambda1752$, shown in the middle panel of Figure~\ref{fig:density}.
The observed flux ratio ($1.6^{+0.3}_{-0.3}$) falls close to a region where the density varies rapidly with line ratio, so that its $1\sigma$ range maps onto a broad range of densities ($n_e = 1.1\times10^3\mbox{--}1.1\times10^8$\,cm$^{-3}$).
This ratio nonetheless rules out at $>3\sigma$ the most extreme densities ($\sim10^9$\,cm$^{-3}$) expected for a BLR.
We also examine the \ion{N}{3}]~$\lambda1754$/\ion{N}{3}]$_{\rm total}$ ratio adopted by \cite{Maiolino2024}, for which we measure $0.11^{+0.02}_{-0.02}$.
As this ratio is double-branched in the line-ratio--density relation, our measured value corresponds to either a low density of $\simeq10^3$\,cm$^{-3}$ or a very high density of $\simeq10^8$\,cm$^{-3}$.
Given the high densities inferred from the \ion{C}{3}] and \ion{N}{4}] ratios, the \ion{N}{3}]-emitting gas is likely also dense ($\simeq10^8$\,cm$^{-3}$), yet still an order of magnitude below BLR densities.

In summary, both the \ion{C}{3}] and \ion{N}{4}] line ratios indicate high densities for the ionized gas in GN-z11, as would be expected for galaxies undergoing intense bursts of star formation at high redshift \citep[e.g.,][]{Vanzella2023,Adamo2024,Topping2025_demographic}.
The nitrogen line ratios point to higher densities, indicating that the nitrogen-emitting gas traces a denser phase that is not necessarily the same component dominating the \ion{C}{3}] emission.
However, we do not find the extreme densities expected for BLRs are required to explain the measured line ratios.

\subsection{Electron Temperature and Gas-Phase Metallicity}\label{sec:phys_TeOH}

We detect two auroral lines in the stacked spectrum of GN-z11 (\ion{O}{3}]$\lambda\lambda$1661,1666, [\ion{O}{3}]$\lambda$4363). To derive the electron temperature and gas-phase oxygen abundances, we combine the auroral line measurements with rest-optical emission line fluxes ([O~{\small III}]~$\lambda5007$) measured from MIRI reported in \citet{Alvarez-Marquez2025}. 
The MIRI [O III] and H$\alpha$ fluxes were extracted from spatial apertures with radii of $0''.25$ and $0''.3$, respectively, corresponding to $1.5\times$ the FWHM of the MIRI MRS channel 1 ([O III]) and channel 2 (H$\alpha$) PSF. Additionally, 
\citet{Alvarez-Marquez2025}  performed an aperture correction to account for lost flux, with a computed aperture-correction factor of $1.64\times$.
Because GN-z11 is spatially unresolved in both NIRCam imaging and the MIRI MRS IFU spectra, and the absolute flux calibration of MIRI and NIRCam agrees to within 5\%\footnote{\url{https://jwst-docs.stsci.edu/jwst-calibration-status/}.} \citep{Law2025}, the aperture-corrected MIRI fluxes and our NIRSpec fluxes, which we matched to the NIRCam photometry, share a common flux scale.
Following \citet{Alvarez-Marquez2025}, we therefore adopt these fluxes directly in the analysis below, without any additional correction between instruments.

We first consider the electron temperature. 
Based on the O~{\small III}]~$\lambda1666$/[O~{\small III}]~$\lambda5007$ ratio, we derive a temperature of $1.3^{+0.1}_{-0.1}\times10^4$~K (assuming the \ion{C}{3}]-based density and zero dust correction).
If we use [\ion{O}{3}]4363/[\ion{O}{3}]5007 ratio, we find a similar temperature of $1.4^{+0.1}_{-0.1}\times10^4$~K.
Because \ion{O}{3}]1666 has a higher SNR, we will use the temperature derived from the \ion{O}{3}]1666/[\ion{O}{3}]5007 flux ratio below.

Using the electron temperatures and densities, we compute the gas-phase oxygen abundance of GN-z11 with the \texttt{PyNeb} code. 
We follow the methodology presented in \citet{Izotov2006} and calculate the oxygen abundance as O/H $\simeq{\rm O}^{2+}/{\rm H}+{\rm O}^+/{\rm H}$.
The abundance of doubly ionized oxygen (O$^{2+}$/H) is inferred from the [O~{\small III}]~$\lambda5007$/H$\beta$ flux ratio. We use the [O~{\small III}]~$\lambda5007$ measured in \citet{Alvarez-Marquez2025} and H$\beta$ flux derived from H$\gamma$ flux assuming case B recombination (H$\gamma$/H$\beta=0.47$). 
For the O$^+$ zone, the temperature and density are likely different from that of the O$^{2+}$ zone. 
Since we do not detect [O~{\small II}]~$\lambda\lambda7320,7330$ auroral lines, we derive the O$^+$ zone temperature following \citet{Campbell1986}: $T({\rm O}^+)=0.7\times T({\rm O}^{2+})+3000$~K. 
For the O$^+$ zone density, since we are not able to deblend the [O~{\small II}]~$\lambda\lambda3727,3729$ doublet due to the resolution of the spectra and directly infer the density, we assume a typical value of the low-ionization zone ($n_{\rm e}\simeq300$~cm$^{-3}$) derived in literature \citep[e.g.,][]{Sanders2024,Topping2025b}. 
The singly ionized oxygen abundances (O$^+$/H) are inferred from the [O~{\small II}]/H$\beta$ flux ratios based on the above O$^+$ zone densities and temperatures. 

Following this approach, we compute a gas-phase oxygen abundance of $12+\log{\rm (O/H)}=7.94^{+0.14}_{-0.14}$. 
This corresponds to a moderately low metallicity of $0.18^{+0.07}_{-0.05}\ Z_{\odot}$ (assuming the solar oxygen abundance is $12+\log{\rm (O/H)}=8.69$; \citealt{Asplund2021}). 
This updated measurement is also consistent with the previous works based on the earlier JADES spectrum (and the same MIRI fluxes) \citep{Bunker2023,Alvarez-Marquez2025}, suggesting that GN-z11 is somewhat chemically enriched.

\subsection{N/O Ratio}

The N/O ratio of GN-z11 was computed using the JADES discovery spectrum. Here we update the measurement of N/O for our deeper spectrum, while considering the impact of the broad \ion{N}{4}] component.
We first compute the (N$^{3+}$+N$^{2+}$)/O$^{2+}$ abundance ratio with \texttt{PyNeb} using the narrow (N~{\small IV}]+N~{\small III}])/O~{\small III}] flux ratio based on the derived electron densities and temperatures. 
To convert (N$^{3+}$+N$^{2+}$)/O$^{2+}$ into N/O, we apply an ionization correction factor (ICF), which can be computed as a function of gas-phase metallicity and the narrow N~{\small IV}]/N~{\small III}] flux ratio \citep{Martinez2025}. 
As the \ion{N}{4} doublet ratio implies a density of $>7.9\times10^5$~cm$^{-3}$, and N/O is insensitive to density up to $1\times10^9$~cm$^{-3}$, we will assume a density of $7.9\times10^5$~cm$^{-3}$ and an O$^{2+}$ temperature of $T=1.3\times10^4$~K. 
We confirm that the N/O does not change if we assume a higher density. 
Using the above methods, we derive a N/O ratio of $0.54^{+0.10}_{-0.08}$ for the narrow-line-emitting gas in GN-z11. 
Adopting the solar N/O ratio ($\log{\rm (N/O)_{\odot}}=-0.86$; \citealt{Asplund2021}), this corresponds to a super-solar N/O ratio of $3.9^{+0.8}_{-0.5}$~N/O$_{\odot}$. 
Compared to earlier measurements based on the JADES spectrum, our narrow-line N/O ratio is consistent with the lower limits reported in \citet{Bunker2023,Cameron2023} (N/O $>0.43-0.56$) in uncertainties, and is in agreement with the measurements of \cite{Senchyna2024}.
Owing to our ability to detect [O III] and resolve the broad component of N IV], we now show cleanly that the narrow \ion{N}{4}] line emitting gas has an N/O abundance enhanced relative to solar.

\subsection{Gas Properties from Photoionization Modeling}\label{sec:phy_modeling}

We perform photoionization modeling with BEAGLE to constrain the ionized gas properties and the bulk stellar population properties.
Our approach broadly follows that adopted in \cite{Chen2026}, where we fit the strong rest-frame optical emission line fluxes ([O~{\sc ii}] $\lambda3728$, [Ne~{\sc iii}] $\lambda3869$, H$\delta$, H$\gamma$, and [O~{\sc iii}] $\lambda4363$ from the NIRSpec spectra, and [O~{\sc iii}] $\lambda5007$ from MIRI) along with the pseudo-NIRCam spectral energy distribution (SED), which is constructed by integrating the full NIRSpec grating spectrum over the standard NIRCam filter set of F150W, F200W, F277W, F356W, F410M, and F444W.
We assume a constant star formation history, allowing the age to vary between 1~Myr and the age of the Universe at the source redshift with a log-uniform prior.
We keep the remaining assumptions and parameter priors identical to those described in \cite{Chen2026}, including the \cite{Chabrier2003} stellar initial mass function with mass limits of 0.1--300~$M_\odot$ and the \cite{Pei1992} SMC dust attenuation curve.

The BEAGLE \cite{Gutkin2016} models successfully reproduce the input rest-frame optical emission line fluxes and their corresponding ratios, with most line fluxes agreeing within $1\sigma$ uncertainties.
As expected from the large O32 and Ne3O2 values, the models prefer a very high ionization parameter of $\log U = -1.82_{-0.09}^{+0.08}$.
The observed Balmer decrements, along with constraints from the continuum slopes of the pseudo-photometry, suggest little to no dust attenuation, yielding a V-band optical depth of $\tau_{\rm V}=0.01_{-0.01}^{+0.01}$ for GN-z11.
To reproduce the full set of emission line ratios, particularly those involving the [O~{\sc iii}] $\lambda4363$ auroral line, the models require low metallicities with $Z_* = Z_{\rm ISM} = 0.15_{-0.02}^{+0.03}$~$Z_\odot$ in GN-z11.
At the fitted dust-to-metal ratio ($\xi_d = 0.18_{-0.05}^{+0.11}$), these ISM metallicities also translate to a low gas-phase oxygen abundance of $12 + \log(\mathrm{O/H}) = 7.92_{-0.09}^{+0.10}$, consistent with the value directly inferred from emission lines using the direct-$T_e$ method in \S~\ref{sec:phys_TeOH}.

\section{Stellar Continuum Modeling} \label{sec:VMS}

The deep rest-UV spectrum for GN-z11 provides us with the first opportunity to compare the latest stellar population models with observations at $z>10$ in detail, offering us a window into the massive star populations in this source. 
In particular, we seek to test whether the observed continuum together with the stellar wind profiles can be reproduced by the models that include modern treatments of very massive star atmospheres, and its implications for their presence in GN-z11.
We first introduce these stellar population models in \S~\ref{sec:VMS_models}, review the stellar wind features predicted by these models in \S~\ref{sec:VMS_expectation}, discuss the impact of the IGM damping wing on our \ion{N}{5} modeling in \S~\ref{sec:VMS_IGM}, and compare the models to the observed spectrum in \S~\ref{sec:VMS_comparison}.

\subsection{Stellar Population Models}\label{sec:VMS_models}

We make use of recent stellar population models presented in \cite{Martins2025} and updated versions of the \cite{Bruzual2003} (Charlot \& Bruzual, in preparation) models, as described in several recent studies \citep{Gutkin2016,Vidal-Garcia2017}.  The models  incorporate developments in the modeling of massive-star populations, including updates to evolutionary tracks, mass-loss prescriptions, and atmosphere modeling.  Below we briefly summarize the stellar population models, and then we describe the addition of nebular continuum emission.

The CB models used here are described in \cite{Plat2019} (hereafter referred to as CB19), updating those presented in \cite{Gutkin2016,Vidal-Garcia2017}, and the reader is directed to these studies for a complete description. 
The CB19 models use updated evolutionary tracks for very massive stars (VMS) up to $300~M_\odot$, computed with the PARSEC code \citep{Chen2015}.
The models adopt a revised stellar mass-loss prescription that accounts for enhanced mass loss at higher stellar Eddington factors \citep{Vink2011}.
Compared to the previous mass-loss treatments (e.g., \citealt{Vink2001}) used by older stellar population models, this recipe produces stronger, more optically thick winds at the highest stellar luminosities, even at low metallicity.
These models use updated spectra for hot massive stars calculated with the WM-basic code \citep{Pauldrach1986} from \cite{Chen2015}.
Emission from Wolf-Rayet (WR) stars is included using the line-blanketed, non-LTE, spherically expanding models from the Potsdam Wolf-Rayet (PoWR) group, spanning stellar effective temperatures of $10^{4.4}\mbox{--}10^{5.3}$~K and metallicities $Z = 0.001\mbox{--}0.014$.
These updates have been shown to provide better agreement with the observed strength of stellar wind lines, particularly broad He~II, in low-metallicity $z\sim0$ galaxies \citep[e.g.,][]{Senchyna2021,Wofford2023}.
We generate a grid of simple stellar population (SSP) CB19 models using the BEAGLE tool \citep{Chevallard2016} to create SSP spectra as a function of age and metallicity, assuming the \cite{Chabrier2003} IMF with the upper mass cutoff of 300~$M_\odot$.

We also use the \cite{Martins2025} stellar population models (hereafter M25), which also adopt updated prescriptions for the evolution and spectroscopic appearance of massive stars.
Here, spectra for VMS with masses up to $300~M_\odot$ are derived from evolutionary tracks computed with the STAREVOL code, using mass-loss recipes thought to be appropriate at these masses \citep{Grafener2021}.
Compared to the \cite{Vink2011} prescription used by the CB19 models, the \citet{Grafener2021} recipe has a steeper scaling of mass-loss rate with Eddington factor, producing stronger winds at fixed stellar mass and metallicity (e.g., 0.3 dex higher for a $250~M_\odot$ track at $0.4~Z_\odot$; \citealt{Martins2022}).
The models consider two treatments of the metallicity dependence of the optically thick winds, motivated by differing results from the still-limited observations of individual massive stars below LMC metallicity.
The first one assumes the metallicity scaling of the mass loss rate for optically thick winds which decreases towards lower stellar metallicity ($\log \dot{M}_{\rm thick} \propto Z$) as found for classical WR stars \citep{Hainich2015}, while the second one adopts no such scaling, where the mass loss rate of optically thick winds $\dot{M}_{\rm thick}$ is held fixed at its value at $Z_{\rm LMC}$ so the winds remain strong down to very low metallicity \citep[e.g.,][]{Smith2023}. We will refer to these models as M25$_{\rm{\dot{M}\propto Z}}$ and M25$_{\rm{\dot{M},LMC}}$, respectively.

In both M25 cases, the atmosphere models and synthetic spectra are computed self-consistently with CMFGEN \citep{Hillier1998}, using the evolutionary models' surface properties at ages of $0$--$2.5$~Myr in steps of $0.5$~Myr, for initial masses of $150$--$300~M_\odot$ in steps of $50~M_\odot$. The SSP spectra from \citet{Martins2025} are synthesized by combining the VMS spectra described above (corresponding to stars with mass $>100~M_\odot$) with those of normal stars below $100~M_\odot$ from the BPASS single star models \citep{Eldridge2017}, normalized to the initial mass function with an upper initial mass function slope of $-2.35$ above $0.5~M_\odot$.
Since the \citet{Martins2025} models are available only at three metallicities ($0.01$, $0.1$, and $0.2~Z_\odot$ \footnote{We note that the \citet{Martins2025} models assume a solar metallicity of $Z_\odot = 0.013446$, which is slightly different from the value ($Z_\odot = 0.01524$) adopted in C\&B models and BEAGLE. As the difference is small, and in order to avoid interpolation when possible, we chose to use the M25 models directly following their own solar metallicity definition.}), we interpolate between these grid points in logarithmic space to obtain models at intermediate metallicities.

To compare with the observed GN-z11 spectrum, we must also add  the nebular continuum, which can contribute significantly to the total continuum in the far-UV at the youngest stellar population ages (e.g., \citealt{Topping2022,Katz2025}). 
Specifically, we compute the nebular continuum with PyNeb \citep{Luridiana2015} from the ionizing photon production rate of the stellar emission, assuming $n_{\rm H} = 2.5\times10^4$~cm$^{-3}$ and $T_e = 1.3\times10^4$~K, consistent with what we measured from the spectrum (see \S~\ref{sec:phys_density} and \ref{sec:phys_TeOH}). 
We then add this component to each CB19 and M25 template.
In addition to these models, we also use the models without very massive stars to investigate the impact of the IMF.
For CB19 models, we will compare to the version synthesized with BEAGLE assuming an IMF upper mass cutoff of 100~$M_\odot$.
For M25 models, we will directly use the BPASS models for single stars with the same upper mass cutoff of 100~$M_\odot$, assuming the same initial mass function slope of $-2.35$ above $0.5~M_\odot$ as used in constructing the M25 population synthesis model.
We use these SSP models to explore the expected stellar wind features as a function of age and metallicity, and to compare with the observed UV continuum.

\subsection{Model Expectations for Stellar Wind Lines}\label{sec:VMS_expectation}

\begin{figure*}
    \centering
    \includegraphics[width=\linewidth]{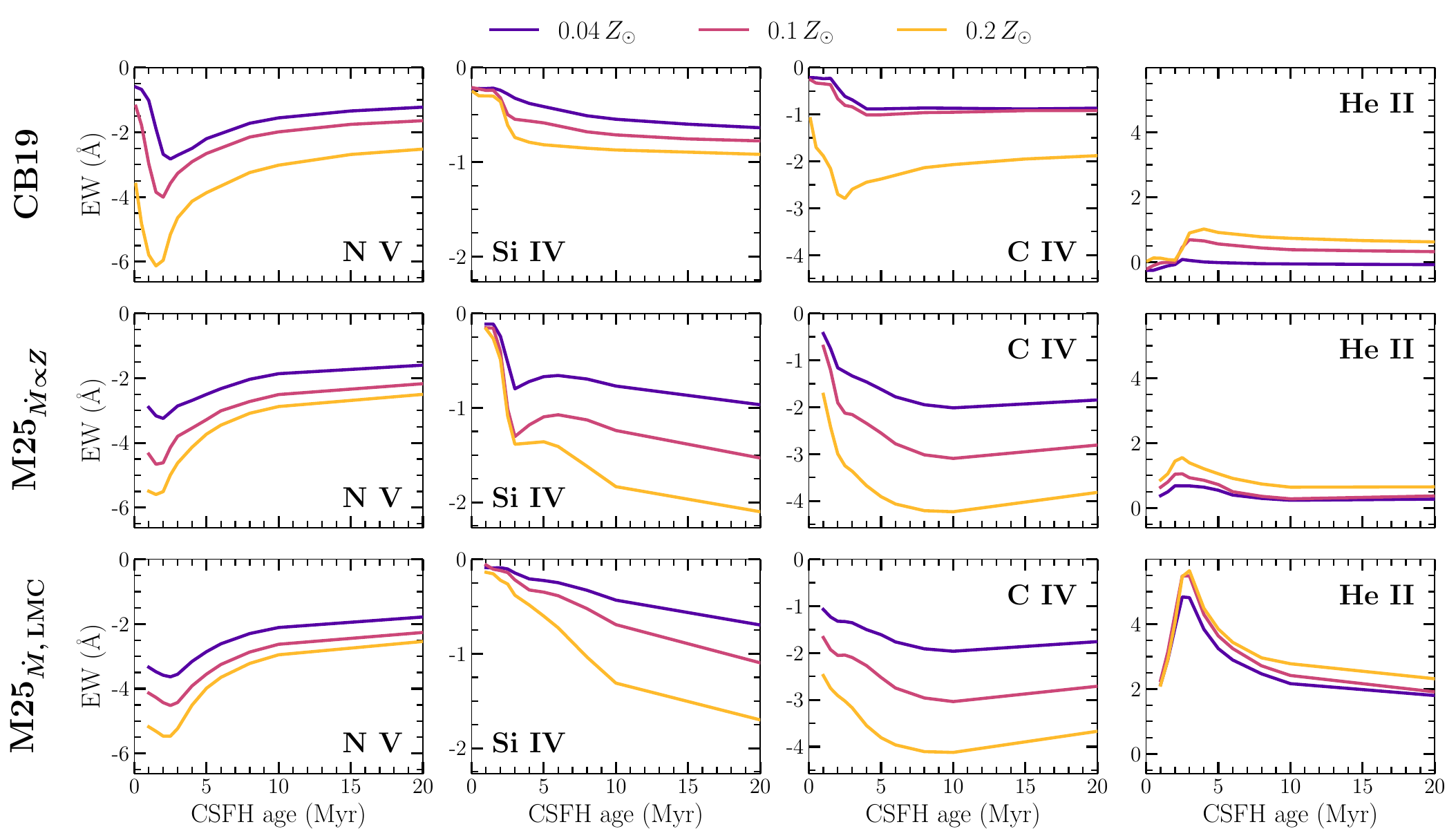}
    \caption{
    Strengths of the N~V, Si~IV, and C~IV P-Cygni absorption features, and the broad stellar He~II emission feature, as a function of stellar population age assuming a constant star formation history (CSFH).
    We show three sets of models: the CB19 stellar population models (top), the \protect\cite{Martins2025} models with metallicity scaling of optically thick mass-loss rates (\ModelMZ{}, middle), and those without such scaling (\ModelMLMC{}, bottom).
    Different colors show three stellar metallicities relevant to this work, $Z=0.04$, $0.1$, and $0.2~Z_\odot$, noting that $Z=0.04$ is the metallicity we would expect for GN-z11 from its gas-phase metallicity assuming Type~II SNe yields ($\simeq0.2$ (O/Fe)$_\odot$).
}
\label{fig:wind_model}
\end{figure*}

To guide understanding of our spectral fitting results (\S~\ref{sec:VMS_comparison}), we first quantify the predicted stellar wind features from the models described in the previous subsection.
We explore the strength of these stellar wind features (in terms of the EW) as a function of stellar population age  and metallicity. We assume a constant star formation history, but note that the true star formation histories inferred will be a combination of different age SSPs. This investigation will help understand how the free parameters in our fit (i.e., metallicity, age, IMF) can be adjusted to reproduce the observed spectrum. 

Our choice of model stellar metallicity  (sensitive primarily to the iron abundance in the O/B photospheres) is informed by the oxygen-based gas-phase metallicity (0.18 Z$_\odot$) derived in \S~\ref{sec:phys_TeOH}. Given the delay in iron production (dominated by Type Ia supernovae) relative to oxygen (dominated by Type II supernovae), we expect high-redshift galaxies to be $\alpha$/Fe-enhanced. Indeed, among $z\simeq2-3$ galaxies, stellar metallicities are typically found to be a factor of 5 lower than gas-phase metallicities \citep[e.g.,][]{Steidel2016,Sanders2020,Topping2020,Runco2021}. This value approaches that expected from Type II supernova yields alone ($\simeq$0.2 (O/Fe)$_\odot$, \citealt{Nomoto2006}). As a result, we adopt models with $Z_*=0.04~Z_\odot$ as those most likely to represent the stellar metallicity of GN-z11. 
We additionally consider models with stellar metallicity $0.1~Z_\odot$ and $Z_*=0.2~Z_\odot$, but consider these unlikely for GN-z11.

We measure the strength of rest-UV stellar wind features in the models described in \S6.1, focusing here on those features covered  by our NIRSpec observations including N~{\sc v}, Si~{\sc iv}, and C~{\sc iv} P-Cygni profiles and the broad He~{\sc ii}~$\lambda1640$ emission.  We measure  equivalent widths by normalizing the model spectrum with a linear fit to the line-free continuum windows given below, then integrating the normalized profile over a velocity interval chosen to span the full blueshifted P-Cygni absorption trough or the broad He~II emission.
For both N~{\sc v}, and C~{\sc iv} P-Cygni features, we integrate the blueshifted absorption trough over $[-3000, 0]$~km~s$^{-1}$ relative to N~{\sc v}~$\lambda1238$ and C~{\sc iv}~$\lambda1548$, respectively, with continuum windows $[1260, 1320]$~\AA{} (N~{\sc v}) and $[1500, 1510]$ and $[1570, 1580]$~\AA{} (C~{\sc iv}).
For Si~{\sc iv} P-Cygni, we integrate the absorption of each component of the doublet over $[-2000, 500]$~km~s$^{-1}$ and sum the two, using continuum windows $[1340, 1360]$ and $[1400, 1450]$~\AA{}.
For the broad He~{\sc ii} emission, we integrate the full profile over $[-2000, 2000]$~km~s$^{-1}$, with continuum windows of $[1600, 1620]$ and $[1670, 1690]$~\AA{}.

We  summarize the range of EWs and their dependence on simple stellar population age and stellar metallicity for three sets of models in Figure~\ref{fig:wind_model}.
Because the N~V P-Cygni profile is limited to the hottest stars, we expect the feature to be a signpost of very young stellar populations \citep[e.g.,][]{Walborn1985,Leitherer1995,Leitherer2010,Crowther2016}.
Indeed, as shown in the  left panels of Figure~\ref{fig:wind_model}, the strongest N~V P-Cygni absorption is seen in stellar populations with CSFH ages of 2--3 Myr. The age dependence of \ion{N}{5} will appear even more pronounced with SSP models, given the rapid disappearance of the most massive stars from the stellar population. 
We also see that \ion{N}{5} wind line is mildly metallicity-sensitive, with the peak absorption EW increasing by a factor of 1.5--2.2 across the three model sets, from the lowest to the largest metallicity we consider (0.04--0.2~$Z_\odot$). 
The Si~IV and C~IV P-Cygni absorption are well-known to trace the stellar metallicity (e.g. \citealt{Rix2004,Leitherer2010,Leitherer2011,Vidal-Garcia2017,Chisholm2019}), with stronger winds as metallicity increases, driving an increase in the median absorption EW by a factor of 2.0--2.3 for Si~IV and 2.4--2.5 for C~IV over the same metallicity range. 
As broad He~II emission traces the optically thick winds from the most massive O stars near the Eddington limit or from WR stars \citep[e.g.,][]{Schaerer1996,Crowther2007,Brinchmann2008,Senchyna2021}, the emission peaks over the first 3--4~Myr after star formation begins. For composite star formation histories, the age dependence can be more complicated, and indeed, we see broad He II remaining visible at slightly older CSFH ages in Figure~\ref{fig:wind_model}.
As has been found in many works \citep[e.g.,][]{Crowther2016,Wofford2023,Upadhyaya2024}, the strength of these features depends on the initial mass function, and including VMS as we consider here leads to stronger P-Cygni profiles and broad He~II emission relative to models without them.
Comparing the M25 and CB19 models, we find differences in the strength of these features at fixed metallicity and age, with the M25 models showing $\approx2\times$ stronger Si~IV and C~IV P-Cygni absorption.
We also see differences between the two M25 models, with \ModelMLMC{} reaching significantly larger ($\approx3\times$) He~II EW than \ModelMZ{}, a result expected from the enhanced optically thick winds at low metallicity in that model.
This variation emphasizes how sensitive the wind lines are to the specific  physics in the models, much of which remains poorly constrained.

\subsection{Impact of IGM Damping Wing on \ion{N}{5}} \label{sec:VMS_IGM}

\begin{figure}
    \centering
    \includegraphics[width=\linewidth]{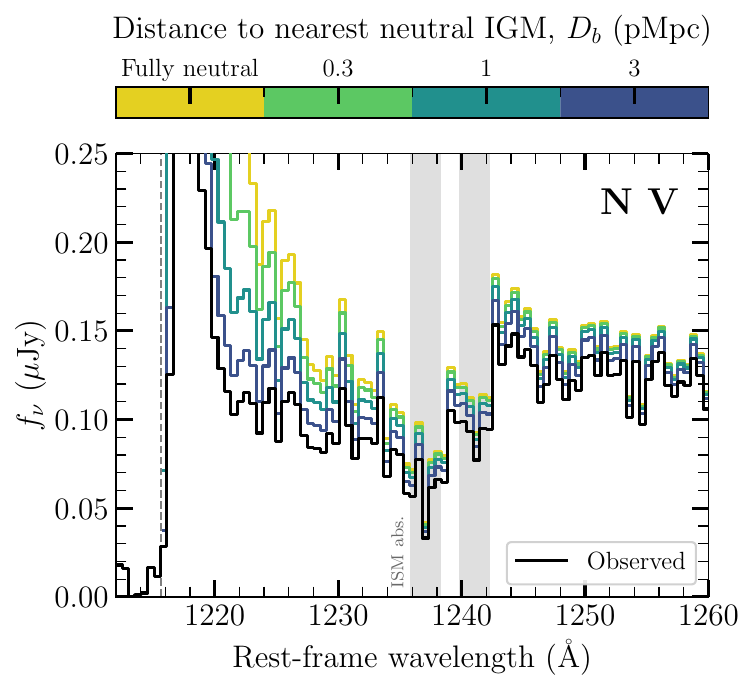}
    \caption{Impact of IGM damping-wing absorption on the N~V stellar-wind profile.
    We show the spectrum corrected for the IGM transmission expected across a range of distances, $D_b$, to the nearest neutral IGM.
    In the case where the IGM is fully ionized, the corrected spectrum is identical to the observed one.
}
\label{fig:igm_nv}
\end{figure}

At the redshift of GN-z11, where the intergalactic medium (IGM) is expected to be largely neutral, strong damping wing absorption may suppress the continuum flux at wavelengths redward of Ly$\alpha$, including across the N~V stellar wind region \citep[e.g.,][]{MiraldaEscude1998}. 
Therefore, to interpret the observed N~V P-Cygni profile and compare it with models, we must account for the damping wing. 
To do so, we first investigate how the intrinsic spectrum would appear considering a range of sizes for the ionized bubble in which GN-z11 resides.

We derive the expected transmission as a function of observed wavelength by integrating the damping wing component of the Ly$\alpha$ optical depth arising from neutral hydrogen along the line of sight, adopting the approach of \cite{Mason2020_bubble,Mason2026}. 
We assume the galaxy sits within an ionized bubble at a distance $D_b$ to the nearest neutral IGM patch, beyond which the IGM is assumed to be fully neutral \citep[which is a reasonable approximation in the early stages of reionization, e.g.,][]{Mesinger2008}.
We integrate the damping wing optical depth from the edge of the ionized region to the redshift at the end of reionization. We adopt a reionization end redshift of $z_{\rm reion} = 5.5$, but note this has little impact on the resulting transmission, since the optical depth is dominated by neutral hydrogen within $\lesssim100$\,cMpc of the galaxy.

In Figure~\ref{fig:igm_nv} we present the spectrum corrected for IGM transmission considering a range of distances from GN-z11 to the nearest neutral IGM, $D_b$.
As expected, the intrinsic N~V P-Cygni profile can deviate from the observed spectrum significantly if GN-z11 resides in very small ionized regions.
In the most extreme case corresponding to a fully neutral IGM ($D_b = 0$~pMpc), the expected transmission at rest-frame 1238~\AA{} (wavelength of the blue component of N~V doublet) is only 81\%.
If we correct the observed spectrum by this transmission, the corresponding intrinsic N~V P-Cygni absorption becomes much weaker, with the resulting absorption EW $\approx 2\times$ lower than what is directly measured from the observed spectrum.
When $D_b$ is small, the broad N~V redshifted emission also becomes stronger in the corrected (intrinsic) spectrum.
The IGM impact on N~V decreases for increasing distances to the neutral IGM, with the corrected spectrum approaching the observed spectrum when $D_b \geq 3$~pMpc.
We note that if $D_b$ is small, the intrinsic flux between Ly$\alpha$ and N~V also increases. This could reflect a larger continuum flux, and potentially contribution from broad Ly$\alpha$ emission -- beyond what is seen in the observed spectrum. In the next subsection we discuss the implications for the Ly$\alpha$ profile given our best-fit continuum models.
While a full, joint fit of the damping wing and stellar populations is left to future work, combining the N~V emission with the other stellar-wind features and the strength of Ly$\alpha$ emission should enable a more precise separation of the IGM attenuation and the intrinsic strength of N~V absorption.

We consider the range of bubble sizes GN-z11 likely resides in, and the resulting implications for our UV continuum modeling, based on recent theoretical work and observational constraints on the IGM at $z>9$. 
The sizes of ionized bubbles depend on the IGM neutral fraction, such that galaxies at the earliest stage of reionization are expected to reside in small ionized bubbles \citep[e.g.,][]{McQuinn2007}.
{\it JWST} observations have demonstrated strong Ly$\alpha$ emission becomes very rare with increasing redshift at $z>8$, indicating decreasing ionized bubble sizes and that the IGM is likely highly neutral at $z=10.6$. Recent estimates based on Ly$\alpha$ visibility suggest the IGM is $\approx90\pm10\%$ neutral at $z>10$ \citep{Tang2024a,Kageura2025}.
For such high neutral fractions, reionization simulations predict small median bubble radii of $0.1-0.3$\,pMpc \citep{Lu2024,Neyer2024}.
UV-bright galaxies, such as GN-z11, are expected to trace overdense regions at these redshifts, which may ionize early and host larger bubbles.
We therefore consider the models of \citet{Lu2024} and \citet{Neyer2024} which estimate bubble sizes around $M_\mathrm{UV} < -21$ galaxies, accounting for their environments. Both models predict UV-bright galaxies sit at a median distance $D_b\approx0.3$\,pMpc from the neutral IGM, with the \citet{Lu2024} models finding a 68\% range of $0.2-0.5$\,pMpc.
This is also consistent with our Ly$\alpha$ measurement, where the small EW and low escape fraction ($f_{\rm esc, Ly\alpha}=0.027_{-0.003}^{+0.003}$) suggest a large bubble is unlikely: assuming $D_b = 0.2-0.5$\,pMpc, the IGM transmission at the Ly$\alpha$ wavelength is 0.3--7\%.
However, for very small bubbles with $D_b<0.1$~pMpc, we find the IGM transmission correction factor at the observed Ly$\alpha$ peak wavelength will be very large, leading to an unphysically high value of the Ly$\alpha$ escape fraction above unity.
Therefore, in our modeling, we will adopt $D_b=0.5$~pMpc as the most likely distance to the neutral IGM, but we will also discuss the impact if the bubble is smaller (down to 0.1~pMpc) or larger.

We also consider damped Ly$\alpha$ absorption (DLA) from high-column-density neutral hydrogen in the vicinity of the stars, finding a DLA is unlikely to significantly affect the continuum near N~{\sc v} of GN-z11.
The weak low-ionization absorption lines (\S~4) imply either the total HI column density in front of the UV continuum is low, or, if present, a very metal-poor DLA.
Low-to-moderate HI column densities have little effect on the continuum near N~{\sc v}, because the DLA optical depth declines more steeply with wavelength offset from Ly$\alpha$ than the IGM damping wing ($\tau_{\lambda, \rm DLA} \propto 1/(\Delta \lambda)^{2}$ versus $\tau_{\lambda, \rm IGM} \propto 1/\Delta \lambda$, \citealt{McQuinn2008}): a $N_{\rm HI} = 10^{20.5}(10^{21.0})$~cm$^{-2}$ DLA still transmits 97\%(92\%) of the flux at rest-frame 1238~\AA{} (wavelength of the blue component of the N~V doublet).
While larger column densities ($N_{\rm HI} \gtrsim 10^{22}$~cm$^{-2}$) could suppress the \ion{N}{5} flux, we find this would imply an unphysical intrinsic spectrum: producing both an unrealistically large continuum flux between Ly$\alpha$ and N~{\sc v} and an extremely high N~{\sc v} emission flux.
Together, this suggests a strong DLA is unlikely in GN-z11 and that the continuum near N~{\sc v} is mainly modulated by the stellar P-Cygni profile and the IGM damping wing, which we focus our modeling on below.

\subsection{Fitting the GN-z11 Rest-UV Spectrum}\label{sec:VMS_comparison}

\begin{figure*}
    \centering
    \includegraphics[width=\linewidth]{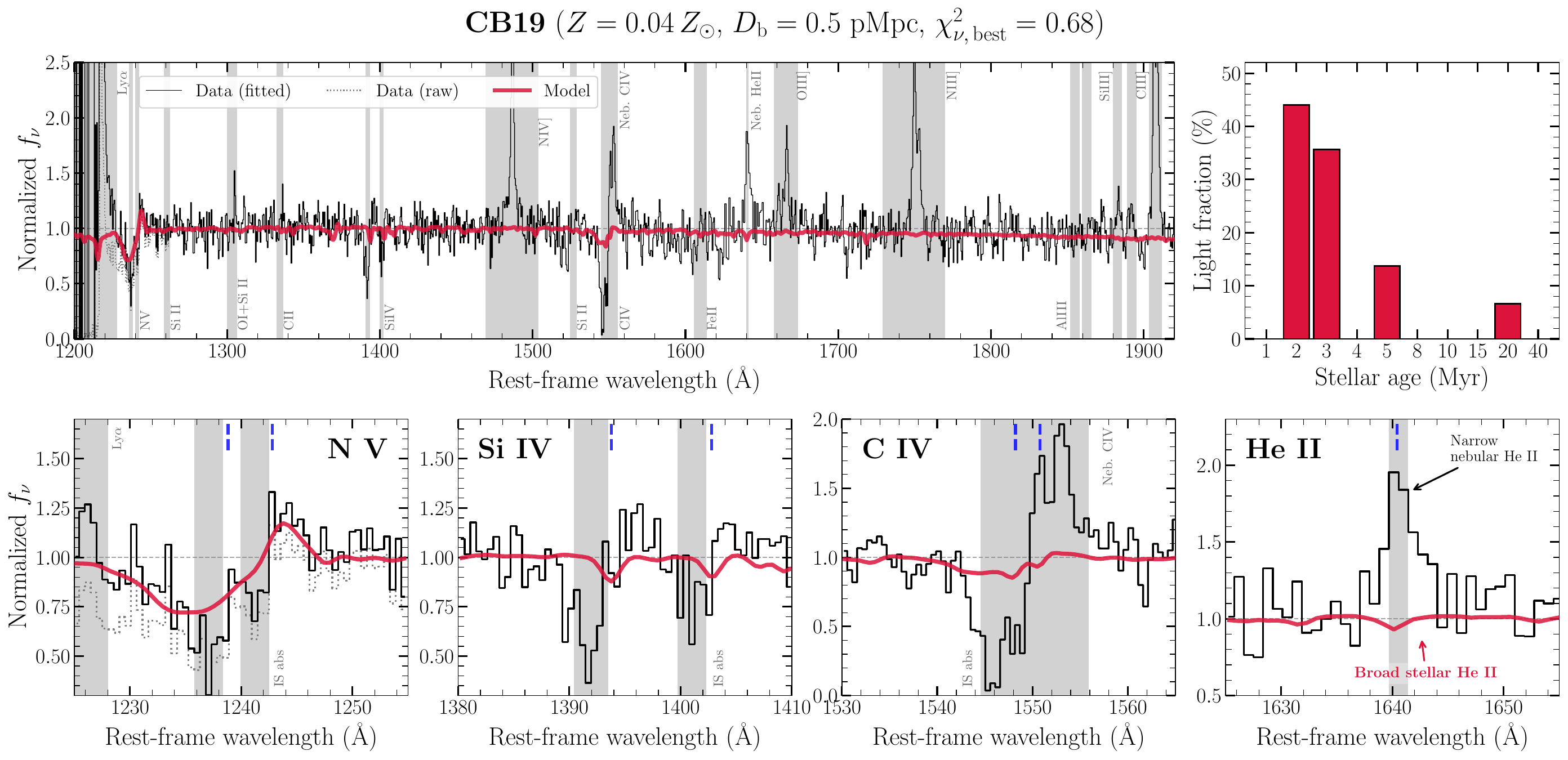}
    \caption{FiCUS fit to the UV continuum of GN-z11 using linear combinations of CB19 SSP models with an IMF upper mass cutoff of 300~$M_\odot$. 
    Shown are the fits to the observed spectrum corrected for IGM transmission, assuming a distance to the nearest neutral HI of 0.5 pMpc, and our preferred low metallicity of $Z_*=0.04~Z_\odot$.
    \textbf{Top left}: The full UV continuum used for fit (black), the original observed spectrum before applying the correction for IGM transmission (dotted blue), and the best-fit model (red). Nebular emission lines and interstellar absorption lines are masked during the fit, which are shaded in gray.
    \textbf{Top right}: The star formation history corresponding to the best-fit model, parameterized as the light fractions of the UV continuum at 1360~\AA{} as a function of SSP age.
    \textbf{Bottom}: Zoomed-in views of the FiCUS fits to the N~V, Si~IV, and C~IV P-Cygni profiles, and broad stellar He~II emission.
    }
    \label{fig:ficus_model_CB19}
\end{figure*}

\begin{figure*}
    \centering
    \includegraphics[width=\linewidth]{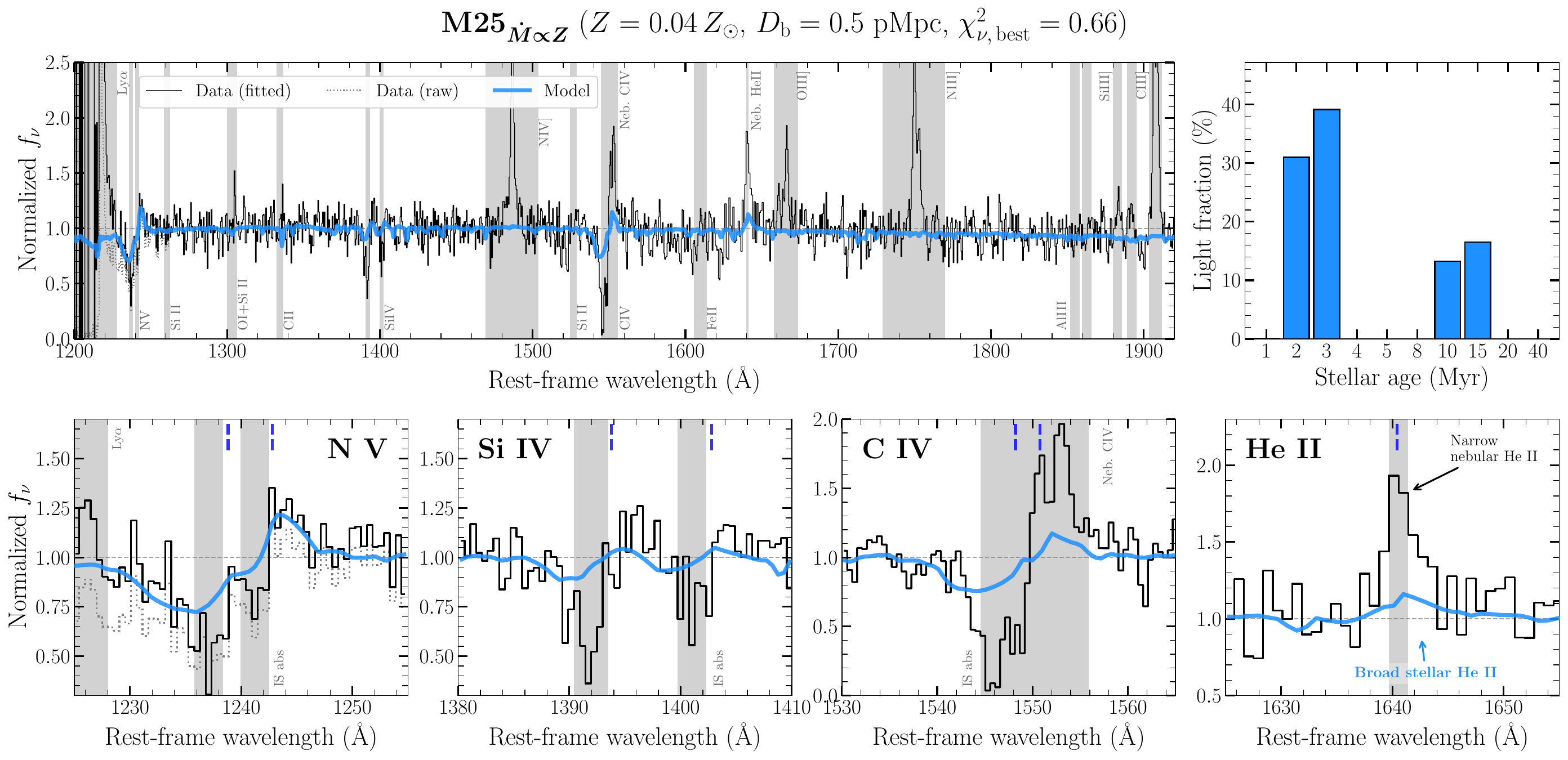}
    \caption{FiCUS fit to the UV continuum of GN-z11 using linear combinations of M25$_{\rm{\dot{M}\propto Z}}$ SSP models. 
    Shown are the fits to the observed spectrum corrected for IGM transmission, assuming a distance to the nearest neutral HI of 0.5 pMpc, and at a low metallicity of $Z*=0.04~Z_\odot$.
    \textbf{Top left}: The full UV continuum used for fit (black), the original observed spectrum before applying the correction for IGM transmission (dotted blue), and the best-fit model (red). Nebular emission lines and interstellar absorption lines are masked during the fit, which are shaded in gray.
    \textbf{Top right}: The star formation history corresponding to the best-fit model, parameterized as the light fractions of the UV continuum  at 1360~\AA{} as a function of SSP age.
    \textbf{Bottom}: Zoomed-in views of the FiCUS fits to the N~V, Si~IV, and C~IV P-Cygni profiles, and broad stellar He~II emission.
    }
    \label{fig:ficus_model_M25}
\end{figure*}

Using the models described above, we now  fit the observed UV continuum spectrum of GN-z11, building on our previous work with the $z=9.3$ galaxy SPURS-A2744-7 \citep{Chen2026}. In this study, we use the FiCUS code \citep{Saldana-Lopez2023}, modeling the UV continuum as a linear combination of simple stellar populations \citep{Chisholm2019}.  We consider several stellar population models (see \S~6.1) and fit the data assuming a range of stellar metallicities (0.04--0.2 Z$_\odot$; see \S6.2), bubble sizes (see \S~6.3), and two upper mass IMF cutoffs  (100 and 300 M$_\odot$). In each case, we fit over an SSP age grid of 1, 2, 3, 4, 5, 8, 10, 15, 20, and 40~Myr. Our goals are twofold. First, we seek to determine whether the current generation of stellar population models are capable of reproducing the UV spectrum of GN-z11. Second, we want to know what the model fits suggest about the properties of the massive star populations in GN-z11, in particular whether the wind lines are best reproduced with the inclusion of VMS as implemented in the CB19 and M25 models.

We fit the rest-frame continuum from 1228 to 1920~\AA{}, where the blue cutoff is chosen to avoid contamination from the red wing of Ly$\alpha$.
We mask nebular emission lines using windows matched to the width of each line.
Specifically, we mask $-3500$ to 3500~km~s$^{-1}$ around the \ion{N}{4}] doublet and \ion{N}{3}] multiplet, $-200$ to 200~km~s$^{-1}$ at the core of He II line that may have a significant contribution from nebular emission, and $-500$ to 500~km~s$^{-1}$ for the remaining lines (including the nebular \ion{C}{4}).
We likewise mask the interstellar absorption, using a $-500$ to $500$~km~s$^{-1}$ window around each absorption centroid identified in Section~4.
We account for dust attenuation, assuming the SMC attenuation curve \citep{Gordon2003}.
FiCUS finds the best-fit model by $\chi^2$ minimization with the \textsc{Lmfit} package \citep{Newville2016}, and estimates uncertainties with a Monte Carlo approach that perturbs the spectrum according to the error array and refits the perturbed spectrum 100 times, adopting the standard deviation of the resulting parameter distributions.

We first consider the results from fits to the GN-z11 spectrum with  CB19  and  M25$_{\rm{\dot{M}\propto Z}}$  models that assume our  preferred low stellar metallicity ($0.04~Z_\odot$, \S 6.2), a small bubble size (0.5~pMpc, \S6.3) and an IMF that extends up to 300 M$_\odot$. We define these as our fiducial model set. 
As shown in Figure~\ref{fig:ficus_model_CB19} and \ref{fig:ficus_model_M25}, we find that both of these fiducial models provide satisfactory fits to the spectrum ($\chi^2_\nu=0.68$ and 0.66, respectively). As expected, the strong \ion{N}{5} P-Cygni profile in GN-z11 forces the models to a very young stellar population. 
In particular, we find that best-fitting SFH is dominated by two SSPs at 2 and 3 Myr in both the CB19 and the M25$_{\rm{\dot{M}\propto Z}}$ fits. 
At such young ages, the \ion{N}{5} wind absorption EW is strong ($-2.5$~\AA\ in CB19 and $-2.2$~\AA\ in M25$_{\rm{\dot{M}\propto Z}}$), providing a good match to the data. Importantly both models also match the \ion{N}{5} P-Cygni emission component. At the low stellar metallicity of these models, the Si IV and \ion{C}{4} P-Cygni profiles are weak but not negligible. We find that the \ion{C}{4} absorption EW in our best fitting models is larger in the M25$_{\rm{\dot{M}\propto Z}}$ ($-1.8$~\AA) than CB19 ($-1.1$~\AA) models. Both stellar wind profiles are consistent with the observations, easily accommodated within the deep interstellar absorption profile. We also find that the broad He II emission is different between the two models: the  M25$_{\rm{\dot{M}\propto Z}}$ best fit has a modest stellar He II EW ($0.7$~\AA) while that from the fiducial CB19 fit is negligible. The observed He II profile is formally consistent with both models since the blend of stellar and nebular emission at $R=1000$ results in ambiguity to the true stellar He II EW. Overall, these fiducial fits confirm that two different VMS models are able to reproduce the GN-z11  continuum and wind features  with a physically-motivated stellar metallicity and bubble size. 
Finally, we note that the \ModelMLMC{} models perform similarly as \ModelMZ{} in reproducing the GN-z11 spectrum (Figure~\ref{fig:ficus_model_M25LMC}). The key difference is that the broad He~II emission component is stronger (EW $= 2.2$\,\AA) than in the \ModelMZ{} models (EW $= 0.7$\,\AA), as expected from the enhanced optically thick winds in the \ModelMLMC{} models at low metallicity. We note that our current measurement of He~II is consistent with both values.

We now explore a wider range of stellar metallicities, with the goal of understanding how our choice of metallicity impacts our ability to reproduce the spectrum.  We first consider CB19 models with a stellar metallicity of $0.1~Z_\odot$ and $0.2~Z_\odot$, 3-5 times larger than our fiducial model. We find that we are able to reproduce the spectrum at these metallicities with a goodness of fit (both with $\chi^2_\nu=0.67$) comparable to our fiducial model. The key change of the larger stellar metallicity is stronger wind lines (see Figure~\ref{fig:ficus_model_CB_Z0p1} and \ref{fig:ficus_model_CB_Z0p2}). 
To reproduce the spectrum, the best-fitting model shifts slightly in its mixture of SSPs, with a single dominant age now at 3 Myr. 
At these higher metallicities, the C\,{\sc iv} absorption EW ranges from comparable to enhanced relative to the fiducial value ($-0.8$~\AA\ at $0.1\,Z_\odot$ and $-1.8$~\AA\ at $0.2\,Z_\odot$) given the additional age dependence, while stellar He~{\sc ii} EW ($1.5$~\AA\ at $0.1\,Z_\odot$ and $2.0$~\AA\ at $0.2\,Z_\odot$) is larger than in the fiducial models described above. Both remain consistent with the observed profiles.
We find similar results when we fit the M25$_{\rm{\dot{M}\propto Z}}$  models at larger metallicities. 
We conclude that the VMS models at stellar metallicity approaching that of the ionized gas are still consistent with the current GN-z11 spectrum.

We now explicitly consider how the fits are impacted by the assumed bubble size. 
While we expect small bubbles to be the norm at $z\simeq 10.6$ (see \S6.3), it is possible that GN-z11 may be situated in an atypically large ionized region of the universe.
As the assumed distance to the neutral IGM increases, the damping wing correction we apply is reduced, resulting in a stronger observed \ion{N}{5} absorption profile. For example, to fit the spectrum with $D_b$=2 pMpc (instead of our fiducial 0.5 pMpc), we find a slight increase in the fraction of light from young SSPs in both CB19 and M25$_{\rm{\dot{M}\propto Z}}$, specifically at ages $\lesssim2$~Myr for CB19 and $\lesssim3$~Myr for M25$_{\rm{\dot{M}\propto Z}}$.
And moving to bubble sizes smaller than our fiducial value shifts the age of the population toward slightly older SSPs. As a limiting case, we consider a highly neutral IGM with a very small ionized bubble ($D_b$=0.1 pMpc) around GN-z11. For both models, the stellar population shifts very slightly to older populations to fit the weaker observed \ion{N}{5} absorption, but the best-fitting models are unable to reproduce the \ion{N}{5} emission profile at a stellar metallicity of $0.04~Z_\odot$. 
Increasing the stellar metallicity (0.1--0.2~$Z_\odot$) helps to match \ion{N}{5} emission at $D_b$=0.1 pMpc, by shifting the stellar population to slightly older ages (now dominated by the SSP at 3 Myr), though we consider these metallicities are unlikely.
Additionally, after correcting for the IGM damping wing, we recover excess flux in the data relative to the best-fit model from Ly$\alpha$ up to $\approx1228$~\AA{}.
This excess flux is suggestive of very broad Ly$\alpha$ emission.
For both the CB19 and \ModelMZ{} models considered (assuming $0.04~Z_\odot$ and $D_b=0.5$\,pMpc), we measure a total EW of 0.8--1.0~\AA{} for this flux at $v>2000$~\kms{}, increasing to 1.3--1.5~\AA{} in the limiting case of a smaller bubble ($D_b=0.1$~pMpc).

We also investigate the role of the upper end of the IMF on the quality of our fit to the spectrum of GN-z11. The models shown above assume an IMF including VMS extending up to 300 M$_\odot$. We now consider both CB19 and M25$_{\rm{\dot{M}\propto Z}}$ models with IMFs that only include stars up to 100 M$_\odot$. In the case of the latter, the non-VMS models are the BPASS single star models with an initial mass function slope of $-2.35$ above $0.5~M_\odot$ as used in constructing the M25 population synthesis models (see \S~\ref{sec:VMS_models}; \citealt{Eldridge2017}).
Without VMS in the models, the stellar wind lines are generally weaker at fixed age and metallicity, and hence the metallicity, ages, and bubble sizes must be adjusted to fit the spectrum. For our fiducial stellar metallicity ($0.04~Z_\odot$) and bubble size (0.5~pMpc), the fits to the stellar wind lines  are considerably worse without VMS for both CB19 and M25$_{\rm{\dot{M}\propto Z}}$, with the \ion{N}{5} emission component significantly underpredicted in both models (see Figure~\ref{fig:ficus_model_CBMup100} and \ref{fig:ficus_model_BPASSMup100}). The non-VMS models with stellar metallicity approaching that of the gas-phase still poorly fit the \ion{N}{5} profile assuming our preferred small bubble size (0.5~pMpc). Only with very large bubble sizes (D$_b>2$ pMpc), which we consider unlikely, can the wind lines be reproduced with the non-VMS models we consider.  

To summarize, we find that the rest-UV continuum and stellar wind lines in the SPURS GN-z11 spectrum can be simultaneously reproduced with stellar population models that have low stellar metallicities ($0.04~Z_\odot$) and small bubble sizes (0.5~pMpc) that we consider most likely. We find that the inclusion of VMS is very important and that the models we consider without stars more massive than 100 M$_\odot$ struggle to fit the observed spectrum without very large  bubble sizes and large stellar metallicities that we deem unlikely. However, we note that even among VMS models, there is a  range of possible wind strengths that are accommodated by our data. Higher spectral resolution observations ($R=2700$) should help distinguish between these scenarios, which in turn should better isolate the physics of the massive stars present in early nitrogen-enriched galaxy populations. Additionally, it is possible that the VMS  included in current models may not appropriately capture the wind properties of the stars that form at low metallicity in dense environments like GN-z11. In the following section, we will consider whether the broad NIV] emission we have detected in GN-z11 may be a spectroscopic signature of such a population 
without clear precedent in population synthesis models or lower-redshift galaxy samples.

\section{Discussion} \label{sec:discussion}

The  depth of the SPURS spectrum provides new insight into two of the most surprising properties of GN-z11: its detectable Ly$\alpha$ emission at $z>10$ and its extreme nitrogen enhancement. In this section, we first discuss what the resolved Ly$\alpha$ profile and interstellar absorption lines reveal about the escape of Ly$\alpha$ photons through the ISM, CGM, and neutral IGM (\S~\ref{sec:discussion_lya}). We then turn to the physical origin of GN-z11's broad N IV] (\S~\ref{sec:discussion_NIV}) before concluding with a broader discussion about the role of massive stars and a possible AGN in shaping its properties (\S~\ref{sec:discussion_nature}).

\subsection{Ly$\alpha$ Escape in $z>10$ Galaxies}\label{sec:discussion_lya}

\begin{figure*}[t]
\centering
\includegraphics[width=\linewidth]{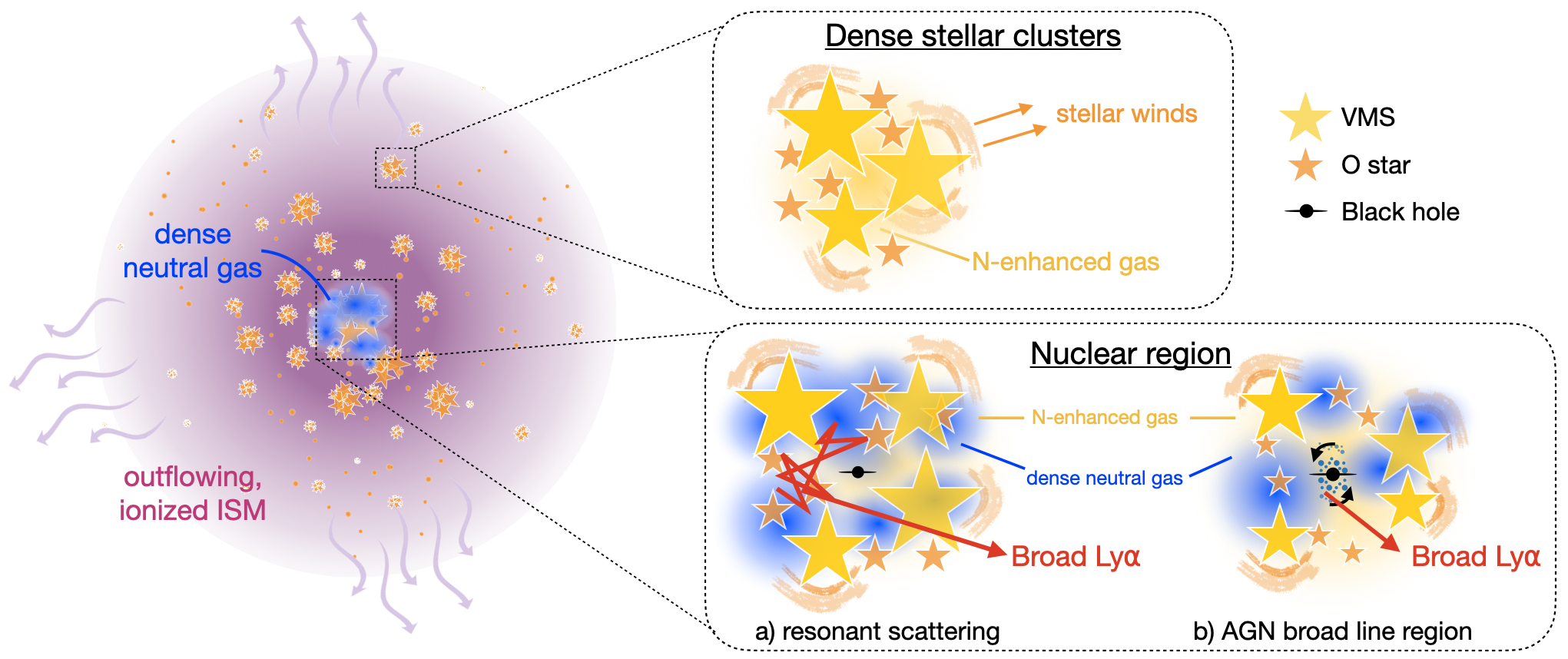}
\caption{Schematic illustrating a potential geometry of GN-z11 that reconciles the observed Ly$\alpha$ emission tail at $>500$\,km~s$^{-1}$ and OI* fluorescent emission with the weak low-ionization absorption lines (\S~\ref{sec:discussion_lya}). 
We propose GN-z11 is composed of multiple dense stellar clusters, hosting young, massive stars, potentially including VMS (\S~\ref{sec:discussion_nature}).
The weak LIS absorption implies the dense neutral gas that is optically thick to Ly$\alpha$ and Ly$\beta$ is concentrated close to ionizing sources in a small volume of GN-z11, potentially around the innermost stellar clusters.
In this nuclear region, Ly$\alpha$ may be broadened by
a) resonant scattering in neutral gas with $N_\mathrm{HI}\gtrsim10^{21}$\,cm$^{-2}$.
b) Keplerian motions in an AGN broad line region, if one is present.
Meanwhile, the majority of the UV continuum is produced in stellar clusters at larger radii that have a low covering fraction of neutral gas.
} 
\label{fig:Lya_cartoon}
\end{figure*}

GN-z11 is one of only two galaxies detected with Ly$\alpha$ emission at $z>10$ \citep{Bunker2023,Witstok2025}.
Prior to the launch of {\it JWST}, the IGM was expected to be almost fully neutral at $z>10$ \citep[e.g.,][]{Mason2018,Jung2020,Bolan2022}.
At such high neutral fractions, ionized regions in the intergalactic medium are predicted to be small \citep[$\lesssim 0.3$\,pMpc, e.g.,][]{McQuinn2007,Lu2024,Neyer2024}, strongly suppressing Ly$\alpha$ flux emerging from galaxies and making these detections surprising.
It has not been clear whether these sources signpost early large ionized bubbles
or reflect Ly$\alpha$ emission properties that facilitate transmission through the neutral IGM \citep[e.g.,][]{Bruton2023,Lu2024,Cain2025,Qin2025,Cohon2026}.

The SPURS spectrum provides new insights into what may boost the visibility of Ly$\alpha$ in GN-z11.
The high SNR Ly$\alpha$ detection reveals a broad red wing, with 44\% of the flux emerging at $>500$\,km\,s$^{-1}$ from line center.
This red wing likely facilitates the transmission of Ly$\alpha$ flux through the neutral IGM.
Due to the strong wavelength dependence of the Ly$\alpha$ damping wing optical depth from the neutral IGM, flux close to line center should be severely suppressed, while Ly$\alpha$ photons emerging from the galaxy at redder wavelengths experience much higher transmission \citep{Dijkstra2011,Mason2018a,Endsley2022,Yuan2025}.
Assuming typical distances of UV-bright galaxies to the neutral IGM at $z>10$ from reionization simulations \citep[$D_b \approx 0.2-0.5$\,pMpc,][see \S~\ref{sec:VMS_IGM}]{Lu2024,Neyer2024}, the IGM damping wing transmission at 500\,km\,s$^{-1}$ is 16--32\%, relative to just 0.3--7\% at Ly$\alpha$ line center.
Given GN-z11's minimal dust attenuation (\S~\ref{sec:phys_dust}) and strong UV and optical emission lines (\S~\ref{sec:results}), based on $z\sim5-6$ samples with similar properties, we would expect strong emergent Ly$\alpha$ emission\footnote{Here we define `emergent' as the emission emerging from the galaxy if it resided in the ionized IGM, excluding the IGM damping wing attenuation.} \citep[EW\,$\gtrsim 25$\,\AA, $f_{\rm esc,Ly\alpha}>0.2$, e.g.,][]{Tang2024}.
However, we observe a low Ly$\alpha$ EW ($5.6_{-0.2}^{+0.2}$\,\AA) and very low Ly$\alpha$ escape fraction ($f_{\rm esc,Ly\alpha}=0.027_{-0.003}^{+0.003}$; \S~\ref{sec:phys_lya}), indicating a large fraction of the emergent Ly$\alpha$ flux may be attenuated by the IGM.
This suggests that GN-z11 need not reside in a very large ($\gtrsim1$\,pMpc) ionized bubble, despite its potential location within an overdensity \citep{Tacchella2023}, but that its Ly$\alpha$ may be detectable in part due to a fraction of its flux emerging from the galaxy at $>500$\,\kms.

The presence of the red Ly$\alpha$ wing is, however, puzzling given the non-detection of low-ionization absorption lines in GN-z11 (\S~\ref{sec:ISMabs}).
Producing significant Ly$\alpha$ flux at $>$500\,km~s$^{-1}$ via resonant scattering requires HI column densities $N_\mathrm{HI}\gtrsim10^{21}$\,cm$^{-2}$ \citep[e.g.,][]{Neufeld1990,Verhamme2006}.
If gas with this high column density covers a large fraction of the UV continuum emitting regions we should also expect metal absorption lines, given these lines are typically optically thick in star-forming galaxies \citep[e.g.,][]{Jones2013,Reddy2016,Reddy2022}.
If instead the metal lines are optically thin, this would imply significantly lower metallicities than the gas-phase metallicity we measure from the emission lines (the \ion{O}{1}~$\lambda$1302 EW lower limit implies [O/H]\,$\lesssim-2.3$, or $12 + \log(O/H) \lesssim 6.39$, assuming a turbulent Doppler parameter of $b\gtrsim20$~\kms), and is therefore unlikely unless the dense neutral gas is far from the \ion{H}{2} regions.
A key challenge is therefore reconciling the evidence for dense neutral gas in GN-z11 with the weak low ionization absorption lines.

The detection of \ion{O}{1}$^*$$\lambda1304$ emission in GN-z11 (Section~\ref{sec:results}) offers an important new insight into where this dense gas is located.
We find the \ion{O}{1}$^*$ emission is most plausibly explained via fluorescent excitation of neutral oxygen atoms by Ly$\beta$, as suggested in AGN \citep{Kwan1981,Matsuoka2007} and LRDs \citep[e.g.,][]{Tripodi2025,Tang2026_spursQSO1,Ji2026_qso1}.
Producing a high \ion{O}{1}$^*$ flux requires both neutral gas which is very optically thick to Ly$\beta$ \citep[$N_\mathrm{HI} > 10^{21}$,][]{Matsuoka2007} -- consistent with the column densities required to produce the Ly$\alpha$ red wing -- and a high Ly$\beta$ flux \citep{Kwan1981}.
This suggests the presence of dense neutral gas very close to at least some of the ionizing sources in GN-z11.
 
One possibility is that only the nuclear star forming region is enshrouded in dense neutral gas, which we illustrate qualitatively in Figure~\ref{fig:Lya_cartoon}.
While we leave a detailed radiative transfer treatment to future work, Ly$\alpha$ photons produced in the nuclear region could resonantly scatter to $>500$\,km~s$^{-1}$ in the high column density neutral gas, and Ly$\beta$ photons could excite a large enough population of OI atoms to produce significant fluorescent emission.
If the majority of the far-UV emission arises from star forming regions exterior to the nuclear region, then the fraction of sightlines passing through the dense neutral gas would be low. 
This could explain the non-detection of low ionization absorption lines in the UV continuum, and also contribute to Ly$\alpha$ flux closer to line center.
A broad-line region (BLR) associated with an AGN in this region could also provide an explanation for the Ly$\alpha$ flux at $>500$\,km~s$^{-1}$. In a BLR Ly$\alpha$ can be broadened to large velocities due to the Keplerian motion of gas. We discuss the potential origins of this geometry and presence of an AGN in GN-z11 further below (\S~\ref{sec:discussion_nature}).

The deep GN-z11 spectrum thus provides critical context for understanding Ly$\alpha$ visibility at $z>10$, including the most distant known Ly$\alpha$ detection, GS-z13-1-LA at $z\simeq13$ \citep{Witstok2025}. GS-z13-1-LA is also expected to have a broad Ly$\alpha$ profile \citep[FWHM\,$\gtrsim600$\,km~s$^{-1}$,][]{Witstok2025,Witstok2026}, based on its detection in the low-resolution prism, but non-detection in current grating spectra.
Our results suggest that these detections of Ly$\alpha$ emission at $z>10$ are likely enhanced by a contribution from broad Ly$\alpha$.
We argue that the origin of this broadened Ly$\alpha$ emission is linked to the extreme nature of $z>10$ galaxies: a consequence of star formation and evolution in very high density environments, and the potential presence of AGN.
This has important implications for using Ly$\alpha$ to infer the onset of reionization.
If such extreme conditions are common in $z>10$ galaxies, the Ly$\alpha$ emerging from the ISM/CGM may be significantly different from that in the baseline $z\sim5-6$ samples used to infer IGM attenuation \citep[e.g.,][]{Tang2024,PrietoLyon2025}.
Accurately inferring the properties of the IGM at the highest redshifts therefore requires a better understanding of the nature of $z>10$ sources.
Progress will require both detailed spectroscopy of analogs in the mostly ionized IGM at $z<6$ \citep[see e.g.,][for an example]{Gagnon-Hartman2026}, and larger samples of $z>10$ sources with deep rest-UV spectroscopy to reveal their detailed gas conditions and obtain a census of Ly$\alpha$ at the highest redshifts.

\begin{figure*}
\centering
\includegraphics[width=\linewidth]{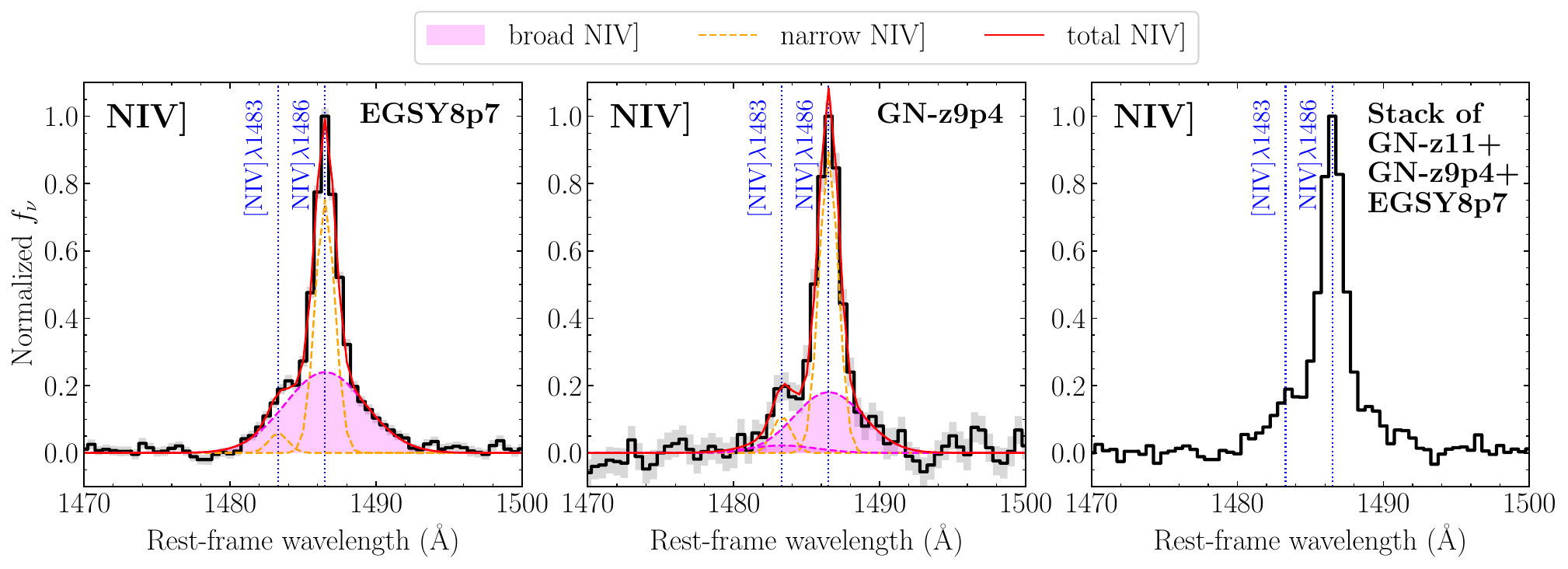}
\caption{\ion{N}{4}] emission line profiles of EGSY8p7 (left), GN-z9p4 (middle), and the stacked \ion{N}{4}] spectrum of GN-z11, EGSY8p7, and GN-z9p4 (right). We also show the best-fit models to the \ion{N}{4}] profiles of EGSY8p7 and GN-z9p4. Similarly to the \ion{N}{4}] profile of GN-z11, the \ion{N}{4}] lines of both EGSY8p7 and GN-z9p4 are best-fitted by narrow \ion{N}{4}] components (orange dashed lines) plus broad \ion{N}{4}] components (magenta shaded regions).}
\label{fig:NIV_comp}
\end{figure*}

\subsection{The Origin of Broad \ion{N}{4}] in GN-z11} \label{sec:discussion_NIV}

\begin{figure*}
    \includegraphics[width=\linewidth]{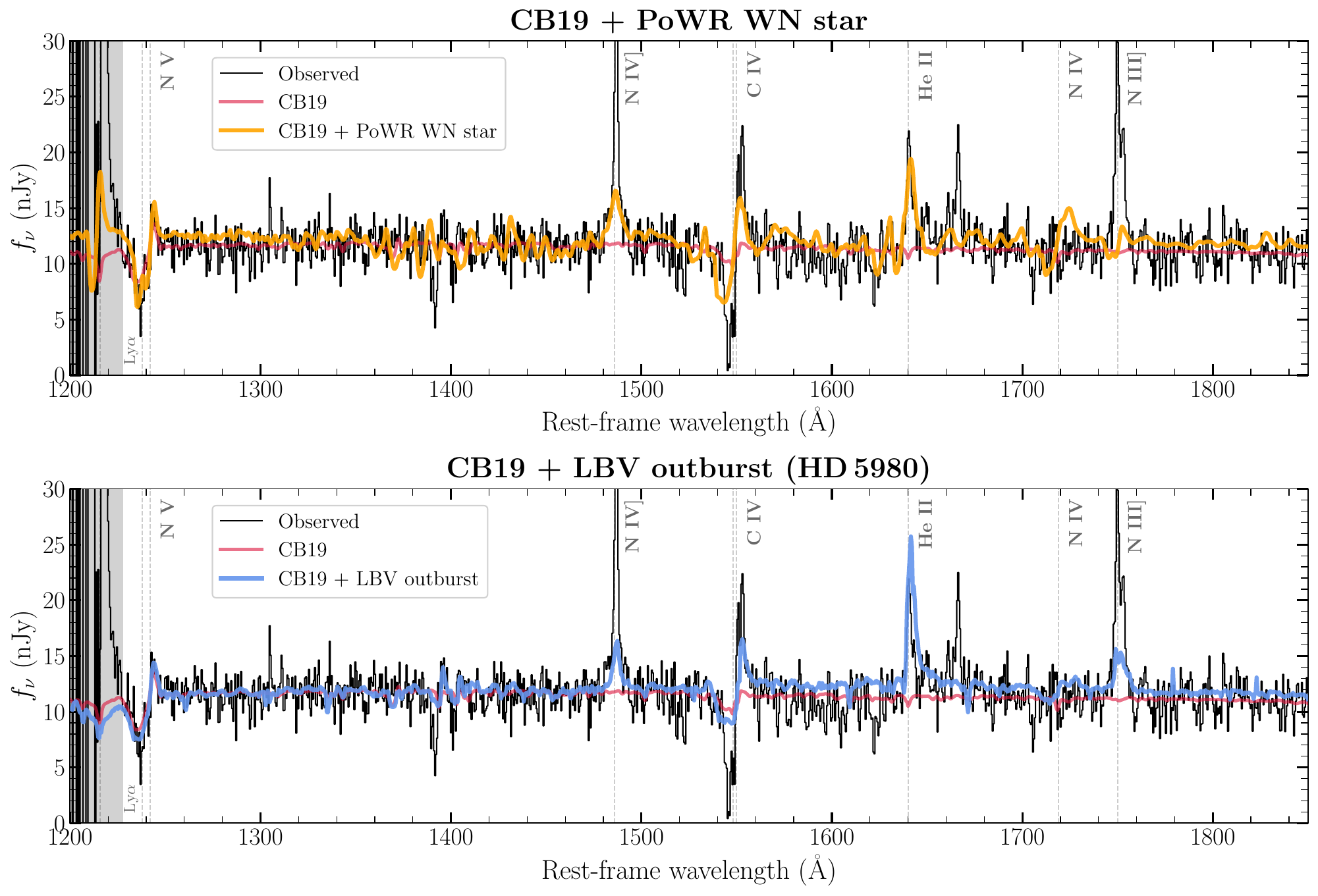}
    \caption{Similarities between the observed GN-z11 spectrum and models adding a WN atmosphere with a particularly prominent \ion{N}{4}]~$\lambda1486$ (top panel), or the outburst of the SMC LBV HD~5980 (bottom panel) to the UV continuum.
    In each panel, we show the GN-z11 spectrum after correcting for damping-wing transmission assuming an ionized bubble of $D_b = 0.5$~pMpc in black, and the fiducial CB19 stellar continuum model ($Z_*=0.04 Z_\odot$) from \S~\ref{sec:VMS_comparison} in red.
    The top panel also shows, in an orange line, a composite model adding to the CB19 model a WN atmosphere (a single atmosphere from the PoWR grids at 40~kK; see text) renormalized to contribute 35\% of the UV continuum.
    The bottom panel presents a second composite model, in a blue line, where we add the UV spectrum of HD~5980 in April~1995, an unusually hot LBV outburst from a massive WN star in a WN+WN+O triple in the SMC, renormalized to contribute 13\% of the UV continuum.
    The addition of either the WN atmosphere or the LBV outburst spectrum produces a prominent broad \ion{N}{4}]~$\lambda1486$ emission in agreement with the data, while the \ion{N}{4}~$\lambda1719$ emission remains weaker than \ion{N}{4}]~$\lambda1486$.
    }
    \label{fig:uv_wpowr_lbv}
\end{figure*}

The physics responsible for nitrogen-enhancements in a subset of early galaxies remains unclear. 
Some have argued the strong  winds of very massive stars \citep[e.g.,][]{Vink2023,Shi2026} or supermassive stars \citep[e.g.,][]{Charbonnel2023,Nagele2023,Gieles2025} may drive nitrogen-enhanced material into the ISM prior to supernovae, but other explanations exist \citep[e.g.,][]{Kobayashi2024,McClymont2026}.
Distinguishing between these pictures will require direct constraints on the stellar populations in nitrogen-enriched galaxies. A first step in this direction was provided by \cite{Berg2026},  with a deep rest-optical spectroscopy revealing strong winds in broad He~II $\lambda$4687 together with strong N~III $\lambda$4642 emission, attributed to a population of WN stars formed following a strong burst of star formation. 

The spectrum of GN-z11 provides another opportunity to study the massive star populations associated with nitrogen emitters. SPURS observations reveal a broad \ion{N}{4}]$\lambda$1486 component which may be related to the physics driving the nitrogen-enhancements. 
In addition to GN-z11, we identify broad \ion{N}{4}] emission in two other $z>8$ galaxies known to have strong narrow \ion{N}{4}] emission: EGSY8p7 ($M_{\rm UV}=-22.1$, $z=8.687$; \citealt{Larson2023,Isobe2023,Marques-Chaves2024,Topping2025_demographic,Whitler2026}), and GNz9p4 ($M_{\rm UV}=-21.0$, $z=9.380$; \citealt{Schaerer2024,Topping2025_demographic}).
In Figure~\ref{fig:NIV_comp}, we show their resolved \ion{N}{4}] profiles based on the SPURS data (Chen et al. in prep, Owens et al. in prep), along with the stacked \ion{N}{5} profile based on all three sources, highlighting their similarities. We find a broad \ion{N}{4}]~$\lambda1486$ component ($1265_{-69}^{+69}$~\kms{} in EGSY8p7 and $1078_{-236}^{+236}$~\kms{} in GN-z9p4). 

The discovery of this broad NIV] component in the first three nitrogen emitters with ultra-deep rest-UV spectroscopy  suggests  
its origin may be somewhat common in strong \ion{N}{4}]-emitting galaxies, motivating a closer investigation of the physics driving this line.
\ion{N}{4}]~$\lambda1486$ is routinely powered in broad Wolf-Rayet emission in dense stellar winds, most prominently in atmospheres of the WN sequence --- from very massive stars on the hydrogen-burning main sequence (and so with significant residual hydrogen; WNh) to hydrogen poor, fully stripped WNE stars.
Its strength is generally a complex function of wind density, ionization structure, and composition.
This line and the permitted \ion{N}{4}~$\lambda 1719$ are the two strongest stellar lines from triply-ionized nitrogen in the UV.
While $\lambda 1719$ is permitted, it is a transition between the excited $2p$ up to the $2p^2$ configuration; whereas while $\lambda 1486$ is semi-forbidden, it is connected to the $2s^2$ ground state of \ion{N}{4}.
Both lines are prominent in nitrogen-rich and moderately hot WR atmospheres, at effective temperatures ranging from $\sim 40$--$80$~kK (above and below this range, \ion{N}{5} and \ion{N}{3} become dominant, respectively; though the resonant \ion{N}{5}~$\lambda\lambda 1238,1242$ is prominent throughout this range as well).

For many WR atmospheres, \ion{N}{4}~$\lambda 1719$ is stronger than \ion{N}{4}]~$\lambda 1486$ \citep[e.g.][]{crowtherFundamentalParametersWolfRayet1995,crowtherFundamentalParametersWolfRayet1997,Hainich2014}.
However, this is not uniformly true, with the ratio between these lines varying substantially across different atmospheres.
In the PoWR WN grids for example, atmospheres at fixed $T_\mathrm{eff}$ and smaller transformed radii $R_t$ (corresponding in turn to some combination of denser, higher $\dot{M}$, and slower winds) tend to have larger $\lambda 1486$ relative to $\lambda 1719$ features (up to a limit above which both \ion{N}{4} features disappear); with $\lambda 1486$ actually becoming dominant for some models at $\sim 40$--$60$~kK and $R_t\sim 5$--$10\, R_\odot$.
To understand why this ratio varies with wind properties at fixed temperature, it is important to first understand how these lines are excited in detail.
The \ion{N}{4}]~$\lambda 1486$ line is directly connected to the ground state and is relatively low-lying, and its emission is predominantly powered by collisional excitation in typical WN atmospheres \citep[and thus also serves as an important coolant, e.g.][]{Hillier1988,Hillier1989,riverogonzalezNitrogenLineSpectroscopy2012a}.
It is typically formed at densities near or above its critical density of $10^{10.5}$~$\mathrm{cm^{-3}}$ within WN atmospheres (and near where emission in \ion{C}{4} originates).
In contrast, \ion{N}{4}~$\lambda 1719$ begins at an upper level significantly above ground ($\sim 23.4$~eV) and has a weak collisional coupling to the $2s^2\, ^1 S$ ground state; and thus while collisional excitation from other levels does likely contribute to and in many cases dominate the population of its upper level, other effects including continuum fluorescence and potentially dielectric recombination also likely contribute to its formation depending on temperature, density, and the radiation field \citep{Hillier1988}.
Crucially, these differences mean that the relative strength of these lines can be sensitive to differences in the radial density and temperature structure of flows in which they are formed.

As a simple experiment, we show in the top panel of Figure~\ref{fig:uv_wpowr_lbv} the effect of adding an additional WN atmosphere with particularly prominent \ion{N}{4}]~$\lambda 1486$ to the fiducial stellar population synthesis model presented in Section~\ref{sec:VMS_comparison}.
The additional WN contribution is a single atmosphere chosen from the PoWR model grids specifically as a visually close approximation of the observed line strengths at relatively low metallicity: an SMC WNe star at $T_\star = 40$~kK and $\dot{M}=10^{-4.5}$~$\mathrm{M_\odot/yr}$, with transformed radius $R_t=10~R_\odot$ near to the upper end of the wind strength distribution for the SMC grid \citep{grafenerLineblanketedModelAtmospheres2002,hamannGridsModelSpectra2004,sanderConsistentTreatmentQuasihydrostatic2015,todtPotsdamWolfRayetModel2015}.
Note that similar models can be found at higher metallicities and could likely be found at lower-$Z$ as well at sufficiently high mass loss rates and wind densities; we plan to more fully explore this parameter space with dedicated modeling in future work.
This is added to the CB19 composite fit by fitting simple scaling factors for the two models and a separate extinction correction for the WN atmosphere, and compared to the IGM-corrected restframe spectrum (assuming the fiducial 0.5~pMpc bubble).
The addition of this atmosphere at a $\sim 35$\% flux contribution in the UV introduces significant broad emission in \ion{N}{4}]~$\lambda 1486$ which matches the observed pedestal closely. 
This likewise significantly strengthens the model profiles of \ion{N}{5}, \ion{C}{4}, and \ion{He}{2}, bringing the emission components of each and the deepest absorption in the latter two into closer agreement with the data.
This is particularly striking for \ion{N}{5}, where the resolved components of the doublet in absorption are reasonably well approximated by the addition of this atmosphere.
Indeed, this wind (even for an essentially fully stripped atmosphere) does also produce some flux in Ly$\alpha$ with similar profile to the observed red wing; though at lower total flux than observed.
The \ion{N}{4}~$\lambda 1719$ in this model is the largest source of tension; while minimal, the predicted P-Cygni does still rise just above the noise in the continuum.

The reasonably good agreement between this very simplified model combination and the data is suggestive of an unusual stellar population.
This model 
suggests
that a strong contribution to the UV from a population of WR atmospheres with particularly dense and relatively low-temperature winds could potentially help
explain the broad \ion{N}{4}], the broad \ion{He}{2}, and the narrower P-Cygni components of \ion{N}{5}.
Such WN atmospheres are broadly expected as products of a variety of processes of stripping and spin-up (particularly in binaries or other interacting systems) which can expose a hydrogen or helium burning core, rich in nitrogen and luminous enough to drive a dense enriched wind.
However, this would require a substantial number of such stars.
The total luminosity of this added component is of-order $10^{10}$--$10^{11}$~$L_\odot$: thus requiring of-order $10^{4}$--$10^5$ `ordinary' ($\lesssim 10^6$~$L_\odot$) massive stars in this phase to explain.
In addition, the winds of these stars would need to be fairly self-similar to produce a composite spectrum with such strong \ion{N}{4}] and coherent low-velocity features in the other lines; as implied by the discussion of the PoWR grid above, much deviation from a fairly narrow temperature and mass loss range and the spectrum would change substantially, weakening \ion{N}{4}]~$\lambda 1486$ and/or strengthening the contribution of other lines.
It is possible that this is the product of catching a VMS population in a common phase, perhaps stripped by a very high rate of interactions in a dense cluster or spun-up to evolve in a chemically homogeneous manner, and that the winds such stars are capable of driving at low metallicity might be confined to relatively low velocities and high densities.
Especially at very low metallicities where strong stellar winds in general require stars close to the Eddington limit, we are increasingly likely to see the impact of the most massive stars (with the highest $L/M$) present in these lines; and models for the evolution and atmospheres of these stars at low metallicity with or without binary interaction remain highly uncertain \citep[e.g.][]{kohlerEvolutionRotatingVery2015,sanderNatureMassiveHelium2020,szecsiLowmetallicityMassiveSingle2025}.

Atmospheres like this model are not common in the Local Group; but interestingly, there is a singular object in the SMC with some remarkable spectral similarities worth drawing attention to.
HD~5980 is an (at least) triple system consisting of two emission-line early WN stars in a close orbit with an O-type tertiary; one of the most luminous systems in the SMC and situated on the outskirts of the standout young star forming region NGC~346 \citep{koenigsbergerHD5980Multiple2014,shenarWolfRayetStarsSmall2016}.
In 1994--1995 one of the WN components of this eclipsing system underwent a luminous outburst generally considered to be a Luminous Blue Variable (LBV) giant eruption, but unusual in both its high luminosity and its exceptionally high temperature \citep[$\sim$WN7--11 at visual maximum, 20-40~kK, much hotter than typical LBV temperatures in-outburst which rarely exceed $\sim 8000$~K; and with estimated luminosity $\sim 10^{6.5}$--$10^7$~$L_\odot$;][]{koenigsbergerRemarkableLongTermChanges1994,koenigsbergerUltravioletObservationsEruption1995,koenigsbergerIUEObservationsHD1997,moffatLuminousEclipsingSMC1998,drissenPhysicalParametersErupting2001,georgievWindStructureLuminosity2011,vinkEtaCarinaeLuminous2012,weisLuminousBlueVariables2020}.
The resulting UV spectrum near peak outburst taken by IUE (April 1995; IUE PID:WRRGK, PI:Koenigsberger) shows some striking similarities to the UV spectrum of GN-z11 (as also noted independently by \citealt{Nakane2026}): with very prominent \ion{N}{4}]~$\lambda 1486$ (and \ion{N}{3}]~$\lambda\lambda 1750$), weak $\lambda 1719$, and low-velocity absorption in \ion{N}{5}, \ion{C}{4}, and \ion{Si}{4} \citep{koenigsbergerUltravioletObservationsEruption1995,georgievWindStructureLuminosity2011}.
Following the same procedure as for the PoWR atmospheres, we add this spectrum, renormalized to contribute 13\% of the UV continuum, to the composite CB spectrum in the bottom panel of Fig.~\ref{fig:uv_wpowr_lbv} to highlight these similarities.
While its properties stand out in the pantheon of LBVs and the precise nature of its outburst cycle remains enigmatic, it is likely that the close binary plays a crucial role; potentially involved in the triggering of the outburst in the first place \citep[e.g.][]{gallagherCloseBinaryModels1989,smithEpisodicMassLoss2011,justhamLuminousBlueVariables2014}, and perhaps responsible for shaping the typically dense and cold outer layers of the outflowing ejecta and explaining its high $T_\mathrm{eff}$ \citep[e.g.][]{koenigsbergerIUEObservationsHD1997,moffatLuminousEclipsingSMC1998,koenigsberger40YearVariability2010}.
While our observations and understanding of LBVs at low metallicity and in close binaries are limited, HD~5980 could reasonably be considered a prototype for LBV-like near Eddington mass eruptions for multiple systems of very massive stars in dense and low metallicity environments, which like `classical' LBV events might be expected to be broadly self-similar \citep[interestingly, arguably the closest analogs to HD~5980 are two events in M33 at sub-LMC metallicity:][]{maryevaHistoryGoesCentury2019,smithNewUnusualLBVlike2020}.
It is suggestive then that spectral features approximated by this event might appear underlying the N-rich nebular spectrum in these high-redshift galaxies. A large number of such luminous eruptive mass loss events, if somehow triggered at very low-$Z$ (potentially aided or driven by interactions with close companions in hardened binaries), would naturally eject very large quantities of N-rich material from CNO cycle hydrogen burning into the ISM at low velocities and temperatures where it might rapidly be integrated into the nebular phase.

Another possibility is that rather than a population of stars with similar winds, we are observing a small number or even a single dense outflow against an extraordinarily luminous central object.
This would need to be an extremely compact system; a simple estimate of an effective radius from the 40~kK effective temperature and implied total luminosity of this component suggests $R_\mathrm{eff}\simeq 10\mbox{--}30$~AU scales ($10^{-5}$--$10^{-4}$~pc), two orders of magnitude smaller than those implied in LRDs \citep[while their luminosities are comparable, the high effective temperature required here necessitates a substantially smaller size; e.g.][]{chisholmLittleRedDots2026} and an order of magnitude smaller than even small measured BLR sizes \citep[of-order light-days or larger; e.g.][]{kaspiRelationshipLuminosityBroadLine2005,Bentz2013,GRAVITYCollaboration2024,Woo2026}.
This is qualitatively consistent with the fact that this broad \ion{N}{4}] emission is superimposed by a very narrow component, which is likely associated given the rarity of the required N/O-enhanced gas. In that case, this narrow component must reside outside this optically thick outflow, and larger extended geometries introduce some tension with the implied kinematics of this near stationary narrow line gas.
Any central engine near the Eddington luminosity would require a mass of at least $10^5$--$10^6$~$M_\odot$, or substantially more luminous \ion{N}{4}] emission than achieved by these stellar models.
It would also naturally require a very high N/C and N/O ratio to explain the observed line ratios.
An extended atmosphere as potentially allowed by such an extreme scenario may also resolve some tension with lines like \ion{N}{4}~$\lambda 1719$, which are likely to be further suppressed in atmospheres where the incident photon density and gas density is further reduced (as is likely the case for HD~5980).
At the implied scales, such a central engine is suggestive of a luminous accretion disk or supermassive star, surrounded by gas which it is driving away in an LBV-like near-Eddington state.

Despite the suggestive compatibility of their velocities, the broad \ion{N}{4}]~$\lambda 1486$ emission may also be unrelated to the broad absorption component of \ion{N}{5} and \ion{C}{4} and the emission in \ion{He}{2}, and instead emerge separately from a highly enriched broad line region in emission.
This would naturally explain the absence of strong \ion{N}{4}~$\lambda 1719$ which is not observed in AGN or generally outside of dense stellar atmospheres.
While \ion{N}{4}]~$\lambda 1486$ is a prominent collisionally excited line in the rare subset of nitrogen loud quasars, these systems also typically show prominent broad emission in \ion{N}{5}~$\lambda\lambda 1238,1242$ and \ion{N}{3}]~$\lambda\lambda 1750$; with the latter two lines comparably or significantly stronger than \ion{N}{4}] \citep{baldwinChemicalAbundancesBroad2003,jiangSampleQuasarsStrong2008,matsuokaChemicalEnrichmentAccretion2017}.
While in GN-z11 the \ion{N}{5} profile is complex, we don't see clear evidence for strong broad emission at similar FWHM on the red side of the P-Cygni profile in excess of the putatively stellar component; and \ion{N}{3}] shows no clear evidence for a strong broad component.
Such a \ion{N}{4}]-dominated BLR might be possible, but would likely require particular temperature and density conditions not achieved in the known nitrogen-loud quasar distribution; and if indeed not associated with the outflow lines, would necessitate even deeper absorption lines from the stellar component to compensate for the additional continuum contribution implied.

\subsection{New Insight into Physical Nature of GN-z11}\label{sec:discussion_nature}

As the most luminous galaxy known at $z>10$, GN-z11 provided our first indications of an extraordinary galaxy population at cosmic dawn already in the HST-era \citep{Oesch2016}. 
Its extreme luminosity makes it our best laboratory to understand the physical conditions in UV-bright galaxies at these redshifts. NIRCam imaging demonstrated its light is dominated by an extremely compact ($r_{1/2} = 64\pm20$\,pc) component containing $\sim10^{8.4}\,M_\odot$ of stars \citep{Tacchella2023}, hinting at an unusually dense mode of star formation. The discovery imaging and spectrum reveal a rising star formation history, with half of the stellar mass building up in the $\simeq 10$ Myr prior to observation \citep{Tacchella2023,Bunker2023}.

The SPURS spectrum sharpens this picture. Detection of multiple stellar P-Cygni wind features provides a direct view of the massive star populations present in GN-z11, with the strong NV feature in particular providing a higher definition view of the recent star formation history. We showed in \S~\ref{sec:VMS_comparison} that reproducing the NV absorption and emission requires a significant fraction of the UV continuum  ($\simeq 2-3$ Myr) arises from very young stellar populations. In the context of these models, the extreme UV continuum luminosity of GN-z11 is driven by the recent burst of star formation. 

The nature of the massive stars in GN-z11 holds important clues as to the origin of the nitrogen-enhanced abundance pattern. In sufficiently dense, low-metallicity stellar systems, VMS are expected to assemble through runaway collisions and stellar mergers (e.g., \citealt{PortegiesZwart2002,Freitag2006,Goswami2012,Fujii2013}). Recent star-by-star simulations of hierarchical cluster assembly support this picture, following the formation of massive clusters through the merging of $\sim1000$ sub-clusters and finding that VMS are efficiently assembled through repeated collisions of massive stars within them \citep{Rantala2024,Rantala2026}. These simulations indicate that the channel operates for cluster half-mass surface densities above $\Sigma_{\rm h}\gtrsim10^{4}-10^{5}\,M_\odot\,{\rm pc^{-2}}$ and metallicities below $Z\lesssim0.2\,Z_\odot$.
While we cannot constrain the densities of individual clusters in GN-z11, its extreme compactness ($r_{1/2}=64$ pc) and low stellar metallicity place it plausibly within this regime. Our analysis in \S~\ref{sec:VMS_comparison} suggests that VMS may be present in GN-z11, powering the stellar wind features observed in the UV spectrum. 
In \S~\ref{sec:discussion_NIV}, we demonstrated that the broad N\textsc{iv}]$\lambda1486$ emission may be powered by the winds of VMS or possibly LBV-like outbursts in GN-z11 and other nitrogen emitters, with the strength of this feature and other narrow wind lines and the absence of N\textsc{iv}$\lambda1719$ offering a first hint that we may be witnessing a stellar population without direct analog at later epochs, plausibly shaped by the lower-metallicity, denser conditions of the early universe.

If VMS are present in GN-z11 and other nitrogen emitters,  simulations predict they may naturally seed IMBHs. In particular, simulations of hierarchically assembling star clusters predict that IMBHs form via stellar collisions in the centers of individual sub-clusters, and are subsequently carried toward the center of the host system as the sub-clusters themselves sink and merge \citep{Fujii2013, Rantala2024, Rantala2026}. The collisionally grown stars collapse after $\simeq2.5-5$ Myr, comparable to the age of the burst inferred in \S~\ref{sec:VMS_comparison}, so IMBH formation may plausibly be underway in GN-z11; their assembly into a single nuclear black hole would follow on the longer timescale over which the cluster system merges. Independently, based on the O~{\sc i}*$\lambda1304$ and broad Ly$\alpha$ features (\S~\ref{sec:discussion_lya}) we have suggested that GN-z11 may harbor a dense reservoir of neutral gas in its nuclear region. This is exactly the environment in which such a black hole would be expected to rapidly grow in mass (e.g., \citealt{Inayoshi2020}).

Previous work has hinted at the possibility that GN-z11 hosts an AGN \citep{Maiolino2024}. We can reconsider this evidence in light of the new observations. 
The SPURS data  confirm the presence of fast ionized outflows in GN-z11, as also noted by \citet{Maiolino2024}.
While the high ionization absorption lines have  contribution from stellar winds, we find that the troughs in the interstellar absorption lines are near $\approx500$\,km~s$^{-1}$ (Section~\ref{sec:ISMabs}), at the extreme end of what is found in star-forming galaxies \citep[e.g.,][]{Chisholm2015,Heckman2016,Xu2022}. In contrast, such high velocities are commonly seen in UV absorption lines of AGN \citep[e.g.,][]{Hainline2011}.
However, the evidence in favor of an AGN is not conclusive.
The electron densities inferred from \ion{N}{4}] and \ion{N}{3}] do not require BLR-like densities, though uncertainties still permit high 
densities from the \ion{N}{3}] ratios 
(Section~\ref{sec:phys_density}). We do not detect the [Ne IV] emission that is present in the discovery spectrum \citep{Maiolino2024,Tang2025}, although we note that the absence of very high ionization lines is common in many AGN discovered by {\it JWST} at very high redshift 
\citep[e.g.,][]{Tang2025,Scholtz2025,Lambrides2026}.
Most importantly, if GN-z11 hosts a broad-line AGN, we should  see broad Balmer lines, which have not been detected in the current NIRSpec and MIRI spectroscopic observations. 
However, as discussed in \S~\ref{sec:res_optical}, if the broad-to-narrow Balmer ratio is similar to that in EGSY8p7, which does show broad rest-frame optical lines \citep{Larson2023,Zamora2025} as well as broad \ion{N}{4}] (Figure~\ref{fig:NIV_comp}), 
both the broad H$\gamma$ and broad H$\alpha$ components would be below the sensitivity limit of current observations.
Deeper rest-frame optical spectroscopy will thus be critical for providing further constraints on the presence of an AGN in GN-z11.

Finally, we consider how these results connect to the LRD population, which shares both nitrogen enhancement (e.g., \citealt{Labbe2024,Morel2025,Tripodi2025_z8}) and evidence for Ly$\beta$-pumped O~{\sc i} emission with GN-z11 (e.g., \citealt{Tripodi2025,Kokorev2026,Tang2026_spursQSO1}). One explanation is simply that both are powered by dense stellar systems in which massive stars are efficiently assembled. But the same conditions may link the two populations more directly: if intense bursts build dense clusters that form VMS, and those VMS in turn seed IMBHs, then the star formation episodes producing objects like GN-z11 may also be efficient black hole seeding sites. Whether such a phase may precede an LRD phase, with the seeded black hole subsequently growing in the dense nuclear gas, or alternatively follow one, after feedback has cleared much of the obscuring material, is unclear. Regardless, the conditions required to seed a central black hole appear to be in place in GN-z11.

\section{Summary} \label{sec:summary}

We present new ultra-deep $R\sim1000$ JWST/NIRSpec spectroscopy of GN-z11, 
the most UV-luminous galaxy confirmed at $z > 10$, 
obtained as part of the SPURS Cycle 4 Large Program. 
The observations provide the deepest rest-UV view yet obtained of a galaxy at $z>10$, and offer new insight into two of GN-z11's most surprising properties: its extreme nitrogen enhancement and its detectable Ly$\alpha$ emission. 
The ultra-deep spectrum provides, for the first time, a direct view of the massive star population responsible for the nitrogen enrichment through stellar wind diagnostics, while the well-resolved Ly$\alpha$ profile sheds new light on how Ly$\alpha$ photons are able to 
be transmitted through a still largely neutral IGM at these redshifts. 
We note, an independent analysis of the SPURS spectrum is presented by \cite{Nakane2026}.
We summarize our findings as follows.

1. The deep medium-resolution G140M spectrum detects and resolves the Ly$\alpha$ velocity profile at high S/N.
The Ly$\alpha$ profile is asymmetric, with the observed flux peak offset from systemic by $418_{-14}^{+12}$~\kms{} and a broad wing at $>500$~\kms{} that contributes 44\% of the total flux.
We measure a small Ly$\alpha$ EW ($5.6_{-0.2}^{+0.2}$~\AA{}) and a low escape fraction ($f_{\rm esc, Ly\alpha}=0.027_{-0.003}^{+0.003}$). The broad red Ly$\alpha$ wing at $>500$\,km~s$^{-1}$ likely boosts the transmission of Ly$\alpha$ flux through the IGM damping wing. 

2. We detect the UV continuum at very high S/N (8.5 in G140M), allowing us to characterize interstellar absorption lines.
We find strong blueshifted absorption from high-ionization lines (N~V, Si~IV, and C~IV), suggesting fast ionized outflows with  velocities centered at $\sim500$~\kms{}.
In contrast, the low-ionization absorption lines are not detected
(typical EW limit of $>-0.3$\,\AA\ at 3$\sigma$), indicating a low covering fraction of neutral gas which should facilitate the escape of Ly$\alpha$ photons through the ISM/CGM. 

3. The broad Ly$\alpha$ emission we detect may be crucial for understanding why Ly$\alpha$ is seen at all at $z>10$. 
However, we point out that the observed flux at $>500$\,\kms{}
is difficult to reconcile with the weak low-ionization absorption lines, which suggest little neutral gas is available to scatter Ly$\alpha$ to these velocities. 
The SPURS spectrum detects the \ion{O}{1}$^*\lambda1304$ fine-structure emission line that offers a resolution.
This line requires dense neutral gas close to a subset of the ionizing sources in GN-z11, that is both optically thick to Ly$\beta$ and exposed to a high Ly$\beta$ flux. Gas at such high column densities will also resonantly scatter Ly$\alpha$ to the large velocities we observe. 
Given the weak LIS absorption,
we suggest this dense neutral gas is confined to a compact nuclear region, with the bulk of the UV continuum arising from surrounding star-forming regions. 
This demonstrates that a more detailed understanding of the gas conditions in early galaxies, and a larger census of $z>10$ Ly$\alpha$, will be important for more accurately inferring properties of the IGM at these redshifts.

4. We detect many emission lines throughout the UV (N~IV], C~IV, He~II, O~III], N~III], Si~III], C~III]), providing constraints on electron density, temperature, and gas-phase abundance pattern. Our results are consistent with those inferred from the JADES discovery spectrum, indicating extreme conditions in the \ion{H}{2} regions of GN-z11.
We find that the ionized gas is characterized by very high electron densities ($2.5\times10^4$~cm$^{-3}$ up to $>7.9\times10^5$~cm$^{-3}$), a hot electron temperature ($1.3_{-0.1}^{+0.1}\times10^4$~K), and a low gas-phase oxygen abundance (12 + log(O/H) = $7.94_{-0.14}^{+0.14}$), along with an enhanced N/O ratio relative to solar ($3.9_{-0.5}^{+0.8}$~(N/O)$_\odot$).

5. The rest-UV continuum spectrum constrains a suite of features (N~V, Si~IV, C~IV, and He~II) tracing winds from massive stars. In particular, we find the N-V P-Cygni feature is very strong, even after accounting for the contribution of IGM damping wing attenuation.  
We perform stellar continuum fitting with a variety of stellar population synthesis models, including those with updated treatments of VMS. 
We find that the far-UV continuum and stellar wind features can be reproduced by models that have the low stellar metallicity ($0.04~Z_\odot$) and  small ionized bubble sizes   ($\lesssim$ 0.5 pMpc) that we consider most likely for GN-z11. 
To reproduce the strong N-V feature, the models require a  significant population of very young stars ($\leq3$~Myr) contributing significantly to the UV continuum. We show that a population of VMS plays an important role in reproducing the observed spectrum.

6.   We show that GN-z11 exhibits broad  \ion{N}{4}] emission, with FWHM = 1670~\kms{} centered at the wavelength of the \ion{N}{4}]~$\lambda$1486 component.
We demonstrate that this broad NIV] component is seen in several other nitrogen emitters at $z\gtrsim9$. 
Broad \ion{N}{4}] emission is often seen in WN stars, but usually alongside a stronger \ion{N}{4}~$\lambda$1719 line, which is not seen in our data.
We demonstrate that strong broad \ion{N}{4}]~$\lambda$1486 with very weak \ion{N}{4}~$\lambda$1719 may nonetheless arise from dense and relatively low-temperature WN-like winds 
or LBV-like outbursts plausibly associated with a population of VMS in a dense cluster environment.
However, a wind driven by an AGN or supermassive star, potentially seeded by an earlier VMS population, cannot be excluded. 
This broad component may trace the mass outflows responsible for GN-z11's nitrogen enhancement.

While prism spectra provided our first indications of the extreme emission features of bright $z>10$ galaxies, UV grating spectra of this depth provide our first direct view of the underlying massive star populations forming under the extreme conditions of the early universe.
Spectra like this of GN-z11 open a unique window onto massive stars at the low metallicities and high densities characteristic of the earliest star-forming clusters, potentially shedding light onto otherwise challenging to constrain stellar astrophysics and providing a stress test for the next generation of stellar population synthesis models. 
Larger samples with deep rest-UV spectroscopy promise to reveal whether such  stellar populations are common among $z>10$ galaxies, and what physical conditions drive their emergence.

\begin{acknowledgments}

The authors thank Andrew Bunker, Alex Cameron, Paul Crowther, Max Gronke, and Andreas Sander for helpful conversations, Stéphane Charlot and Jacopo Chevallard for providing access to the \beagle{} tool, and Fabrice Martins for sharing their population synthesis models.
ZC acknowledges support by VILLUM FONDEN under grant 37459.
CAM acknowledges support by the European Union ERC grant RISES (101163035), Carlsberg Foundation (CF22-1322), and VILLUM FONDEN (37459).
KVGC is supported by NASA through the STScI grants JWST-GO-03777, JWST-GO-04265, and JWST-GO-05974.
LW acknowledges support from the Gavin Boyle Fellowship at the Kavli Institute for Cosmology, Cambridge and from the Kavli Foundation. 
VG acknowledges support by the Carlsberg Foundation (CF22-1322).
The Cosmic Dawn Center (DAWN) is funded by the Danish National Research Foundation under grant DNRF140.

This work is based in part on observations made with the NASA/ESA/CSA JWST. The data were obtained from the Mikulski Archive for Space Telescopes at the Space Telescope Science Institute, which is operated by the Association of Universities for Research in Astronomy, Inc., under NASA contract NAS 5-03127 for JWST. 
These observations are associated with program GTO 1181 (JADES; \citealt{Rieke2023JADES}) and GO 9214 (SPURS).
The authors acknowledge the JADES team for developing their observing program.
The authors thank the GO 9214 program coordinator, Christian Soto, and NIRSpec reviewer, Diane Karakla.
Some of the data products presented herein were retrieved from the Dawn JWST Archive (DJA). DJA is an initiative of the Cosmic Dawn Center (DAWN), which is funded by the Danish National Research Foundation under grant DNRF140.
The Tycho supercomputer hosted at the SCIENCE HPC center at the University of Copenhagen was used for supporting this work.

\end{acknowledgments}





%
\facilities{JWST(NIRSpec)}

\software{{\sc Astropy}:\footnote{\url{http://www.astropy.org}} a community-developed core Python package and an ecosystem of tools and resources for astronomy \citep{astropy:2013, astropy:2018, astropy:2022}; \beagle{} \citep{Chevallard2016}; {\sc Emcee} \citep{Foreman-Mackey2013}; {\sc FiCUS} \citep{Saldana-Lopez2023}; {\sc Jupyter} \citep{Kluyver2016}; {\sc Matplotlib} \citep{Hunter:2007}; {\sc Msaexp} \citep{Brammer2023}; {\sc Numpy} \citep{harris2020array}; {\sc Scipy} \citep{2020SciPy-NMeth}.
}


\appendix
\restartappendixnumbering

\section{BIC Analysis of Emission Line Fitting} \label{app:BIC}

The \ion{N}{4}], \ion{N}{3}], \ion{C}{3}], and \ion{O}{3}] emission lines of GN-z11 show complex profiles with multiple components. 
To interpret each of these lines, we fit the line profile with a few different models. 
We evaluate the goodness of fit of each model using the Bayesian Information Criterion (BIC; \citealt{Schwarz1978,Liddle2007}). 
According to \citet{Liddle2007}, the BIC of each model fit is defined as
\begin{equation}
{\rm BIC} \equiv -2\ln{\mathcal{L}}+k\ln{N},
\end{equation}
where $\mathcal{L}$ is the maximum likelihood achievable by the model, $k$ is the number of parameters of the model, and $N$ is the number of data points used in the fit.
To calculate the maximum likelihood, we assume independent Gaussian uncertainties, obtaining
\begin{equation}
\mathcal{L} = \prod^N_{i=1} \frac{1}{\sqrt{2\pi}\sigma_i}\cdot\exp{[-\frac{(f_i-m_i)^2}{2\sigma^2_i}]},
\end{equation}
where $f_i$ and $\sigma_i$ are the observed flux density and its uncertainty of the $i$th data point in the spectrum to fit, and $m_i$ is the best-fit model value of the $i$th data point. 
When comparing between two models, we take $\Delta$BIC $>10$ as very strong evidence that the model with the lower BIC provides a better fit to the data than the other. 

The SPURS deep G140M spectrum of GN-z11 reveals the detection of a broad \ion{N}{4}]$\lambda1486$ component. 
Motivated by this, we simultaneously fit the entire \ion{N}{4}] profile with two narrow Gaussians and one broad Gaussian. 
Here we fix the centroids of the two narrow Gaussians to the rest-frame wavelengths of [\ion{N}{4}]$\lambda1483$, \ion{N}{4}]$\lambda1486$, respectively. 
For the broad component, we fix the centroid to the rest-frame wavelength of \ion{N}{4}]$\lambda1486$. 
In this fitting, we allow the peak fluxes of three Gaussians and the widths of the narrow and broad lines to vary as free parameters, where we require the line widths of narrow [\ion{N}{4}]$\lambda1483$ and \ion{N}{4}]$\lambda1486$ to be the same. 
The fitting results are presented in Section~3.2.
This model can well reproduce the observed \ion{N}{4}] profile with a reduced chi-square ($\chi^2_{\nu}$) value of $0.7$ and a BIC of $49$.  

If we fit the \ion{N}{4}] profile by adding a second broad component with centroid fixed at the rest-frame wavelength of [\ion{N}{4}]~$\lambda1483$, we will find a similar result ($\chi^2_{\nu}=0.7$, BIC $=52$). 
The derived broad [\ion{N}{4}]~$\lambda1483$ line flux is negligible, and hence we place a $3\sigma$ limiting flux of $<2.0\times10^{-19}$~erg~s$^{-1}$~cm$^{-2}$. 
This indicates a large \ion{N}{4}]~$\lambda1486$/[\ion{N}{4}]~$\lambda1483$ flux ratio to the broad line ($>3.0$).

We also consider whether the \ion{N}{4}] profile of GN-z11 can be fitted by two narrow Gaussians only. 
With this model, we derive a $\chi^2_{\nu}$ of $1.2$ and a BIC of $66$, both larger than that of fitting with a broad component. 
Notably, the difference between BIC values of these two fits is significant ($\Delta$BIC $=17$), indicating that the model including a broad component indeed provides a better fit to the \ion{N}{4}] profile. 
The \ion{N}{4}]/[\ion{N}{4}] flux ratio inferred from two narrow Gaussian model fitting is $8.2$, consistent with that inferred from fitting with models including both narrow and broad Gaussians.

We consider whether the \ion{N}{3}] multiplet profile shows a broad component. 
We simultaneously fit \ion{N}{3}] with five narrow Gaussians and a broad Gaussian, where we fix the line width of the broad component to that of the broad \ion{N}{4}] (FWHM $=1670$~km~s$^{-1}$).
This fit is statistically marginally worse than fitting with five narrow Gaussian model, with a larger BIC ($\Delta$BIC $=4$). 
The inferred broad \ion{N}{3}] component has SNR $<3$. 
The fitted \ion{N}{3}]$\lambda1750$/\ion{N}{3}]$\lambda1752$ ratio is similar to the narrow Gaussian fit, with $1.7^{+0.4}_{-0.4}$.

Given the presence of a broad \ion{N}{4}], we examine whether there is a broad component in \ion{C}{3}] as well. 
To do so, we fit the \ion{C}{3}] profile with two different models: a combination of two narrow Gaussians (centered on the rest-frame wavelengths of [\ion{C}{3}]$\lambda1907$ and \ion{C}{3}]$\lambda1909$), and a combination of two narrow Gaussians and two broad Gaussians. 
In the latter model, we fix the line width of each broad component to be the same as the broad \ion{N}{4}] (FWHM $=1670$~km~s$^{-1}$). 
We then perform the same BIC and reduced chi-square test as for \ion{N}{4}]. 
When fitting the \ion{C}{3}] profile with two narrow Gaussians, we find a BIC of 53 and $\chi^2_{\nu}=1.0$. 
By adding broad components, the fitting is almost equal to two narrow Gaussian fitting with $\Delta$BIC $=2$ and a similar $\chi^2_{\nu}$ ($=0.9$). 
Both the inferred broad [\ion{C}{3}] and \ion{C}{3}] components have SNRs below $3$. 
This demonstrates that we do not clearly detect a broad component underlying \ion{C}{3}].

We similarly test whether there is a broad component in \ion{O}{3}]$\lambda1666$. 
We fit the \ion{O}{3}]$\lambda1666$ profile with two models: a single narrow Gaussian model and a combination of a narrow and a broad Gaussian, where the line width of the broad Gaussian is fixed to that of the broad \ion{N}{4}]. 
We then perform the BIC test for \ion{O}{3}]. 
There is no significant difference between fitting \ion{O}{3}] with both models ($\Delta$BIC $=1$), suggesting that there is no clear broad component underlying \ion{O}{3}].

\section{Additional Stellar Continuum Fitting Models}

\begin{figure*}
    \centering
    \includegraphics[width=\linewidth]{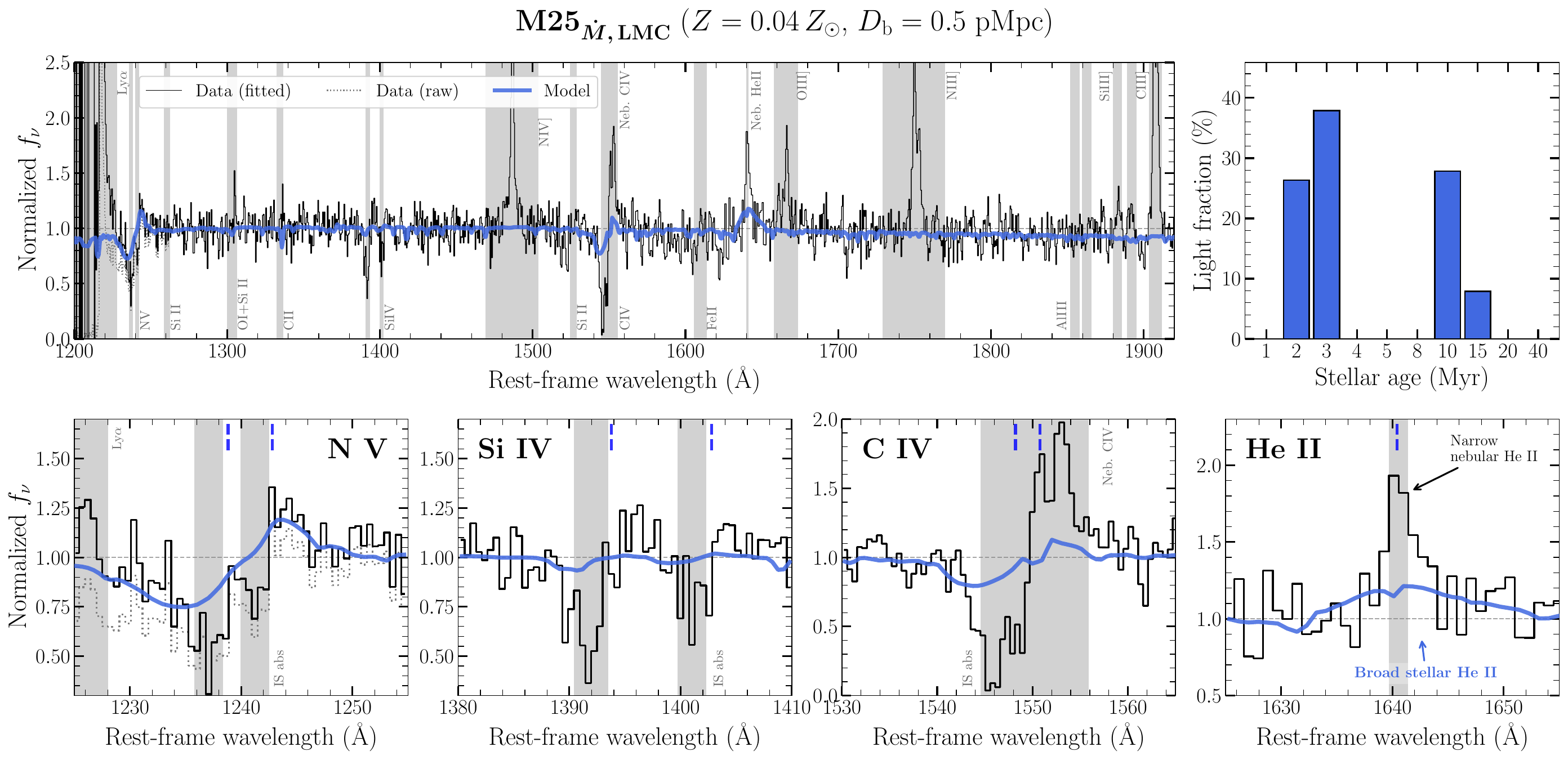}
    \caption{FiCUS fit to the UV continuum of GN-z11 using linear combinations of \ModelMLMC{} SSP models, assuming the fiducial low metallicity ($0.04~Z_\odot$) and a smaller ionized bubble ($D_b = 0.5$~pMpc).
    As in Figure~\ref{fig:ficus_model_CB19}, we show the fit to the full UV continuum (\textbf{Top left}), the best-fit star formation history (\textbf{Top right}), and zoomed-in views of the N~V, Si~IV, C~IV, and He~II regions (\textbf{Bottom} panels).
    }
    \label{fig:ficus_model_M25LMC}
\end{figure*}

\begin{figure*}
    \centering
    \includegraphics[width=\linewidth]{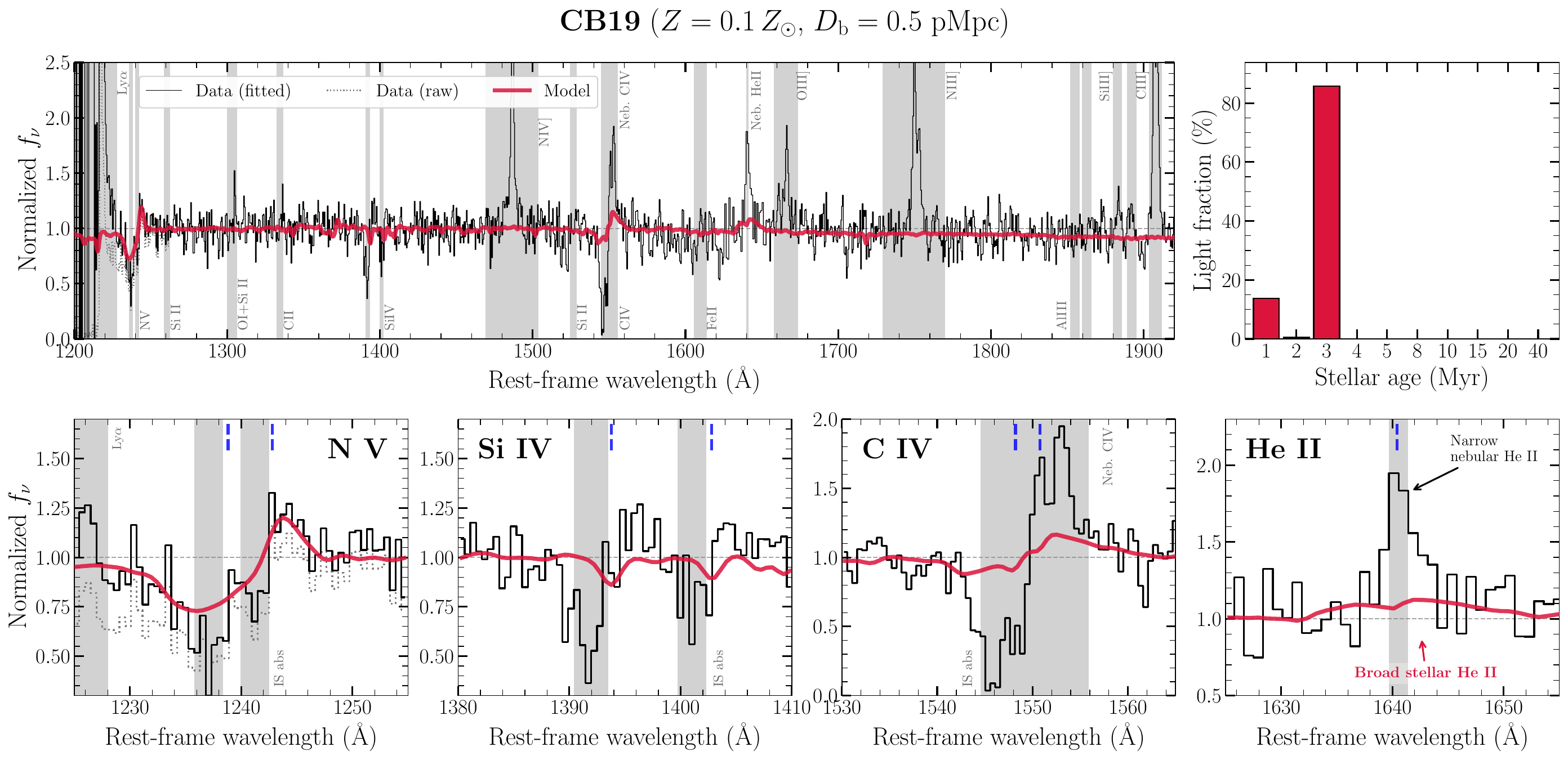}
    \caption{FiCUS fit to the UV continuum of GN-z11 using linear combinations of CB19 SSP models with a higher metallicity of $0.1~Z_\odot$ than the fiducial CB19 model in Figure~\ref{fig:ficus_model_CB19}, plotted in the same format.
    Such a high stellar metallicity is 2.5$\times$ higher than what we expect, and we consider it unlikely.
    }
    \label{fig:ficus_model_CB_Z0p1}
\end{figure*}

\begin{figure*}
    \centering
    \includegraphics[width=\linewidth]{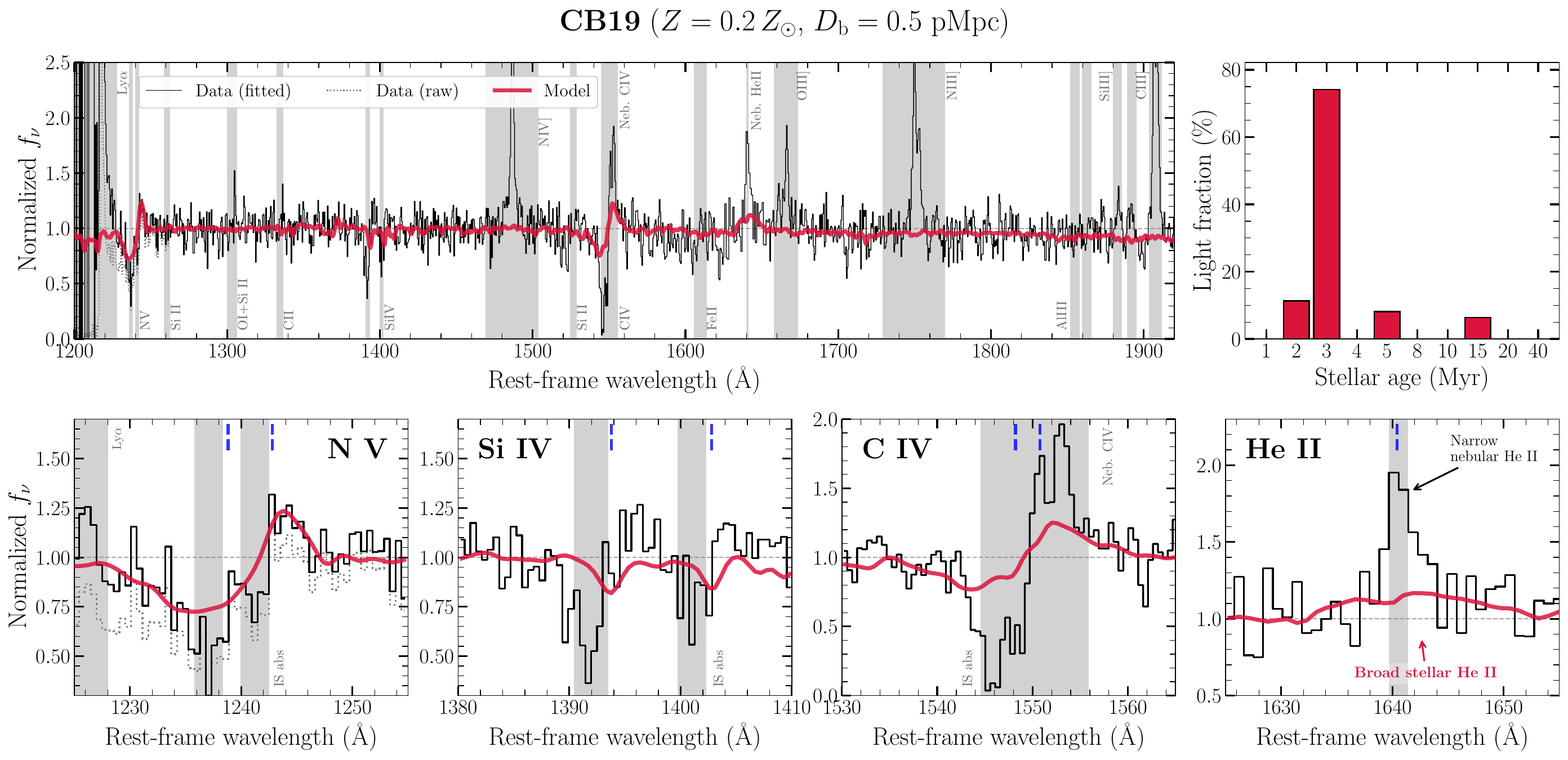}
    \caption{FiCUS fit to the UV continuum of GN-z11 using linear combinations of CB19 SSP models with a higher metallicity of $0.2~Z_\odot$ than the fiducial CB19 model in Figure~\ref{fig:ficus_model_CB19}, plotted in the same format.
    As in Figure~\ref{fig:ficus_model_CB19}, we show the fit to the full UV continuum (\textbf{Top left}), the best-fit star formation history (\textbf{Top right}), and zoomed-in views of the N~V, Si~IV, C~IV, and He~II regions (\textbf{Bottom} panels).
    However, we note such a high stellar metallicity is unlikely.
    }
    \label{fig:ficus_model_CB_Z0p2}
\end{figure*}

\begin{figure*}
    \centering
    \includegraphics[width=\linewidth]{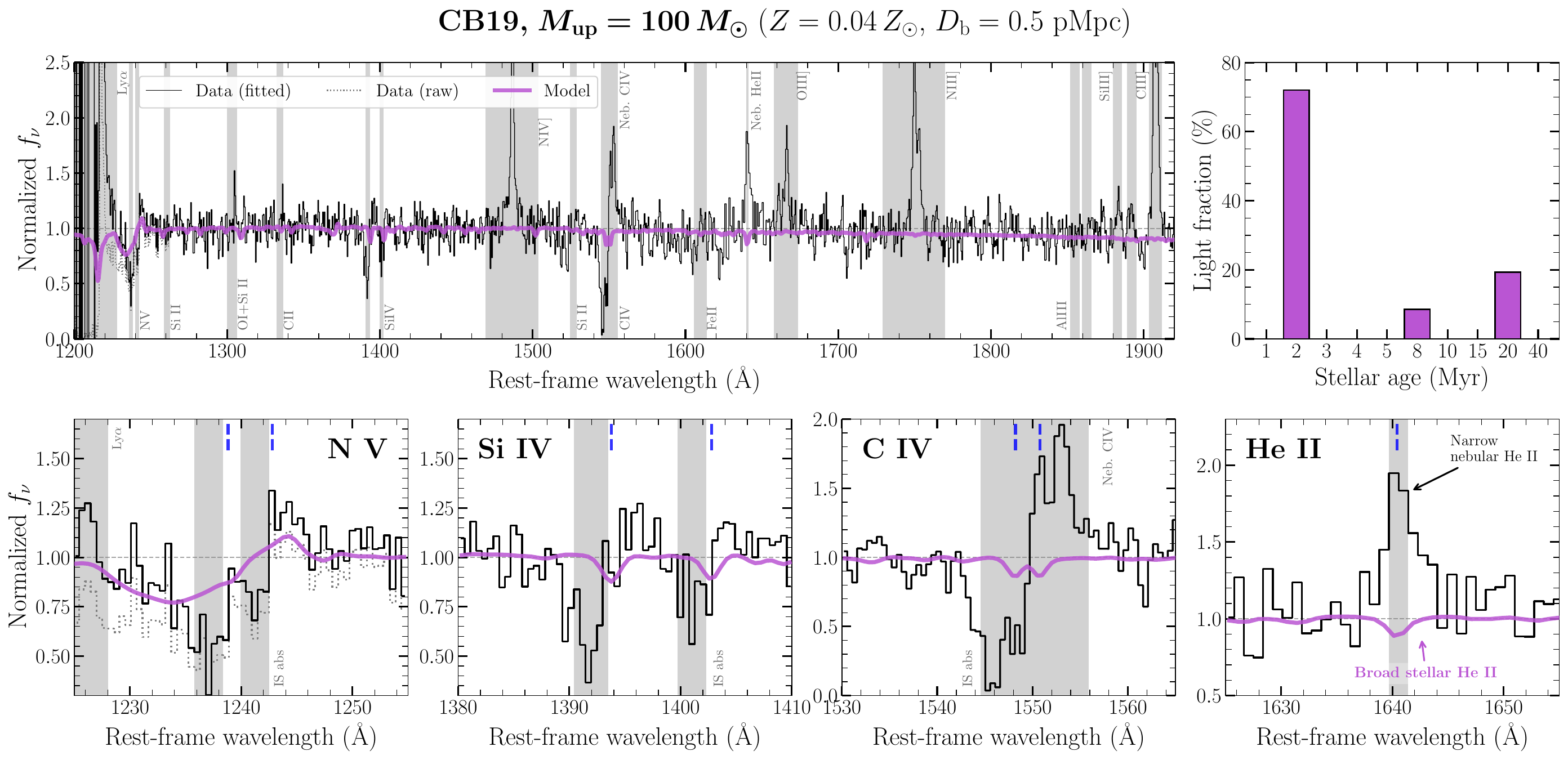}
    \caption{FiCUS fit to the UV continuum of GN-z11 using linear combinations of CB19 SSP models with an upper mass cutoff of 100~$M_\odot$, plotted in the same format as Figure~\ref{fig:ficus_model_CB19}.
    For the fiducial low metallicity ($0.04~Z_\odot$) and a small ionized bubble ($D_b = 0.5$~pMpc), the model underpredicts the P-Cygni emission components compared to the CB19 models that include very massive stars (Figure~\ref{fig:ficus_model_CB19}).
    }
    \label{fig:ficus_model_CBMup100}
\end{figure*}

\begin{figure*}
    \centering
    \includegraphics[width=\linewidth]{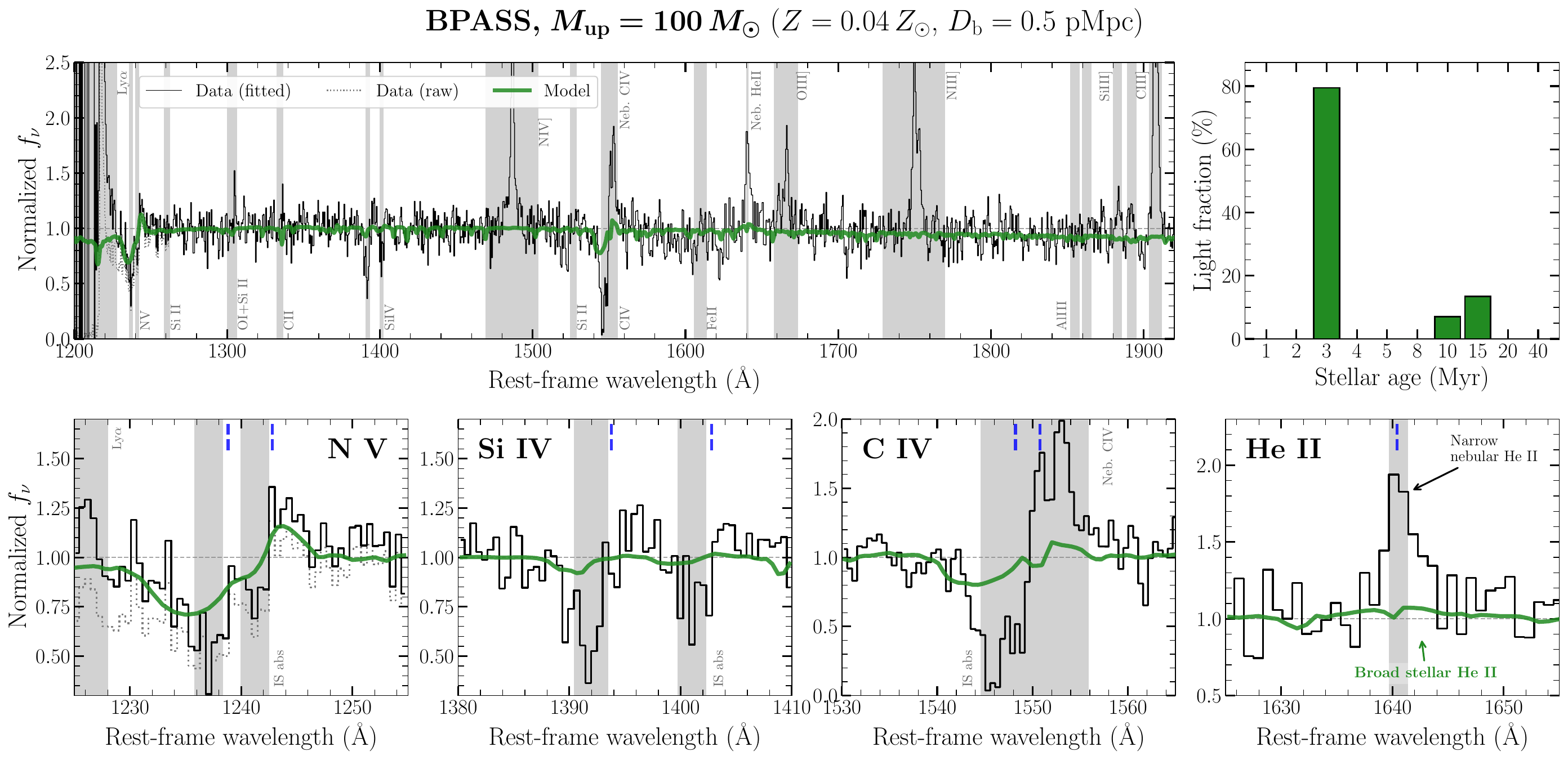}
    \caption{FiCUS fit to the UV continuum of GN-z11 using linear combinations of BPASS SSP models with an upper mass cutoff of 100~$M_\odot$, plotted in the same format as  Figure~\ref{fig:ficus_model_CBMup100}.
    Compared to the M25 models that include very massive stars (Figure~\ref{fig:ficus_model_M25}), this model underpredicts the P-Cygni emission component for the fiducial low metallicity ($0.04~Z_\odot$) and a small ionized bubble ($D_b = 0.5$~pMpc).
    }
    \label{fig:ficus_model_BPASSMup100}
\end{figure*}

We present the additional stellar continuum fitting models discussed in \S~\ref{sec:VMS_comparison}.
In Figure~\ref{fig:ficus_model_M25LMC}, we show the continuum fit based on the \ModelMLMC{} models, assuming the fiducial low metallicity (0.04~$Z_\odot$) and small ionized bubble size (0.5~pMPc).
In Figures~\ref{fig:ficus_model_CB_Z0p1} and \ref{fig:ficus_model_CB_Z0p2}, we show also the best-fit model with the CB19 models (0.1 and $0.2~Z_\odot$) at higher metallicities than our fiducial value, though we consider such high metallicities are unlikely for GN-z11 (see \S~\ref{sec:VMS_expectation}).
In Figures~\ref{fig:ficus_model_CBMup100} and \ref{fig:ficus_model_BPASSMup100}, we show the best-fit model without very massive stars (i.e., assuming an IMF upper mass cutoff of $100~M_\odot$) based on the CB19 and BPASS SSP models, respectively.
For both models, the fit to \ion{N}{5} is worse than the models including very massive stars (Figure~\ref{fig:ficus_model_CB19} and \ref{fig:ficus_model_M25}), underpredicting the P-Cygni emission components.


\bibliography{SPURS_GNz11}{}

@ARTICLE{Xu2023,
       author = {{Xu}, Xinfeng and {Henry}, Alaina and {Heckman}, Timothy and {Chisholm}, John and {Marques-Chaves}, Rui and {Leclercq}, Floriane and {Berg}, Danielle A. and {Jaskot}, Anne and {Schaerer}, Daniel and {Worseck}, G{\'a}bor and {Amor{\'\i}n}, Ricardo O. and {Atek}, Hakim and {Hayes}, Matthew and {Ji}, Zhiyuan and {{\"O}stlin}, G{\"o}ran and {Saldana-Lopez}, Alberto and {Thuan}, Trinh},
        title = "{The Low-redshift Lyman Continuum Survey: Optically Thin and Thick Mg II Lines as Probes of Lyman Continuum Escape}",
      journal = {\apj},
         year = 2023,
        month = feb,
       volume = {943},
       number = {2},
          eid = {94},
        pages = {94},
          doi = {10.3847/1538-4357/aca89a},
archivePrefix = {arXiv},
       eprint = {2301.04087},
 primaryClass = {astro-ph.GA},
       adsurl = {https://ui.adsabs.harvard.edu/abs/2023ApJ...943...94X}
}

@ARTICLE{steidel2010,
       author = {{Steidel}, Charles C. and {Erb}, Dawn K. and {Shapley}, Alice E. and {Pettini}, Max and {Reddy}, Naveen and {Bogosavljevi{\'c}}, Milan and {Rudie}, Gwen C. and {Rakic}, Olivera},
        title = "{The Structure and Kinematics of the Circumgalactic Medium from Far-ultraviolet Spectra of z \raisebox{-0.5ex}\textasciitilde= 2-3 Galaxies}",
      journal = {\apj},
         year = 2010,
        month = jul,
       volume = {717},
       number = {1},
        pages = {289-322},
          doi = {10.1088/0004-637X/717/1/289},
archivePrefix = {arXiv},
       eprint = {1003.0679},
 primaryClass = {astro-ph.CO},
       adsurl = {https://ui.adsabs.harvard.edu/abs/2010ApJ...717..289S}
}

@ARTICLE{Gardner2023,
       author = {{Gardner}, Jonathan P. and {Mather}, John C. and {Abbott}, Randy and {Abell}, James S. and {Abernathy}, Mark and {Abney}, Faith E. and {Abraham}, John G. and {Abraham}, Roberto and {Abul-Huda}, Yasin M. and {Acton}, Scott and {Adams}, Cynthia K. and {Adams}, Evan and {Adler}, David S. and {Adriaensen}, Maarten and {Aguilar}, Jonathan Albert and {Ahmed}, Mansoor and {Ahmed}, Nasif S. and {Ahmed}, Tanjira and {Albat}, R{\"u}deger and {Albert}, Lo{\"\i}c and {Alberts}, Stacey and {Aldridge}, David and {Allen}, Mary Marsha and {Allen}, Shaune S. and {Altenburg}, Martin and {Altunc}, Serhat and {Alvarez}, Jose Lorenzo and {{\'A}lvarez-M{\'a}rquez}, Javier and {Alves de Oliveira}, Catarina and {Ambrose}, Leslie L. and {Anandakrishnan}, Satya M. and {Andersen}, Gregory C. and {Anderson}, Harry James and {Anderson}, Jay and {Anderson}, Kristen and {Anderson}, Sara M. and {Aprea}, Julio and {Archer}, Benita J. and {Arenberg}, Jonathan W. and {Argyriou}, Ioannis and {Arribas}, Santiago and {Artigau}, {\'E}tienne and {Arvai}, Amanda Rose and {Atcheson}, Paul and {Atkinson}, Charles B. and {Averbukh}, Jesse and {Aymergen}, Cagatay and {Bacinski}, John J. and {Baggett}, Wayne E. and {Bagnasco}, Giorgio and {Baker}, Lynn L. and {Balzano}, Vicki Ann and {Banks}, Kimberly A. and {Baran}, David A. and {Barker}, Elizabeth A. and {Barrett}, Larry K. and {Barringer}, Bruce O. and {Barto}, Allison and {Bast}, William and {Baudoz}, Pierre and {Baum}, Stefi and {Beatty}, Thomas G. and {Beaulieu}, Mathilde and {Bechtold}, Kathryn and {Beck}, Tracy and {Beddard}, Megan M. and {Beichman}, Charles and {Bellagama}, Larry and {Bely}, Pierre and {Berger}, Timothy W. and {Bergeron}, Louis E. and {Bernier}, Antoine-Darveau and {Bertch}, Maria D. and {Beskow}, Charlotte and {Betz}, Laura E. and {Biagetti}, Carl P. and {Birkmann}, Stephan and {Bjorklund}, Kurt F. and {Blackwood}, James D. and {Blazek}, Ronald Paul and {Blossfeld}, Stephen and {Bluth}, Marcel and {Boccaletti}, Anthony and {Boegner}, Jr., Martin E. and {Bohlin}, Ralph C. and {Boia}, John Joseph and {B{\"o}ker}, Torsten and {Bonaventura}, N. and {Bond}, Nicholas A. and {Bosley}, Kari Ann and {Boucarut}, Rene A. and {Bouchet}, Patrice and {Bouwman}, Jeroen and {Bower}, Gary and {Bowers}, Ariel S. and {Bowers}, Charles W. and {Boyce}, Leslye A. and {Boyer}, Christine T. and {Boyer}, Martha L. and {Boyer}, Michael and {Boyer}, Robert and {Bradley}, Larry D. and {Brady}, Gregory R. and {Brandl}, Bernhard R. and {Brannen}, Judith L. and {Breda}, David and {Bremmer}, Harold G. and {Brennan}, David and {Bresnahan}, Pamela A. and {Bright}, Stacey N. and {Broiles}, Brian J. and {Bromenschenkel}, Asa and {Brooks}, Brian H. and {Brooks}, Keira J. and {Brown}, Bob and {Brown}, Bruce and {Brown}, Thomas M. and {Bruce}, Barry W. and {Bryson}, Jonathan G. and {Bujanda}, Edwin D. and {Bullock}, Blake M. and {Bunker}, A.~J. and {Bureo}, Rafael and {Burt}, Irving J. and {Bush}, James Aaron and {Bushouse}, Howard A. and {Bussman}, Marie C. and {Cabaud}, Olivier and {Cale}, Steven and {Calhoon}, Charles D. and {Calvani}, Humberto and {Canipe}, Alicia M. and {Caputo}, Francis M. and {Cara}, Mihai and {Carey}, Larkin and {Case}, Michael Eli and {Cesari}, Thaddeus and {Cetorelli}, Lee D. and {Chance}, Don R. and {Chandler}, Lynn and {Chaney}, Dave and {Chapman}, George N. and {Charlot}, S. and {Chayer}, Pierre and {Cheezum}, Jeffrey I. and {Chen}, Bin and {Chen}, Christine H. and {Cherinka}, Brian and {Chichester}, Sarah C. and {Chilton}, Zachary S. and {Chittiraibalan}, Dharini and {Clampin}, Mark and {Clark}, Charles R. and {Clark}, Kerry W. and {Clark}, Stephanie M. and {Claybrooks}, Edward E. and {Cleveland}, Keith A. and {Cohen}, Andrew L. and {Cohen}, Lester M. and {Col{\'o}n}, Knicole D. and {Coleman}, Benee L. and {Colina}, Luis and {Comber}, Brian J. and {Comeau}, Thomas M. and {Comer}, Thomas and {Conde Reis}, Alain and {Connolly}, Dennis C. and {Conroy}, Kyle E. and {Contos}, Adam R. and {Contreras}, James and {Cook}, Neil J. and {Cooper}, James L. and {Cooper}, Rachel Aviva and {Correia}, Michael F. and {Correnti}, Matteo and {Cossou}, Christophe and {Costanza}, Brian F. and {Coulais}, Alain and {Cox}, Colin R. and {Coyle}, Ray T. and {Cracraft}, Misty M. and {Crew}, Keith A. and {Curtis}, Gary J. and {Cusveller}, Bianca and {Da Costa Maciel}, Cleyciane and {Dailey}, Christopher T. and {Daugeron}, Fr{\'e}d{\'e}ric and {Davidson}, Greg S. and {Davies}, James E. and {Davis}, Katherine Anne and {Davis}, Michael S. and {Day}, Ratna and {de Chambure}, Daniel and {de Jong}, Pauline and {De Marchi}, Guido and {Dean}, Bruce H. and {Decker}, John E. and {Delisa}, Amy S. and {Dell}, Lawrence C. and {Dellagatta}, Gail},
        title = "{The James Webb Space Telescope Mission}",
      journal = {\pasp},
         year = 2023,
        month = jun,
       volume = {135},
       number = {1048},
          eid = {068001},
        pages = {068001},
          doi = {10.1088/1538-3873/acd1b5},
archivePrefix = {arXiv},
       eprint = {2304.04869},
 primaryClass = {astro-ph.IM},
       adsurl = {https://ui.adsabs.harvard.edu/abs/2023PASP..135f8001G}
}

@ARTICLE{Rigby2023,
       author = {{Rigby}, Jane and {Perrin}, Marshall and {McElwain}, Michael and {Kimble}, Randy and {Friedman}, Scott and {Lallo}, Matt and {Doyon}, Ren{\'e} and {Feinberg}, Lee and {Ferruit}, Pierre and {Glasse}, Alistair and {Rieke}, Marcia and {Rieke}, George and {Wright}, Gillian and {Willott}, Chris and {Colon}, Knicole and {Milam}, Stefanie and {Neff}, Susan and {Stark}, Christopher and {Valenti}, Jeff and {Abell}, Jim and {Abney}, Faith and {Abul-Huda}, Yasin and {Acton}, D. Scott and {Adams}, Evan and {Adler}, David and {Aguilar}, Jonathan and {Ahmed}, Nasif and {Albert}, Lo{\"\i}c and {Alberts}, Stacey and {Aldridge}, David and {Allen}, Marsha and {Altenburg}, Martin and {{\'A}lvarez-M{\'a}rquez}, Javier and {Alves de Oliveira}, Catarina and {Andersen}, Greg and {Anderson}, Harry and {Anderson}, Sara and {Argyriou}, Ioannis and {Armstrong}, Amber and {Arribas}, Santiago and {Artigau}, Etienne and {Arvai}, Amanda and {Atkinson}, Charles and {Bacon}, Gregory and {Bair}, Thomas and {Banks}, Kimberly and {Barrientes}, Jaclyn and {Barringer}, Bruce and {Bartosik}, Peter and {Bast}, William and {Baudoz}, Pierre and {Beatty}, Thomas and {Bechtold}, Katie and {Beck}, Tracy and {Bergeron}, Eddie and {Bergkoetter}, Matthew and {Bhatawdekar}, Rachana and {Birkmann}, Stephan and {Blazek}, Ronald and {Blome}, Claire and {Boccaletti}, Anthony and {B{\"o}ker}, Torsten and {Boia}, John and {Bonaventura}, Nina and {Bond}, Nicholas and {Bosley}, Kari and {Boucarut}, Ray and {Bourque}, Matthew and {Bouwman}, Jeroen and {Bower}, Gary and {Bowers}, Charles and {Boyer}, Martha and {Bradley}, Larry and {Brady}, Greg and {Braun}, Hannah and {Breda}, David and {Bresnahan}, Pamela and {Bright}, Stacey and {Britt}, Christopher and {Bromenschenkel}, Asa and {Brooks}, Brian and {Brooks}, Keira and {Brown}, Bob and {Brown}, Matthew and {Brown}, Patricia and {Bunker}, Andy and {Burger}, Matthew and {Bushouse}, Howard and {Cale}, Steven and {Cameron}, Alex and {Cameron}, Peter and {Canipe}, Alicia and {Caplinger}, James and {Caputo}, Francis and {Cara}, Mihai and {Carey}, Larkin and {Carniani}, Stefano and {Carrasquilla}, Maria and {Carruthers}, Margaret and {Case}, Michael and {Catherine}, Riggs and {Chance}, Don and {Chapman}, George and {Charlot}, St{\'e}phane and {Charlow}, Brian and {Chayer}, Pierre and {Chen}, Bin and {Cherinka}, Brian and {Chichester}, Sarah and {Chilton}, Zack and {Chonis}, Taylor and {Clampin}, Mark and {Clark}, Charles and {Clark}, Kerry and {Coe}, Dan and {Coleman}, Benee and {Comber}, Brian and {Comeau}, Tom and {Connolly}, Dennis and {Cooper}, James and {Cooper}, Rachel and {Coppock}, Eric and {Correnti}, Matteo and {Cossou}, Christophe and {Coulais}, Alain and {Coyle}, Laura and {Cracraft}, Misty and {Curti}, Mirko and {Cuturic}, Steven and {Davis}, Katherine and {Davis}, Michael and {Dean}, Bruce and {DeLisa}, Amy and {deMeester}, Wim and {Dencheva}, Nadia and {Dencheva}, Nadezhda and {DePasquale}, Joseph and {Deschenes}, Jeremy and {Hunor Detre}, {\"O}rs and {Diaz}, Rosa and {Dicken}, Dan and {DiFelice}, Audrey and {Dillman}, Matthew and {Dixon}, William and {Doggett}, Jesse and {Donaldson}, Tom and {Douglas}, Rob and {DuPrie}, Kimberly and {Dupuis}, Jean and {Durning}, John and {Easmin}, Nilufar and {Eck}, Weston and {Edeani}, Chinwe and {Egami}, Eiichi and {Ehrenwinkler}, Ralf and {Eisenhamer}, Jonathan and {Eisenhower}, Michael and {Elie}, Michelle and {Elliott}, James and {Elliott}, Kyle and {Ellis}, Tracy and {Engesser}, Michael and {Espinoza}, Nestor and {Etienne}, Odessa and {Etxaluze}, Mireya and {Falini}, Patrick and {Feeney}, Matthew and {Ferry}, Malcolm and {Filippazzo}, Joseph and {Fincham}, Brian and {Fix}, Mees and {Flagey}, Nicolas and {Florian}, Michael and {Flynn}, Jim and {Fontanella}, Erin and {Ford}, Terrance and {Forshay}, Peter and {Fox}, Ori and {Franz}, David and {Fu}, Henry and {Fullerton}, Alexander and {Galkin}, Sergey and {Galyer}, Anthony and {Garc{\'\i}a Mar{\'\i}n}, Macarena and {Gardner}, Jonathan P. and {Gardner}, Lisa and {Garland}, Dennis and {Garrett}, Bruce and {Gasman}, Danny and {Gaspar}, Andras and {Gaudreau}, Daniel and {Gauthier}, Peter and {Geers}, Vincent and {Geithner}, Paul and {Gennaro}, Mario and {Giardino}, Giovanna and {Girard}, Julien and {Giuliano}, Mark and {Glassmire}, Kirk and {Glauser}, Adrian},
        title = "{The Science Performance of JWST as Characterized in Commissioning}",
      journal = {\pasp},
         year = 2023,
        month = apr,
       volume = {135},
       number = {1046},
          eid = {048001},
        pages = {048001},
          doi = {10.1088/1538-3873/acb293},
archivePrefix = {arXiv},
       eprint = {2207.05632},
 primaryClass = {astro-ph.IM},
       adsurl = {https://ui.adsabs.harvard.edu/abs/2023PASP..135d8001R}
}

@INPROCEEDINGS{Stark2026,
       author = {{Stark}, Daniel P. and {Topping}, Michael W. and {Endsley}, Ryan and {Tang}, Mengtao},
        title = "{Observations of the first galaxies in the Era of JWST}",
    booktitle = {Encyclopedia of Astrophysics, Volume 4},
         year = 2026,
       volume = {4},
        month = jan,
        pages = {453-499},
          doi = {10.1016/B978-0-443-21439-4.00128-0},
archivePrefix = {arXiv},
       eprint = {2501.17078},
 primaryClass = {astro-ph.GA},
       adsurl = {https://ui.adsabs.harvard.edu/abs/2026enap....4..453S}
}

@ARTICLE{Bunker2023,
       author = {{Bunker}, Andrew J. and {Saxena}, Aayush and {Cameron}, Alex J. and {Willott}, Chris J. and {Curtis-Lake}, Emma and {Jakobsen}, Peter and {Carniani}, Stefano and {Smit}, Renske and {Maiolino}, Roberto and {Witstok}, Joris and {Curti}, Mirko and {D'Eugenio}, Francesco and {Jones}, Gareth C. and {Ferruit}, Pierre and {Arribas}, Santiago and {Charlot}, Stephane and {Chevallard}, Jacopo and {Giardino}, Giovanna and {de Graaff}, Anna and {Looser}, Tobias J. and {L{\"u}tzgendorf}, Nora and {Maseda}, Michael V. and {Rawle}, Tim and {Rix}, Hans-Walter and {Del Pino}, Bruno Rodr{\'\i}guez and {Alberts}, Stacey and {Egami}, Eiichi and {Eisenstein}, Daniel J. and {Endsley}, Ryan and {Hainline}, Kevin and {Hausen}, Ryan and {Johnson}, Benjamin D. and {Rieke}, George and {Rieke}, Marcia and {Robertson}, Brant E. and {Shivaei}, Irene and {Stark}, Daniel P. and {Sun}, Fengwu and {Tacchella}, Sandro and {Tang}, Mengtao and {Williams}, Christina C. and {Willmer}, Christopher N.~A. and {Baker}, William M. and {Baum}, Stefi and {Bhatawdekar}, Rachana and {Bowler}, Rebecca and {Boyett}, Kristan and {Chen}, Zuyi and {Circosta}, Chiara and {Helton}, Jakob M. and {Ji}, Zhiyuan and {Kumari}, Nimisha and {Lyu}, Jianwei and {Nelson}, Erica and {Parlanti}, Eleonora and {Perna}, Michele and {Sandles}, Lester and {Scholtz}, Jan and {Suess}, Katherine A. and {Topping}, Michael W. and {{\"U}bler}, Hannah and {Wallace}, Imaan E.~B. and {Whitler}, Lily},
        title = "{JADES NIRSpec Spectroscopy of GN-z11: Lyman-{\ensuremath{\alpha}} emission and possible enhanced nitrogen abundance in a z = 10.60 luminous galaxy}",
      journal = {\aap},
         year = 2023,
        month = sep,
       volume = {677},
          eid = {A88},
        pages = {A88},
          doi = {10.1051/0004-6361/202346159},
archivePrefix = {arXiv},
       eprint = {2302.07256},
 primaryClass = {astro-ph.GA},
       adsurl = {https://ui.adsabs.harvard.edu/abs/2023A&A...677A..88B}
}

@ARTICLE{vasan2026,
       author = {{Vasan G. C}. and {Senchyna}, Peter and {Mason}, Charlotte A. and {Chen}, Zuyi and {Stark}, Daniel P. and {Jones}, Tucker and {Whitler}, Lily and {Glazer}, Kelsey S. and {Aravena}, Manuel and {Gonzalez-Lopez}, Jorge and {Endsley}, Ryan and {Gelli}, Viola and {Tang}, Mengtao and {Topping}, Michael W.},
        title = "{JWST absorption line spectroscopy with SPURS: ISM covering fractions and kinematics in individual galaxies at $z=5-9$}",
      journal = {arXiv e-prints},
         year = 2026,
        month = jun,
          eid = {arXiv:2606.06609},
        pages = {arXiv:2606.06609},
          doi = {10.48550/arXiv.2606.06609},
archivePrefix = {arXiv},
       eprint = {2606.06609},
 primaryClass = {astro-ph.GA},
       adsurl = {https://ui.adsabs.harvard.edu/abs/2026arXiv260606609K}
}

@ARTICLE{Bouwens2015,
       author = {{Bouwens}, R.~J. and {Illingworth}, G.~D. and {Oesch}, P.~A. and {Trenti}, M. and {Labb{\'e}}, I. and {Bradley}, L. and {Carollo}, M. and {van Dokkum}, P.~G. and {Gonzalez}, V. and {Holwerda}, B. and {Franx}, M. and {Spitler}, L. and {Smit}, R. and {Magee}, D.},
        title = "{UV Luminosity Functions at Redshifts z {\ensuremath{\sim}} 4 to z {\ensuremath{\sim}} 10: 10,000 Galaxies from HST Legacy Fields}",
      journal = {\apj},
         year = 2015,
        month = apr,
       volume = {803},
       number = {1},
          eid = {34},
        pages = {34},
          doi = {10.1088/0004-637X/803/1/34},
archivePrefix = {arXiv},
       eprint = {1403.4295},
 primaryClass = {astro-ph.CO},
       adsurl = {https://ui.adsabs.harvard.edu/abs/2015ApJ...803...34B}
}

@ARTICLE{Bouwens2021,
       author = {{Bouwens}, R.~J. and {Oesch}, P.~A. and {Stefanon}, M. and {Illingworth}, G. and {Labb{\'e}}, I. and {Reddy}, N. and {Atek}, H. and {Montes}, M. and {Naidu}, R. and {Nanayakkara}, T. and {Nelson}, E. and {Wilkins}, S.},
        title = "{New Determinations of the UV Luminosity Functions from z   9 to 2 Show a Remarkable Consistency with Halo Growth and a Constant Star Formation Efficiency}",
      journal = {\aj},
         year = 2021,
        month = aug,
       volume = {162},
       number = {2},
          eid = {47},
        pages = {47},
          doi = {10.3847/1538-3881/abf83e},
archivePrefix = {arXiv},
       eprint = {2102.07775},
 primaryClass = {astro-ph.GA},
       adsurl = {https://ui.adsabs.harvard.edu/abs/2021AJ....162...47B}
}

@ARTICLE{Finkelstein2015,
       author = {{Finkelstein}, Steven L. and {Ryan}, Jr., Russell E. and {Papovich}, Casey and {Dickinson}, Mark and {Song}, Mimi and {Somerville}, Rachel S. and {Ferguson}, Henry C. and {Salmon}, Brett and {Giavalisco}, Mauro and {Koekemoer}, Anton M. and {Ashby}, Matthew L.~N. and {Behroozi}, Peter and {Castellano}, Marco and {Dunlop}, James S. and {Faber}, Sandy M. and {Fazio}, Giovanni G. and {Fontana}, Adriano and {Grogin}, Norman A. and {Hathi}, Nimish and {Jaacks}, Jason and {Kocevski}, Dale D. and {Livermore}, Rachael and {McLure}, Ross J. and {Merlin}, Emiliano and {Mobasher}, Bahram and {Newman}, Jeffrey A. and {Rafelski}, Marc and {Tilvi}, Vithal and {Willner}, S.~P.},
        title = "{The Evolution of the Galaxy Rest-frame Ultraviolet Luminosity Function over the First Two Billion Years}",
      journal = {\apj},
         year = 2015,
        month = sep,
       volume = {810},
       number = {1},
          eid = {71},
        pages = {71},
          doi = {10.1088/0004-637X/810/1/71},
archivePrefix = {arXiv},
       eprint = {1410.5439},
 primaryClass = {astro-ph.GA},
       adsurl = {https://ui.adsabs.harvard.edu/abs/2015ApJ...810...71F}
}

@ARTICLE{Oesch2018,
       author = {{Oesch}, P.~A. and {Bouwens}, R.~J. and {Illingworth}, G.~D. and {Labb{\'e}}, I. and {Stefanon}, M.},
        title = "{The Dearth of z {\ensuremath{\sim}} 10 Galaxies in All HST Legacy Fields{\textemdash}The Rapid Evolution of the Galaxy Population in the First 500 Myr}",
      journal = {\apj},
         year = 2018,
        month = mar,
       volume = {855},
       number = {2},
          eid = {105},
        pages = {105},
          doi = {10.3847/1538-4357/aab03f},
archivePrefix = {arXiv},
       eprint = {1710.11131},
 primaryClass = {astro-ph.GA},
       adsurl = {https://ui.adsabs.harvard.edu/abs/2018ApJ...855..105O}
}

@ARTICLE{Bowler2020,
       author = {{Bowler}, R.~A.~A. and {Jarvis}, M.~J. and {Dunlop}, J.~S. and {McLure}, R.~J. and {McLeod}, D.~J. and {Adams}, N.~J. and {Milvang-Jensen}, B. and {McCracken}, H.~J.},
        title = "{A lack of evolution in the very bright end of the galaxy luminosity function from z ≃ 8 to 10}",
      journal = {\mnras},
         year = 2020,
        month = apr,
       volume = {493},
       number = {2},
        pages = {2059-2084},
          doi = {10.1093/mnras/staa313},
archivePrefix = {arXiv},
       eprint = {1911.12832},
 primaryClass = {astro-ph.GA},
       adsurl = {https://ui.adsabs.harvard.edu/abs/2020MNRAS.493.2059B}
}

@ARTICLE{Castellano2022,
       author = {{Castellano}, Marco and {Fontana}, Adriano and {Treu}, Tommaso and {Santini}, Paola and {Merlin}, Emiliano and {Leethochawalit}, Nicha and {Trenti}, Michele and {Vanzella}, Eros and {Mestric}, Uros and {Bonchi}, Andrea and {Belfiori}, Davide and {Nonino}, Mario and {Paris}, Diego and {Polenta}, Gianluca and {Roberts-Borsani}, Guido and {Boyett}, Kristan and {Brada{\v{c}}}, Maru{\v{s}}a and {Calabr{\`o}}, Antonello and {Glazebrook}, Karl and {Grillo}, Claudio and {Mascia}, Sara and {Mason}, Charlotte and {Mercurio}, Amata and {Morishita}, Takahiro and {Nanayakkara}, Themiya and {Pentericci}, Laura and {Rosati}, Piero and {Vulcani}, Benedetta and {Wang}, Xin and {Yang}, Lilan},
        title = "{Early Results from GLASS-JWST. III. Galaxy Candidates at z  9-15}",
      journal = {\apjl},
         year = 2022,
        month = oct,
       volume = {938},
       number = {2},
          eid = {L15},
        pages = {L15},
          doi = {10.3847/2041-8213/ac94d0},
archivePrefix = {arXiv},
       eprint = {2207.09436},
 primaryClass = {astro-ph.GA},
       adsurl = {https://ui.adsabs.harvard.edu/abs/2022ApJ...938L..15C}
}

@ARTICLE{Naidu2022,
       author = {{Naidu}, Rohan P. and {Oesch}, Pascal A. and {van Dokkum}, Pieter and {Nelson}, Erica J. and {Suess}, Katherine A. and {Brammer}, Gabriel and {Whitaker}, Katherine E. and {Illingworth}, Garth and {Bouwens}, Rychard and {Tacchella}, Sandro and {Matthee}, Jorryt and {Allen}, Natalie and {Bezanson}, Rachel and {Conroy}, Charlie and {Labbe}, Ivo and {Leja}, Joel and {Leonova}, Ecaterina and {Magee}, Dan and {Price}, Sedona H. and {Setton}, David J. and {Strait}, Victoria and {Stefanon}, Mauro and {Toft}, Sune and {Weaver}, John R. and {Weibel}, Andrea},
        title = "{Two Remarkably Luminous Galaxy Candidates at z {\ensuremath{\approx}} 10-12 Revealed by JWST}",
      journal = {\apjl},
         year = 2022,
        month = nov,
       volume = {940},
       number = {1},
          eid = {L14},
        pages = {L14},
          doi = {10.3847/2041-8213/ac9b22},
archivePrefix = {arXiv},
       eprint = {2207.09434},
 primaryClass = {astro-ph.GA},
       adsurl = {https://ui.adsabs.harvard.edu/abs/2022ApJ...940L..14N}
}

@ARTICLE{Adams2023,
       author = {{Adams}, N.~J. and {Conselice}, C.~J. and {Ferreira}, L. and {Austin}, D. and {Trussler}, J.~A.~A. and {Juod{\v{z}}balis}, I. and {Wilkins}, S.~M. and {Caruana}, J. and {Dayal}, P. and {Verma}, A. and {Vijayan}, A.~P.},
        title = "{Discovery and properties of ultra-high redshift galaxies (9 < z < 12) in the JWST ERO SMACS 0723 Field}",
      journal = {\mnras},
         year = 2023,
        month = jan,
       volume = {518},
       number = {3},
        pages = {4755-4766},
          doi = {10.1093/mnras/stac3347},
archivePrefix = {arXiv},
       eprint = {2207.11217},
 primaryClass = {astro-ph.GA},
       adsurl = {https://ui.adsabs.harvard.edu/abs/2023MNRAS.518.4755A}
}

@ARTICLE{Finkelstein2023,
       author = {{Finkelstein}, Steven L. and {Bagley}, Micaela B. and {Ferguson}, Henry C. and {Wilkins}, Stephen M. and {Kartaltepe}, Jeyhan S. and {Papovich}, Casey and {Yung}, L.~Y. Aaron and {Arrabal Haro}, Pablo and {Behroozi}, Peter and {Dickinson}, Mark and {Kocevski}, Dale D. and {Koekemoer}, Anton M. and {Larson}, Rebecca L. and {Le Bail}, Aur{\'e}lien and {Morales}, Alexa M. and {P{\'e}rez-Gonz{\'a}lez}, Pablo G. and {Burgarella}, Denis and {Dav{\'e}}, Romeel and {Hirschmann}, Michaela and {Somerville}, Rachel S. and {Wuyts}, Stijn and {Bromm}, Volker and {Casey}, Caitlin M. and {Fontana}, Adriano and {Fujimoto}, Seiji and {Gardner}, Jonathan P. and {Giavalisco}, Mauro and {Grazian}, Andrea and {Grogin}, Norman A. and {Hathi}, Nimish P. and {Hutchison}, Taylor A. and {Jha}, Saurabh W. and {Jogee}, Shardha and {Kewley}, Lisa J. and {Kirkpatrick}, Allison and {Long}, Arianna S. and {Lotz}, Jennifer M. and {Pentericci}, Laura and {Pierel}, Justin D.~R. and {Pirzkal}, Nor and {Ravindranath}, Swara and {Ryan}, Russell E. and {Trump}, Jonathan R. and {Yang}, Guang and {Bhatawdekar}, Rachana and {Bisigello}, Laura and {Buat}, V{\'e}ronique and {Calabr{\`o}}, Antonello and {Castellano}, Marco and {Cleri}, Nikko J. and {Cooper}, M.~C. and {Croton}, Darren and {Daddi}, Emanuele and {Dekel}, Avishai and {Elbaz}, David and {Franco}, Maximilien and {Gawiser}, Eric and {Holwerda}, Benne W. and {Huertas-Company}, Marc and {Jaskot}, Anne E. and {Leung}, Gene C.~K. and {Lucas}, Ray A. and {Mobasher}, Bahram and {Pandya}, Viraj and {Tacchella}, Sandro and {Weiner}, Benjamin J. and {Zavala}, Jorge A.},
        title = "{CEERS Key Paper. I. An Early Look into the First 500 Myr of Galaxy Formation with JWST}",
      journal = {\apjl},
         year = 2023,
        month = mar,
       volume = {946},
       number = {1},
          eid = {L13},
        pages = {L13},
          doi = {10.3847/2041-8213/acade4},
archivePrefix = {arXiv},
       eprint = {2211.05792},
 primaryClass = {astro-ph.GA},
       adsurl = {https://ui.adsabs.harvard.edu/abs/2023ApJ...946L..13F}
}

@ARTICLE{Harikane2023a,
       author = {{Harikane}, Yuichi and {Ouchi}, Masami and {Oguri}, Masamune and {Ono}, Yoshiaki and {Nakajima}, Kimihiko and {Isobe}, Yuki and {Umeda}, Hiroya and {Mawatari}, Ken and {Zhang}, Yechi},
        title = "{A Comprehensive Study of Galaxies at z   9-16 Found in the Early JWST Data: Ultraviolet Luminosity Functions and Cosmic Star Formation History at the Pre-reionization Epoch}",
      journal = {\apjs},
         year = 2023,
        month = mar,
       volume = {265},
       number = {1},
          eid = {5},
        pages = {5},
          doi = {10.3847/1538-4365/acaaa9},
archivePrefix = {arXiv},
       eprint = {2208.01612},
 primaryClass = {astro-ph.GA},
       adsurl = {https://ui.adsabs.harvard.edu/abs/2023ApJS..265....5H}
}

@ARTICLE{Donnan2024,
       author = {{Donnan}, C.~T. and {McLure}, R.~J. and {Dunlop}, J.~S. and {McLeod}, D.~J. and {Magee}, D. and {Arellano-C{\'o}rdova}, K.~Z. and {Barrufet}, L. and {Begley}, R. and {Bowler}, R.~A.~A. and {Carnall}, A.~C. and {Cullen}, F. and {Ellis}, R.~S. and {Fontana}, A. and {Illingworth}, G.~D. and {Grogin}, N.~A. and {Hamadouche}, M.~L. and {Koekemoer}, A.~M. and {Liu}, F. -Y. and {Mason}, C. and {Santini}, P. and {Stanton}, T.~M.},
        title = "{JWST PRIMER: a new multifield determination of the evolving galaxy UV luminosity function at redshifts z ≃ 9 - 15}",
      journal = {\mnras},
         year = 2024,
        month = sep,
       volume = {533},
       number = {3},
        pages = {3222-3237},
          doi = {10.1093/mnras/stae2037},
archivePrefix = {arXiv},
       eprint = {2403.03171},
 primaryClass = {astro-ph.GA},
       adsurl = {https://ui.adsabs.harvard.edu/abs/2024MNRAS.533.3222D}
}

@ARTICLE{Robertson2024,
       author = {{Robertson}, Brant and {Johnson}, Benjamin D. and {Tacchella}, Sandro and {Eisenstein}, Daniel J. and {Hainline}, Kevin and {Arribas}, Santiago and {Baker}, William M. and {Bunker}, Andrew J. and {Carniani}, Stefano and {Cargile}, Phillip A. and {Carreira}, Courtney and {Charlot}, Stephane and {Chevallard}, Jacopo and {Curti}, Mirko and {Curtis-Lake}, Emma and {D'Eugenio}, Francesco and {Egami}, Eiichi and {Hausen}, Ryan and {Helton}, Jakob M. and {Jakobsen}, Peter and {Ji}, Zhiyuan and {Jones}, Gareth C. and {Maiolino}, Roberto and {Maseda}, Michael V. and {Nelson}, Erica and {P{\'e}rez-Gonz{\'a}lez}, Pablo G. and {Pusk{\'a}s}, D{\'a}vid and {Rieke}, Marcia and {Smit}, Renske and {Sun}, Fengwu and {{\"U}bler}, Hannah and {Whitler}, Lily and {Williams}, Christina C. and {Willmer}, Christopher N.~A. and {Willott}, Chris and {Witstok}, Joris},
        title = "{Earliest Galaxies in the JADES Origins Field: Luminosity Function and Cosmic Star Formation Rate Density 300 Myr after the Big Bang}",
      journal = {\apj},
         year = 2024,
        month = jul,
       volume = {970},
       number = {1},
          eid = {31},
        pages = {31},
          doi = {10.3847/1538-4357/ad463d},
archivePrefix = {arXiv},
       eprint = {2312.10033},
 primaryClass = {astro-ph.GA},
       adsurl = {https://ui.adsabs.harvard.edu/abs/2024ApJ...970...31R}
}

@ARTICLE{Whitler2025,
       author = {{Whitler}, Lily and {Stark}, Daniel P. and {Topping}, Michael W. and {Robertson}, Brant and {Rieke}, Marcia and {Hainline}, Kevin N. and {Endsley}, Ryan and {Chen}, Zuyi and {Baker}, William M. and {Bhatawdekar}, Rachana and {Bunker}, Andrew J. and {Carniani}, Stefano and {Charlot}, St{\'e}phane and {Chevallard}, Jacopo and {Curtis-Lake}, Emma and {Egami}, Eiichi and {Eisenstein}, Daniel J. and {Helton}, Jakob M. and {Ji}, Zhiyuan and {Johnson}, Benjamin D. and {P{\'e}rez-Gonz{\'a}lez}, Pablo G. and {Rinaldi}, Pierluigi and {Tacchella}, Sandro and {Williams}, Christina C. and {Willmer}, Christopher N.~A. and {Willott}, Chris and {Witstok}, Joris},
        title = "{The z {\ensuremath{\gtrsim}} 9 Galaxy UV Luminosity Function from the JWST Advanced Deep Extragalactic Survey: Insights into Early Galaxy Evolution and Reionization}",
      journal = {\apj},
         year = 2025,
        month = oct,
       volume = {992},
       number = {1},
          eid = {63},
        pages = {63},
          doi = {10.3847/1538-4357/adfddc},
archivePrefix = {arXiv},
       eprint = {2501.00984},
 primaryClass = {astro-ph.GA},
       adsurl = {https://ui.adsabs.harvard.edu/abs/2025ApJ...992...63W}
}

@ARTICLE{Dekel2023,
       author = {{Dekel}, Avishai and {Sarkar}, Kartick C. and {Birnboim}, Yuval and {Mandelker}, Nir and {Li}, Zhaozhou},
        title = "{Efficient formation of massive galaxies at cosmic dawn by feedback-free starbursts}",
      journal = {\mnras},
         year = 2023,
        month = aug,
       volume = {523},
       number = {3},
        pages = {3201-3218},
          doi = {10.1093/mnras/stad1557},
archivePrefix = {arXiv},
       eprint = {2303.04827},
 primaryClass = {astro-ph.GA},
       adsurl = {https://ui.adsabs.harvard.edu/abs/2023MNRAS.523.3201D}
}

@ARTICLE{Qin2023,
       author = {{Qin}, Yuxiang and {Balu}, Sreedhar and {Wyithe}, J. Stuart B.},
        title = "{Implications of z {\ensuremath{\gtrsim}} 12 JWST galaxies for galaxy formation at high redshift}",
      journal = {\mnras},
         year = 2023,
        month = nov,
       volume = {526},
       number = {1},
        pages = {1324-1342},
          doi = {10.1093/mnras/stad2448},
archivePrefix = {arXiv},
       eprint = {2305.17959},
 primaryClass = {astro-ph.GA},
       adsurl = {https://ui.adsabs.harvard.edu/abs/2023MNRAS.526.1324Q}
}

@ARTICLE{Feldmann2025,
       author = {{Feldmann}, Robert and {Boylan-Kolchin}, Michael and {Bullock}, James S. and {{\c{C}}atmabacak}, Onur and {Faucher-Gigu{\`e}re}, Claude-Andr{\'e} and {Hayward}, Christopher C. and {Kere{\v{s}}}, Du{\v{s}}an and {Lazar}, Alexandres and {Liang}, Lichen and {Moreno}, Jorge and {Oesch}, Pascal A. and {Quataert}, Eliot and {Shen}, Xuejian and {Sun}, Guochao},
        title = "{Elevated UV luminosity density at Cosmic Dawn explained by non-evolving, weakly mass-dependent star formation efficiency}",
      journal = {\mnras},
         year = 2025,
        month = jan,
       volume = {536},
       number = {1},
        pages = {988-1016},
          doi = {10.1093/mnras/stae2633},
archivePrefix = {arXiv},
       eprint = {2407.02674},
 primaryClass = {astro-ph.CO},
       adsurl = {https://ui.adsabs.harvard.edu/abs/2025MNRAS.536..988F}
}

@ARTICLE{Somerville2025,
       author = {{Somerville}, Rachel S. and {Yung}, L.~Y. Aaron and {Lancaster}, Lachlan and {Menon}, Shyam and {Sommovigo}, Laura and {Finkelstein}, Steven L.},
        title = "{Density-modulated star formation efficiency: implications for the observed abundance of ultraviolet luminous galaxies at z > 10}",
      journal = {\mnras},
         year = 2025,
        month = dec,
       volume = {544},
       number = {4},
        pages = {3774-3798},
          doi = {10.1093/mnras/staf1824},
archivePrefix = {arXiv},
       eprint = {2505.05442},
 primaryClass = {astro-ph.GA},
       adsurl = {https://ui.adsabs.harvard.edu/abs/2025MNRAS.544.3774S}
}

@ARTICLE{Mason2023,
       author = {{Mason}, Charlotte A. and {Trenti}, Michele and {Treu}, Tommaso},
        title = "{The brightest galaxies at cosmic dawn}",
      journal = {\mnras},
         year = 2023,
        month = may,
       volume = {521},
       number = {1},
        pages = {497-503},
          doi = {10.1093/mnras/stad035},
archivePrefix = {arXiv},
       eprint = {2207.14808},
 primaryClass = {astro-ph.GA},
       adsurl = {https://ui.adsabs.harvard.edu/abs/2023MNRAS.521..497M}
}

@ARTICLE{Mirocha2023,
       author = {{Mirocha}, Jordan and {Furlanetto}, Steven R.},
        title = "{Balancing the efficiency and stochasticity of star formation with dust extinction in z {\ensuremath{\gtrsim}} 10 galaxies observed by JWST}",
      journal = {\mnras},
         year = 2023,
        month = feb,
       volume = {519},
       number = {1},
        pages = {843-853},
          doi = {10.1093/mnras/stac3578},
archivePrefix = {arXiv},
       eprint = {2208.12826},
 primaryClass = {astro-ph.GA},
       adsurl = {https://ui.adsabs.harvard.edu/abs/2023MNRAS.519..843M}
}

@ARTICLE{Shen2023,
       author = {{Shen}, Xuejian and {Vogelsberger}, Mark and {Boylan-Kolchin}, Michael and {Tacchella}, Sandro and {Kannan}, Rahul},
        title = "{The impact of UV variability on the abundance of bright galaxies at z {\ensuremath{\geq}} 9}",
      journal = {\mnras},
         year = 2023,
        month = nov,
       volume = {525},
       number = {3},
        pages = {3254-3261},
          doi = {10.1093/mnras/stad2508},
archivePrefix = {arXiv},
       eprint = {2305.05679},
 primaryClass = {astro-ph.GA},
       adsurl = {https://ui.adsabs.harvard.edu/abs/2023MNRAS.525.3254S}
}

@ARTICLE{Kravtsov2024,
       author = {{Kravtsov}, Andrey and {Belokurov}, Vasily},
        title = "{Stochastic star formation and the abundance of $z>10$ UV-bright galaxies}",
      journal = {arXiv e-prints},
         year = 2024,
        month = may,
          eid = {arXiv:2405.04578},
        pages = {arXiv:2405.04578},
          doi = {10.48550/arXiv.2405.04578},
archivePrefix = {arXiv},
       eprint = {2405.04578},
 primaryClass = {astro-ph.GA},
       adsurl = {https://ui.adsabs.harvard.edu/abs/2024arXiv240504578K}
}

@ARTICLE{Ferrara2025,
       author = {{Ferrara}, A. and {Pallottini}, A. and {Sommovigo}, L.},
        title = "{Blue monsters at z > 10: Where all their dust has gone}",
      journal = {\aap},
         year = 2025,
        month = feb,
       volume = {694},
          eid = {A286},
        pages = {A286},
          doi = {10.1051/0004-6361/202452707},
archivePrefix = {arXiv},
       eprint = {2410.19042},
 primaryClass = {astro-ph.GA},
       adsurl = {https://ui.adsabs.harvard.edu/abs/2025A&A...694A.286F}
}

@ARTICLE{Chon2022,
       author = {{Chon}, Sunmyon and {Ono}, Haruka and {Omukai}, Kazuyuki and {Schneider}, Raffaella},
        title = "{Impact of the cosmic background radiation on the initial mass function of metal-poor stars}",
      journal = {\mnras},
         year = 2022,
        month = aug,
       volume = {514},
       number = {3},
        pages = {4639-4654},
          doi = {10.1093/mnras/stac1549},
archivePrefix = {arXiv},
       eprint = {2205.15328},
 primaryClass = {astro-ph.GA},
       adsurl = {https://ui.adsabs.harvard.edu/abs/2022MNRAS.514.4639C}
}

@ARTICLE{Harikane2023b,
       author = {{Harikane}, Yuichi and {Zhang}, Yechi and {Nakajima}, Kimihiko and {Ouchi}, Masami and {Isobe}, Yuki and {Ono}, Yoshiaki and {Hatano}, Shun and {Xu}, Yi and {Umeda}, Hiroya},
        title = "{A JWST/NIRSpec First Census of Broad-line AGNs at z = 4-7: Detection of 10 Faint AGNs with M $_{BH}$ {}10$^{6}$-{}10$^{8}$ M $_{☉}$ and Their Host Galaxy Properties}",
      journal = {\apj},
         year = 2023,
        month = dec,
       volume = {959},
       number = {1},
          eid = {39},
        pages = {39},
          doi = {10.3847/1538-4357/ad029e},
archivePrefix = {arXiv},
       eprint = {2303.11946},
 primaryClass = {astro-ph.GA},
       adsurl = {https://ui.adsabs.harvard.edu/abs/2023ApJ...959...39H}
}

@ARTICLE{Hegde2024,
       author = {{Hegde}, Sahil and {Wyatt}, Michael M. and {Furlanetto}, Steven R.},
        title = "{A hidden population of active galactic nuclei can explain the overabundance of luminous z > 10 objects observed by JWST}",
      journal = {\jcap},
         year = 2024,
        month = aug,
       volume = {2024},
       number = {8},
          eid = {025},
        pages = {025},
          doi = {10.1088/1475-7516/2024/08/025},
archivePrefix = {arXiv},
       eprint = {2405.01629},
 primaryClass = {astro-ph.GA},
       adsurl = {https://ui.adsabs.harvard.edu/abs/2024JCAP...08..025H}
}

@ARTICLE{Steinhardt2023,
       author = {{Steinhardt}, Charles L. and {Kokorev}, Vasily and {Rusakov}, Vadim and {Garcia}, Ethan and {Sneppen}, Albert},
        title = "{Templates for Fitting Photometry of Ultra-high-redshift Galaxies}",
      journal = {\apjl},
         year = 2023,
        month = jul,
       volume = {951},
       number = {2},
          eid = {L40},
        pages = {L40},
          doi = {10.3847/2041-8213/acdef6},
archivePrefix = {arXiv},
       eprint = {2208.07879},
 primaryClass = {astro-ph.GA},
       adsurl = {https://ui.adsabs.harvard.edu/abs/2023ApJ...951L..40S}
}

@ARTICLE{Ventura2024,
       author = {{Ventura}, Emanuele M. and {Qin}, Yuxiang and {Balu}, Sreedhar and {Wyithe}, J. Stuart B.},
        title = "{Semi-analytic modelling of Pop. III star formation and metallicity evolution - I. Impact on the UV luminosity functions at z = 9-16}",
      journal = {\mnras},
         year = 2024,
        month = mar,
       volume = {529},
       number = {1},
        pages = {628-646},
          doi = {10.1093/mnras/stae567},
archivePrefix = {arXiv},
       eprint = {2401.07396},
 primaryClass = {astro-ph.GA},
       adsurl = {https://ui.adsabs.harvard.edu/abs/2024MNRAS.529..628V}
}

@ARTICLE{Isobe2023,
       author = {{Isobe}, Yuki and {Ouchi}, Masami and {Tominaga}, Nozomu and {Watanabe}, Kuria and {Nakajima}, Kimihiko and {Umeda}, Hiroya and {Yajima}, Hidenobu and {Harikane}, Yuichi and {Fukushima}, Hajime and {Xu}, Yi and {Ono}, Yoshiaki and {Zhang}, Yechi},
        title = "{JWST Identification of Extremely Low C/N Galaxies with [N/O] {\ensuremath{\gtrsim}} 0.5 at z 6-10 Evidencing the Early CNO-cycle Enrichment and a Connection with Globular Cluster Formation}",
      journal = {\apj},
         year = 2023,
        month = dec,
       volume = {959},
       number = {2},
          eid = {100},
        pages = {100},
          doi = {10.3847/1538-4357/ad09be},
archivePrefix = {arXiv},
       eprint = {2307.00710},
 primaryClass = {astro-ph.GA},
       adsurl = {https://ui.adsabs.harvard.edu/abs/2023ApJ...959..100I}
}

@ARTICLE{Senchyna2024,
       author = {{Senchyna}, Peter and {Plat}, Adele and {Stark}, Daniel P. and {Rudie}, Gwen C. and {Berg}, Danielle and {Charlot}, St{\'e}phane and {James}, Bethan L. and {Mingozzi}, Matilde},
        title = "{GN-z11 in Context: Possible Signatures of Globular Cluster Precursors at Redshift 10}",
      journal = {\apj},
         year = 2024,
        month = may,
       volume = {966},
       number = {1},
          eid = {92},
        pages = {92},
          doi = {10.3847/1538-4357/ad235e},
archivePrefix = {arXiv},
       eprint = {2303.04179},
 primaryClass = {astro-ph.GA},
       adsurl = {https://ui.adsabs.harvard.edu/abs/2024ApJ...966...92S}
}

@ARTICLE{Maiolino2024,
       author = {{Maiolino}, Roberto and {Scholtz}, Jan and {Witstok}, Joris and {Carniani}, Stefano and {D'Eugenio}, Francesco and {de Graaff}, Anna and {{\"U}bler}, Hannah and {Tacchella}, Sandro and {Curtis-Lake}, Emma and {Arribas}, Santiago and {Bunker}, Andrew and {Charlot}, St{\'e}phane and {Chevallard}, Jacopo and {Curti}, Mirko and {Looser}, Tobias J. and {Maseda}, Michael V. and {Rawle}, Timothy D. and {Rodr{\'\i}guez del Pino}, Bruno and {Willott}, Chris J. and {Egami}, Eiichi and {Eisenstein}, Daniel J. and {Hainline}, Kevin N. and {Robertson}, Brant and {Williams}, Christina C. and {Willmer}, Christopher N.~A. and {Baker}, William M. and {Boyett}, Kristan and {DeCoursey}, Christa and {Fabian}, Andrew C. and {Helton}, Jakob M. and {Ji}, Zhiyuan and {Jones}, Gareth C. and {Kumari}, Nimisha and {Laporte}, Nicolas and {Nelson}, Erica J. and {Perna}, Michele and {Sandles}, Lester and {Shivaei}, Irene and {Sun}, Fengwu},
        title = "{A small and vigorous black hole in the early Universe}",
      journal = {\nat},
         year = 2024,
        month = mar,
       volume = {627},
       number = {8002},
        pages = {59-63},
          doi = {10.1038/s41586-024-07052-5},
archivePrefix = {arXiv},
       eprint = {2305.12492},
 primaryClass = {astro-ph.GA},
       adsurl = {https://ui.adsabs.harvard.edu/abs/2024Natur.627...59M}
}

@ARTICLE{Alvarez-Marquez2025,
       author = {{{\'A}lvarez-M{\'a}rquez}, J. and {Crespo G{\'o}mez}, A. and {Colina}, L. and {Langeroodi}, D. and {Marques-Chaves}, R. and {Prieto-Jim{\'e}nez}, C. and {Bik}, A. and {Alonso-Herrero}, A. and {Boogaard}, L. and {Costantin}, L. and {Garc{\'\i}a-Mar{\'\i}n}, M. and {Gillman}, S. and {Hjorth}, J. and {Iani}, E. and {Jermann}, I. and {Labiano}, A. and {Melinder}, J. and {Meyer}, R. and {{\"O}stlin}, G. and {P{\'e}rez-Gonz{\'a}lez}, P.~G. and {Rinaldi}, P. and {Walter}, F. and {van der Werf}, P. and {Wright}, G.},
        title = "{Insight into the starburst nature of Galaxy GN-z11 with JWST MIRI spectroscopy}",
      journal = {\aap},
         year = 2025,
        month = mar,
       volume = {695},
          eid = {A250},
        pages = {A250},
          doi = {10.1051/0004-6361/202451731},
archivePrefix = {arXiv},
       eprint = {2412.12826},
 primaryClass = {astro-ph.GA},
       adsurl = {https://ui.adsabs.harvard.edu/abs/2025A&A...695A.250A}
}

@ARTICLE{Asplund2021,
       author = {{Asplund}, M. and {Amarsi}, A.~M. and {Grevesse}, N.},
        title = "{The chemical make-up of the Sun: A 2020 vision}",
      journal = {\aap},
         year = 2021,
        month = sep,
       volume = {653},
          eid = {A141},
        pages = {A141},
          doi = {10.1051/0004-6361/202140445},
archivePrefix = {arXiv},
       eprint = {2105.01661},
 primaryClass = {astro-ph.SR},
       adsurl = {https://ui.adsabs.harvard.edu/abs/2021A&A...653A.141A}
}

@ARTICLE{FroeseFischer1985,
       author = {{Froese Fischer}, Charlotte and {Saha}, H.~P.},
        title = "{Multiconfiguration Hartree-Fock Results with Breit-Pauli Corrections for Transitions in the Carbon Sequence}",
      journal = {\physscr},
         year = 1985,
        month = sep,
       volume = {32},
       number = {3},
        pages = {181-194},
          doi = {10.1088/0031-8949/32/3/004},
       adsurl = {https://ui.adsabs.harvard.edu/abs/1985PhyS...32..181F}
}

@ARTICLE{Luridiana2015,
       author = {{Luridiana}, V. and {Morisset}, C. and {Shaw}, R.~A.},
        title = "{PyNeb: a new tool for analyzing emission lines. I. Code description and validation of results}",
      journal = {\aap},
         year = 2015,
        month = jan,
       volume = {573},
          eid = {A42},
        pages = {A42},
          doi = {10.1051/0004-6361/201323152},
archivePrefix = {arXiv},
       eprint = {1410.6662},
 primaryClass = {astro-ph.IM},
       adsurl = {https://ui.adsabs.harvard.edu/abs/2015A&A...573A..42L}
}

@BOOK{Osterbrock2006,
       author = {{Osterbrock}, Donald E. and {Ferland}, Gary J.},
        title = "{Astrophysics of gaseous nebulae and active galactic nuclei}",
         year = 2006,
       adsurl = {https://ui.adsabs.harvard.edu/abs/2006agna.book.....O}
}

@ARTICLE{Izotov2006,
       author = {{Izotov}, Y.~I. and {Stasi{\'n}ska}, G. and {Meynet}, G. and {Guseva}, N.~G. and {Thuan}, T.~X.},
        title = "{The chemical composition of metal-poor emission-line galaxies in the Data Release 3 of the Sloan Digital Sky Survey}",
      journal = {\aap},
         year = 2006,
        month = mar,
       volume = {448},
       number = {3},
        pages = {955-970},
          doi = {10.1051/0004-6361:20053763},
archivePrefix = {arXiv},
       eprint = {astro-ph/0511644},
 primaryClass = {astro-ph},
       adsurl = {https://ui.adsabs.harvard.edu/abs/2006A&A...448..955I}
}

@ARTICLE{Campbell1986,
       author = {{Campbell}, Alison and {Terlevich}, Roberto and {Melnick}, Jorge},
        title = "{The stellar populations and evolution of H II galaxies - I. High signal-to-noise optical spectroscopy.}",
      journal = {\mnras},
         year = 1986,
        month = dec,
       volume = {223},
        pages = {811-825},
          doi = {10.1093/mnras/223.4.811},
       adsurl = {https://ui.adsabs.harvard.edu/abs/1986MNRAS.223..811C}
}

@ARTICLE{Sanders2024,
       author = {{Sanders}, Ryan L. and {Shapley}, Alice E. and {Topping}, Michael W. and {Reddy}, Naveen A. and {Brammer}, Gabriel B.},
        title = "{Direct T $_{e}$-based Metallicities of z = 2{\textendash}9 Galaxies with JWST/NIRSpec: Empirical Metallicity Calibrations Applicable from Reionization to Cosmic Noon}",
      journal = {\apj},
         year = 2024,
        month = feb,
       volume = {962},
       number = {1},
          eid = {24},
        pages = {24},
          doi = {10.3847/1538-4357/ad15fc},
archivePrefix = {arXiv},
       eprint = {2303.08149},
 primaryClass = {astro-ph.GA},
       adsurl = {https://ui.adsabs.harvard.edu/abs/2024ApJ...962...24S}
}

@ARTICLE{Topping2025b,
       author = {{Topping}, Michael W. and {Sanders}, Ryan L. and {Shapley}, Alice E. and {Pahl}, Anthony J. and {Reddy}, Naveen A. and {Stark}, Daniel P. and {Berg}, Danielle A. and {Clarke}, Leonardo and {Cullen}, Fergus and {Dunlop}, James S. and {Ellis}, Richard S. and {Schreiber}, N.~M. F{\"o}rster and {Illingworth}, Garth D. and {Jones}, Tucker and {Narayanan}, Desika and {Pettini}, Max and {Schaerer}, Daniel},
        title = "{The AURORA survey: the evolution of multiphase electron densities at high redshift}",
      journal = {\mnras},
         year = 2025,
        month = aug,
       volume = {541},
       number = {2},
        pages = {1707-1721},
          doi = {10.1093/mnras/staf903},
archivePrefix = {arXiv},
       eprint = {2502.08712},
 primaryClass = {astro-ph.GA},
       adsurl = {https://ui.adsabs.harvard.edu/abs/2025MNRAS.541.1707T}
}

@ARTICLE{Martinez2025,
       author = {{Martinez}, Zorayda and {Berg}, Danielle A. and {James}, Bethan L. and {Arellano-C{\'o}rdova}, Karla Z. and {Stark}, Daniel P. and {Senchyna}, Peter and {Skillman}, Evan D. and {Rogers}, Noah S.~J. and {Chisholm}, John},
        title = "{Under Pressure: Decoding the Effect of High Densities on Derived Nebular Properties}",
      journal = {\apj},
         year = 2025,
        month = dec,
       volume = {995},
       number = {2},
          eid = {204},
        pages = {204},
          doi = {10.3847/1538-4357/ae17c6},
archivePrefix = {arXiv},
       eprint = {2510.21960},
 primaryClass = {astro-ph.GA},
       adsurl = {https://ui.adsabs.harvard.edu/abs/2025ApJ...995..204M}
}

@ARTICLE{Topping2024,
       author = {{Topping}, Michael W. and {Stark}, Daniel P. and {Senchyna}, Peter and {Plat}, Adele and {Zitrin}, Adi and {Endsley}, Ryan and {Charlot}, St{\'e}phane and {Furtak}, Lukas J. and {Maseda}, Michael V. and {Smit}, Renske and {Mainali}, Ramesh and {Chevallard}, Jacopo and {Molyneux}, Stephen and {Rigby}, Jane R.},
        title = "{Metal-poor star formation at z > 6 with JWST: new insight into hard radiation fields and nitrogen enrichment on 20 pc scales}",
      journal = {\mnras},
         year = 2024,
        month = apr,
       volume = {529},
       number = {4},
        pages = {3301-3322},
          doi = {10.1093/mnras/stae682},
archivePrefix = {arXiv},
       eprint = {2401.08764},
 primaryClass = {astro-ph.GA},
       adsurl = {https://ui.adsabs.harvard.edu/abs/2024MNRAS.529.3301T}
}

@ARTICLE{Larson2023,
       author = {{Larson}, Rebecca L. and {Finkelstein}, Steven L. and {Kocevski}, Dale D. and {Hutchison}, Taylor A. and {Trump}, Jonathan R. and {Arrabal Haro}, Pablo and {Bromm}, Volker and {Cleri}, Nikko J. and {Dickinson}, Mark and {Fujimoto}, Seiji and {Kartaltepe}, Jeyhan S. and {Koekemoer}, Anton M. and {Papovich}, Casey and {Pirzkal}, Nor and {Tacchella}, Sandro and {Zavala}, Jorge A. and {Bagley}, Micaela and {Behroozi}, Peter and {Champagne}, Jaclyn B. and {Cole}, Justin W. and {Jung}, Intae and {Morales}, Alexa M. and {Yang}, Guang and {Zhang}, Haowen and {Zitrin}, Adi and {Amor{\'\i}n}, Ricardo O. and {Burgarella}, Denis and {Casey}, Caitlin M. and {Ch{\'a}vez Ortiz}, {\'O}scar A. and {Cox}, Isabella G. and {Chworowsky}, Katherine and {Fontana}, Adriano and {Gawiser}, Eric and {Grazian}, Andrea and {Grogin}, Norman A. and {Harish}, Santosh and {Hathi}, Nimish P. and {Hirschmann}, Michaela and {Holwerda}, Benne W. and {Juneau}, St{\'e}phanie and {Leung}, Gene C.~K. and {Lucas}, Ray A. and {McGrath}, Elizabeth J. and {P{\'e}rez-Gonz{\'a}lez}, Pablo G. and {Rigby}, Jane R. and {Seill{\'e}}, Lise-Marie and {Simons}, Raymond C. and {de La Vega}, Alexander and {Weiner}, Benjamin J. and {Wilkins}, Stephen M. and {Yung}, L.~Y. Aaron and {Ceers Team}},
        title = "{A CEERS Discovery of an Accreting Supermassive Black Hole 570 Myr after the Big Bang: Identifying a Progenitor of Massive z > 6 Quasars}",
      journal = {\apjl},
         year = 2023,
        month = aug,
       volume = {953},
       number = {2},
          eid = {L29},
        pages = {L29},
          doi = {10.3847/2041-8213/ace619},
archivePrefix = {arXiv},
       eprint = {2303.08918},
 primaryClass = {astro-ph.GA},
       adsurl = {https://ui.adsabs.harvard.edu/abs/2023ApJ...953L..29L}
}

@ARTICLE{Zamora2025,
       author = {{Zamora}, Sandra and {Carniani}, Stefano and {Bertola}, Elena and {Parlanti}, Eleonora and {P{\'e}rez-Gonz{\'a}lez}, Pablo G. and {Arribas}, Santiago and {B{\"o}ker}, Torsten and {Bunker}, Andrew J. and {D'Eugenio}, Francesco and {Maiolino}, Roberto and {Perna}, Michele and {Rodr{\'\i}guez Del Pino}, Bruno and {{\"U}bler}, Hannah and {Cresci}, Giovanni and {Jones}, Gareth C. and {Lamperti}, Isabella and {Scholtz}, Jan and {Trefoloni}, Bartolomeo and {Venturi}, Giacomo},
        title = "{GA-NIFS: Understanding the ionization nature of EGSY8p7/CEERS-1019. Evidence for a star formation-driven outflow at z = 8.6}",
      journal = {arXiv e-prints},
         year = 2025,
        month = dec,
          eid = {arXiv:2512.09022},
        pages = {arXiv:2512.09022},
          doi = {10.48550/arXiv.2512.09022},
archivePrefix = {arXiv},
       eprint = {2512.09022},
 primaryClass = {astro-ph.GA},
       adsurl = {https://ui.adsabs.harvard.edu/abs/2025arXiv251209022Z}
}

@ARTICLE{Tang2025,
       author = {{Tang}, Mengtao and {Stark}, Daniel P. and {Plat}, Ad{\`e}le and {Feltre}, Anna and {Katz}, Harley and {Senchyna}, Peter and {Mason}, Charlotte A. and {Whitler}, Lily and {Chen}, Zuyi and {Topping}, Michael W.},
        title = "{JWST/NIRSpec Observations of High-ionization Emission Lines in Galaxies at High Redshift}",
      journal = {\apj},
         year = 2025,
        month = oct,
       volume = {991},
       number = {2},
          eid = {217},
        pages = {217},
          doi = {10.3847/1538-4357/adfd57},
archivePrefix = {arXiv},
       eprint = {2505.06359},
 primaryClass = {astro-ph.GA},
       adsurl = {https://ui.adsabs.harvard.edu/abs/2025ApJ...991..217T}
}

@ARTICLE{Whitler2026,
       author = {{Whitler}, Lily and {Stark}, Daniel P. and {Mason}, Charlotte A. and {Tang}, Mengtao and {Chen}, Zuyi and {Lu}, Ting-Yi and {Prieto-Lyon}, Gonzalo and {Hutter}, Anne},
        title = "{Deep JWST spectroscopy of galaxies in a candidate ionized bubble at z = 8.7: probing reionization at pMpc scales with Ly {\ensuremath{\alpha}} emission}",
      journal = {\mnras},
         year = 2026,
        month = may,
       volume = {548},
       number = {2},
          eid = {stag639},
        pages = {stag639},
          doi = {10.1093/mnras/stag639},
archivePrefix = {arXiv},
       eprint = {2510.12019},
 primaryClass = {astro-ph.GA},
       adsurl = {https://ui.adsabs.harvard.edu/abs/2026MNRAS.548ag639W}
}

@ARTICLE{Witstok2025,
       author = {{Witstok}, Joris and {Jakobsen}, Peter and {Maiolino}, Roberto and {Helton}, Jakob M. and {Johnson}, Benjamin D. and {Robertson}, Brant E. and {Tacchella}, Sandro and {Cameron}, Alex J. and {Smit}, Renske and {Bunker}, Andrew J. and {Saxena}, Aayush and {Sun}, Fengwu and {Alberts}, Stacey and {Arribas}, Santiago and {Baker}, William M. and {Bhatawdekar}, Rachana and {Boyett}, Kristan and {Cargile}, Phillip A. and {Carniani}, Stefano and {Charlot}, St{\'e}phane and {Chevallard}, Jacopo and {Curti}, Mirko and {Curtis-Lake}, Emma and {D'Eugenio}, Francesco and {Eisenstein}, Daniel J. and {Hainline}, Kevin N. and {Jones}, Gareth C. and {Kumari}, Nimisha and {Maseda}, Michael V. and {P{\'e}rez-Gonz{\'a}lez}, Pablo G. and {Rinaldi}, Pierluigi and {Scholtz}, Jan and {{\"U}bler}, Hannah and {Williams}, Christina C. and {Willmer}, Christopher N.~A. and {Willott}, Chris and {Zhu}, Yongda},
        title = "{Witnessing the onset of reionization through Lyman-{\ensuremath{\alpha}} emission at redshift 13}",
      journal = {\nat},
         year = 2025,
        month = mar,
       volume = {639},
       number = {8056},
        pages = {897-901},
          doi = {10.1038/s41586-025-08779-5},
archivePrefix = {arXiv},
       eprint = {2408.16608},
 primaryClass = {astro-ph.GA},
       adsurl = {https://ui.adsabs.harvard.edu/abs/2025Natur.639..897W}
}

@ARTICLE{Tang2024,
       author = {{Tang}, Mengtao and {Stark}, Daniel P. and {Ellis}, Richard S. and {Sun}, Fengwu and {Topping}, Michael and {Robertson}, Brant and {Tacchella}, Sandro and {Arribas}, Santiago and {Baker}, William M. and {Bhatawdekar}, Rachana and {Boyett}, Kristan and {Bunker}, Andrew J. and {Charlot}, St{\'e}phane and {Chen}, Zuyi and {Chevallard}, Jacopo and {Jones}, Gareth C. and {Kumari}, Nimisha and {Lyu}, Jianwei and {Maiolino}, Roberto and {Maseda}, Michael V. and {Saxena}, Aayush and {Whitler}, Lily and {Williams}, Christina C. and {Willott}, Chris and {Witstok}, Joris},
        title = "{Ly{\ensuremath{\alpha}} emission in galaxies at z ≃ 5-6: new insight from JWST into the statistical distributions of Ly{\ensuremath{\alpha}} properties at the end of reionization}",
      journal = {\mnras},
         year = 2024,
        month = jun,
       volume = {531},
       number = {2},
        pages = {2701-2730},
          doi = {10.1093/mnras/stae1338},
archivePrefix = {arXiv},
       eprint = {2402.06070},
 primaryClass = {astro-ph.GA},
       adsurl = {https://ui.adsabs.harvard.edu/abs/2024MNRAS.531.2701T}
}

@ARTICLE{Shapley2003,
       author = {{Shapley}, Alice E. and {Steidel}, Charles C. and {Pettini}, Max and {Adelberger}, Kurt L.},
        title = "{Rest-Frame Ultraviolet Spectra of z\raisebox{-0.5ex}\textasciitilde3 Lyman Break Galaxies}",
      journal = {\apj},
         year = 2003,
        month = may,
       volume = {588},
       number = {1},
        pages = {65-89},
          doi = {10.1086/373922},
archivePrefix = {arXiv},
       eprint = {astro-ph/0301230},
 primaryClass = {astro-ph},
       adsurl = {https://ui.adsabs.harvard.edu/abs/2003ApJ...588...65S}
}

@ARTICLE{Chen2026,
       author = {{Chen}, Zuyi and {Stark}, Daniel P. and {Mason}, Charlotte A. and {Plat}, Adele and {Gelli}, Viola and {Senchyna}, Peter and {Keerthi Vasan G.}, C. and {Endsley}, Ryan and {Tang}, Mengtao and {Topping}, Michael W. and {Whitler}, Lily},
        title = "{SPURS: Bursty Star Formation in an Extremely Luminous Weak Emission Line Galaxy at $z=9.3$}",
      journal = {arXiv e-prints},
         year = 2026,
        month = apr,
          eid = {arXiv:2604.21516},
        pages = {arXiv:2604.21516},
          doi = {10.48550/arXiv.2604.21516},
archivePrefix = {arXiv},
       eprint = {2604.21516},
 primaryClass = {astro-ph.GA},
       adsurl = {https://ui.adsabs.harvard.edu/abs/2026arXiv260421516C}
}

@ARTICLE{Castellano2024,
       author = {{Castellano}, Marco and {Napolitano}, Lorenzo and {Fontana}, Adriano and {Roberts-Borsani}, Guido and {Treu}, Tommaso and {Vanzella}, Eros and {Zavala}, Jorge A. and {Arrabal Haro}, Pablo and {Calabr{\`o}}, Antonello and {Llerena}, Mario and {Mascia}, Sara and {Merlin}, Emiliano and {Paris}, Diego and {Pentericci}, Laura and {Santini}, Paola and {Bakx}, Tom J.~L.~C. and {Bergamini}, Pietro and {Cupani}, Guido and {Dickinson}, Mark and {Filippenko}, Alexei V. and {Glazebrook}, Karl and {Grillo}, Claudio and {Kelly}, Patrick L. and {Malkan}, Matthew A. and {Mason}, Charlotte A. and {Morishita}, Takahiro and {Nanayakkara}, Themiya and {Rosati}, Piero and {Sani}, Eleonora and {Wang}, Xin and {Yoon}, Ilsang},
        title = "{JWST NIRSpec Spectroscopy of the Remarkable Bright Galaxy GHZ2/GLASS-z12 at Redshift 12.34}",
      journal = {\apj},
         year = 2024,
        month = sep,
       volume = {972},
       number = {2},
          eid = {143},
        pages = {143},
          doi = {10.3847/1538-4357/ad5f88},
archivePrefix = {arXiv},
       eprint = {2403.10238},
 primaryClass = {astro-ph.GA},
       adsurl = {https://ui.adsabs.harvard.edu/abs/2024ApJ...972..143C}
}

@ARTICLE{Oesch2016,
       author = {{Oesch}, P.~A. and {Brammer}, G. and {van Dokkum}, P.~G. and {Illingworth}, G.~D. and {Bouwens}, R.~J. and {Labb{\'e}}, I. and {Franx}, M. and {Momcheva}, I. and {Ashby}, M.~L.~N. and {Fazio}, G.~G. and {Gonzalez}, V. and {Holden}, B. and {Magee}, D. and {Skelton}, R.~E. and {Smit}, R. and {Spitler}, L.~R. and {Trenti}, M. and {Willner}, S.~P.},
        title = "{A Remarkably Luminous Galaxy at z=11.1 Measured with Hubble Space Telescope Grism Spectroscopy}",
      journal = {\apj},
         year = 2016,
        month = mar,
       volume = {819},
       number = {2},
          eid = {129},
        pages = {129},
          doi = {10.3847/0004-637X/819/2/129},
archivePrefix = {arXiv},
       eprint = {1603.00461},
 primaryClass = {astro-ph.GA},
       adsurl = {https://ui.adsabs.harvard.edu/abs/2016ApJ...819..129O}
}

@ARTICLE{Tacchella2023,
       author = {{Tacchella}, Sandro and {Eisenstein}, Daniel J. and {Hainline}, Kevin and {Johnson}, Benjamin D. and {Baker}, William M. and {Helton}, Jakob M. and {Robertson}, Brant and {Suess}, Katherine A. and {Chen}, Zuyi and {Nelson}, Erica and {Pusk{\'a}s}, D{\'a}vid and {Sun}, Fengwu and {Alberts}, Stacey and {Egami}, Eiichi and {Hausen}, Ryan and {Rieke}, George and {Rieke}, Marcia and {Shivaei}, Irene and {Williams}, Christina C. and {Willmer}, Christopher N.~A. and {Bunker}, Andrew and {Cameron}, Alex J. and {Carniani}, Stefano and {Charlot}, Stephane and {Curti}, Mirko and {Curtis-Lake}, Emma and {Looser}, Tobias J. and {Maiolino}, Roberto and {Maseda}, Michael V. and {Rawle}, Tim and {Rix}, Hans-Walter and {Smit}, Renske and {{\"U}bler}, Hannah and {Willott}, Chris and {Witstok}, Joris and {Baum}, Stefi and {Bhatawdekar}, Rachana and {Boyett}, Kristan and {Danhaive}, A. Lola and {de Graaff}, Anna and {Endsley}, Ryan and {Ji}, Zhiyuan and {Lyu}, Jianwei and {Sandles}, Lester and {Saxena}, Aayush and {Scholtz}, Jan and {Topping}, Michael W. and {Whitler}, Lily},
        title = "{JADES Imaging of GN-z11: Revealing the Morphology and Environment of a Luminous Galaxy 430 Myr after the Big Bang}",
      journal = {\apj},
         year = 2023,
        month = jul,
       volume = {952},
       number = {1},
          eid = {74},
        pages = {74},
          doi = {10.3847/1538-4357/acdbc6},
archivePrefix = {arXiv},
       eprint = {2302.07234},
 primaryClass = {astro-ph.GA},
       adsurl = {https://ui.adsabs.harvard.edu/abs/2023ApJ...952...74T}
}

@ARTICLE{deGraaff2025,
       author = {{de Graaff}, Anna and {Brammer}, Gabriel and {Weibel}, Andrea and {Lewis}, Zach and {Maseda}, Michael V. and {Oesch}, Pascal A. and {Bezanson}, Rachel and {Boogaard}, Leindert A. and {Cleri}, Nikko J. and {Cooper}, Olivia R. and {Gottumukkala}, Rashmi and {Greene}, Jenny E. and {Hirschmann}, Michaela and {Hviding}, Raphael E. and {Katz}, Harley and {Labb{\'e}}, Ivo and {Leja}, Joel and {Matthee}, Jorryt and {McConachie}, Ian and {Miller}, Tim B. and {Naidu}, Rohan P. and {Price}, Sedona H. and {Rix}, Hans-Walter and {Setton}, David J. and {Suess}, Katherine A. and {Wang}, Bingjie and {Whitaker}, Katherine E. and {Williams}, Christina C.},
        title = "{RUBIES: A complete census of the bright and red distant Universe with JWST/NIRSpec}",
      journal = {\aap},
         year = 2025,
        month = may,
       volume = {697},
          eid = {A189},
        pages = {A189},
          doi = {10.1051/0004-6361/202452186},
archivePrefix = {arXiv},
       eprint = {2409.05948},
 primaryClass = {astro-ph.GA},
       adsurl = {https://ui.adsabs.harvard.edu/abs/2025A&A...697A.189D}
}

@ARTICLE{Heintz2025,
       author = {{Heintz}, K.~E. and {Brammer}, G.~B. and {Watson}, D. and {Oesch}, P.~A. and {Keating}, L.~C. and {Hayes}, M.~J. and {Abdurro'uf} and {Arellano-C{\'o}rdova}, K.~Z. and {Carnall}, A.~C. and {Christiansen}, C.~R. and {Cullen}, F. and {Dav{\'e}}, R. and {Dayal}, P. and {Ferrara}, A. and {Finlator}, K. and {Fynbo}, J.~P.~U. and {Flury}, S.~R. and {Gelli}, V. and {Gillman}, S. and {Gottumukkala}, R. and {Gould}, K. and {Greve}, T.~R. and {Hardin}, S.~E. and {Hsiao}, T.~Y.-Y. and {Hutter}, A. and {Jakobsson}, P. and {Killi}, M. and {Khosravaninezhad}, N. and {Laursen}, P. and {Lee}, M.~M. and {Magdis}, G.~E. and {Matthee}, J. and {Naidu}, R.~P. and {Narayanan}, D. and {Pollock}, C. and {Prescott}, M.~K.~M. and {Rusakov}, V. and {Shuntov}, M. and {Sneppen}, A. and {Smit}, R. and {Tanvir}, N.~R. and {Terp}, C. and {Toft}, S. and {Valentino}, F. and {Vijayan}, A.~P. and {Weaver}, J.~R. and {Wise}, J.~H. and {Witstok}, J.},
        title = "{The JWST-PRIMAL archival survey: A JWST/NIRSpec reference sample for the physical properties and Lyman-{\ensuremath{\alpha}} absorption and emission of {\ensuremath{\sim}}600 galaxies at z = 5.0 {\ensuremath{-}} 13.4}",
      journal = {\aap},
         year = 2025,
        month = jan,
       volume = {693},
          eid = {A60},
        pages = {A60},
          doi = {10.1051/0004-6361/202450243},
archivePrefix = {arXiv},
       eprint = {2404.02211},
 primaryClass = {astro-ph.GA},
       adsurl = {https://ui.adsabs.harvard.edu/abs/2025A&A...693A..60H}
}

@ARTICLE{Valentino2025,
       author = {{Valentino}, F. and {Heintz}, K.~E. and {Brammer}, G. and {Ito}, K. and {Kokorev}, V. and {Whitaker}, K.~E. and {Gallazzi}, A. and {de Graaff}, A. and {Weibel}, A. and {Frye}, B.~L. and {Kamieneski}, P.~S. and {Jin}, S. and {Ceverino}, D. and {Faisst}, A. and {Farcy}, M. and {Fujimoto}, S. and {Gillman}, S. and {Gottumukkala}, R. and {Hamadouche}, M. and {Harrington}, K.~C. and {Hirschmann}, M. and {Jespersen}, C.~K. and {Kakimoto}, T. and {Kubo}, M. and {Lagos}, C. d. P. and {Lee}, M. and {Magdis}, G.~E. and {Man}, A.~W.~S. and {Onodera}, M. and {Rizzo}, F. and {Shimakawa}, R. and {Setton}, D.~J. and {Tanaka}, M. and {Toft}, S. and {Wu}, P.-F. and {Zhu}, P.},
        title = "{Gas outflows in two recently quenched galaxies at z = 4 and 7}",
      journal = {\aap},
         year = 2025,
        month = jul,
       volume = {699},
          eid = {A358},
        pages = {A358},
          doi = {10.1051/0004-6361/202553908},
archivePrefix = {arXiv},
       eprint = {2503.01990},
 primaryClass = {astro-ph.GA},
       adsurl = {https://ui.adsabs.harvard.edu/abs/2025A&A...699A.358V}
}

@ARTICLE{Chevallard2016,
       author = {{Chevallard}, Jacopo and {Charlot}, St{\'e}phane},
        title = "{Modelling and interpreting spectral energy distributions of galaxies with BEAGLE}",
      journal = {\mnras},
         year = 2016,
        month = oct,
       volume = {462},
       number = {2},
        pages = {1415-1443},
          doi = {10.1093/mnras/stw1756},
archivePrefix = {arXiv},
       eprint = {1603.03037},
 primaryClass = {astro-ph.GA},
       adsurl = {https://ui.adsabs.harvard.edu/abs/2016MNRAS.462.1415C}
}

@ARTICLE{Gutkin2016,
       author = {{Gutkin}, Julia and {Charlot}, St{\'e}phane and {Bruzual}, Gustavo},
        title = "{Modelling the nebular emission from primeval to present-day star-forming galaxies}",
      journal = {\mnras},
         year = 2016,
        month = oct,
       volume = {462},
       number = {2},
        pages = {1757-1774},
          doi = {10.1093/mnras/stw1716},
archivePrefix = {arXiv},
       eprint = {1607.06086},
 primaryClass = {astro-ph.GA},
       adsurl = {https://ui.adsabs.harvard.edu/abs/2016MNRAS.462.1757G}
}

@ARTICLE{Bruzual2003,
       author = {{Bruzual}, G. and {Charlot}, S.},
        title = "{Stellar population synthesis at the resolution of 2003}",
      journal = {\mnras},
         year = 2003,
        month = oct,
       volume = {344},
       number = {4},
        pages = {1000-1028},
          doi = {10.1046/j.1365-8711.2003.06897.x},
archivePrefix = {arXiv},
       eprint = {astro-ph/0309134},
 primaryClass = {astro-ph},
       adsurl = {https://ui.adsabs.harvard.edu/abs/2003MNRAS.344.1000B}
}

@ARTICLE{Chabrier2003,
       author = {{Chabrier}, Gilles},
        title = "{Galactic Stellar and Substellar Initial Mass Function}",
      journal = {\pasp},
         year = 2003,
        month = jul,
       volume = {115},
       number = {809},
        pages = {763-795},
          doi = {10.1086/376392},
archivePrefix = {arXiv},
       eprint = {astro-ph/0304382},
 primaryClass = {astro-ph},
       adsurl = {https://ui.adsabs.harvard.edu/abs/2003PASP..115..763C}
}

@ARTICLE{Chen2015,
       author = {{Chen}, Yang and {Bressan}, Alessandro and {Girardi}, L{\'e}o and {Marigo}, Paola and {Kong}, Xu and {Lanza}, Antonio},
        title = "{PARSEC evolutionary tracks of massive stars up to 350 M$_{{\ensuremath{\odot}}}$ at metallicities 0.0001 {\ensuremath{\leq}} Z {\ensuremath{\leq}} 0.04}",
      journal = {\mnras},
         year = 2015,
        month = sep,
       volume = {452},
       number = {1},
        pages = {1068-1080},
          doi = {10.1093/mnras/stv1281},
archivePrefix = {arXiv},
       eprint = {1506.01681},
 primaryClass = {astro-ph.SR},
       adsurl = {https://ui.adsabs.harvard.edu/abs/2015MNRAS.452.1068C}
}

@ARTICLE{Pei1992,
       author = {{Pei}, Yichuan C.},
        title = "{Interstellar Dust from the Milky Way to the Magellanic Clouds}",
      journal = {\apj},
         year = 1992,
        month = aug,
       volume = {395},
        pages = {130},
          doi = {10.1086/171637},
       adsurl = {https://ui.adsabs.harvard.edu/abs/1992ApJ...395..130P}
}

@ARTICLE{Schwarz1978,
       author = {{Schwarz}, Gideon},
        title = "{Estimating the Dimension of a Model}",
      journal = {Annals of Statistics},
         year = 1978,
        month = jul,
       volume = {6},
       number = {2},
        pages = {461-464},
       adsurl = {https://ui.adsabs.harvard.edu/abs/1978AnSta...6..461S}
}

@ARTICLE{Liddle2007,
       author = {{Liddle}, Andrew R.},
        title = "{Information criteria for astrophysical model selection}",
      journal = {\mnras},
         year = 2007,
        month = may,
       volume = {377},
       number = {1},
        pages = {L74-L78},
          doi = {10.1111/j.1745-3933.2007.00306.x},
archivePrefix = {arXiv},
       eprint = {astro-ph/0701113},
 primaryClass = {astro-ph},
       adsurl = {https://ui.adsabs.harvard.edu/abs/2007MNRAS.377L..74L}
}

@ARTICLE{Chang2024,
       author = {{Chang}, Seok-Jun and {Gronke}, Max},
        title = "{Probing cold gas with Mg II and Ly {\ensuremath{\alpha}} radiative transfer}",
      journal = {\mnras},
         year = 2024,
        month = aug,
       volume = {532},
       number = {3},
        pages = {3526-3555},
          doi = {10.1093/mnras/stae1664},
archivePrefix = {arXiv},
       eprint = {2403.11524},
 primaryClass = {astro-ph.GA},
       adsurl = {https://ui.adsabs.harvard.edu/abs/2024MNRAS.532.3526C}
}

@ARTICLE{Henry2018,
       author = {{Henry}, Alaina and {Berg}, Danielle A. and {Scarlata}, Claudia and {Verhamme}, Anne and {Erb}, Dawn},
        title = "{A Close Relationship between Ly{\ensuremath{\alpha}} and Mg II in Green Pea Galaxies}",
      journal = {\apj},
         year = 2018,
        month = mar,
       volume = {855},
       number = {2},
          eid = {96},
        pages = {96},
          doi = {10.3847/1538-4357/aab099},
archivePrefix = {arXiv},
       eprint = {1803.10243},
 primaryClass = {astro-ph.GA},
       adsurl = {https://ui.adsabs.harvard.edu/abs/2018ApJ...855...96H}
}

@ARTICLE{Chisholm2020,
       author = {{Chisholm}, J. and {Prochaska}, J.~X. and {Schaerer}, D. and {Gazagnes}, S. and {Henry}, A.},
        title = "{Optically thin spatially resolved Mg II emission maps the escape of ionizing photons}",
      journal = {\mnras},
         year = 2020,
        month = oct,
       volume = {498},
       number = {2},
        pages = {2554-2574},
          doi = {10.1093/mnras/staa2470},
archivePrefix = {arXiv},
       eprint = {2008.06059},
 primaryClass = {astro-ph.GA},
       adsurl = {https://ui.adsabs.harvard.edu/abs/2020MNRAS.498.2554C}
}

@ARTICLE{Marques-Chaves2024,
       author = {{Marques-Chaves}, R. and {Schaerer}, D. and {Kuruvanthodi}, A. and {Korber}, D. and {Prantzos}, N. and {Charbonnel}, C. and {Weibel}, A. and {Izotov}, Y.~I. and {Messa}, M. and {Brammer}, G. and {Dessauges-Zavadsky}, M. and {Oesch}, P.},
        title = "{Extreme N-emitters at high redshift: Possible signatures of supermassive stars and globular cluster or black hole formation in action}",
      journal = {\aap},
         year = 2024,
        month = jan,
       volume = {681},
          eid = {A30},
        pages = {A30},
          doi = {10.1051/0004-6361/202347411},
archivePrefix = {arXiv},
       eprint = {2307.04234},
 primaryClass = {astro-ph.GA},
       adsurl = {https://ui.adsabs.harvard.edu/abs/2024A&A...681A..30M}
}

@ARTICLE{Martins2025,
       author = {{Martins}, F. and {Palacios}, A. and {Schaerer}, D. and {Marques-Chaves}, R.},
        title = "{Very massive stars at low metallicity: Evolution, synthetic spectroscopy, and impact on the integrated light of starbursts}",
      journal = {\aap},
         year = 2025,
        month = jun,
       volume = {698},
          eid = {A262},
        pages = {A262},
          doi = {10.1051/0004-6361/202554421},
archivePrefix = {arXiv},
       eprint = {2505.02993},
 primaryClass = {astro-ph.SR},
       adsurl = {https://ui.adsabs.harvard.edu/abs/2025A&A...698A.262M}
}

@ARTICLE{Grafener2021,
       author = {{Gr{\"a}fener}, G{\"o}tz},
        title = "{Physics and evolution of the most massive stars in 30 Doradus. Mass loss, envelope inflation, and a variable upper stellar mass limit}",
      journal = {\aap},
         year = 2021,
        month = mar,
       volume = {647},
          eid = {A13},
        pages = {A13},
          doi = {10.1051/0004-6361/202040037},
archivePrefix = {arXiv},
       eprint = {2101.03837},
 primaryClass = {astro-ph.SR},
       adsurl = {https://ui.adsabs.harvard.edu/abs/2021A&A...647A..13G}
}

@ARTICLE{Eldridge2017,
       author = {{Eldridge}, J.~J. and {Stanway}, E.~R. and {Xiao}, L. and {McClelland}, L.~A.~S. and {Taylor}, G. and {Ng}, M. and {Greis}, S.~M.~L. and {Bray}, J.~C.},
        title = "{Binary Population and Spectral Synthesis Version 2.1: Construction, Observational Verification, and New Results}",
      journal = {\pasa},
         year = 2017,
        month = nov,
       volume = {34},
          eid = {e058},
        pages = {e058},
          doi = {10.1017/pasa.2017.51},
archivePrefix = {arXiv},
       eprint = {1710.02154},
 primaryClass = {astro-ph.SR},
       adsurl = {https://ui.adsabs.harvard.edu/abs/2017PASA...34...58E}
}

@ARTICLE{Saldana-Lopez2023,
       author = {{Saldana-Lopez}, A. and {Schaerer}, D. and {Chisholm}, J. and {Calabr{\`o}}, A. and {Pentericci}, L. and {Cullen}, F. and {Saxena}, A. and {Amor{\'\i}n}, R. and {Carnall}, A.~C. and {Fontanot}, F. and {Fynbo}, J.~P.~U. and {Guaita}, L. and {Hathi}, N.~P. and {Hibon}, P. and {Ji}, Z. and {McLeod}, D.~J. and {Pompei}, E. and {Zamorani}, G.},
        title = "{The VANDELS survey: the ionizing properties of star-forming galaxies at 3 {\ensuremath{\leq}} z {\ensuremath{\leq}} 5 using deep rest-frame ultraviolet spectroscopy}",
      journal = {\mnras},
         year = 2023,
        month = jul,
       volume = {522},
       number = {4},
        pages = {6295-6325},
          doi = {10.1093/mnras/stad1283},
archivePrefix = {arXiv},
       eprint = {2211.01351},
 primaryClass = {astro-ph.GA},
       adsurl = {https://ui.adsabs.harvard.edu/abs/2023MNRAS.522.6295S}
}

@software{Newville2016,
       author = {{Newville}, Matthew and {Stensitzki}, Till and {Allen}, Daniel B. and {Rawlik}, Michal and {Ingargiola}, Antonino and {Nelson}, Andrew},
        title = "{Lmfit: Non-Linear Least-Square Minimization and Curve-Fitting for Python}",
 howpublished = {Astrophysics Source Code Library, record ascl:1606.014},
         year = 2016,
        month = jun,
          eid = {ascl:1606.014},
archivePrefix = {ascl},
       eprint = {1606.014},
       adsurl = {https://ui.adsabs.harvard.edu/abs/2016ascl.soft06014N}
}

@ARTICLE{Scholtz2024,
       author = {{Scholtz}, Jan and {Witten}, Callum and {Laporte}, Nicolas and {{\"U}bler}, Hannah and {Perna}, Michele and {Maiolino}, Roberto and {Arribas}, Santiago and {Baker}, William M. and {Bennett}, Jake S. and {D'Eugenio}, Francesco and {Simmonds}, Charlotte and {Tacchella}, Sandro and {Witstok}, Joris and {Bunker}, Andrew J. and {Carniani}, Stefano and {Charlot}, St{\'e}phane and {Cresci}, Giovanni and {Curtis-Lake}, Emma and {Eisenstein}, Daniel J. and {Kumari}, Nimisha and {Robertson}, Brant and {Rodr{\'\i}guez Del Pino}, Bruno and {Smit}, Renske and {Venturi}, Giacomo and {Williams}, Christina C. and {Willmer}, Christopher N.~A.},
        title = "{GN-z11: The environment of an active galactic nucleus at z = 10.603. New insights into the most distant Ly{\ensuremath{\alpha}} detection}",
      journal = {\aap},
         year = 2024,
        month = jul,
       volume = {687},
          eid = {A283},
        pages = {A283},
          doi = {10.1051/0004-6361/202347187},
archivePrefix = {arXiv},
       eprint = {2306.09142},
 primaryClass = {astro-ph.GA},
       adsurl = {https://ui.adsabs.harvard.edu/abs/2024A&A...687A.283S}
}

@ARTICLE{Ji2025,
       author = {{Ji}, Xihan and {Maiolino}, Roberto and {Ferland}, Gary and {D'Eugenio}, Francesco and {Bhatawdekar}, Rachana and {Charlot}, St{\'e}phane and {Chevallard}, Jacopo and {Curti}, Mirko and {Curtis-Lake}, Emma and {Hainline}, Kevin and {Ji}, Zhiyuan and {Robertson}, Brant and {Rodr{\'\i}guez Del Pino}, Bruno and {Scholtz}, Jan and {Tacchella}, Sandro and {Williams}, Christina C. and {Witstok}, Joris},
        title = "{JADES ─ the small blue bump in GN-z11: insights into the nuclear region of a galaxy at z = 10.6}",
      journal = {\mnras},
         year = 2025,
        month = aug,
       volume = {541},
       number = {3},
        pages = {2134-2161},
          doi = {10.1093/mnras/staf1083},
archivePrefix = {arXiv},
       eprint = {2405.05772},
 primaryClass = {astro-ph.GA},
       adsurl = {https://ui.adsabs.harvard.edu/abs/2025MNRAS.541.2134J}
}

@ARTICLE{Tang2026_spursQSO1,
       author = {{Tang}, Mengtao and {Stark}, Daniel P. and {Mason}, Charlotte A. and {Chen}, Zuyi and {Katz}, Harley and {Gronke}, Max and {Furtak}, Lukas J. and {Chang}, Seok-Jun and {Matthee}, Jorryt and {Whitler}, Lily and {Zitrin}, Adi and {Endsley}, Ryan and {Gelli}, Viola and {Roychowdhury}, Tamojeet and {Senchyna}, Peter and {Topping}, Michael W. and {Zhang}, Meng},
        title = "{SPURS: Evidence for Clumpy Neutral Envelopes and Ionized IGM Surrounding Little Red Dots in Abell 2744 from Ultra-Deep Rest-UV Spectroscopy}",
      journal = {arXiv e-prints},
         year = 2026,
        month = apr,
          eid = {arXiv:2604.03563},
        pages = {arXiv:2604.03563},
          doi = {10.48550/arXiv.2604.03563},
archivePrefix = {arXiv},
       eprint = {2604.03563},
 primaryClass = {astro-ph.GA},
       adsurl = {https://ui.adsabs.harvard.edu/abs/2026arXiv260403563T}
}

@ARTICLE{Endsley2022_bubble,
       author = {{Endsley}, Ryan and {Stark}, Daniel P.},
        title = "{Strong Lyman-{\ensuremath{\alpha}} emission in an overdense region at z = 6.8: a very large (R   3 physical Mpc) ionized bubble in COSMOS?}",
      journal = {\mnras},
         year = 2022,
        month = apr,
       volume = {511},
       number = {4},
        pages = {6042-6054},
          doi = {10.1093/mnras/stac524},
archivePrefix = {arXiv},
       eprint = {2112.14779},
 primaryClass = {astro-ph.GA},
       adsurl = {https://ui.adsabs.harvard.edu/abs/2022MNRAS.511.6042E}
}

@ARTICLE{Nomoto2006,
       author = {{Nomoto}, Ken'ichi and {Tominaga}, Nozomu and {Umeda}, Hideyuki and {Kobayashi}, Chiaki and {Maeda}, Keiichi},
        title = "{Nucleosynthesis yields of core-collapse supernovae and hypernovae, and galactic chemical evolution}",
      journal = {\nphysa},
         year = 2006,
        month = oct,
       volume = {777},
        pages = {424-458},
          doi = {10.1016/j.nuclphysa.2006.05.008},
archivePrefix = {arXiv},
       eprint = {astro-ph/0605725},
 primaryClass = {astro-ph},
       adsurl = {https://ui.adsabs.harvard.edu/abs/2006NuPhA.777..424N}
}

@ARTICLE{Santos2004,
       author = {{Santos}, Michael R.},
        title = "{Probing reionization with Lyman {\ensuremath{\alpha}} emission lines}",
      journal = {\mnras},
         year = 2004,
        month = apr,
       volume = {349},
       number = {3},
        pages = {1137-1152},
          doi = {10.1111/j.1365-2966.2004.07594.x},
archivePrefix = {arXiv},
       eprint = {astro-ph/0308196},
 primaryClass = {astro-ph},
       adsurl = {https://ui.adsabs.harvard.edu/abs/2004MNRAS.349.1137S}
}

@ARTICLE{Xu2024,
       author = {{Xu}, Yi and {Ouchi}, Masami and {Yajima}, Hidenobu and {Fukushima}, Hajime and {Harikane}, Yuichi and {Isobe}, Yuki and {Nakajima}, Kimihiko and {Nakane}, Minami and {Ono}, Yoshiaki and {Umeda}, Hiroya and {Yanagisawa}, Hiroto and {Zhang}, Yechi},
        title = "{Dynamics of a Galaxy at z > 10 Explored by JWST Integral Field Spectroscopy: Hints of Rotating Disk Suggesting Weak Feedback}",
      journal = {\apj},
         year = 2024,
        month = nov,
       volume = {976},
       number = {1},
          eid = {142},
        pages = {142},
          doi = {10.3847/1538-4357/ad82dd},
archivePrefix = {arXiv},
       eprint = {2404.16963},
 primaryClass = {astro-ph.GA},
       adsurl = {https://ui.adsabs.harvard.edu/abs/2024ApJ...976..142X}
}

@ARTICLE{Cameron2023,
       author = {{Cameron}, Alex J. and {Katz}, Harley and {Rey}, Martin P. and {Saxena}, Aayush},
        title = "{Nitrogen enhancements 440 Myr after the big bang: supersolar N/O, a tidal disruption event, or a dense stellar cluster in GN-z11?}",
      journal = {\mnras},
         year = 2023,
        month = aug,
       volume = {523},
       number = {3},
        pages = {3516-3525},
          doi = {10.1093/mnras/stad1579},
archivePrefix = {arXiv},
       eprint = {2302.10142},
 primaryClass = {astro-ph.GA},
       adsurl = {https://ui.adsabs.harvard.edu/abs/2023MNRAS.523.3516C}
}

@ARTICLE{CrespoGomez2026,
       author = {{Crespo G{\'o}mez}, A. and {Colina}, L. and {P{\'e}rez-Gonz{\'a}lez}, P.~G. and {{\'A}lvarez-M{\'a}rquez}, J. and {Garc{\'\i}a-Mar{\'\i}n}, M. and {Alonso-Herrero}, A. and {Annunziatella}, M. and {Bik}, A. and {Bosman}, S. and {Bunker}, A.~J. and {Labiano}, A. and {Langeroodi}, D. and {Rinaldi}, P. and {{\"O}stlin}, G. and {Boogaard}, L. and {Gillman}, S. and {Barro}, G. and {Finkelstein}, S.~L. and {Leung}, G.~C.~K.},
        title = "{MIRI spectrophotometry of GN-z11: Detection and nature of an optical red continuum component}",
      journal = {\aap},
         year = 2026,
        month = jan,
       volume = {706},
          eid = {A46},
        pages = {A46},
          doi = {10.1051/0004-6361/202556814},
archivePrefix = {arXiv},
       eprint = {2512.02997},
 primaryClass = {astro-ph.GA},
       adsurl = {https://ui.adsabs.harvard.edu/abs/2026A&A...706A..46C}
}

@ARTICLE{Vidal-Garcia2017,
       author = {{Vidal-Garc{\'\i}a}, A. and {Charlot}, S. and {Bruzual}, G. and {Hubeny}, I.},
        title = "{Modelling ultraviolet-line diagnostics of stars, the ionized and the neutral interstellar medium in star-forming galaxies}",
      journal = {\mnras},
         year = 2017,
        month = sep,
       volume = {470},
       number = {3},
        pages = {3532-3556},
          doi = {10.1093/mnras/stx1324},
archivePrefix = {arXiv},
       eprint = {1705.10320},
 primaryClass = {astro-ph.GA},
       adsurl = {https://ui.adsabs.harvard.edu/abs/2017MNRAS.470.3532V}
}

@ARTICLE{Senchyna2022,
       author = {{Senchyna}, Peter and {Stark}, Daniel P. and {Charlot}, St{\'e}phane and {Plat}, Adele and {Chevallard}, Jacopo and {Chen}, Zuyi and {Jones}, Tucker and {Sanders}, Ryan L. and {Rudie}, Gwen C. and {Cooper}, Thomas J. and {Bruzual}, Gustavo},
        title = "{Direct Constraints on the Extremely Metal-poor Massive Stars Underlying Nebular C IV Emission from Ultra-deep HST/COS Ultraviolet Spectroscopy}",
      journal = {\apj},
         year = 2022,
        month = may,
       volume = {930},
       number = {2},
          eid = {105},
        pages = {105},
          doi = {10.3847/1538-4357/ac5d38},
archivePrefix = {arXiv},
       eprint = {2111.11508},
 primaryClass = {astro-ph.GA},
       adsurl = {https://ui.adsabs.harvard.edu/abs/2022ApJ...930..105S}
}

@ARTICLE{Hainich2014,
       author = {{Hainich}, R. and {R{\"u}hling}, U. and {Todt}, H. and {Oskinova}, L.~M. and {Liermann}, A. and {Gr{\"a}fener}, G. and {Foellmi}, C. and {Schnurr}, O. and {Hamann}, W.-R.},
        title = "{The Wolf-Rayet stars in the Large Magellanic Cloud. A comprehensive analysis of the WN class}",
      journal = {\aap},
         year = 2014,
        month = may,
       volume = {565},
          eid = {A27},
        pages = {A27},
          doi = {10.1051/0004-6361/201322696},
archivePrefix = {arXiv},
       eprint = {1401.5474},
 primaryClass = {astro-ph.SR},
       adsurl = {https://ui.adsabs.harvard.edu/abs/2014A&A...565A..27H}
}

@Article{Park2021,
  author           = {Park, Hyunbae and Jung, Intae and Song, Hyunmi and Ocvirk, Pierre and Shapiro, Paul R. and Dawoodbhoy, Taha and Iliev, Ilian T. and Ahn, Kyungjin and Bianco, Michele and Kim, Hyo Jeong},
  journal          = {\apj},
  title            = {Crucial Factors for Ly{\ensuremath{\alpha}} Transmission in the Reionizing Intergalactic Medium: Infall Motion, H II Bubble Size, and Self-shielded Systems},
  year             = {2021},
  month            = dec,
  number           = {2},
  pages            = {263},
  volume           = {922},
  archiveprefix    = {arXiv},
  doi              = {10.3847/1538-4357/ac2f4b},
  eid              = {263},
  eprint           = {2105.10770},
  modificationdate = {2026-07-14T18:22:26},
  primaryclass     = {astro-ph.GA},
  url              = {https://ui.adsabs.harvard.edu/abs/2021ApJ...922..263P},
}

@ARTICLE{Robertson2026,
       author = {{Robertson}, Brant E. and {Johnson}, Benjamin D. and {Tacchella}, Sandro and {Eisenstein}, Daniel J. and {Hainline}, Kevin and {Alberts}, Stacey and {Arribas}, Santiago and {Baker}, William M. and {Bunker}, Andrew J. and {Cameron}, Alex J. and {Carniani}, Stefano and {Carreira}, Courtney and {Chevallard}, Jacopo and {Circosta}, Chiara and {Curtis-Lake}, Emma and {Danhaive}, A. Lola and {Duan}, Qiao and {Egami}, Eiichi and {Hausen}, Ryan and {Helton}, Jakob M. and {Ji}, Zhiyuan and {Maiolino}, Roberto and {P{\'e}rez-Gonz{\'a}lez}, Pablo G. and {Pusk{\'a}s}, D{\'a}vid and {Rieke}, Marcia and {Rinaldi}, Pierluigi and {Sun}, Fengwu and {Sun}, Yang and {{\"U}bler}, Hannah and {Trussler}, James A.~A. and {Villanueva}, Natalia C. and {Whitler}, Lily and {Williams}, Christina C. and {Willmer}, Christopher N.~A. and {Willott}, Chris and {Wu}, Zihao and {Zhu}, Yongda},
        title = "{JWST Advanced Deep Extragalactic Survey (JADES) Data Release 5: Photometric Catalog}",
      journal = {arXiv e-prints},
         year = 2026,
        month = jan,
          eid = {arXiv:2601.15956},
        pages = {arXiv:2601.15956},
          doi = {10.48550/arXiv.2601.15956},
archivePrefix = {arXiv},
       eprint = {2601.15956},
 primaryClass = {astro-ph.GA},
       adsurl = {https://ui.adsabs.harvard.edu/abs/2026arXiv260115956R}
}

@ARTICLE{Mason2018,
       author = {{Mason}, Charlotte A. and {Treu}, Tommaso and {Dijkstra}, Mark and {Mesinger}, Andrei and {Trenti}, Michele and {Pentericci}, Laura and {de Barros}, Stephane and {Vanzella}, Eros},
        title = "{The Universe Is Reionizing at z {\ensuremath{\sim}} 7: Bayesian Inference of the IGM Neutral Fraction Using Ly{\ensuremath{\alpha}} Emission from Galaxies}",
      journal = {\apj},
         year = 2018,
        month = mar,
       volume = {856},
       number = {1},
          eid = {2},
        pages = {2},
          doi = {10.3847/1538-4357/aab0a7},
archivePrefix = {arXiv},
       eprint = {1709.05356},
 primaryClass = {astro-ph.CO},
       adsurl = {https://ui.adsabs.harvard.edu/abs/2018ApJ...856....2M}
}

@ARTICLE{Vink2001,
       author = {{Vink}, Jorick S. and {de Koter}, A. and {Lamers}, H.~J.~G.~L.~M.},
        title = "{Mass-loss predictions for O and B stars as a function of metallicity}",
      journal = {\aap},
         year = 2001,
        month = apr,
       volume = {369},
        pages = {574-588},
          doi = {10.1051/0004-6361:20010127},
archivePrefix = {arXiv},
       eprint = {astro-ph/0101509},
 primaryClass = {astro-ph},
       adsurl = {https://ui.adsabs.harvard.edu/abs/2001A&A...369..574V}
}

@ARTICLE{Lu2024,
       author = {{Lu}, Ting-Yi and {Mason}, Charlotte A. and {Hutter}, Anne and {Mesinger}, Andrei and {Qin}, Yuxiang and {Stark}, Daniel P. and {Endsley}, Ryan},
        title = "{The reionizing bubble size distribution around galaxies}",
      journal = {\mnras},
         year = 2024,
        month = mar,
       volume = {528},
       number = {3},
        pages = {4872-4890},
          doi = {10.1093/mnras/stae266},
archivePrefix = {arXiv},
       eprint = {2304.11192},
 primaryClass = {astro-ph.GA},
       adsurl = {https://ui.adsabs.harvard.edu/abs/2024MNRAS.528.4872L}
}

@ARTICLE{Topping2022,
       author = {{Topping}, Michael W. and {Stark}, Daniel P. and {Endsley}, Ryan and {Plat}, Adele and {Whitler}, Lily and {Chen}, Zuyi and {Charlot}, St{\'e}phane},
        title = "{Searching for Extremely Blue UV Continuum Slopes at z = 7-11 in JWST/NIRCam Imaging: Implications for Stellar Metallicity and Ionizing Photon Escape in Early Galaxies}",
      journal = {\apj},
         year = 2022,
        month = dec,
       volume = {941},
       number = {2},
          eid = {153},
        pages = {153},
          doi = {10.3847/1538-4357/aca522},
archivePrefix = {arXiv},
       eprint = {2208.01610},
 primaryClass = {astro-ph.GA},
       adsurl = {https://ui.adsabs.harvard.edu/abs/2022ApJ...941..153T}
}

@ARTICLE{Vink2011,
       author = {{Vink}, Jorick S. and {Muijres}, L.~E. and {Anthonisse}, B. and {de Koter}, A. and {Gr{\"a}fener}, G. and {Langer}, N.},
        title = "{Wind modelling of very massive stars up to 300 solar masses}",
      journal = {\aap},
         year = 2011,
        month = jul,
       volume = {531},
          eid = {A132},
        pages = {A132},
          doi = {10.1051/0004-6361/201116614},
archivePrefix = {arXiv},
       eprint = {1105.0556},
 primaryClass = {astro-ph.SR},
       adsurl = {https://ui.adsabs.harvard.edu/abs/2011A&A...531A.132V}
}

@ARTICLE{PrietoLyon2025,
       author = {{Prieto-Lyon}, Gonzalo and {Mason}, Charlotte A. and {Strait}, Victoria and {Brammer}, Gabriel and {Naidu}, Rohan P. and {Meyer}, Romain A. and {Oesch}, Pascal and {Tacchella}, Sandro and {Covelo-Paz}, Alba and {Giovinazzo}, Emma and {Xiao}, Mengyuan},
        title = "{Lyman-alpha emission at the end of reionization: line strengths and profiles from MMT and JWST observations at z\raisebox{-0.5ex}\textasciitilde5-6}",
      journal = {arXiv e-prints},
         year = 2025,
        month = sep,
          eid = {arXiv:2509.18302},
        pages = {arXiv:2509.18302},
          doi = {10.48550/arXiv.2509.18302},
archivePrefix = {arXiv},
       eprint = {2509.18302},
 primaryClass = {astro-ph.GA},
       adsurl = {https://ui.adsabs.harvard.edu/abs/2025arXiv250918302P}
}

@ARTICLE{Pentericci2018,
       author = {{Pentericci}, L. and {Vanzella}, E. and {Castellano}, M. and {Fontana}, A. and {De Barros}, S. and {Grazian}, A. and {Marchi}, F. and {Bradac}, M. and {Conselice}, C.~J. and {Cristiani}, S. and {Dickinson}, M. and {Finkelstein}, S.~L. and {Giallongo}, E. and {Guaita}, L. and {Koekemoer}, A.~M. and {Maiolino}, R. and {Santini}, P. and {Tilvi}, V.},
        title = "{CANDELSz7: a large spectroscopic survey of CANDELS galaxies in the reionization epoch}",
      journal = {\aap},
         year = 2018,
        month = nov,
       volume = {619},
          eid = {A147},
        pages = {A147},
          doi = {10.1051/0004-6361/201732465},
archivePrefix = {arXiv},
       eprint = {1808.01847},
 primaryClass = {astro-ph.GA},
       adsurl = {https://ui.adsabs.harvard.edu/abs/2018A&A...619A.147P}
}

@ARTICLE{Hayes2021,
       author = {{Hayes}, Matthew J. and {Runnholm}, Axel and {Gronke}, Max and {Scarlata}, Claudia},
        title = "{Spectral Shapes of the Ly{\ensuremath{\alpha}} Emission from Galaxies. I. Blueshifted Emission and Intrinsic Invariance with Redshift}",
      journal = {\apj},
         year = 2021,
        month = feb,
       volume = {908},
       number = {1},
          eid = {36},
        pages = {36},
          doi = {10.3847/1538-4357/abd246},
archivePrefix = {arXiv},
       eprint = {2006.03232},
 primaryClass = {astro-ph.GA},
       adsurl = {https://ui.adsabs.harvard.edu/abs/2021ApJ...908...36H}
}

@ARTICLE{deGraaff2024,
       author = {{de Graaff}, Anna and {Rix}, Hans-Walter and {Carniani}, Stefano and {Suess}, Katherine A. and {Charlot}, St{\'e}phane and {Curtis-Lake}, Emma and {Arribas}, Santiago and {Baker}, William M. and {Boyett}, Kristan and {Bunker}, Andrew J. and {Cameron}, Alex J. and {Chevallard}, Jacopo and {Curti}, Mirko and {Eisenstein}, Daniel J. and {Franx}, Marijn and {Hainline}, Kevin and {Hausen}, Ryan and {Ji}, Zhiyuan and {Johnson}, Benjamin D. and {Jones}, Gareth C. and {Maiolino}, Roberto and {Maseda}, Michael V. and {Nelson}, Erica and {Parlanti}, Eleonora and {Rawle}, Tim and {Robertson}, Brant and {Tacchella}, Sandro and {{\"U}bler}, Hannah and {Williams}, Christina C. and {Willmer}, Christopher N.~A. and {Willott}, Chris},
        title = "{Ionised gas kinematics and dynamical masses of z {\ensuremath{\gtrsim}} 6 galaxies from JADES/NIRSpec high-resolution spectroscopy}",
      journal = {\aap},
         year = 2024,
        month = apr,
       volume = {684},
          eid = {A87},
        pages = {A87},
          doi = {10.1051/0004-6361/202347755},
archivePrefix = {arXiv},
       eprint = {2308.09742},
 primaryClass = {astro-ph.GA},
       adsurl = {https://ui.adsabs.harvard.edu/abs/2024A&A...684A..87D}
}

@ARTICLE{Shajib2025,
       author = {{Shajib}, Anowar J. and {Treu}, Tommaso and {Melo}, Alejandra and {Roberts-Borsani}, Guido and {Knabel}, Shawn and {Cappellari}, Michele and {Frieman}, Joshua A.},
        title = "{An accurate measurement of the spectral resolution of the JWST Near Infrared Spectrograph}",
      journal = {\aap},
         year = 2025,
        month = oct,
       volume = {702},
          eid = {L12},
        pages = {L12},
          doi = {10.1051/0004-6361/202556281},
archivePrefix = {arXiv},
       eprint = {2507.03746},
 primaryClass = {astro-ph.IM},
       adsurl = {https://ui.adsabs.harvard.edu/abs/2025A&A...702L..12S}
}

@ARTICLE{Stark2011,
       author = {{Stark}, Daniel P. and {Ellis}, Richard S. and {Ouchi}, Masami},
        title = "{Keck Spectroscopy of Faint 3>z>7 Lyman Break Galaxies: A High Fraction of Line Emitters at Redshift Six}",
      journal = {\apjl},
         year = 2011,
        month = feb,
       volume = {728},
       number = {1},
          eid = {L2},
        pages = {L2},
          doi = {10.1088/2041-8205/728/1/L2},
archivePrefix = {arXiv},
       eprint = {1009.5471},
 primaryClass = {astro-ph.CO},
       adsurl = {https://ui.adsabs.harvard.edu/abs/2011ApJ...728L...2S}
}

@ARTICLE{Stark2010,
       author = {{Stark}, Daniel P. and {Ellis}, Richard S. and {Chiu}, Kuenley and {Ouchi}, Masami and {Bunker}, Andrew},
        title = "{Keck spectroscopy of faint 3 < z < 7 Lyman break galaxies - I. New constraints on cosmic reionization from the luminosity and redshift-dependent fraction of Lyman {\ensuremath{\alpha}} emission}",
      journal = {\mnras},
         year = 2010,
        month = nov,
       volume = {408},
       number = {3},
        pages = {1628-1648},
          doi = {10.1111/j.1365-2966.2010.17227.x},
archivePrefix = {arXiv},
       eprint = {1003.5244},
 primaryClass = {astro-ph.CO},
       adsurl = {https://ui.adsabs.harvard.edu/abs/2010MNRAS.408.1628S}
}

@ARTICLE{Pentericci2014,
       author = {{Pentericci}, L. and {Vanzella}, E. and {Fontana}, A. and {Castellano}, M. and {Treu}, T. and {Mesinger}, A. and {Dijkstra}, M. and {Grazian}, A. and {Brada{\v{c}}}, M. and {Conselice}, C. and {Cristiani}, S. and {Dunlop}, J. and {Galametz}, A. and {Giavalisco}, M. and {Giallongo}, E. and {Koekemoer}, A. and {McLure}, R. and {Maiolino}, R. and {Paris}, D. and {Santini}, P.},
        title = "{New Observations of z \raisebox{-0.5ex}\textasciitilde 7 Galaxies: Evidence for a Patchy Reionization}",
      journal = {\apj},
         year = 2014,
        month = oct,
       volume = {793},
       number = {2},
          eid = {113},
        pages = {113},
          doi = {10.1088/0004-637X/793/2/113},
archivePrefix = {arXiv},
       eprint = {1403.5466},
 primaryClass = {astro-ph.CO},
       adsurl = {https://ui.adsabs.harvard.edu/abs/2014ApJ...793..113P}
}

@ARTICLE{Treu2012,
       author = {{Treu}, Tommaso and {Trenti}, Michele and {Stiavelli}, Massimo and {Auger}, Matthew W. and {Bradley}, Larry D.},
        title = "{Inferences on the Distribution of Ly{\ensuremath{\alpha}} Emission of z \raisebox{-0.5ex}\textasciitilde 7 and z \raisebox{-0.5ex}\textasciitilde 8 Galaxies}",
      journal = {\apj},
         year = 2012,
        month = mar,
       volume = {747},
       number = {1},
          eid = {27},
        pages = {27},
          doi = {10.1088/0004-637X/747/1/27},
archivePrefix = {arXiv},
       eprint = {1201.0016},
 primaryClass = {astro-ph.CO},
       adsurl = {https://ui.adsabs.harvard.edu/abs/2012ApJ...747...27T}
}

@ARTICLE{Mason2026,
       author = {{Mason}, Charlotte A. and {Chen}, Zuyi and {Stark}, Daniel P. and {Yi Lu}, Ting and {Topping}, Michael and {Tang}, Mengtao},
        title = "{Constraints on the z {\ensuremath{\sim}} 6{\ensuremath{-}}13 intergalactic medium from JWST spectroscopy of Lyman-alpha damping wings in galaxies}",
      journal = {\aap},
         year = 2026,
        month = jan,
       volume = {705},
          eid = {A114},
        pages = {A114},
          doi = {10.1051/0004-6361/202553820},
archivePrefix = {arXiv},
       eprint = {2501.11702},
 primaryClass = {astro-ph.GA},
       adsurl = {https://ui.adsabs.harvard.edu/abs/2026A&A...705A.114M}
}

@ARTICLE{Keating2024,
       author = {{Keating}, Laura C. and {Bolton}, James S. and {Cullen}, Fergus and {Haehnelt}, Martin G. and {Puchwein}, Ewald and {Kulkarni}, Girish},
        title = "{JWST observations of galaxy-damping wings during reionization interpreted with cosmological simulations}",
      journal = {\mnras},
         year = 2024,
        month = aug,
       volume = {532},
       number = {2},
        pages = {1646-1658},
          doi = {10.1093/mnras/stae1530},
archivePrefix = {arXiv},
       eprint = {2308.05800},
 primaryClass = {astro-ph.GA},
       adsurl = {https://ui.adsabs.harvard.edu/abs/2024MNRAS.532.1646K}
}

@ARTICLE{Tang2024a,
       author = {{Tang}, Mengtao and {Stark}, Daniel P. and {Topping}, Michael W. and {Mason}, Charlotte and {Ellis}, Richard S.},
        title = "{JWST/NIRSpec Observations of Lyman {\ensuremath{\alpha}} Emission in Star-forming Galaxies at 6.5 {\ensuremath{\lesssim}} z {\ensuremath{\lesssim}} 13}",
      journal = {\apj},
         year = 2024,
        month = nov,
       volume = {975},
       number = {2},
          eid = {208},
        pages = {208},
          doi = {10.3847/1538-4357/ad7eb7},
archivePrefix = {arXiv},
       eprint = {2408.01507},
 primaryClass = {astro-ph.GA},
       adsurl = {https://ui.adsabs.harvard.edu/abs/2024ApJ...975..208T}
}

@ARTICLE{Kageura2025,
       author = {{Kageura}, Yuta and {Ouchi}, Masami and {Nakane}, Minami and {Umeda}, Hiroya and {Harikane}, Yuichi and {Yoshiura}, Shintaro and {Nakajima}, Kimihiko and {Yajima}, Hidenobu and {Thai}, Tran Thi},
        title = "{Census of Ly{\ensuremath{\alpha}} Emission from {\ensuremath{\sim}}600 Galaxies at z = 5─14: Evolution of the Ly{\ensuremath{\alpha}} Luminosity Function and a Late Sharp Cosmic Reionization}",
      journal = {\apjs},
         year = 2025,
        month = jun,
       volume = {278},
       number = {2},
          eid = {33},
        pages = {33},
          doi = {10.3847/1538-4365/adc690},
archivePrefix = {arXiv},
       eprint = {2501.05834},
 primaryClass = {astro-ph.GA},
       adsurl = {https://ui.adsabs.harvard.edu/abs/2025ApJS..278...33K}
}

@ARTICLE{Tripodi2025,
       author = {{Tripodi}, Roberta and {Brada{\v{c}}}, Maru{\v{s}}a and {D'Eugenio}, Francesco and {Martis}, Nicholas and {Rihtar{\v{s}}i{\v{c}}}, Gregor and {Willott}, Chris and {Pentericci}, Laura and {Moreschini}, Bianca and {Markevitch}, Maxim and {Asada}, Yoshihisa and {Calabr{\'o}}, Antonello and {Desprez}, Guillaume and {Felicioni}, Giordano and {Gaspar}, Gaia and {Gonzalez}, Anthony H. and {Harshan}, Anishya and {Ji}, Xihan and {Jude{\v{z}}}, Jon and {Lemaux}, Brian C. and {Marconi}, Alessandro and {Markov}, Vladan and {Merida}, Rosa M. and {Napolitano}, Lorenzo and {Noirot}, Ga{\"e}l and {Parente}, Massimiliano and {Peter}, Annika H.~G. and {Robbins}, Luke and {Robertson}, Andrew and {Sarrouh}, Ghassan T.~E. and {Sawicki}, Marcin},
        title = "{A Deep Dive down the Broad-line Region: Permitted O I, Ca II, and Fe II Emission in an Active Galactic Nucleus Little Red Dot at z = 5.3}",
      journal = {\apjl},
         year = 2025,
        month = nov,
       volume = {994},
       number = {1},
          eid = {L6},
        pages = {L6},
          doi = {10.3847/2041-8213/ae13a9},
archivePrefix = {arXiv},
       eprint = {2507.20684},
 primaryClass = {astro-ph.GA},
       adsurl = {https://ui.adsabs.harvard.edu/abs/2025ApJ...994L...6T}
}

@ARTICLE{Kwan1981,
       author = {{Kwan}, J. and {Krolik}, J.~H.},
        title = "{The formation of emission lines in quasars and Seyfert nuclei}",
      journal = {\apj},
         year = 1981,
        month = nov,
       volume = {250},
        pages = {478-507},
          doi = {10.1086/159395},
       adsurl = {https://ui.adsabs.harvard.edu/abs/1981ApJ...250..478K}
}

@ARTICLE{Matsuoka2007,
       author = {{Matsuoka}, Y. and {Oyabu}, S. and {Tsuzuki}, Y. and {Kawara}, K.},
        title = "{Observations of O I and Ca II Emission Lines in Quasars: Implications for the Site of Fe II Line Emission}",
      journal = {\apj},
         year = 2007,
        month = jul,
       volume = {663},
       number = {2},
        pages = {781-798},
          doi = {10.1086/518399},
archivePrefix = {arXiv},
       eprint = {astro-ph/0703659},
 primaryClass = {astro-ph},
       adsurl = {https://ui.adsabs.harvard.edu/abs/2007ApJ...663..781M}
}

@ARTICLE{Senchyna2021,
       author = {{Senchyna}, Peter and {Stark}, Daniel P. and {Charlot}, St{\'e}phane and {Chevallard}, Jacopo and {Bruzual}, Gustavo and {Vidal-Garc{\'\i}a}, Alba},
        title = "{Ultraviolet spectra of extreme nearby star-forming regions: Evidence for an overabundance of very massive stars}",
      journal = {\mnras},
         year = 2021,
        month = jun,
       volume = {503},
       number = {4},
        pages = {6112-6135},
          doi = {10.1093/mnras/stab884},
archivePrefix = {arXiv},
       eprint = {2008.09780},
 primaryClass = {astro-ph.GA},
       adsurl = {https://ui.adsabs.harvard.edu/abs/2021MNRAS.503.6112S}
}

@ARTICLE{Wang2020,
       author = {{Wang}, Bingjie and {Heckman}, Timothy M. and {Zhu}, Guangtun and {Norman}, Colin A.},
        title = "{A Systematic Study of Galactic Outflows via Fluorescence Emission: Implications for Their Size and Structure}",
      journal = {\apj},
         year = 2020,
        month = may,
       volume = {894},
       number = {2},
          eid = {149},
        pages = {149},
          doi = {10.3847/1538-4357/ab88b4},
archivePrefix = {arXiv},
       eprint = {2004.06105},
 primaryClass = {astro-ph.GA},
       adsurl = {https://ui.adsabs.harvard.edu/abs/2020ApJ...894..149W}
}

@ARTICLE{Kornei2013,
       author = {{Kornei}, Katherine A. and {Shapley}, Alice E. and {Martin}, Crystal L. and {Coil}, Alison L. and {Lotz}, Jennifer M. and {Weiner}, Benjamin J.},
        title = "{Fine-structure Fe II* Emission and Resonant Mg II Emission in z \raisebox{-0.5ex}\textasciitilde 1 Star-forming Galaxies}",
      journal = {\apj},
         year = 2013,
        month = sep,
       volume = {774},
       number = {1},
          eid = {50},
        pages = {50},
          doi = {10.1088/0004-637X/774/1/50},
archivePrefix = {arXiv},
       eprint = {1302.6997},
 primaryClass = {astro-ph.CO},
       adsurl = {https://ui.adsabs.harvard.edu/abs/2013ApJ...774...50K}
}

@ARTICLE{Verhamme2006,
       author = {{Verhamme}, A. and {Schaerer}, D. and {Maselli}, A.},
        title = "{3D Ly{\ensuremath{\alpha}} radiation transfer. I. Understanding Ly{\ensuremath{\alpha}} line profile morphologies}",
      journal = {\aap},
         year = 2006,
        month = dec,
       volume = {460},
       number = {2},
        pages = {397-413},
          doi = {10.1051/0004-6361:20065554},
archivePrefix = {arXiv},
       eprint = {astro-ph/0608075},
 primaryClass = {astro-ph},
       adsurl = {https://ui.adsabs.harvard.edu/abs/2006A&A...460..397V}
}

@ARTICLE{Verhamme2008,
       author = {{Verhamme}, A. and {Schaerer}, D. and {Atek}, H. and {Tapken}, C.},
        title = "{3D Ly{\ensuremath{\alpha}} radiation transfer. III. Constraints on gas and stellar properties of z \raisebox{-0.5ex}\textasciitilde 3 Lyman break galaxies (LBG) and implications for high-z LBGs and Ly{\ensuremath{\alpha}} emitters}",
      journal = {\aap},
         year = 2008,
        month = nov,
       volume = {491},
       number = {1},
        pages = {89-111},
          doi = {10.1051/0004-6361:200809648},
archivePrefix = {arXiv},
       eprint = {0805.3601},
 primaryClass = {astro-ph},
       adsurl = {https://ui.adsabs.harvard.edu/abs/2008A&A...491...89V}
}

@ARTICLE{Neufeld1990,
       author = {{Neufeld}, David A.},
        title = "{The Transfer of Resonance-Line Radiation in Static Astrophysical Media}",
      journal = {\apj},
         year = 1990,
        month = feb,
       volume = {350},
        pages = {216},
          doi = {10.1086/168375},
       adsurl = {https://ui.adsabs.harvard.edu/abs/1990ApJ...350..216N}
}

@ARTICLE{Erb2012,
       author = {{Erb}, Dawn K. and {Quider}, Anna M. and {Henry}, Alaina L. and {Martin}, Crystal L.},
        title = "{Galactic Outflows in Absorption and Emission: Near-ultraviolet Spectroscopy of Galaxies at 1 < z < 2}",
      journal = {\apj},
         year = 2012,
        month = nov,
       volume = {759},
       number = {1},
          eid = {26},
        pages = {26},
          doi = {10.1088/0004-637X/759/1/26},
archivePrefix = {arXiv},
       eprint = {1209.4903},
 primaryClass = {astro-ph.CO},
       adsurl = {https://ui.adsabs.harvard.edu/abs/2012ApJ...759...26E}
}

@ARTICLE{Kewley2019,
       author = {{Kewley}, Lisa J. and {Nicholls}, David C. and {Sutherland}, Ralph S.},
        title = "{Understanding Galaxy Evolution Through Emission Lines}",
      journal = {\araa},
         year = 2019,
        month = aug,
       volume = {57},
        pages = {511-570},
          doi = {10.1146/annurev-astro-081817-051832},
archivePrefix = {arXiv},
       eprint = {1910.09730},
 primaryClass = {astro-ph.GA},
       adsurl = {https://ui.adsabs.harvard.edu/abs/2019ARA&A..57..511K}
}

@ARTICLE{Steidel2016,
       author = {{Steidel}, Charles C. and {Strom}, Allison L. and {Pettini}, Max and {Rudie}, Gwen C. and {Reddy}, Naveen A. and {Trainor}, Ryan F.},
        title = "{Reconciling the Stellar and Nebular Spectra of High-redshift Galaxies}",
      journal = {\apj},
         year = 2016,
        month = aug,
       volume = {826},
       number = {2},
          eid = {159},
        pages = {159},
          doi = {10.3847/0004-637X/826/2/159},
archivePrefix = {arXiv},
       eprint = {1605.07186},
 primaryClass = {astro-ph.GA},
       adsurl = {https://ui.adsabs.harvard.edu/abs/2016ApJ...826..159S}
}

@ARTICLE{Pahl2020,
       author = {{Pahl}, Anthony J. and {Shapley}, Alice and {Faisst}, Andreas L. and {Capak}, Peter L. and {Du}, Xinnan and {Reddy}, Naveen A. and {Laursen}, Peter and {Topping}, Michael W.},
        title = "{The redshift evolution of rest-UV spectroscopic properties to z {\ensuremath{\sim}} 5}",
      journal = {\mnras},
         year = 2020,
        month = apr,
       volume = {493},
       number = {3},
        pages = {3194-3211},
          doi = {10.1093/mnras/staa355},
archivePrefix = {arXiv},
       eprint = {1910.04179},
 primaryClass = {astro-ph.GA},
       adsurl = {https://ui.adsabs.harvard.edu/abs/2020MNRAS.493.3194P}
}

@ARTICLE{Du2018,
       author = {{Du}, Xinnan and {Shapley}, Alice E. and {Reddy}, Naveen A. and {Jones}, Tucker and {Stark}, Daniel P. and {Steidel}, Charles C. and {Strom}, Allison L. and {Rudie}, Gwen C. and {Erb}, Dawn K. and {Ellis}, Richard S. and {Pettini}, Max},
        title = "{The Redshift Evolution of Rest-UV Spectroscopic Properties in Lyman-break Galaxies at z {\ensuremath{\sim}} 2-4}",
      journal = {\apj},
         year = 2018,
        month = jun,
       volume = {860},
       number = {1},
          eid = {75},
        pages = {75},
          doi = {10.3847/1538-4357/aabfcf},
archivePrefix = {arXiv},
       eprint = {1803.05912},
 primaryClass = {astro-ph.GA},
       adsurl = {https://ui.adsabs.harvard.edu/abs/2018ApJ...860...75D}
}

@ARTICLE{Jones2013,
       author = {{Jones}, Tucker A. and {Ellis}, Richard S. and {Schenker}, Matthew A. and {Stark}, Daniel P.},
        title = "{Keck Spectroscopy of Gravitationally Lensed z \raisebox{-0.5ex}\textasciitilde= 4 Galaxies: Improved Constraints on the Escape Fraction of Ionizing Photons}",
      journal = {\apj},
         year = 2013,
        month = dec,
       volume = {779},
       number = {1},
          eid = {52},
        pages = {52},
          doi = {10.1088/0004-637X/779/1/52},
archivePrefix = {arXiv},
       eprint = {1304.7015},
 primaryClass = {astro-ph.CO},
       adsurl = {https://ui.adsabs.harvard.edu/abs/2013ApJ...779...52J}
}

@ARTICLE{Berg2022,
       author = {{Berg}, Danielle A. and {James}, Bethan L. and {King}, Teagan and {McDonald}, Meaghan and {Chen}, Zuyi and {Chisholm}, John and {Heckman}, Timothy and {Martin}, Crystal L. and {Stark}, Dan P. and {Aloisi}, Alessandra and {Amor{\'\i}n}, Ricardo O. and {Arellano-C{\'o}rdova}, Karla Z. and {Bayliss}, Matthew and {Bordoloi}, Rongmon and {Brinchmann}, Jarle and {Charlot}, St{\'e}phane and {Chevallard}, Jacopo and {Clark}, Ilyse and {Erb}, Dawn K. and {Feltre}, Anna and {Gronke}, Max and {Hayes}, Matthew and {Henry}, Alaina and {Hernandez}, Svea and {Jaskot}, Anne and {Jones}, Tucker and {Kewley}, Lisa J. and {Kumari}, Nimisha and {Leitherer}, Claus and {Llerena}, Mario and {Maseda}, Michael and {Mingozzi}, Matilde and {Nanayakkara}, Themiya and {Ouchi}, Masami and {Plat}, Adele and {Pogge}, Richard W. and {Ravindranath}, Swara and {Rigby}, Jane R. and {Sanders}, Ryan and {Scarlata}, Claudia and {Senchyna}, Peter and {Skillman}, Evan D. and {Steidel}, Charles C. and {Strom}, Allison L. and {Sugahara}, Yuma and {Wilkins}, Stephen M. and {Wofford}, Aida and {Xu}, Xinfeng and {Classy Team}},
        title = "{The COS Legacy Archive Spectroscopy Survey (CLASSY) Treasury Atlas}",
      journal = {\apjs},
         year = 2022,
        month = aug,
       volume = {261},
       number = {2},
          eid = {31},
        pages = {31},
          doi = {10.3847/1538-4365/ac6c03},
archivePrefix = {arXiv},
       eprint = {2203.07357},
 primaryClass = {astro-ph.GA},
       adsurl = {https://ui.adsabs.harvard.edu/abs/2022ApJS..261...31B}
}

@ARTICLE{Law2025,
       author = {{Law}, David R. and {Argyriou}, Ioannis and {Gordon}, Karl D. and {Sloan}, G.~C. and {Gasman}, Danny and {Glasse}, Alistair and {Larson}, Kirsten and {Fletcher}, Leigh N. and {Labiano}, Alvaro and {Noriega-Crespo}, Alberto},
        title = "{The James Webb Space Telescope Absolute Flux Calibration. III. Mid-infrared Instrument Medium Resolution Integral Field Unit Spectrometer}",
      journal = {\aj},
         year = 2025,
        month = feb,
       volume = {169},
       number = {2},
          eid = {67},
        pages = {67},
          doi = {10.3847/1538-3881/ad9685},
archivePrefix = {arXiv},
       eprint = {2409.15435},
 primaryClass = {astro-ph.IM},
       adsurl = {https://ui.adsabs.harvard.edu/abs/2025AJ....169...67L}
}

@ARTICLE{Pettini2000,
       author = {{Pettini}, Max and {Steidel}, Charles C. and {Adelberger}, Kurt L. and {Dickinson}, Mark and {Giavalisco}, Mauro},
        title = "{The Ultraviolet Spectrum of MS 1512-CB58: An Insight into Lyman-Break Galaxies}",
      journal = {\apj},
         year = 2000,
        month = jan,
       volume = {528},
       number = {1},
        pages = {96-107},
          doi = {10.1086/308176},
archivePrefix = {arXiv},
       eprint = {astro-ph/9908007},
 primaryClass = {astro-ph},
       adsurl = {https://ui.adsabs.harvard.edu/abs/2000ApJ...528...96P}
}

@ARTICLE{Pettini2002,
       author = {{Pettini}, Max and {Rix}, Samantha A. and {Steidel}, Charles C. and {Adelberger}, Kurt L. and {Hunt}, Matthew P. and {Shapley}, Alice E.},
        title = "{New Observations of the Interstellar Medium in the Lyman Break Galaxy MS 1512-cB58}",
      journal = {\apj},
         year = 2002,
        month = apr,
       volume = {569},
       number = {2},
        pages = {742-757},
          doi = {10.1086/339355},
archivePrefix = {arXiv},
       eprint = {astro-ph/0110637},
 primaryClass = {astro-ph},
       adsurl = {https://ui.adsabs.harvard.edu/abs/2002ApJ...569..742P}
}

@ARTICLE{Mainali2023,
       author = {{Mainali}, Ramesh and {Stark}, Daniel P. and {Jones}, Tucker and {Ellis}, Richard S. and {Hezaveh}, Yashar D. and {Rigby}, Jane R.},
        title = "{Spectroscopy of CASSOWARY gravitationally lensed galaxies in SDSS: characterization of an extremely bright reionization-era analogue at z = 1.42}",
      journal = {\mnras},
         year = 2023,
        month = apr,
       volume = {520},
       number = {3},
        pages = {4037-4056},
          doi = {10.1093/mnras/stad387},
archivePrefix = {arXiv},
       eprint = {2301.11264},
 primaryClass = {astro-ph.GA},
       adsurl = {https://ui.adsabs.harvard.edu/abs/2023MNRAS.520.4037M}
}

@ARTICLE{Christensen2012,
       author = {{Christensen}, Lise and {Richard}, Johan and {Hjorth}, Jens and {Milvang-Jensen}, Bo and {Laursen}, Peter and {Limousin}, Marceau and {Dessauges-Zavadsky}, Miroslava and {Grillo}, Claudio and {Ebeling}, Harald},
        title = "{The low-mass end of the fundamental relation for gravitationally lensed star-forming galaxies at 1 < z < 6}",
      journal = {\mnras},
         year = 2012,
        month = dec,
       volume = {427},
       number = {3},
        pages = {1953-1972},
          doi = {10.1111/j.1365-2966.2012.22006.x},
archivePrefix = {arXiv},
       eprint = {1209.0767},
 primaryClass = {astro-ph.CO},
       adsurl = {https://ui.adsabs.harvard.edu/abs/2012MNRAS.427.1953C}
}

@ARTICLE{Topping2025_demographic,
       author = {{Topping}, Michael W. and {Stark}, Daniel P. and {Senchyna}, Peter and {Chen}, Zuyi and {Zitrin}, Adi and {Endsley}, Ryan and {Charlot}, St{\'e}phane and {Furtak}, Lukas J. and {Maseda}, Michael V. and {Plat}, Adele and {Smit}, Renske and {Mainali}, Ramesh and {Chevallard}, Jacopo and {Molyneux}, Stephen and {Rigby}, Jane R.},
        title = "{Deep Rest-UV JWST/NIRSpec Spectroscopy of Early Galaxies: The Demographics of C IV and N-emitters in the Reionization Era}",
      journal = {\apj},
         year = 2025,
        month = feb,
       volume = {980},
       number = {2},
          eid = {225},
        pages = {225},
          doi = {10.3847/1538-4357/ada95c},
archivePrefix = {arXiv},
       eprint = {2407.19009},
 primaryClass = {astro-ph.GA},
       adsurl = {https://ui.adsabs.harvard.edu/abs/2025ApJ...980..225T}
}

@ARTICLE{Mingozzi2022,
       author = {{Mingozzi}, Matilde and {James}, Bethan L. and {Arellano-C{\'o}rdova}, Karla Z. and {Berg}, Danielle A. and {Senchyna}, Peter and {Chisholm}, John and {Brinchmann}, Jarle and {Aloisi}, Alessandra and {Amor{\'\i}n}, Ricardo O. and {Charlot}, St{\'e}phane and {Feltre}, Anna and {Hayes}, Matthew and {Heckman}, Timothy and {Henry}, Alaina and {Hernandez}, Svea and {Kumari}, Nimisha and {Leitherer}, Claus and {Llerena}, Mario and {Martin}, Crystal L. and {Nanayakkara}, Themiya and {Ravindranath}, Swara and {Skillman}, Evan D. and {Sugahara}, Yuma and {Wofford}, Aida and {Xu}, Xinfeng},
        title = "{CLASSY IV. Exploring UV Diagnostics of the Interstellar Medium in Local High-z Analogs at the Dawn of the JWST Era}",
      journal = {\apj},
         year = 2022,
        month = nov,
       volume = {939},
       number = {2},
          eid = {110},
        pages = {110},
          doi = {10.3847/1538-4357/ac952c},
archivePrefix = {arXiv},
       eprint = {2209.09047},
 primaryClass = {astro-ph.GA},
       adsurl = {https://ui.adsabs.harvard.edu/abs/2022ApJ...939..110M}
}

@ARTICLE{James2014,
       author = {{James}, Bethan L. and {Pettini}, Max and {Christensen}, Lise and {Auger}, Matthew W. and {Becker}, George D. and {King}, Lindsay J. and {Quider}, Anna M. and {Shapley}, Alice E. and {Steidel}, Charles C.},
        title = "{Testing metallicity indicators at z {\ensuremath{\sim}} 1.4 with the gravitationally lensed galaxy CASSOWARY 20}",
      journal = {\mnras},
         year = 2014,
        month = may,
       volume = {440},
       number = {2},
        pages = {1794-1809},
          doi = {10.1093/mnras/stu287},
archivePrefix = {arXiv},
       eprint = {1311.5092},
 primaryClass = {astro-ph.GA},
       adsurl = {https://ui.adsabs.harvard.edu/abs/2014MNRAS.440.1794J}
}

@ARTICLE{Maseda2017,
       author = {{Maseda}, Michael V. and {Brinchmann}, Jarle and {Franx}, Marijn and {Bacon}, Roland and {Bouwens}, Rychard J. and {Schmidt}, Kasper B. and {Boogaard}, Leindert A. and {Contini}, Thierry and {Feltre}, Anna and {Inami}, Hanae and {Kollatschny}, Wolfram and {Marino}, Raffaella A. and {Richard}, Johan and {Verhamme}, Anne and {Wisotzki}, Lutz},
        title = "{The MUSE Hubble Ultra Deep Field Survey. IV. Global properties of C III] emitters}",
      journal = {\aap},
         year = 2017,
        month = dec,
       volume = {608},
          eid = {A4},
        pages = {A4},
          doi = {10.1051/0004-6361/201730985},
archivePrefix = {arXiv},
       eprint = {1710.06432},
 primaryClass = {astro-ph.GA},
       adsurl = {https://ui.adsabs.harvard.edu/abs/2017A&A...608A...4M}
}

@ARTICLE{Adamo2024,
       author = {{Adamo}, Angela and {Bradley}, Larry D. and {Vanzella}, Eros and {Claeyssens}, Ad{\'e}la{\"\i}de and {Welch}, Brian and {Diego}, Jose M. and {Mahler}, Guillaume and {Oguri}, Masamune and {Sharon}, Keren and {Abdurro'uf} and {Hsiao}, Tiger Yu-Yang and {Xu}, Xinfeng and {Messa}, Matteo and {Lassen}, Augusto E. and {Zackrisson}, Erik and {Brammer}, Gabriel and {Coe}, Dan and {Kokorev}, Vasily and {Ricotti}, Massimo and {Zitrin}, Adi and {Fujimoto}, Seiji and {Inoue}, Akio K. and {Resseguier}, Tom and {Rigby}, Jane R. and {Jim{\'e}nez-Teja}, Yolanda and {Windhorst}, Rogier A. and {Hashimoto}, Takuya and {Tamura}, Yoichi},
        title = "{Bound star clusters observed in a lensed galaxy 460 Myr after the Big Bang}",
      journal = {\nat},
         year = 2024,
        month = aug,
       volume = {632},
       number = {8025},
        pages = {513-516},
          doi = {10.1038/s41586-024-07703-7},
archivePrefix = {arXiv},
       eprint = {2401.03224},
 primaryClass = {astro-ph.GA},
       adsurl = {https://ui.adsabs.harvard.edu/abs/2024Natur.632..513A}
}

@ARTICLE{Vanzella2023,
       author = {{Vanzella}, Eros and {Claeyssens}, Ad{\'e}la{\"\i}de and {Welch}, Brian and {Adamo}, Angela and {Coe}, Dan and {Diego}, Jose M. and {Mahler}, Guillaume and {Khullar}, Gourav and {Kokorev}, Vasily and {Oguri}, Masamune and {Ravindranath}, Swara and {Furtak}, Lukas J. and {Hsiao}, Tiger Yu-Yang and {Abdurro'uf} and {Mandelker}, Nir and {Brammer}, Gabriel and {Bradley}, Larry D. and {Brada{\v{c}}}, Maru{\v{s}}a and {Conselice}, Christopher J. and {Dayal}, Pratika and {Nonino}, Mario and {Andrade-Santos}, Felipe and {Windhorst}, Rogier A. and {Pirzkal}, Nor and {Sharon}, Keren and {de Mink}, S.~E. and {Fujimoto}, Seiji and {Zitrin}, Adi and {Eldridge}, Jan J. and {Norman}, Colin},
        title = "{JWST/NIRCam Probes Young Star Clusters in the Reionization Era Sunrise Arc}",
      journal = {\apj},
         year = 2023,
        month = mar,
       volume = {945},
       number = {1},
          eid = {53},
        pages = {53},
          doi = {10.3847/1538-4357/acb59a},
archivePrefix = {arXiv},
       eprint = {2211.09839},
 primaryClass = {astro-ph.GA},
       adsurl = {https://ui.adsabs.harvard.edu/abs/2023ApJ...945...53V}
}

@ARTICLE{Trainor2015,
       author = {{Trainor}, Ryan F. and {Steidel}, Charles C. and {Strom}, Allison L. and {Rudie}, Gwen C.},
        title = "{The Spectroscopic Properties of Ly{\ensuremath{\alpha}}-Emitters at z {\ensuremath{\sim}}2.7: Escaping Gas and Photons from Faint Galaxies}",
      journal = {\apj},
         year = 2015,
        month = aug,
       volume = {809},
       number = {1},
          eid = {89},
        pages = {89},
          doi = {10.1088/0004-637X/809/1/89},
archivePrefix = {arXiv},
       eprint = {1506.08205},
 primaryClass = {astro-ph.GA},
       adsurl = {https://ui.adsabs.harvard.edu/abs/2015ApJ...809...89T}
}

@ARTICLE{Trainor2019,
       author = {{Trainor}, Ryan F. and {Strom}, Allison L. and {Steidel}, Charles C. and {Rudie}, Gwen C. and {Chen}, Yuguang and {Theios}, Rachel L.},
        title = "{Predicting Ly{\ensuremath{\alpha}} Emission from Galaxies via Empirical Markers of Production and Escape in the KBSS}",
      journal = {\apj},
         year = 2019,
        month = dec,
       volume = {887},
       number = {1},
          eid = {85},
        pages = {85},
          doi = {10.3847/1538-4357/ab4993},
archivePrefix = {arXiv},
       eprint = {1908.04794},
 primaryClass = {astro-ph.GA},
       adsurl = {https://ui.adsabs.harvard.edu/abs/2019ApJ...887...85T}
}

@ARTICLE{Glazer2025,
       author = {{Glazer}, Kelsey S. and {Jones}, Tucker and {Chen}, Yuguang and {Sanders}, Ryan L. and {Brada{\v{c}}}, Maru{\v{s}}a and {Pahl}, Anthony J. and {Shapley}, Alice E. and {Ellis}, Richard S. and {Topping}, Michael W. and {Reddy}, Naveen A.},
        title = "{Stacking PANCAKEZ: Spectroscopic Analysis with NIRSpec Stacks in the Epoch of Reionization. Weak Interstellar Medium Absorption and Implications for Ionizing Photon Escape at z {\ensuremath{\sim}} 7}",
      journal = {\apj},
         year = 2025,
        month = oct,
       volume = {992},
       number = {2},
          eid = {191},
        pages = {191},
          doi = {10.3847/1538-4357/ae0194},
archivePrefix = {arXiv},
       eprint = {2504.21080},
 primaryClass = {astro-ph.GA},
       adsurl = {https://ui.adsabs.harvard.edu/abs/2025ApJ...992..191G}
}

@ARTICLE{Reddy2022,
       author = {{Reddy}, Naveen A. and {Topping}, Michael W. and {Shapley}, Alice E. and {Steidel}, Charles C. and {Sanders}, Ryan L. and {Du}, Xinnan and {Coil}, Alison L. and {Mobasher}, Bahram and {Price}, Sedona H. and {Shivaei}, Irene},
        title = "{The Effects of Stellar Population and Gas Covering Fraction on the Emergent Ly{\ensuremath{\alpha}} Emission of High-redshift Galaxies}",
      journal = {\apj},
         year = 2022,
        month = feb,
       volume = {926},
       number = {1},
          eid = {31},
        pages = {31},
          doi = {10.3847/1538-4357/ac3b4c},
archivePrefix = {arXiv},
       eprint = {2108.05363},
 primaryClass = {astro-ph.GA},
       adsurl = {https://ui.adsabs.harvard.edu/abs/2022ApJ...926...31R}
}

@ARTICLE{Ji2024,
       author = {{Ji}, Xihan and {{\"U}bler}, Hannah and {Maiolino}, Roberto and {D'Eugenio}, Francesco and {Arribas}, Santiago and {Bunker}, Andrew J. and {Charlot}, St{\'e}phane and {Perna}, Michele and {Rodr{\'\i}guez Del Pino}, Bruno and {B{\"o}ker}, Torsten and {Cresci}, Giovanni and {Curti}, Mirko and {Kumari}, Nimisha and {Lamperti}, Isabella},
        title = "{GA-NIFS: an extremely nitrogen-loud and chemically stratified galaxy at z   5.55}",
      journal = {\mnras},
         year = 2024,
        month = nov,
       volume = {535},
       number = {1},
        pages = {881-908},
          doi = {10.1093/mnras/stae2375},
archivePrefix = {arXiv},
       eprint = {2404.04148},
 primaryClass = {astro-ph.GA},
       adsurl = {https://ui.adsabs.harvard.edu/abs/2024MNRAS.535..881J}
}

@ARTICLE{Sanders2020,
       author = {{Sanders}, Ryan L. and {Shapley}, Alice E. and {Reddy}, Naveen A. and {Kriek}, Mariska and {Siana}, Brian and {Coil}, Alison L. and {Mobasher}, Bahram and {Shivaei}, Irene and {Freeman}, William R. and {Azadi}, Mojegan and {Price}, Sedona H. and {Leung}, Gene and {Fetherolf}, Tara and {de Groot}, Laura and {Zick}, Tom and {Fornasini}, Francesca M. and {Barro}, Guillermo},
        title = "{The MOSDEF survey: direct-method metallicities and ISM conditions at z {\ensuremath{\sim}} 1.5-3.5}",
      journal = {\mnras},
         year = 2020,
        month = jan,
       volume = {491},
       number = {1},
        pages = {1427-1455},
          doi = {10.1093/mnras/stz3032},
archivePrefix = {arXiv},
       eprint = {1907.00013},
 primaryClass = {astro-ph.GA},
       adsurl = {https://ui.adsabs.harvard.edu/abs/2020MNRAS.491.1427S}
}

@Article{Topping2020,
  author           = {Topping, Michael W. and Shapley, Alice E. and Reddy, Naveen A. and Sanders, Ryan L. and Coil, Alison L. and Kriek, Mariska and Mobasher, Bahram and Siana, Brian},
  journal          = {\mnras},
  title            = {The MOSDEF-LRIS Survey: the interplay between massive stars and ionized gas in high-redshift star-forming galaxies},
  year             = {2020},
  month            = jul,
  number           = {4},
  pages            = {4430-4444},
  volume           = {495},
  archiveprefix    = {arXiv},
  doi              = {10.1093/mnras/staa1410},
  eprint           = {1912.10243},
  modificationdate = {2026-01-30T21:04:09},
  primaryclass     = {astro-ph.GA},
  url              = {https://ui.adsabs.harvard.edu/abs/2020MNRAS.495.4430T},
}

@Article{Runco2021,
  author           = {Runco, Jordan N and Shapley, Alice E and Sanders, Ryan L and Topping, Michael W and Kriek, Mariska and Reddy, Naveen A and Coil, Alison L and Mobasher, Bahram and Siana, Brian and Freeman, William R and Shivaei, Irene and Azadi, Mojegan and Price, Sedona H and Leung, Gene C K and Fetherolf, Tara and de Groot, Laura and Zick, Tom and Fornasini, Francesca M and Barro, Guillermo},
  journal          = {Monthly Notices of the Royal Astronomical Society},
  title            = {The MOSDEF survey: a comprehensive analysis of the rest-optical emission-line properties of z ∼ 2.3 star-forming galaxies},
  year             = {2021},
  issn             = {1365-2966},
  month            = jan,
  number           = {2},
  pages            = {2600--2614},
  volume           = {502},
  doi              = {10.1093/mnras/stab119},
  modificationdate = {2026-01-30T21:05:22},
  publisher        = {Oxford University Press (OUP)},
}

@ARTICLE{Wofford2023,
       author = {{Wofford}, Aida and {Sixtos}, Andr{\'e}s and {Charlot}, Stephane and {Bruzual}, Gustavo and {Cullen}, Fergus and {Stanton}, Thomas M. and {Hern{\'a}ndez}, Svea and {Smith}, Linda J. and {Hayes}, Matthew},
        title = "{Extreme broad He II emission at high and low redshifts: the dominant role of VMS in NGC 3125-A1 and CDFS131717}",
      journal = {\mnras},
         year = 2023,
        month = aug,
       volume = {523},
       number = {3},
        pages = {3949-3966},
          doi = {10.1093/mnras/stad1622},
archivePrefix = {arXiv},
       eprint = {2305.14563},
 primaryClass = {astro-ph.GA},
       adsurl = {https://ui.adsabs.harvard.edu/abs/2023MNRAS.523.3949W}
}

@ARTICLE{Mason2018a,
       author = {{Mason}, Charlotte A. and {Treu}, Tommaso and {de Barros}, Stephane and {Dijkstra}, Mark and {Fontana}, Adriano and {Mesinger}, Andrei and {Pentericci}, Laura and {Trenti}, Michele and {Vanzella}, Eros},
        title = "{Beacons into the Cosmic Dark Ages: Boosted Transmission of Ly{\ensuremath{\alpha}} from UV Bright Galaxies at z {\ensuremath{\gtrsim}} 7}",
      journal = {\apjl},
         year = 2018,
        month = apr,
       volume = {857},
       number = {2},
          eid = {L11},
        pages = {L11},
          doi = {10.3847/2041-8213/aabbab},
archivePrefix = {arXiv},
       eprint = {1801.01891},
 primaryClass = {astro-ph.CO},
       adsurl = {https://ui.adsabs.harvard.edu/abs/2018ApJ...857L..11M}
}

@ARTICLE{Dijkstra2011,
       author = {{Dijkstra}, Mark and {Mesinger}, Andrei and {Wyithe}, J. Stuart B.},
        title = "{The detectability of Ly{\ensuremath{\alpha}} emission from galaxies during the epoch of reionization}",
      journal = {\mnras},
         year = 2011,
        month = jul,
       volume = {414},
       number = {3},
        pages = {2139-2147},
          doi = {10.1111/j.1365-2966.2011.18530.x},
archivePrefix = {arXiv},
       eprint = {1101.5160},
 primaryClass = {astro-ph.CO},
       adsurl = {https://ui.adsabs.harvard.edu/abs/2011MNRAS.414.2139D}
}

@ARTICLE{Reddy2016,
       author = {{Reddy}, Naveen A. and {Steidel}, Charles C. and {Pettini}, Max and {Bogosavljevi{\'c}}, Milan and {Shapley}, Alice E.},
        title = "{The Connection Between Reddening, Gas Covering Fraction, and the Escape of Ionizing Radiation at High Redshift}",
      journal = {\apj},
         year = 2016,
        month = sep,
       volume = {828},
       number = {2},
          eid = {108},
        pages = {108},
          doi = {10.3847/0004-637X/828/2/108},
archivePrefix = {arXiv},
       eprint = {1606.03452},
 primaryClass = {astro-ph.GA},
       adsurl = {https://ui.adsabs.harvard.edu/abs/2016ApJ...828..108R}
}

@ARTICLE{Yuan2025,
       author = {{Yuan}, Yuxuan and {Martin-Alvarez}, Sergio and {Haehnelt}, Martin G. and {Garel}, Thibault and {Keating}, Laura and {Witstok}, Joris and {Sijacki}, Debora},
        title = "{Extended red wings and the visibility of reionization-epoch Lyman-{\ensuremath{\alpha}} emitters}",
      journal = {\mnras},
         year = 2025,
        month = sep,
       volume = {542},
       number = {2},
        pages = {762-789},
          doi = {10.1093/mnras/staf1252},
archivePrefix = {arXiv},
       eprint = {2412.07970},
 primaryClass = {astro-ph.GA},
       adsurl = {https://ui.adsabs.harvard.edu/abs/2025MNRAS.542..762Y}
}

@ARTICLE{Pascale2024,
       author = {{Pascale}, Massimo and {Dai}, Liang},
        title = "{A Young Super Star Cluster Powering a Nebula of Retained Massive Star Ejecta}",
      journal = {\apj},
         year = 2024,
        month = dec,
       volume = {976},
       number = {2},
          eid = {166},
        pages = {166},
          doi = {10.3847/1538-4357/ad7732},
archivePrefix = {arXiv},
       eprint = {2404.10755},
 primaryClass = {astro-ph.GA},
       adsurl = {https://ui.adsabs.harvard.edu/abs/2024ApJ...976..166P}
}

@ARTICLE{Bolan2022,
       author = {{Bolan}, Patricia and {Lemaux}, Brian C. and {Mason}, Charlotte and {Brada{\v{c}}}, Maru{\v{s}}a and {Treu}, Tommaso and {Strait}, Victoria and {Pelliccia}, Debora and {Pentericci}, Laura and {Malkan}, Matthew},
        title = "{Inferring the intergalactic medium neutral fraction at z   6-8 with low-luminosity Lyman break galaxies}",
      journal = {\mnras},
         year = 2022,
        month = dec,
       volume = {517},
       number = {3},
        pages = {3263-3274},
          doi = {10.1093/mnras/stac1963},
archivePrefix = {arXiv},
       eprint = {2111.14912},
 primaryClass = {astro-ph.GA},
       adsurl = {https://ui.adsabs.harvard.edu/abs/2022MNRAS.517.3263B}
}

@ARTICLE{Cain2025,
       author = {{Cain}, Christopher and {Lopez}, Garett and {D'Aloisio}, Anson and {Mu{\~n}oz}, Julian B. and {Jansen}, Rolf A. and {Windhorst}, Rogier A. and {Gangolli}, Nakul},
        title = "{Chasing the Beginning of Reionization in the JWST Era}",
      journal = {\apj},
         year = 2025,
        month = feb,
       volume = {980},
       number = {1},
          eid = {83},
        pages = {83},
          doi = {10.3847/1538-4357/ada152},
archivePrefix = {arXiv},
       eprint = {2409.02989},
 primaryClass = {astro-ph.CO},
       adsurl = {https://ui.adsabs.harvard.edu/abs/2025ApJ...980...83C}
}

@ARTICLE{Endsley2022,
       author = {{Endsley}, Ryan and {Stark}, Daniel P. and {Bouwens}, Rychard J. and {Schouws}, Sander and {Smit}, Renske and {Stefanon}, Mauro and {Inami}, Hanae and {Bowler}, Rebecca A.~A. and {Oesch}, Pascal and {Gonzalez}, Valentino and {Aravena}, Manuel and {da Cunha}, Elisabete and {Dayal}, Pratika and {Ferrara}, Andrea and {Graziani}, Luca and {Nanayakkara}, Themiya and {Pallottini}, Andrea and {Schneider}, Raffaella and {Sommovigo}, Laura and {Topping}, Michael and {van der Werf}, Paul and {Hutter}, Anne},
        title = "{The REBELS ALMA Survey: efficient Ly {\ensuremath{\alpha}} transmission of UV-bright z ≃ 7 galaxies from large velocity offsets and broad line widths}",
      journal = {\mnras},
         year = 2022,
        month = dec,
       volume = {517},
       number = {4},
        pages = {5642-5659},
          doi = {10.1093/mnras/stac3064},
archivePrefix = {arXiv},
       eprint = {2202.01219},
 primaryClass = {astro-ph.GA},
       adsurl = {https://ui.adsabs.harvard.edu/abs/2022MNRAS.517.5642E}
}

@ARTICLE{Gelli2024,
       author = {{Gelli}, Viola and {Mason}, Charlotte and {Hayward}, Christopher C.},
        title = "{The Impact of Mass-dependent Stochasticity at Cosmic Dawn}",
      journal = {\apj},
         year = 2024,
        month = nov,
       volume = {975},
       number = {2},
          eid = {192},
        pages = {192},
          doi = {10.3847/1538-4357/ad7b36},
archivePrefix = {arXiv},
       eprint = {2405.13108},
 primaryClass = {astro-ph.GA},
       adsurl = {https://ui.adsabs.harvard.edu/abs/2024ApJ...975..192G}
}

@ARTICLE{Hutter2025,
       author = {{Hutter}, Anne and {Cueto}, Elie R. and {Dayal}, Pratika and {Gottl{\"o}ber}, Stefan and {Trebitsch}, Maxime and {Yepes}, Gustavo},
        title = "{ASTRAEUS: X. Indications of a top-heavy initial mass function in highly star-forming galaxies from JWST observations at z > 10}",
      journal = {\aap},
         year = 2025,
        month = feb,
       volume = {694},
          eid = {A254},
        pages = {A254},
          doi = {10.1051/0004-6361/202452460},
archivePrefix = {arXiv},
       eprint = {2410.00730},
 primaryClass = {astro-ph.GA},
       adsurl = {https://ui.adsabs.harvard.edu/abs/2025A&A...694A.254H}
}

@ARTICLE{Katz2025,
       author = {{Katz}, Harley and {Cameron}, Alex J. and {Saxena}, Aayush and {Barrufet}, Laia and {Choustikov}, Nichloas and {Cleri}, Nikko J. and {de Graff}, Anna and {Ellis}, Richard S. and {Fosbury}, Robert A.~E. and {Heintz}, Kasper E. and {Maseda}, Michael and {Matthee}, Jorryt and {McConachie}, Ian and {Oesch}, Pascal A.},
        title = "{21 Balmer Jump Street: The Nebular Continuum at High Redshift and Implications for the Bright Galaxy Problem, UV Continuum Slopes, and Early Stellar Populations}",
      journal = {The Open Journal of Astrophysics},
         year = 2025,
        month = jul,
       volume = {8},
          eid = {104},
        pages = {104},
          doi = {10.33232/001c.142570},
archivePrefix = {arXiv},
       eprint = {2408.03189},
 primaryClass = {astro-ph.GA},
       adsurl = {https://ui.adsabs.harvard.edu/abs/2025OJAp....8E.104K}
}

@ARTICLE{Hainich2015,
       author = {{Hainich}, R. and {Pasemann}, D. and {Todt}, H. and {Shenar}, T. and {Sander}, A. and {Hamann}, W.-R.},
        title = "{Wolf-Rayet stars in the Small Magellanic Cloud. I. Analysis of the single WN stars}",
      journal = {\aap},
         year = 2015,
        month = sep,
       volume = {581},
          eid = {A21},
        pages = {A21},
          doi = {10.1051/0004-6361/201526241},
archivePrefix = {arXiv},
       eprint = {1507.04000},
 primaryClass = {astro-ph.SR},
       adsurl = {https://ui.adsabs.harvard.edu/abs/2015A&A...581A..21H}
}

@ARTICLE{Smith2023,
       author = {{Smith}, Linda J. and {Oey}, M.~S. and {Hernandez}, Svea and {Ryon}, Jenna and {Leitherer}, Claus and {Charlot}, Stephane and {Bruzual}, Gustavo and {Calzetti}, Daniela and {Chu}, You-Hua and {Hayes}, Matthew J. and {James}, Bethan L. and {Jaskot}, Anne E. and {{\"O}stlin}, G{\"o}ran},
        title = "{HST FUV Spectroscopy of Super Star Cluster A in the Green Pea Analog Mrk 71: Revealing the Presence of Very Massive Stars}",
      journal = {\apj},
         year = 2023,
        month = dec,
       volume = {958},
       number = {2},
          eid = {194},
        pages = {194},
          doi = {10.3847/1538-4357/ad00b4},
archivePrefix = {arXiv},
       eprint = {2310.03413},
 primaryClass = {astro-ph.GA},
       adsurl = {https://ui.adsabs.harvard.edu/abs/2023ApJ...958..194S}
}

@ARTICLE{Hillier1998,
       author = {{Hillier}, D. John and {Miller}, D.~L.},
        title = "{The Treatment of Non-LTE Line Blanketing in Spherically Expanding Outflows}",
      journal = {\apj},
         year = 1998,
        month = mar,
       volume = {496},
       number = {1},
        pages = {407-427},
          doi = {10.1086/305350},
       adsurl = {https://ui.adsabs.harvard.edu/abs/1998ApJ...496..407H}
}

@ARTICLE{Chisholm2019,
       author = {{Chisholm}, J. and {Rigby}, J.~R. and {Bayliss}, M. and {Berg}, D.~A. and {Dahle}, H. and {Gladders}, M. and {Sharon}, K.},
        title = "{Constraining the Metallicities, Ages, Star Formation Histories, and Ionizing Continua of Extragalactic Massive Star Populations}",
      journal = {\apj},
         year = 2019,
        month = sep,
       volume = {882},
       number = {2},
          eid = {182},
        pages = {182},
          doi = {10.3847/1538-4357/ab3104},
archivePrefix = {arXiv},
       eprint = {1905.04314},
 primaryClass = {astro-ph.GA},
       adsurl = {https://ui.adsabs.harvard.edu/abs/2019ApJ...882..182C}
}

@ARTICLE{Gordon2003,
       author = {{Gordon}, Karl D. and {Clayton}, Geoffrey C. and {Misselt}, K.~A. and {Landolt}, Arlo U. and {Wolff}, Michael J.},
        title = "{A Quantitative Comparison of the Small Magellanic Cloud, Large Magellanic Cloud, and Milky Way Ultraviolet to Near-Infrared Extinction Curves}",
      journal = {\apj},
         year = 2003,
        month = sep,
       volume = {594},
       number = {1},
        pages = {279-293},
          doi = {10.1086/376774},
archivePrefix = {arXiv},
       eprint = {astro-ph/0305257},
 primaryClass = {astro-ph},
       adsurl = {https://ui.adsabs.harvard.edu/abs/2003ApJ...594..279G}
}

@ARTICLE{Plat2019,
       author = {{Plat}, A. and {Charlot}, S. and {Bruzual}, G. and {Feltre}, A. and {Vidal-Garc{\'\i}a}, A. and {Morisset}, C. and {Chevallard}, J. and {Todt}, H.},
        title = "{Constraints on the production and escape of ionizing radiation from the emission-line spectra of metal-poor star-forming galaxies}",
      journal = {\mnras},
         year = 2019,
        month = nov,
       volume = {490},
       number = {1},
        pages = {978-1009},
          doi = {10.1093/mnras/stz2616},
archivePrefix = {arXiv},
       eprint = {1909.07386},
 primaryClass = {astro-ph.GA},
       adsurl = {https://ui.adsabs.harvard.edu/abs/2019MNRAS.490..978P}
}

@ARTICLE{Pauldrach1986,
       author = {{Pauldrach}, A. and {Puls}, J. and {Kudritzki}, R.~P.},
        title = "{Radiation-driven winds of hot luminous stars. Improvements of the theory and first results.}",
      journal = {\aap},
         year = 1986,
        month = aug,
       volume = {164},
        pages = {86-100},
       adsurl = {https://ui.adsabs.harvard.edu/abs/1986A&A...164...86P}
}

@ARTICLE{McQuinn2008,
       author = {{McQuinn}, Matthew and {Lidz}, Adam and {Zaldarriaga}, Matias and {Hernquist}, Lars and {Dutta}, Suvendra},
        title = "{Probing the neutral fraction of the IGM with GRBs during the epoch of reionization}",
      journal = {\mnras},
         year = 2008,
        month = aug,
       volume = {388},
       number = {3},
        pages = {1101-1110},
          doi = {10.1111/j.1365-2966.2008.13271.x},
archivePrefix = {arXiv},
       eprint = {0710.1018},
 primaryClass = {astro-ph},
       adsurl = {https://ui.adsabs.harvard.edu/abs/2008MNRAS.388.1101M}
}

@ARTICLE{Boyett2024_z9p3,
       author = {{Boyett}, Kristan and {Trenti}, Michele and {Leethochawalit}, Nicha and {Calabr{\'o}}, Antonello and {Metha}, Benjamin and {Roberts-Borsani}, Guido and {Dalmasso}, Nicol{\'o} and {Yang}, Lilan and {Santini}, Paola and {Treu}, Tommaso and {Jones}, Tucker and {Henry}, Alaina and {Mason}, Charlotte A. and {Morishita}, Takahiro and {Nanayakkara}, Themiya and {Roy}, Namrata and {Wang}, Xin and {Fontana}, Adriano and {Merlin}, Emiliano and {Castellano}, Marco and {Paris}, Diego and {Brada{\v{c}}}, Maru{\v{s}}a and {Malkan}, Matt and {Marchesini}, Danilo and {Mascia}, Sara and {Glazebrook}, Karl and {Pentericci}, Laura and {Vanzella}, Eros and {Vulcani}, Benedetta},
        title = "{A massive interacting galaxy 510 million years after the Big Bang}",
      journal = {Nature Astronomy},
         year = 2024,
        month = may,
       volume = {8},
        pages = {657-672},
          doi = {10.1038/s41550-024-02218-7},
archivePrefix = {arXiv},
       eprint = {2303.00306},
 primaryClass = {astro-ph.GA},
       adsurl = {https://ui.adsabs.harvard.edu/abs/2024NatAs...8..657B}
}

@ARTICLE{Carniani2024_z14,
       author = {{Carniani}, Stefano and {Hainline}, Kevin and {D'Eugenio}, Francesco and {Eisenstein}, Daniel J. and {Jakobsen}, Peter and {Witstok}, Joris and {Johnson}, Benjamin D. and {Chevallard}, Jacopo and {Maiolino}, Roberto and {Helton}, Jakob M. and {Willott}, Chris and {Robertson}, Brant and {Alberts}, Stacey and {Arribas}, Santiago and {Baker}, William M. and {Bhatawdekar}, Rachana and {Boyett}, Kristan and {Bunker}, Andrew J. and {Cameron}, Alex J. and {Cargile}, Phillip A. and {Charlot}, St{\'e}phane and {Curti}, Mirko and {Curtis-Lake}, Emma and {Egami}, Eiichi and {Giardino}, Giovanna and {Isaak}, Kate and {Ji}, Zhiyuan and {Jones}, Gareth C. and {Kumari}, Nimisha and {Maseda}, Michael V. and {Parlanti}, Eleonora and {P{\'e}rez-Gonz{\'a}lez}, Pablo G. and {Rawle}, Tim and {Rieke}, George and {Rieke}, Marcia and {Del Pino}, Bruno Rodr{\'\i}guez and {Saxena}, Aayush and {Scholtz}, Jan and {Smit}, Renske and {Sun}, Fengwu and {Tacchella}, Sandro and {{\"U}bler}, Hannah and {Venturi}, Giacomo and {Williams}, Christina C. and {Willmer}, Christopher N.~A.},
        title = "{Spectroscopic confirmation of two luminous galaxies at a redshift of 14}",
      journal = {\nat},
         year = 2024,
        month = sep,
       volume = {633},
       number = {8029},
        pages = {318-322},
          doi = {10.1038/s41586-024-07860-9},
archivePrefix = {arXiv},
       eprint = {2405.18485},
 primaryClass = {astro-ph.GA},
       adsurl = {https://ui.adsabs.harvard.edu/abs/2024Natur.633..318C}
}

@Article{Naidu2026_MoMz14,
  author           = {Naidu, Rohan P. and Oesch, Pascal A. and Brammer, Gabriel and Weibel, Andrea and Li, Yijia and Matthee, Jorryt and Chisolm, John and Pollock, Clara L. and Heintz, Kasper E. and Johnson, Benjamin D. and Shan, Xuejian and Hviding, Raphael E. and Leja, Joel and Tacchella, Sandro and Ganguly, Arpita and Witten, Callum and Atek, Hakim and Belli, Siro and Bose, Sownak and Bouwens, Rychard and Dayal, Pratika and Decarli, Roberto and de Graaff, Anna and Fudamoto, Yoshinobu and Giovinazzo, Emma and Greene, Jenny E. and Illingworth, Garth and Inoue, Akio K. and Kane, Sarah G. and Labbe, Ivo and Leonova, Ecaterina and Marques-Chaves, Rui and Meyer, Roman A. and Nelson, Erica J. and Roberts-Borsani, Guido and Schaerer, Daniel and Simcoe, Robert A. and Stefanon, Mauro and Sugahara, Yuma and Toft, Sune and van der Wel, Arjen and van Dokkum, Pieter and Walter, Fabian and Watson, Darrach and Weaver, John R. and Whitaker, Katherine E.},
  journal          = {The Open Journal of Astrophysics},
  title            = {A Cosmic Miracle: A Remarkably Luminous Galaxy at zspec = 14.44 Confirmed with JWST},
  year             = {2026},
  month            = jan,
  pages            = {56033},
  volume           = {9},
  archiveprefix    = {arXiv},
  doi              = {10.33232/001c.156033},
  eprint           = {2505.11263},
  modificationdate = {2026-02-15T14:11:55},
  primaryclass     = {astro-ph.GA},
  url              = {https://ui.adsabs.harvard.edu/abs/2026OJAp....956033N},
}

@Article{Donnan2026,
  author           = {Donnan, Callum T. and McLeod, Derek J. and McLure, Ross J. and Dunlop, James S. and Cullen, Fergus and Dickinson, Mark and Haro, Pablo Arrabal and Taylor, Anthony J. and Bondestam, Cecilia and Liu, Feng-Yuan and Arellano-Córdova, Karla Z. and Barrufet, Laia and Begley, Ryan and Carnall, Adam C. and Golawska, Hanna and Leung, Ho-Hin and Scholte, Dirk and Stanton, Thomas M.},
  title            = {Spectroscopic confirmation of a large and luminous galaxy with weak emission lines at $\mathbf{z = 13.53}$},
  year             = {2026},
  month            = jan,
  archiveprefix    = {arXiv},
  copyright        = {Creative Commons Attribution 4.0 International},
  doi              = {10.48550/ARXIV.2601.11515},
  eprint           = {2601.11515},
  modificationdate = {2026-01-20T17:20:48},
  primaryclass     = {astro-ph.GA},
  publisher        = {arXiv},
}

@ARTICLE{MiraldaEscude1998,
       author = {{Miralda-Escud{\'e}}, Jordi},
        title = "{Reionization of the Intergalactic Medium and the Damping Wing of the Gunn-Peterson Trough}",
      journal = {\apj},
         year = 1998,
        month = jul,
       volume = {501},
       number = {1},
        pages = {15-22},
          doi = {10.1086/305799},
archivePrefix = {arXiv},
       eprint = {astro-ph/9708253},
 primaryClass = {astro-ph},
       adsurl = {https://ui.adsabs.harvard.edu/abs/1998ApJ...501...15M}
}

@ARTICLE{Mason2020_bubble,
       author = {{Mason}, Charlotte A. and {Gronke}, Max},
        title = "{Measuring the properties of reionized bubbles with resolved Ly{\ensuremath{\alpha}} spectra}",
      journal = {\mnras},
         year = 2020,
        month = nov,
       volume = {499},
       number = {1},
        pages = {1395-1405},
          doi = {10.1093/mnras/staa2910},
archivePrefix = {arXiv},
       eprint = {2004.13065},
 primaryClass = {astro-ph.GA},
       adsurl = {https://ui.adsabs.harvard.edu/abs/2020MNRAS.499.1395M}
}

@ARTICLE{Mesinger2008,
       author = {{Mesinger}, Andrei and {Furlanetto}, Steven R.},
        title = "{Ly{\ensuremath{\alpha}} damping wing constraints on inhomogeneous reionization}",
      journal = {\mnras},
         year = 2008,
        month = apr,
       volume = {385},
       number = {3},
        pages = {1348-1358},
          doi = {10.1111/j.1365-2966.2007.12836.x},
archivePrefix = {arXiv},
       eprint = {0710.0371},
 primaryClass = {astro-ph},
       adsurl = {https://ui.adsabs.harvard.edu/abs/2008MNRAS.385.1348M}
}

@ARTICLE{Martins2022,
       author = {{Martins}, F. and {Palacios}, A.},
        title = "{Spectroscopic evolution of very massive stars at Z = 1/2.5 Z$_{☉}$}",
      journal = {\aap},
         year = 2022,
        month = mar,
       volume = {659},
          eid = {A163},
        pages = {A163},
          doi = {10.1051/0004-6361/202243048},
archivePrefix = {arXiv},
       eprint = {2202.13703},
 primaryClass = {astro-ph.SR},
       adsurl = {https://ui.adsabs.harvard.edu/abs/2022A&A...659A.163M}
}

@ARTICLE{Schaerer1996,
       author = {{Schaerer}, D.},
        title = "{Combined stellar structure and atmosphere models for massive stars. Wolf-Rayet models with spherically outflowing envelopes.}",
      journal = {\aap},
         year = 1996,
        month = may,
       volume = {309},
        pages = {129-143},
          doi = {10.48550/arXiv.astro-ph/9509079},
archivePrefix = {arXiv},
       eprint = {astro-ph/9509079},
 primaryClass = {astro-ph},
       adsurl = {https://ui.adsabs.harvard.edu/abs/1996A&A...309..129S}
}

@ARTICLE{Brinchmann2008,
       author = {{Brinchmann}, Jarle and {Pettini}, Max and {Charlot}, St{\'e}phane},
        title = "{New insights into the stellar content and physical conditions of star-forming galaxies at z = 2-3 from spectral modelling}",
      journal = {\mnras},
         year = 2008,
        month = apr,
       volume = {385},
       number = {2},
        pages = {769-782},
          doi = {10.1111/j.1365-2966.2008.12914.x},
archivePrefix = {arXiv},
       eprint = {0801.1678},
 primaryClass = {astro-ph},
       adsurl = {https://ui.adsabs.harvard.edu/abs/2008MNRAS.385..769B}
}

@ARTICLE{Chandar2004,
       author = {{Chandar}, Rupali and {Leitherer}, Claus and {Tremonti}, Christy A.},
        title = "{NGC 3125-1: The Most Extreme Wolf-Rayet Star Cluster Known in the Local Universe}",
      journal = {\apj},
         year = 2004,
        month = mar,
       volume = {604},
       number = {1},
        pages = {153-166},
          doi = {10.1086/381723},
       adsurl = {https://ui.adsabs.harvard.edu/abs/2004ApJ...604..153C}
}

@ARTICLE{Nanayakkara2019,
       author = {{Nanayakkara}, Themiya and {Brinchmann}, Jarle and {Boogaard}, Leindert and {Bouwens}, Rychard and {Cantalupo}, Sebastiano and {Feltre}, Anna and {Kollatschny}, Wolfram and {Marino}, Raffaella Anna and {Maseda}, Michael and {Matthee}, Jorryt and {Paalvast}, Mieke and {Richard}, Johan and {Verhamme}, Anne},
        title = "{Exploring He II {\ensuremath{\lambda}}1640 emission line properties at z {\ensuremath{\sim}}2-4}",
      journal = {\aap},
         year = 2019,
        month = apr,
       volume = {624},
          eid = {A89},
        pages = {A89},
          doi = {10.1051/0004-6361/201834565},
archivePrefix = {arXiv},
       eprint = {1902.05960},
 primaryClass = {astro-ph.GA},
       adsurl = {https://ui.adsabs.harvard.edu/abs/2019A&A...624A..89N}
}

@ARTICLE{Chisholm2015,
       author = {{Chisholm}, John and {Tremonti}, Christy A. and {Leitherer}, Claus and {Chen}, Yanmei and {Wofford}, Aida and {Lundgren}, Britt},
        title = "{Scaling Relations Between Warm Galactic Outflows and Their Host Galaxies}",
      journal = {\apj},
         year = 2015,
        month = oct,
       volume = {811},
       number = {2},
          eid = {149},
        pages = {149},
          doi = {10.1088/0004-637X/811/2/149},
archivePrefix = {arXiv},
       eprint = {1412.2139},
 primaryClass = {astro-ph.GA},
       adsurl = {https://ui.adsabs.harvard.edu/abs/2015ApJ...811..149C}
}

@ARTICLE{Heckman2016,
       author = {{Heckman}, Timothy M. and {Borthakur}, Sanchayeeta},
        title = "{The Implications of Extreme Outflows from Extreme Starbursts}",
      journal = {\apj},
         year = 2016,
        month = may,
       volume = {822},
       number = {1},
          eid = {9},
        pages = {9},
          doi = {10.3847/0004-637X/822/1/9},
archivePrefix = {arXiv},
       eprint = {1603.03036},
 primaryClass = {astro-ph.GA},
       adsurl = {https://ui.adsabs.harvard.edu/abs/2016ApJ...822....9H}
}

@ARTICLE{Xu2022,
       author = {{Xu}, Xinfeng and {Heckman}, Timothy and {Henry}, Alaina and {Berg}, Danielle A. and {Chisholm}, John and {James}, Bethan L. and {Martin}, Crystal L. and {Stark}, Daniel P. and {Aloisi}, Alessandra and {Amor{\'\i}n}, Ricardo O. and {Arellano-C{\'o}rdova}, Karla Z. and {Bordoloi}, Rongmon and {Charlot}, St{\'e}phane and {Chen}, Zuyi and {Hayes}, Matthew and {Mingozzi}, Matilde and {Sugahara}, Yuma and {Kewley}, Lisa J. and {Ouchi}, Masami and {Scarlata}, Claudia and {Steidel}, Charles C.},
        title = "{CLASSY III. The Properties of Starburst-driven Warm Ionized Outflows}",
      journal = {\apj},
         year = 2022,
        month = jul,
       volume = {933},
       number = {2},
          eid = {222},
        pages = {222},
          doi = {10.3847/1538-4357/ac6d56},
archivePrefix = {arXiv},
       eprint = {2204.09181},
 primaryClass = {astro-ph.GA},
       adsurl = {https://ui.adsabs.harvard.edu/abs/2022ApJ...933..222X}
}

@ARTICLE{Hainline2011,
       author = {{Hainline}, Kevin N. and {Shapley}, Alice E. and {Greene}, Jenny E. and {Steidel}, Charles C.},
        title = "{The Rest-frame Ultraviolet Spectra of UV-selected Active Galactic Nuclei at z \raisebox{-0.5ex}\textasciitilde 2-3}",
      journal = {\apj},
         year = 2011,
        month = may,
       volume = {733},
       number = {1},
          eid = {31},
        pages = {31},
          doi = {10.1088/0004-637X/733/1/31},
archivePrefix = {arXiv},
       eprint = {1012.0075},
 primaryClass = {astro-ph.CO},
       adsurl = {https://ui.adsabs.harvard.edu/abs/2011ApJ...733...31H}
}

@ARTICLE{Lambrides2026,
       author = {{Lambrides}, Erini and {Larson}, Rebecca L. and {Garofali}, Kristen and {Ptak}, Andrew and {Chiaberge}, Marco and {Long}, Arianna S. and {Hutchison}, Taylor A. and {Norman}, Colin and {McKinney}, Jed and {Akins}, Hollis B. and {Berg}, Danielle A. and {Chisholm}, John and {Civano}, Francesca and {Cloonan}, Aidan P. and {Endsley}, Ryan and {Faisst}, Andreas L. and {Gilli}, Roberto and {Gillman}, Steven and {Hirschmann}, Michaela and {Kartaltepe}, Jeyhan S. and {Kocevski}, Dale D. and {Kokorev}, Vasily and {Pacucci}, Fabio and {Richardson}, Chris T. and {Stiavelli}, Massimo and {Whalen}, Kelly E.},
        title = "{The case for super-Eddington accretion in JWST broad-line active galactic nuclei during the first billion years}",
      journal = {Nature Astronomy},
         year = 2026,
        month = jun,
       volume = {10},
        pages = {868-879},
          doi = {10.1038/s41550-026-02813-w},
archivePrefix = {arXiv},
       eprint = {2409.13047},
 primaryClass = {astro-ph.HE},
       adsurl = {https://ui.adsabs.harvard.edu/abs/2026NatAs..10..868L}
}

@ARTICLE{Scholtz2025,
       author = {{Scholtz}, Jan and {Maiolino}, Roberto and {D'Eugenio}, Francesco and {Curtis-Lake}, Emma and {Carniani}, Stefano and {Charlot}, Stephane and {Curti}, Mirko and {Silcock}, Maddie S. and {Arribas}, Santiago and {Baker}, William and {Bhatawdekar}, Rachana and {Boyett}, Kristan and {Bunker}, Andrew J. and {Chevallard}, Jacopo and {Circosta}, Chiara and {Eisenstein}, Daniel J. and {Hainline}, Kevin and {Hausen}, Ryan and {Ji}, Xihan and {Ji}, Zhiyuan and {Johnson}, Benjamin D. and {Kumari}, Nimisha and {Looser}, Tobias J. and {Lyu}, Jianwei and {Maseda}, Michael V. and {Parlanti}, Eleonora and {Perna}, Michele and {Rieke}, Marcia and {Robertson}, Brant and {Del Pino}, Bruno Rodr{\'\i}guez and {Sun}, Fengwu and {Tacchella}, Sandro and {{\"U}bler}, Hannah and {Venturi}, Giacomo and {Williams}, Christina C. and {Willmer}, Christopher N.~A. and {Willott}, Chris and {Witstok}, Joris},
        title = "{JADES: A large population of obscured, narrow-line active galactic nuclei at high redshift}",
      journal = {\aap},
         year = 2025,
        month = may,
       volume = {697},
          eid = {A175},
        pages = {A175},
          doi = {10.1051/0004-6361/202348804},
archivePrefix = {arXiv},
       eprint = {2311.18731},
 primaryClass = {astro-ph.GA},
       adsurl = {https://ui.adsabs.harvard.edu/abs/2025A&A...697A.175S}
}

@ARTICLE{Cohon2026,
       author = {{Cohon}, Joshua and {Cain}, Christopher and {Windhorst}, Rogier and {D'Aloisio}, Anson and {Carleton}, Timothy and {Zhu}, Yongda},
        title = "{A long time ago in an LAE far, far away: A signpost of early reionisation or a nascent AGN at z = 13?}",
      journal = {\pasa},
         year = 2026,
        month = mar,
       volume = {43},
          eid = {e046},
        pages = {e046},
          doi = {10.1017/pasa.2026.10166},
archivePrefix = {arXiv},
       eprint = {2508.05739},
 primaryClass = {astro-ph.GA},
       adsurl = {https://ui.adsabs.harvard.edu/abs/2026PASA...43...46C}
}

@ARTICLE{Bruton2023,
       author = {{Bruton}, Sean and {Lin}, Yu-Heng and {Scarlata}, Claudia and {Hayes}, Matthew J.},
        title = "{The Universe is at Most 88\% Neutral at z = 10.6}",
      journal = {\apjl},
         year = 2023,
        month = jun,
       volume = {949},
       number = {2},
          eid = {L40},
        pages = {L40},
          doi = {10.3847/2041-8213/acd5d0},
archivePrefix = {arXiv},
       eprint = {2303.03419},
 primaryClass = {astro-ph.GA},
       adsurl = {https://ui.adsabs.harvard.edu/abs/2023ApJ...949L..40B}
}

@ARTICLE{Qin2025,
       author = {{Qin}, Yuxiang and {Wyithe}, J. Stuart B.},
        title = "{Reionization morphology and intrinsic velocity offsets allow transmission of Lyman-{\ensuremath{\alpha}} emission from JADES-GS-z13-1-LA}",
      journal = {\mnras},
         year = 2025,
        month = mar,
       volume = {538},
       number = {1},
        pages = {L16-L23},
          doi = {10.1093/mnrasl/slaf001},
archivePrefix = {arXiv},
       eprint = {2409.07356},
 primaryClass = {astro-ph.CO},
       adsurl = {https://ui.adsabs.harvard.edu/abs/2025MNRAS.538L..16Q}
}

@ARTICLE{Jung2020,
       author = {{Jung}, Intae and {Finkelstein}, Steven L. and {Dickinson}, Mark and {Hutchison}, Taylor A. and {Larson}, Rebecca L. and {Papovich}, Casey and {Pentericci}, Laura and {Straughn}, Amber N. and {Guo}, Yicheng and {Malhotra}, Sangeeta and {Rhoads}, James and {Song}, Mimi and {Tilvi}, Vithal and {Wold}, Isak},
        title = "{Texas Spectroscopic Search for Ly{\ensuremath{\alpha}} Emission at the End of Reionization. III. The Ly{\ensuremath{\alpha}} Equivalent-width Distribution and Ionized Structures at z > 7}",
      journal = {\apj},
         year = 2020,
        month = dec,
       volume = {904},
       number = {2},
          eid = {144},
        pages = {144},
          doi = {10.3847/1538-4357/abbd44},
archivePrefix = {arXiv},
       eprint = {2009.10092},
 primaryClass = {astro-ph.GA},
       adsurl = {https://ui.adsabs.harvard.edu/abs/2020ApJ...904..144J}
}

@ARTICLE{Neyer2024,
       author = {{Neyer}, Meredith and {Smith}, Aaron and {Kannan}, Rahul and {Vogelsberger}, Mark and {Garaldi}, Enrico and {Gal{\'a}rraga-Espinosa}, Daniela and {Borrow}, Josh and {Hernquist}, Lars and {Pakmor}, R{\"u}diger and {Springel}, Volker},
        title = "{The THESAN project: connecting ionized bubble sizes to their local environments during the Epoch of Reionization}",
      journal = {\mnras},
         year = 2024,
        month = jul,
       volume = {531},
       number = {3},
        pages = {2943-2957},
          doi = {10.1093/mnras/stae1325},
archivePrefix = {arXiv},
       eprint = {2310.03783},
 primaryClass = {astro-ph.GA},
       adsurl = {https://ui.adsabs.harvard.edu/abs/2024MNRAS.531.2943N}
}

@ARTICLE{McQuinn2007,
       author = {{McQuinn}, Matthew and {Lidz}, Adam and {Zahn}, Oliver and {Dutta}, Suvendra and {Hernquist}, Lars and {Zaldarriaga}, Matias},
        title = "{The morphology of HII regions during reionization}",
      journal = {\mnras},
         year = 2007,
        month = may,
       volume = {377},
       number = {3},
        pages = {1043-1063},
          doi = {10.1111/j.1365-2966.2007.11489.x},
archivePrefix = {arXiv},
       eprint = {astro-ph/0610094},
 primaryClass = {astro-ph},
       adsurl = {https://ui.adsabs.harvard.edu/abs/2007MNRAS.377.1043M}
}

@ARTICLE{Choe2025,
       author = {{Choe}, S. and {Emil Rivera-Thorsen}, T. and {Dahle}, H. and {Sharon}, K. and {Owens}, M. Riley and {Rigby}, J.~R. and {Bayliss}, M.~B. and {Hayes}, M.~J. and {Hutchison}, T. and {Welch}, B. and {Chisholm}, J. and {Gladders}, M.~D. and {Khullar}, G. and {Kim}, K.},
        title = "{The Sunburst Arc with JWST: II. Observations of an Eta Carinae analog at z = 2.37}",
      journal = {\aap},
         year = 2025,
        month = jun,
       volume = {698},
          eid = {A16},
        pages = {A16},
          doi = {10.1051/0004-6361/202450685},
archivePrefix = {arXiv},
       eprint = {2405.06953},
 primaryClass = {astro-ph.GA},
       adsurl = {https://ui.adsabs.harvard.edu/abs/2025A&A...698A..16C}
}

@ARTICLE{Gagnon-Hartman2026,
       author = {{Gagnon-Hartman}, Samuel and {Mesinger}, Andrei and {Nikoli{\'c}}, Ivan and {Parlanti}, Eleonora and {Venturi}, Giacomo},
        title = "{Characterizing Lyman alpha emission from high-redshift galaxies}",
      journal = {arXiv e-prints},
         year = 2026,
        month = feb,
          eid = {arXiv:2602.13389},
        pages = {arXiv:2602.13389},
          doi = {10.48550/arXiv.2602.13389},
archivePrefix = {arXiv},
       eprint = {2602.13389},
 primaryClass = {astro-ph.GA},
       adsurl = {https://ui.adsabs.harvard.edu/abs/2026arXiv260213389G}
}

@ARTICLE{Berg2021,
       author = {{Berg}, Danielle A. and {Chisholm}, John and {Erb}, Dawn K. and {Skillman}, Evan D. and {Pogge}, Richard W. and {Olivier}, Grace M.},
        title = "{Characterizing Extreme Emission-line Galaxies. I. A Four-zone Ionization Model for Very High-ionization Emission}",
      journal = {\apj},
         year = 2021,
        month = dec,
       volume = {922},
       number = {2},
          eid = {170},
        pages = {170},
          doi = {10.3847/1538-4357/ac141b},
archivePrefix = {arXiv},
       eprint = {2105.12765},
 primaryClass = {astro-ph.GA},
       adsurl = {https://ui.adsabs.harvard.edu/abs/2021ApJ...922..170B}
}

@ARTICLE{Crowther2007,
       author = {{Crowther}, Paul A.},
        title = "{Physical Properties of Wolf-Rayet Stars}",
      journal = {\araa},
         year = 2007,
        month = sep,
       volume = {45},
       number = {1},
        pages = {177-219},
          doi = {10.1146/annurev.astro.45.051806.110615},
archivePrefix = {arXiv},
       eprint = {astro-ph/0610356},
 primaryClass = {astro-ph},
       adsurl = {https://ui.adsabs.harvard.edu/abs/2007ARA&A..45..177C}
}

@ARTICLE{Upadhyaya2024,
       author = {{Upadhyaya}, A. and {Marques-Chaves}, R. and {Schaerer}, D. and {Martins}, F. and {P{\'e}rez-Fournon}, I. and {Palacios}, A. and {Stanway}, E.~R.},
        title = "{Evidence for very massive stars in extremely UV-bright star-forming galaxies at z {\ensuremath{\sim}} 2.2-3.6}",
      journal = {\aap},
         year = 2024,
        month = jun,
       volume = {686},
          eid = {A185},
        pages = {A185},
          doi = {10.1051/0004-6361/202449184},
archivePrefix = {arXiv},
       eprint = {2401.16165},
 primaryClass = {astro-ph.GA},
       adsurl = {https://ui.adsabs.harvard.edu/abs/2024A&A...686A.185U}
}

@ARTICLE{Crowther2016,
       author = {{Crowther}, Paul A. and {Caballero-Nieves}, S.~M. and {Bostroem}, K.~A. and {Ma{\'\i}z Apell{\'a}niz}, J. and {Schneider}, F.~R.~N. and {Walborn}, N.~R. and {Angus}, C.~R. and {Brott}, I. and {Bonanos}, A. and {de Koter}, A. and {de Mink}, S.~E. and {Evans}, C.~J. and {Gr{\"a}fener}, G. and {Herrero}, A. and {Howarth}, I.~D. and {Langer}, N. and {Lennon}, D.~J. and {Puls}, J. and {Sana}, H. and {Vink}, J.~S.},
        title = "{The R136 star cluster dissected with Hubble Space Telescope/STIS. I. Far-ultraviolet spectroscopic census and the origin of He II {\ensuremath{\lambda}}1640 in young star clusters}",
      journal = {\mnras},
         year = 2016,
        month = may,
       volume = {458},
       number = {1},
        pages = {624-659},
          doi = {10.1093/mnras/stw273},
archivePrefix = {arXiv},
       eprint = {1603.04994},
 primaryClass = {astro-ph.SR},
       adsurl = {https://ui.adsabs.harvard.edu/abs/2016MNRAS.458..624C}
}

@ARTICLE{Leitherer2010,
       author = {{Leitherer}, Claus and {Ortiz Ot{\'a}lvaro}, Paula A. and {Bresolin}, Fabio and {Kudritzki}, Rolf-Peter and {Lo Faro}, Barbara and {Pauldrach}, Adalbert W.~A. and {Pettini}, Max and {Rix}, Samantha A.},
        title = "{A Library of Theoretical Ultraviolet Spectra of Massive, Hot Stars for Evolutionary Synthesis}",
      journal = {\apjs},
         year = 2010,
        month = aug,
       volume = {189},
       number = {2},
        pages = {309-335},
          doi = {10.1088/0067-0049/189/2/309},
archivePrefix = {arXiv},
       eprint = {1006.5624},
 primaryClass = {astro-ph.SR},
       adsurl = {https://ui.adsabs.harvard.edu/abs/2010ApJS..189..309L}
}

@ARTICLE{Leitherer1995,
       author = {{Leitherer}, Claus and {Robert}, Carmelle and {Heckman}, Timothy M.},
        title = "{Atlas of Synthetic Ultraviolet Spectra of Massive Star Populations}",
      journal = {\apjs},
         year = 1995,
        month = jul,
       volume = {99},
        pages = {173},
          doi = {10.1086/192183},
       adsurl = {https://ui.adsabs.harvard.edu/abs/1995ApJS...99..173L}
}

@ARTICLE{Rix2004,
       author = {{Rix}, Samantha A. and {Pettini}, Max and {Leitherer}, Claus and {Bresolin}, Fabio and {Kudritzki}, Rolf-Peter and {Steidel}, Charles C.},
        title = "{Spectral Modeling of Star-forming Regions in the Ultraviolet: Stellar Metallicity Diagnostics for High-Redshift Galaxies}",
      journal = {\apj},
         year = 2004,
        month = nov,
       volume = {615},
       number = {1},
        pages = {98-117},
          doi = {10.1086/424031},
archivePrefix = {arXiv},
       eprint = {astro-ph/0407296},
 primaryClass = {astro-ph},
       adsurl = {https://ui.adsabs.harvard.edu/abs/2004ApJ...615...98R}
}

@ARTICLE{Walborn1985,
       author = {{Walborn}, N.~R. and {Nichols-Bohlin}, J. and {Panek}, R.~J.},
        title = "{International Ultraviolet Explorer Atlas of O-type Spectra from 1200 to 1900 {\r{A}}.}",
      journal = {NASA Reference Publication},
         year = 1985,
        month = dec,
       volume = {1155},
       adsurl = {https://ui.adsabs.harvard.edu/abs/1985NASRP1155.....W}
}

@ARTICLE{Leitherer2011,
       author = {{Leitherer}, Claus and {Tremonti}, Christy A. and {Heckman}, Timothy M. and {Calzetti}, Daniela},
        title = "{An Ultraviolet Spectroscopic Atlas of Local Starbursts and Star-forming Galaxies: The Legacy of FOS and GHRS}",
      journal = {\aj},
         year = 2011,
        month = feb,
       volume = {141},
       number = {2},
          eid = {37},
        pages = {37},
          doi = {10.1088/0004-6256/141/2/37},
archivePrefix = {arXiv},
       eprint = {1011.0385},
 primaryClass = {astro-ph.CO},
       adsurl = {https://ui.adsabs.harvard.edu/abs/2011AJ....141...37L}
}

@ARTICLE{Hillier1988,
       author = {{Hillier}, D.~J.},
        title = "{The Formation of Nitrogen and Carbon Emission Lines in HD 50896 (WN5)}",
      journal = {\apj},
         year = 1988,
        month = apr,
       volume = {327},
        pages = {822},
          doi = {10.1086/166240},
       adsurl = {https://ui.adsabs.harvard.edu/abs/1988ApJ...327..822H}
}

@ARTICLE{Hillier1989,
       author = {{Hillier}, D.~J.},
        title = "{WC Stars: Hot Stars with Cold Winds}",
      journal = {\apj},
         year = 1989,
        month = dec,
       volume = {347},
        pages = {392},
          doi = {10.1086/168127},
       adsurl = {https://ui.adsabs.harvard.edu/abs/1989ApJ...347..392H}
}

@software{Brammer2023,
       author = {{Brammer}, Gabriel},
        title = "{msaexp: NIRSpec analyis tools}",
         year = 2023,
        month = sep,
          eid = {10.5281/zenodo.7299500},
          doi = {10.5281/zenodo.7299500},
      version = {0.9.18},
    publisher = {Zenodo},
       adsurl = {https://ui.adsabs.harvard.edu/abs/2022zndo...7299500B}
}

@article{astropy:2013,
Adsurl = {http://adsabs.harvard.edu/abs/2013A%26A...558A..33A},
Archiveprefix = {arXiv},
Author = {{Astropy Collaboration} and {Robitaille}, T.~P. and {Tollerud}, E.~J. and {Greenfield}, P. and {Droettboom}, M. and {Bray}, E. and {Aldcroft}, T. and {Davis}, M. and {Ginsburg}, A. and {Price-Whelan}, A.~M. and {Kerzendorf}, W.~E. and {Conley}, A. and {Crighton}, N. and {Barbary}, K. and {Muna}, D. and {Ferguson}, H. and {Grollier}, F. and {Parikh}, M.~M. and {Nair}, P.~H. and {Unther}, H.~M. and {Deil}, C. and {Woillez}, J. and {Conseil}, S. and {Kramer}, R. and {Turner}, J.~E.~H. and {Singer}, L. and {Fox}, R. and {Weaver}, B.~A. and {Zabalza}, V. and {Edwards}, Z.~I. and {Azalee Bostroem}, K. and {Burke}, D.~J. and {Casey}, A.~R. and {Crawford}, S.~M. and {Dencheva}, N. and {Ely}, J. and {Jenness}, T. and {Labrie}, K. and {Lim}, P.~L. and {Pierfederici}, F. and {Pontzen}, A. and {Ptak}, A. and {Refsdal}, B. and {Servillat}, M. and {Streicher}, O.},
Doi = {10.1051/0004-6361/201322068},
Eid = {A33},
Eprint = {1307.6212},
Journal = {\aap},
Month = oct,
Pages = {A33},
Primaryclass = {astro-ph.IM},
Title = {{Astropy: A community Python package for astronomy}},
Volume = 558,
Year = 2013}

@ARTICLE{astropy:2018,
       author = {{Astropy Collaboration} and {Price-Whelan}, A.~M. and
         {Sip{\H{o}}cz}, B.~M. and {G{\"u}nther}, H.~M. and {Lim}, P.~L. and
         {Crawford}, S.~M. and {Conseil}, S. and {Shupe}, D.~L. and
         {Craig}, M.~W. and {Dencheva}, N. and {Ginsburg}, A. and {Vand
        erPlas}, J.~T. and {Bradley}, L.~D. and {P{\'e}rez-Su{\'a}rez}, D. and
         {de Val-Borro}, M. and {Aldcroft}, T.~L. and {Cruz}, K.~L. and
         {Robitaille}, T.~P. and {Tollerud}, E.~J. and {Ardelean}, C. and
         {Babej}, T. and {Bach}, Y.~P. and {Bachetti}, M. and {Bakanov}, A.~V. and
         {Bamford}, S.~P. and {Barentsen}, G. and {Barmby}, P. and
         {Baumbach}, A. and {Berry}, K.~L. and {Biscani}, F. and {Boquien}, M. and
         {Bostroem}, K.~A. and {Bouma}, L.~G. and {Brammer}, G.~B. and
         {Bray}, E.~M. and {Breytenbach}, H. and {Buddelmeijer}, H. and
         {Burke}, D.~J. and {Calderone}, G. and {Cano Rodr{\'\i}guez}, J.~L. and
         {Cara}, M. and {Cardoso}, J.~V.~M. and {Cheedella}, S. and {Copin}, Y. and
         {Corrales}, L. and {Crichton}, D. and {D'Avella}, D. and {Deil}, C. and
         {Depagne}, {\'E}. and {Dietrich}, J.~P. and {Donath}, A. and
         {Droettboom}, M. and {Earl}, N. and {Erben}, T. and {Fabbro}, S. and
         {Ferreira}, L.~A. and {Finethy}, T. and {Fox}, R.~T. and
         {Garrison}, L.~H. and {Gibbons}, S.~L.~J. and {Goldstein}, D.~A. and
         {Gommers}, R. and {Greco}, J.~P. and {Greenfield}, P. and
         {Groener}, A.~M. and {Grollier}, F. and {Hagen}, A. and {Hirst}, P. and
         {Homeier}, D. and {Horton}, A.~J. and {Hosseinzadeh}, G. and {Hu}, L. and
         {Hunkeler}, J.~S. and {Ivezi{\'c}}, {\v{Z}}. and {Jain}, A. and
         {Jenness}, T. and {Kanarek}, G. and {Kendrew}, S. and {Kern}, N.~S. and
         {Kerzendorf}, W.~E. and {Khvalko}, A. and {King}, J. and {Kirkby}, D. and
         {Kulkarni}, A.~M. and {Kumar}, A. and {Lee}, A. and {Lenz}, D. and
         {Littlefair}, S.~P. and {Ma}, Z. and {Macleod}, D.~M. and
         {Mastropietro}, M. and {McCully}, C. and {Montagnac}, S. and
         {Morris}, B.~M. and {Mueller}, M. and {Mumford}, S.~J. and {Muna}, D. and
         {Murphy}, N.~A. and {Nelson}, S. and {Nguyen}, G.~H. and
         {Ninan}, J.~P. and {N{\"o}the}, M. and {Ogaz}, S. and {Oh}, S. and
         {Parejko}, J.~K. and {Parley}, N. and {Pascual}, S. and {Patil}, R. and
         {Patil}, A.~A. and {Plunkett}, A.~L. and {Prochaska}, J.~X. and
         {Rastogi}, T. and {Reddy Janga}, V. and {Sabater}, J. and
         {Sakurikar}, P. and {Seifert}, M. and {Sherbert}, L.~E. and
         {Sherwood-Taylor}, H. and {Shih}, A.~Y. and {Sick}, J. and
         {Silbiger}, M.~T. and {Singanamalla}, S. and {Singer}, L.~P. and
         {Sladen}, P.~H. and {Sooley}, K.~A. and {Sornarajah}, S. and
         {Streicher}, O. and {Teuben}, P. and {Thomas}, S.~W. and
         {Tremblay}, G.~R. and {Turner}, J.~E.~H. and {Terr{\'o}n}, V. and
         {van Kerkwijk}, M.~H. and {de la Vega}, A. and {Watkins}, L.~L. and
         {Weaver}, B.~A. and {Whitmore}, J.~B. and {Woillez}, J. and
         {Zabalza}, V. and {Astropy Contributors}},
        title = "{The Astropy Project: Building an Open-science Project and Status of the v2.0 Core Package}",
      journal = {\aj},
         year = 2018,
        month = sep,
       volume = {156},
       number = {3},
          eid = {123},
        pages = {123},
          doi = {10.3847/1538-3881/aabc4f},
archivePrefix = {arXiv},
       eprint = {1801.02634},
 primaryClass = {astro-ph.IM},
       adsurl = {https://ui.adsabs.harvard.edu/abs/2018AJ....156..123A}
}

@ARTICLE{astropy:2022,
       author = {{Astropy Collaboration} and {Price-Whelan}, Adrian M. and {Lim}, Pey Lian and {Earl}, Nicholas and {Starkman}, Nathaniel and {Bradley}, Larry and {Shupe}, David L. and {Patil}, Aarya A. and {Corrales}, Lia and {Brasseur}, C.~E. and {N{"o}the}, Maximilian and {Donath}, Axel and {Tollerud}, Erik and {Morris}, Brett M. and {Ginsburg}, Adam and {Vaher}, Eero and {Weaver}, Benjamin A. and {Tocknell}, James and {Jamieson}, William and {van Kerkwijk}, Marten H. and {Robitaille}, Thomas P. and {Merry}, Bruce and {Bachetti}, Matteo and {G{"u}nther}, H. Moritz and {Aldcroft}, Thomas L. and {Alvarado-Montes}, Jaime A. and {Archibald}, Anne M. and {B{'o}di}, Attila and {Bapat}, Shreyas and {Barentsen}, Geert and {Baz{'a}n}, Juanjo and {Biswas}, Manish and {Boquien}, M{'e}d{'e}ric and {Burke}, D.~J. and {Cara}, Daria and {Cara}, Mihai and {Conroy}, Kyle E. and {Conseil}, Simon and {Craig}, Matthew W. and {Cross}, Robert M. and {Cruz}, Kelle L. and {D'Eugenio}, Francesco and {Dencheva}, Nadia and {Devillepoix}, Hadrien A.~R. and {Dietrich}, J{"o}rg P. and {Eigenbrot}, Arthur Davis and {Erben}, Thomas and {Ferreira}, Leonardo and {Foreman-Mackey}, Daniel and {Fox}, Ryan and {Freij}, Nabil and {Garg}, Suyog and {Geda}, Robel and {Glattly}, Lauren and {Gondhalekar}, Yash and {Gordon}, Karl D. and {Grant}, David and {Greenfield}, Perry and {Groener}, Austen M. and {Guest}, Steve and {Gurovich}, Sebastian and {Handberg}, Rasmus and {Hart}, Akeem and {Hatfield-Dodds}, Zac and {Homeier}, Derek and {Hosseinzadeh}, Griffin and {Jenness}, Tim and {Jones}, Craig K. and {Joseph}, Prajwel and {Kalmbach}, J. Bryce and {Karamehmetoglu}, Emir and {Ka{l}uszy{'n}ski}, Miko{l}aj and {Kelley}, Michael S.~P. and {Kern}, Nicholas and {Kerzendorf}, Wolfgang E. and {Koch}, Eric W. and {Kulumani}, Shankar and {Lee}, Antony and {Ly}, Chun and {Ma}, Zhiyuan and {MacBride}, Conor and {Maljaars}, Jakob M. and {Muna}, Demitri and {Murphy}, N.~A. and {Norman}, Henrik and {O'Steen}, Richard and {Oman}, Kyle A. and {Pacifici}, Camilla and {Pascual}, Sergio and {Pascual-Granado}, J. and {Patil}, Rohit R. and {Perren}, Gabriel I. and {Pickering}, Timothy E. and {Rastogi}, Tanuj and {Roulston}, Benjamin R. and {Ryan}, Daniel F. and {Rykoff}, Eli S. and {Sabater}, Jose and {Sakurikar}, Parikshit and {Salgado}, Jes{'u}s and {Sanghi}, Aniket and {Saunders}, Nicholas and {Savchenko}, Volodymyr and {Schwardt}, Ludwig and {Seifert-Eckert}, Michael and {Shih}, Albert Y. and {Jain}, Anany Shrey and {Shukla}, Gyanendra and {Sick}, Jonathan and {Simpson}, Chris and {Singanamalla}, Sudheesh and {Singer}, Leo P. and {Singhal}, Jaladh and {Sinha}, Manodeep and {Sip{H{o}}cz}, Brigitta M. and {Spitler}, Lee R. and {Stansby}, David and {Streicher}, Ole and {{{S}}umak}, Jani and {Swinbank}, John D. and {Taranu}, Dan S. and {Tewary}, Nikita and {Tremblay}, Grant R. and {Val-Borro}, Miguel de and {Van Kooten}, Samuel J. and {Vasovi{'c}}, Zlatan and {Verma}, Shresth and {de Miranda Cardoso}, Jos{'e} Vin{'i}cius and {Williams}, Peter K.~G. and {Wilson}, Tom J. and {Winkel}, Benjamin and {Wood-Vasey}, W.~M. and {Xue}, Rui and {Yoachim}, Peter and {Zhang}, Chen and {Zonca}, Andrea and {Astropy Project Contributors}},
        title = "{The Astropy Project: Sustaining and Growing a Community-oriented Open-source Project and the Latest Major Release (v5.0) of the Core Package}",
      journal = {apj},
         year = 2022,
        month = aug,
       volume = {935},
       number = {2},
          eid = {167},
        pages = {167},
          doi = {10.3847/1538-4357/ac7c74},
archivePrefix = {arXiv},
       eprint = {2206.14220},
 primaryClass = {astro-ph.IM},
       adsurl = {https://ui.adsabs.harvard.edu/abs/2022ApJ...935..167A}
}

@Article{Hunter:2007,
  Author    = {Hunter, J. D.},
  Title     = {Matplotlib: A 2D graphics environment},
  Journal   = {Computing in Science \& Engineering},
  Volume    = {9},
  Number    = {3},
  Pages     = {90--95},
  publisher = {IEEE COMPUTER SOC},
  doi       = {10.1109/MCSE.2007.55},
  year      = 2007
}

@Article{         harris2020array,
 title         = {Array programming with {NumPy}},
 author        = {Charles R. Harris and K. Jarrod Millman and St{\'{e}}fan J.
                 van der Walt and Ralf Gommers and Pauli Virtanen and David
                 Cournapeau and Eric Wieser and Julian Taylor and Sebastian
                 Berg and Nathaniel J. Smith and Robert Kern and Matti Picus
                 and Stephan Hoyer and Marten H. van Kerkwijk and Matthew
                 Brett and Allan Haldane and Jaime Fern{\'{a}}ndez del
                 R{\'{i}}o and Mark Wiebe and Pearu Peterson and Pierre
                 G{\'{e}}rard-Marchant and Kevin Sheppard and Tyler Reddy and
                 Warren Weckesser and Hameer Abbasi and Christoph Gohlke and
                 Travis E. Oliphant},
 year          = {2020},
 month         = sep,
 journal       = {Nature},
 volume        = {585},
 number        = {7825},
 pages         = {357--362},
 doi           = {10.1038/s41586-020-2649-2},
 publisher     = {Springer Science and Business Media {LLC}},
 url           = {https://doi.org/10.1038/s41586-020-2649-2}
}

@ARTICLE{2020SciPy-NMeth,
  author  = {Virtanen, Pauli and Gommers, Ralf and Oliphant, Travis E. and
            Haberland, Matt and Reddy, Tyler and Cournapeau, David and
            Burovski, Evgeni and Peterson, Pearu and Weckesser, Warren and
            Bright, Jonathan and {van der Walt}, St{\'e}fan J. and
            Brett, Matthew and Wilson, Joshua and Millman, K. Jarrod and
            Mayorov, Nikolay and Nelson, Andrew R. J. and Jones, Eric and
            Kern, Robert and Larson, Eric and Carey, C J and
            Polat, {\.I}lhan and Feng, Yu and Moore, Eric W. and
            {VanderPlas}, Jake and Laxalde, Denis and Perktold, Josef and
            Cimrman, Robert and Henriksen, Ian and Quintero, E. A. and
            Harris, Charles R. and Archibald, Anne M. and
            Ribeiro, Ant{\^o}nio H. and Pedregosa, Fabian and
            {van Mulbregt}, Paul and {SciPy 1.0 Contributors}},
  title   = {{{SciPy} 1.0: Fundamental Algorithms for Scientific
            Computing in Python}},
  journal = {Nature Methods},
  year    = {2020},
  volume  = {17},
  pages   = {261--272},
  adsurl  = {https://rdcu.be/b08Wh},
  doi     = {10.1038/s41592-019-0686-2},
}

@ARTICLE{Foreman-Mackey2013,
       author = {{Foreman-Mackey}, Daniel and {Hogg}, David W. and {Lang}, Dustin and {Goodman}, Jonathan},
        title = "{emcee: The MCMC Hammer}",
      journal = {\pasp},
         year = 2013,
        month = mar,
       volume = {125},
       number = {925},
        pages = {306},
          doi = {10.1086/670067},
archivePrefix = {arXiv},
       eprint = {1202.3665},
 primaryClass = {astro-ph.IM},
       adsurl = {https://ui.adsabs.harvard.edu/abs/2013PASP..125..306F}
}

@INCOLLECTION{Kluyver2016,
       author = {{Kluyver}, Thomas and {Ragan-Kelley}, Benjain and {P{\'e}rez}, Fernando and {Granger}, Brian and {Bussonnier}, Matthias and {Frederic}, Jonathan and {Kelley}, Kyle and {Hamrick}, Jessica and {Grout}, Jason and {Corlay}, Sylvain and {Ivanov}, Paul and {Avila}, Dami{\'a}n and {Abdalla}, Safia and {Willing}, Carol and {Jupyter Development Team}},
        title = "{Jupyter Notebooks{\textemdash}a publishing format for reproducible computational workflows}",
    booktitle = {IOS Press},
         year = 2016,
        pages = {87-90},
          doi = {10.3233/978-1-61499-649-1-87},
       adsurl = {https://ui.adsabs.harvard.edu/abs/2016ppap.book...87K}
}

@ARTICLE{Witstok2026,
       author = {{Witstok}, Joris and {Carniani}, Stefano and {Jakobsen}, Peter and {Bunker}, Andrew J. and {Cameron}, Alex J. and {D'Eugenio}, Francesco and {Hainline}, Kevin and {Helton}, Jakob M. and {Looser}, Tobias J. and {Rinaldi}, Pierluigi and {Robertson}, Brant and {Baker}, William M. and {Charlot}, St{\'e}phane and {Johnson}, Benjamin D. and {Jones}, Gareth C. and {Kumari}, Nimisha and {Maiolino}, Roberto and {Scholtz}, Jan and {Tacchella}, Sandro and {Willmer}, Christopher N.~A. and {Willott}, Chris and {Wu}, Zihao},
        title = "{An OASIS of Lyman-$α$ within a neutral intergalactic desert: reaffirmed line and blue continuum reveal efficient ionising agents at $z = 13$}",
      journal = {arXiv e-prints},
         year = 2026,
        month = mar,
          eid = {arXiv:2603.18775},
        pages = {arXiv:2603.18775},
          doi = {10.48550/arXiv.2603.18775},
archivePrefix = {arXiv},
       eprint = {2603.18775},
 primaryClass = {astro-ph.GA},
       adsurl = {https://ui.adsabs.harvard.edu/abs/2026arXiv260318775W}
}

@misc{Rieke2023JADES,
  author       = {{Rieke}, Marcia and {Robertson}, Brant and {Tacchella}, Sandro and {Willmer}, Christopher and {Johnson}, Ben and {Carniani}, Stefano and {Bunker}, Andy and {Willott}, Chris and {Cameron}, Alex and {Curtis-Lake}, Emma and {D'Eugenio}, Francesco},
  title        = {{Data from the JWST Advanced Deep Extragalactic Survey (JADES)}},
  year         = {2023},
  publisher    = {STScI/MAST},
  doi          = {10.17909/8tdj-8n28},
  url          = {https://doi.org/10.17909/8tdj-8n28}
}

@ARTICLE{Rantala2026,
       author = {{Rantala}, Antti and {Naab}, Thorsten and {Lah{\'e}n}, Natalia and {Reuter}, Klaus and {Rampp}, Markus and {Chru{\'s}li{\'n}ska}, Martyna and {Reinoso}, Basti{\'a}n},
        title = "{FROST-CLUSTERS ─ III. Metallicity-dependent intermediate-mass black hole formation by runaway collisions in dense star clusters}",
      journal = {\mnras},
         year = 2026,
        month = jul,
       volume = {549},
       number = {4},
          eid = {stag986},
        pages = {stag986},
          doi = {10.1093/mnras/stag986},
archivePrefix = {arXiv},
       eprint = {2601.07917},
 primaryClass = {astro-ph.GA},
       adsurl = {https://ui.adsabs.harvard.edu/abs/2026MNRAS.549ag986R}
}

@ARTICLE{Rantala2024,
       author = {{Rantala}, Antti and {Naab}, Thorsten and {Lah{\'e}n}, Natalia},
        title = "{FROST-CLUSTERS - I. Hierarchical star cluster assembly boosts intermediate-mass black hole formation}",
      journal = {\mnras},
         year = 2024,
        month = jul,
       volume = {531},
       number = {3},
        pages = {3770-3799},
          doi = {10.1093/mnras/stae1413},
archivePrefix = {arXiv},
       eprint = {2403.10602},
 primaryClass = {astro-ph.GA},
       adsurl = {https://ui.adsabs.harvard.edu/abs/2024MNRAS.531.3770R}
}

@article{crowtherFundamentalParametersWolfRayet1995,
  title = {Fundamental Parameters of {{Wolf-Rayet}} Stars. {{IV}}. {{Weak-lined WNE}} Stars.},
  author = {Crowther, P. A. and Smith, L. J. and Hillier, D. J.},
  year = 1995,
  month = oct,
  journal = {Astronomy and Astrophysics},
  volume = {302},
  pages = {457},
  issn = {0004-6361},
  urldate = {2023-12-19}
}

@article{crowtherFundamentalParametersWolfRayet1997,
  title = {Fundamental Parameters of {{Wolf-Rayet}} Stars. {{VI}}. {{Large Magellanic Cloud WNL}} Stars.},
  author = {Crowther, P. A. and Smith, L. J.},
  year = 1997,
  month = apr,
  journal = {Astronomy and Astrophysics},
  volume = {320},
  pages = {500--524},
  issn = {0004-6361},
  urldate = {2023-05-01}
}

@article{riverogonzalezNitrogenLineSpectroscopy2012a,
  title = {Nitrogen Line Spectroscopy in {{O-stars}}. {{III}}. {{The}} Earliest {{O-stars}}},
  author = {Rivero Gonz{\'a}lez, J. G. and Puls, J. and Massey, P. and Najarro, F.},
  year = 2012,
  month = jul,
  journal = {Astronomy and Astrophysics},
  volume = {543},
  pages = {A95},
  issn = {0004-6361},
  doi = {10.1051/0004-6361/201218955},
  urldate = {2024-02-14}
}

@article{grafenerLineblanketedModelAtmospheres2002,
  title = {Line-blanketed model atmospheres for WR stars},
  author = {Gräfener, G. and Koesterke, L. and Hamann, W. -R.},
  year = 2002,
  month = may,
  journal = {\aap},
  volume = {387},
  pages = {244-257},
  doi = {10.1051/0004-6361:20020269}
}

@article{hamannGridsModelSpectra2004,
  title = {Grids of Model Spectra for {{WN}} Stars, Ready for Use},
  author = {Hamann, W.-R. and Gr{\"a}fener, G.},
  year = 2004,
  month = nov,
  journal = {Astronomy and Astrophysics},
  volume = {427},
  pages = {697--704},
  issn = {0004-6361},
  doi = {10.1051/0004-6361:20040506},
  urldate = {2020-04-02}
}

@article{sanderConsistentTreatmentQuasihydrostatic2015,
  title = {On the Consistent Treatment of the Quasi-Hydrostatic Layers in Hot Star Atmospheres},
  author = {Sander, A. and Shenar, T. and Hainich, R. and {G{\'i}menez-Garc{\'i}a}, A. and Todt, H. and Hamann, W.-R.},
  year = 2015,
  month = may,
  journal = {Astronomy and Astrophysics},
  volume = {577},
  pages = {A13},
  doi = {10.1051/0004-6361/201425356},
  urldate = {2019-07-19},
  langid = {english}
}

@article{todtPotsdamWolfRayetModel2015,
  title = {Potsdam Wolf-Rayet model atmosphere grids for WN stars},
  author = {Todt, H. and Sander, A. and Hainich, R. and Hamann, W.-R. and Quade, M. and Shenar, T.},
  year = 2015,
  month = jul,
  journal = {\aap},
  volume = {579},
  pages = {A75},
  doi = {10.1051/0004-6361/201526253}
}

@article{kohlerEvolutionRotatingVery2015,
  title = {The Evolution of Rotating Very Massive Stars with {{LMC}} Composition},
  author = {K{\"o}hler, K. and Langer, N. and {de Koter}, A. and {de Mink}, S. E. and Crowther, P. A. and Evans, C. J. and Gr{\"a}fener, G. and Sana, H. and Sanyal, D. and Schneider, F. R. N. and Vink, J. S.},
  year = 2015,
  month = jan,
  journal = {Astronomy and Astrophysics},
  volume = {573},
  pages = {A71},
  issn = {0004-6361},
  doi = {10.1051/0004-6361/201424356},
  urldate = {2020-03-05}
}

@article{sanderNatureMassiveHelium2020,
  title = {On the Nature of Massive Helium Star Winds and {{Wolf-Rayet-type}} Mass-Loss},
  author = {Sander, Andreas A. C. and Vink, Jorick S.},
  year = 2020,
  month = nov,
  journal = {Monthly Notices of the Royal Astronomical Society},
  volume = {499},
  pages = {873--892},
  issn = {0035-8711},
  doi = {10.1093/mnras/staa2712},
  urldate = {2021-04-07}
}

@article{jiangSampleQuasarsStrong2008,
  title = {A {{Sample}} of {{Quasars}} with {{Strong Nitrogen Emission Lines}} from the {{Sloan Digital Sky Survey}}},
  author = {Jiang, Linhua and Fan, Xiaohui and Vestergaard, M.},
  year = 2008,
  month = jun,
  journal = {The Astrophysical Journal},
  volume = {679},
  pages = {962--966},
  issn = {0004-637X},
  doi = {10.1086/587868},
  urldate = {2023-05-18}
}
\bibliographystyle{aasjournalv7}



\end{document}